\documentclass[11pt,a4paper]{article}
\usepackage[utf8]{inputenc}
\usepackage[T1]{fontenc}
\usepackage{amsmath,amssymb,amsthm,mathrsfs,bm}
\usepackage[margin=2.4cm]{geometry}
\usepackage{xcolor}
\usepackage{tikz}
\usetikzlibrary{arrows.meta,decorations.markings,calc,shapes.misc}
\usepackage{hyperref}
\hypersetup{colorlinks=true,linkcolor=blue!60!black,citecolor=blue!60!black,urlcolor=blue!60!black}
\newtheorem{theorem}{Theorem}[section]
\newtheorem{proposition}[theorem]{Proposition}

\newtheorem{lemma}[theorem]{Lemma}
\theoremstyle{definition}
\newtheorem{definition}[theorem]{Definition}
\theoremstyle{remark}
\newtheorem{remark}[theorem]{Remark}
\newtheorem{example}[theorem]{Example}

\newcommand{\tr}{\mathrm{tr}}
\newcommand{\Tr}{\mathrm{Tr}}
\newcommand{\trD}{\mathrm{tr}_D}
\newcommand{\trG}{\mathrm{tr}_G}
\newcommand{\trH}{\mathrm{tr}_H}
\newcommand{\Sym}{\mathrm{Sym}}
\newcommand{\sgn}{\mathrm{sgn}}
\newcommand{\Pexp}{\mathrm{Pexp}}
\newcommand{\ch}{\mathrm{ch}}
\newcommand{\Td}{\mathrm{Td}}
\newcommand{\ind}{\mathrm{index}\,}
\newcommand{\sA}{\mathscr{A}}
\newcommand{\hx}{\hat x}
\newcommand{\hp}{\hat p}
\newcommand{\hpi}{\hat\pi}
\newcommand{\hQ}{\hat Q}
\newcommand{\hG}{\hat G}
\newcommand{\hX}{\hat X}
\newcommand{\hY}{\hat Y}
\newcommand{\hO}{\hat O}
\newcommand{\bG}{\mathbf G}
\newcommand{\bQ}{\mathbf Q}
\newcommand{\R}{\mathbb R}
\newcommand{\Z}{\mathbb Z}
\newcommand{\C}{\mathbb C}
\newcommand{\cD}{\mathcal D}
\newcommand{\cN}{\mathcal N}
\newcommand{\cM}{\mathcal M}
\newcommand{\cT}{\mathcal T}
\newcommand{\cK}{\mathcal K}
\newcommand{\corr}[1]{#1}
\newenvironment{GPTcorrection}{\begingroup}{\endgroup}
\newcommand{\corrB}[1]{#1}
\DeclareRobustCommand{\reviewfix}[1]{#1}

\newenvironment{reviewchange}{\begingroup}{\endgroup}
\DeclareRobustCommand{\GPT}[1]{#1}
\title{Covariant Wigner--Weyl calculus, non-dissipative transport and the chiral anomaly in inhomogeneous
systems with nontrivial momentum-space topology}
\author{M.~A. Zubkov\\[1ex]\small Physics Department, Ariel University, Ariel 40700, Israel}
\date{}

\begin{document}
\maketitle

\begin{abstract}
\reviewfix{We review, systematize and extend the covariant Wigner--Weyl calculus for fermions in non-Abelian gauge fields
and its applications to systems inhomogeneous in coordinate space with nontrivial momentum-space topology: the
non-Abelian chiral separation and quantum Hall effects, their stability under interactions, the Fermi-surface
invariant of inhomogeneous systems and the non-elliptic models that carry it, the factorization of the chiral anomaly
into the instanton number and the momentum-space invariant $N_3$, and the noncommutative-geometric form of the
invariants. }
\end{abstract}

\tableofcontents

%======================================================================
\section{Introduction}

Topological invariants expressed through the one-particle Green function $G(p)$ are the standard tool for the
description of the stability of Fermi points and Fermi surfaces \cite{Volovik} and of quantized transport
\cite{TKNN,Ishikawa,Volovik,ZZJPA}. For a homogeneous system in $D=4$ Euclidean dimensions the invariant responsible
for the stability of a Fermi point is
\begin{equation}
\corr{N_3^0=-\frac{1}{24\pi^2}\epsilon_{\mu\nu\rho\lambda}}\tr\int_S dS_\mu\,G\,\partial_\nu G^{-1}\,G\,\partial_\rho
G^{-1}\,G\,\partial_\lambda G^{-1},
\label{eq:N3hom}
\end{equation}
\corr{$S$ an outward-oriented three-surface enclosing the point, with $\epsilon_{1234}=1$. Here $G$ is the inverse chiral block; the minus sign fixes $N_3^0=1$ for $o=p_4+i\vec\sigma\cdot\vec p$. The full Dirac expression instead includes $\gamma^5$ and the factor $1/(48\pi^2)$. Under the chiral and response assumptions specified below, this charge enters the conductivity} of the chiral separation effect (CSE)
\cite{SonZhit,MetZhit}, $\sigma_{\rm CSE}=N_3/2\pi^2$ \cite{KhaikinZubkov,SuleymanovZubkov,ZA}, and, as shown for
lattice models in \cite{ZubkovKhaidukov2017} and for inhomogeneous systems in \cite{XZPLB}, the chiral anomaly. For
an inhomogeneous system the Green function is not diagonal in momentum and
one has to pass to the Wigner--Weyl calculus \cite{Wigner,Moyal,Zachos}: the invariants keep their form with
$G(p)\to G_W(x,p)$, products $\to$ Moyal products and an average over coordinate space
\cite{ZubkovWu,ZZAnn,SuleymanovZubkov,SuleymanovZubkovNPB}. \reviewfix{Throughout, $N_3$ denotes this Moyal (star-product) expression built from the full Weyl symbol, and $N_3^0$ its star-free counterpart, the degree of the principal symbol on a large sphere in momentum space (\eqref{eq:N3hom} for a homogeneous system); the two coincide for $x$-independent symbols and, under the conditions of Sec.~\ref{sec:stokes}, in general.} When the external field is non-Abelian the ordinary Wigner transform is not
gauge covariant, and a covariant version is needed; this was constructed in \cite{XZPRD}, following the
gauge-covariant Wigner functions of the non-Abelian kinetic theory \cite{Heinz83,Heinz85,EGV86,EGVgluon,VGE87,ElzeHeinz}
and the magnetic Weyl calculus \cite{Muller,MantoiuPurice}.

The present review collects, with the derivations, the content of a series of works on fermionic systems that are
inhomogeneous in coordinate space and have nontrivial topology in momentum space: the covariant Wigner--Weyl
calculus for non-Abelian gauge fields and the resulting non-Abelian CSE and quantum Hall effect (QHE) \cite{XZPRD};
the factorization of the chiral anomaly into the instanton number and the momentum-space invariant $N_3$
\cite{XZPLB}; the effect of interactions on the topological expressions for the QHE \cite{ZZAnn,ZZJPA} and for the
CSE \cite{ZA}; the Fermi-surface invariant of non-homogeneous systems \cite{ZubJETP}; and the
noncommutative-geometric form of the invariants \cite{KMMZZ,KMMZZindex}. The main results are the following. For
finite-order operators with gauge-singlet coefficients the covariant Weyl symbol is independent of the gauge field if
and only if the covariant derivatives are completely symmetrized; a change of ordering does not affect the integrated
anomaly as long as it preserves the elliptic chiral symbol class. On a closed four-manifold with vanishing rational
first Pontryagin class the anomaly obeys a factorization formula with background and mixed terms; for $c_1(E)=0$ the
factorization into the instanton number and $N_3^0$ holds precisely when the untwisted index vanishes. We state the
analytic assumptions under which the CSE invariant equals the winding of the principal symbol, derive the
Luttinger--Ward identity that justifies the current with the renormalized velocity, transfer the non-renormalization
results from the lattice to the continuum, and describe the index-pairing form of the invariants for lattice
symbols. Throughout we cite the earlier results close to these: the gauge-covariant Wigner functions of Heinz and of
Elze, Gyulassy and Vasak, the magnetic Weyl calculus, the non-Abelian chiral kinetic theory, the induced non-Abelian
Chern--Simons terms, the $U(1)\times SU(2)$ Hall effects of Fr\"ohlich and Studer, the anomaly of generalized Dirac
operators of Yee and Yi, the non-Abelian anomalies of multi-Weyl semimetals, and the index theory of Atiyah--Singer,
Fedosov and Connes.

Section~\ref{sec:WW} develops the covariant calculus and identifies the class of operators whose
symbol is independent of the gauge field (Sec.~\ref{sec:class}). Sections~\ref{sec:currents}--\ref{sec:groenewold}
give the currents \reviewfix{(by the Noether procedure and as the response of $\log Z$ to the gauge field, with the
comparison of the two definitions)} and the iterative solution of the Groenewold equation. Sections~\ref{sec:CSE}
\corr{and~\ref{sec:QHE} treat the non-Abelian CSE and QHE, the conditions for their stability under interactions, and the Fermi-surface}
models\reviewfix{ -- the latter in Sec.~\ref{sec:CME}: the non-elliptic models of \cite{ZubJETP}, their invariant, the
equilibrium current of the conserved charge (a \reviewfix{chiral magnetic like effect}), and the chiral separation effect and anomaly
of the model with spherical Fermi surfaces}; Sec.~\ref{sec:latcont} transfers the non-renormalization results, originally obtained for lattice models with
the approximate lattice Wigner--Weyl calculus, to the continuum, where the calculus is exact, and lists the conditions
at $|p|\to\infty$ that replace the compactness of the Brillouin zone. Section~\ref{sec:anomaly} treats the chiral anomaly: the point-splitting derivation, the factorization
theorem beyond the homotopy hypothesis, the identification of the Moyal $N_3$ with the degree $N_3^0$ of the principal
symbol, \reviewfix{the quasi-elliptic operators (Sec.~\ref{sec:quasiell}),} and the non-elliptic models. Section~\ref{sec:NCG} gives the noncommutative-geometric form.
Section~\ref{sec:summary} summarizes the results and lists open problems. Definitions of the mathematical and
physical notions used throughout -- principal symbol and ellipticity, Fredholm index, degree of a map, Chern
character and Todd class, $K$-theory and fibre integration, Moyal product and Weyl ordering, Luttinger--Ward
functional, Landau levels, cyclic cocycles and Fredholm modules -- are collected in Appendix~\ref{app:defs}; in
particular, $N_3^0$ in \eqref{eq:N3hom} is the degree (winding number) of the map $p\mapsto G(p)$ from $S$ to the
invertible matrices (Appendix~\ref{app:degree}).

\paragraph{Conventions.} Euclidean $D$-dimensional space ($D=4$ unless stated), $\hbar=c=1$; operators
$\hx_\mu,\hp_\mu$ with $[\hx_\mu,\hp_\nu]=i\delta_{\mu\nu}$, $\langle x|p\rangle=(2\pi)^{-D/2}e^{ipx}$; external
gauge field $A_\mu=A^a_\mu t^a$ of $G\subset U(N)$, covariant momentum $\hpi_\mu=\hp_\mu-A_\mu(\hx)$, covariant
derivative $D_\mu=\partial_\mu-iA_\mu$, $-iD_\mu\leftrightarrow\hpi_\mu$;
$F_{\mu\nu}=\partial_\mu A_\nu-\partial_\nu A_\mu-i[A_\mu,A_\nu]$, $[D_\mu,D_\nu]=-iF_{\mu\nu}$,
$[\hpi_\mu,\hpi_\nu]=iF_{\mu\nu}(\hx)$; $\cD_\mu=\partial_\mu-i[A_\mu,\cdot]$ in the adjoint. Traces: $\trD$ over
spinor and other internal indices, $\trG$ over gauge indices, $\trH$ over the Hilbert space; the trace per unit
volume $\frac1V\int d^3x$ in the topological expressions is defined by periodization (after \eqref{eq:CSE});
$\gamma^5=\mathrm{diag}(1_2,-1_2)$. Euclidean components: $A_4=-iA_0$, $x_4=ix_0$, $p_4=\omega=-ip_0$,
$F_{ij}=\epsilon_{ijk}B_k$, $F_{4j}=-iE_j$ for static fields. Equation numbers of the original papers are quoted as
[Ref.~\cite{XZPRD}, Eq.~(n)] and [Ref.~\cite{XZPLB}, Eq.~(n)].
\corr{Antisymmetrization over two indices is unnormalized: $T_{[\mu\nu]}=T_{\mu\nu}-T_{\nu\mu}$.
Momentum-space orientation in four dimensions is $dp_1\wedge dp_2\wedge dp_3\wedge dp_4$.
The CSE slice difference uses $dp_1\wedge dp_2\wedge dp_3$ on the upper slice and its negative on the lower slice;
this is the inward boundary orientation of the intervening slab. Gauge generators are normalized by
$\trG(t^at^b)=T_R\delta^{ab}$; an adjoint component response includes this factor.}
\begin{reviewchange}
The transform below is defined on $\R^D$ (or on an explicitly periodic realization); on a general manifold it is a
local construction, and the global index arguments of Sec.~\ref{sec:anomaly} use the symbol class rather than a
globally defined Moyal product. Unitarity of the Wilson lines requires Hermitian Euclidean $A_\mu$; the
continuation $A_4=-iA_0$ and a real chemical potential are analytic continuations, which need not preserve it.
\end{reviewchange}

%======================================================================
\section{Covariant Wigner--Weyl calculus}
\label{sec:WW}

\subsection{Parallel transporters and the covariant transform}

Let $U(w,x)=\Pexp\big(i\int_{x\to w}dz\,A(z)\big)$ be the path-ordered exponential along the straight segment from
$x$ to $w$, defined as the limit of the ordered product [Ref.~\cite{XZPRD}, Eqs.~(52)--(53)]
\begin{equation}
U(w,x)=\lim_{N\to\infty}f_N(N)\cdots f_N(0),\qquad f_N(n)=1+\frac iN(w-x)^\mu A_\mu\Big(x+\frac nN(w-x)\Big),
\end{equation}
so that $A(x)$ stands on the right, $A(w)$ on the left, $U$ is unitary and $U(x,w)^{-1}=U(w,x)$.

\begin{lemma}[Translation by the covariant derivative]
For $\psi$ in the fundamental and $\Psi$ in the adjoint representation [Ref.~\cite{XZPRD}, Eqs.~(50), (54)],
\begin{equation}
e^{yD}\psi(x)=U(x,x+y)\psi(x+y),\qquad e^{y\cD}\Psi(x)=U(x,x+y)\Psi(x+y)U(x+y,x).
\label{eq:transl}
\end{equation}
\end{lemma}
\begin{proof}
Write $e^{yD}=\lim_N\prod_{n=1}^N(1+\frac yN D)$ and $1+\frac yND=\big(1+\frac yN\partial\big)\big(1-i\frac yNA(x)\big)
+O(N^{-2})=T(y/N)\big(1-i\frac yNA(x)\big)$ with $T(\varepsilon)\psi(x)=\psi(x+\varepsilon)$. Moving all
translations to the left produces $\psi(x+y)$ multiplied on the left by the ordered product
$\prod_n\big(1-i\frac yNA(x+\frac nNy)\big)$, which is $U(x,x+y)$. The adjoint case follows by applying the
fundamental one to both indices.
\end{proof}

In bra-ket form, with $\hpi=\hp-A(\hx)$ representing $-iD$,
\begin{equation}
e^{-iy\hpi}|x\rangle=|x+y\rangle U(x+y,x),\qquad \langle x|e^{iy\hpi}=U(x,x+y)\langle x+y| .
\label{eq:braket}
\end{equation}

\begin{definition}
The covariant Wigner transform (covariant Weyl symbol) of an operator $\hX$ is [Ref.~\cite{XZPRD}, Eqs.~(44), (62); Ref.~\cite{XZPLB}, Eqs.~(15), (17)]
\begin{equation}
X_W(x,p)=\int d^Dy\,e^{ipy}\,\langle x|e^{-\frac i2y\hpi}\hX e^{-\frac i2y\hpi}|x\rangle
=\int d^Dy\,e^{ipy}\,U(x,x-\tfrac y2)\,\langle x-\tfrac y2|\hX|x+\tfrac y2\rangle\,U(x+\tfrac y2,x),
\label{eq:WT}
\end{equation}
the two forms being equal by \eqref{eq:braket}.
\end{definition}
The definition goes back, for Abelian fields, to Vasak, Gyulassy and Elze \cite{VGE87}; the non-Abelian Wigner
operator with straight transporters is that of Heinz \cite{Heinz83,Heinz85} and Elze, Gyulassy and Vasak
\cite{EGV86,EGVgluon}, reviewed in \cite{ElzeHeinz}; in the Abelian case the resulting calculus is the magnetic Weyl
calculus of M\"uller \cite{Muller} and M\u{a}ntoiu and Purice \cite{MantoiuPurice}.

\paragraph{Gauge covariance.} Under $\psi\to\Omega\psi$, $A\to\Omega A\Omega^\dagger+i\Omega\partial\Omega^\dagger$,
the matrix element transforms as $\langle x-\frac y2|\hX|x+\frac y2\rangle\to\Omega(x-\frac y2)(\dots)\Omega^\dagger
(x+\frac y2)$ and $U(w,x)\to\Omega(w)U(w,x)\Omega^\dagger(x)$, so that $X_W(x,p)\to\Omega(x)X_W(x,p)\Omega(x)^\dagger$:
the symbol lives in the adjoint (two-index) representation at the single point $x$.

\paragraph{Trace formula.} Setting $\hX\to\hX\hY$ and integrating \eqref{eq:WT} over $p$ gives $\delta(y)$, the
transporters become $1$, and [Ref.~\cite{XZPRD}, Eq.~(11); Ref.~\cite{XZPLB}, Eq.~(26)]
\begin{equation}
\trG\trH(\hX\hY)=\int\frac{d^Dx\,d^Dp}{(2\pi)^D}\trG\big(X_W\star Y_W\big)=\int\frac{d^Dx\,d^Dp}{(2\pi)^D}\trG\big(X_WY_W\big),
\label{eq:trace}
\end{equation}
the second equality (and the cyclic property) holding whenever the integrals converge [Ref.~\cite{XZPLB}, Eq.~(27)].

\paragraph{Exact symbols.} (i) For a gauge-singlet function $c(\hx)$: $\langle x-\frac y2|c(\hx)|x+\frac y2\rangle
=c(x)\delta(y)$ and the transporters cancel, so $[c(\hx)]_W=c(x)$ exactly. (ii) For a gauge-covariant function such as
$F_{\mu\nu}(\hx)$ the same holds, $[F(\hx)]_W=F(x)$, because at $y=0$ the transporters are $1$. (iii)
$[\hpi_\mu]_W=p_\mu$: see Sec.~\ref{sec:cons}.

\subsection{The inverse (Weyl) transform}
\label{sec:inverse}

\begin{proposition}[Ref.~\cite{XZPRD}, Eq.~(45); Ref.~\cite{XZPLB}, Eq.~(23)]
The covariant Weyl quantization
\begin{equation}
\hX=(2\pi)^{-2D}\int dq\,dy\,dx\,dp\;e^{\frac i2y(\hpi-p)}\,e^{iq(\hx-x)}\,X_W(x,p)\,e^{\frac i2y(\hpi-p)}
\label{eq:inv}
\end{equation}
is the two-sided inverse of \eqref{eq:WT}. For $A=0$ it is the ordinary Weyl quantization.
\end{proposition}
\begin{proof}
(a) Substitute \eqref{eq:inv} into \eqref{eq:WT}. The matrix element to be computed is
\[
\langle x|e^{-\frac i2y\hpi}e^{\frac i2y'\hpi}e^{iq'\hx}X_W(x',p')e^{\frac i2y'\hpi}e^{-\frac i2y\hpi}|x\rangle .
\]
By \eqref{eq:braket}, $\langle x|e^{-\frac i2y\hpi}e^{\frac i2y'\hpi}=U(x,x-\frac y2)U(x-\frac y2,x-\frac y2+\frac{y'}2)
\langle x-\frac y2+\frac{y'}2|$ and similarly on the right; the middle factor $e^{iq'\hx}X_W(x',p')$ is a function
of $\hx$ (the gauge matrix $X_W(x',p')$ is a constant matrix at fixed $x',p'$), so its matrix element is
$e^{iq'x}X_W(x',p')\delta(y-y')$; at $y=y'$ the transporters run along a single segment back and forth and cancel.
The remaining integrals $\int dq'e^{iq'(x-x')}=(2\pi)^D\delta(x-x')$ and $\int dy\,e^{iy(p-p')}=(2\pi)^D\delta(p-p')$
return $X_W(x,p)$. (b) Conversely, substitute \eqref{eq:WT} into \eqref{eq:inv} and integrate over $p$ and $q$:
one obtains $\int dy\,dx\,e^{\frac i2y\hpi}|x\rangle\,U(x,x-\frac y2)\langle x-\frac y2|\hX|x+\frac y2\rangle
U(x+\frac y2,x)\,\langle x|e^{\frac i2y\hpi}$; by \eqref{eq:braket} $e^{\frac i2y\hpi}|x\rangle U(x,x-\frac y2)
=|x-\frac y2\rangle$ and $U(x+\frac y2,x)\langle x|e^{\frac i2y\hpi}=\langle x+\frac y2|$, whence
$\int dy\,dx\,|x-\frac y2\rangle\langle x-\frac y2|\hX|x+\frac y2\rangle\langle x+\frac y2|=\hX$.
\end{proof}

\subsection{Consequences: what is quantized into what}
\label{sec:cons}

For a multi-index $\alpha$ with $|\alpha|=n$ realized by $(\mu_1,\dots,\mu_n)$ (so $p^\alpha=p_{\mu_1}\cdots
p_{\mu_n}$) define the completely symmetrized product
\begin{equation}
\Sym(\hpi^\alpha):=\frac1{n!}\sum_{\sigma\in S_n}\hpi_{\mu_{\sigma(1)}}\cdots\hpi_{\mu_{\sigma(n)}} .
\end{equation}

\begin{proposition}\label{prop:quant}
(a) The quantization \eqref{eq:inv} of $p^\alpha$ is $\Sym(\hpi^\alpha)$; hence $[\Sym(\hpi^\alpha)]_W=p^\alpha$
exactly. (b) The quantization of $c(x)p^\alpha$ with a gauge singlet $c$ is the symmetrically split operator
\begin{equation}
\mathrm{Weyl}[c\,p^\alpha]=\Big[(-i\partial_y)^\alpha\,e^{\frac i2y\hpi}c(\hx)e^{\frac i2y\hpi}\Big]_{y=0}
=\frac1{2^n}\sum_{S\subseteq\{1,\dots,n\}}\Sym(\hpi_S)\,c(\hx)\,\Sym(\hpi_{S^c}),
\label{eq:split}
\end{equation}
where $\hpi_S$ denotes the sub-multi-index of the factors with labels in $S$. (c) The quantization of
$c(x)e^{ib\cdot p}$ is $e^{\frac i2b\hpi}c(\hx)e^{\frac i2b\hpi}$.
\end{proposition}
\begin{proof}
In \eqref{eq:inv} the $x$- and $q$-integrals of $e^{iq(\hx-x)}c(x)$ give $c(\hx)$; the $p$-integral of
$p^\alpha e^{-iyp}$ gives $(i\partial_y)^\alpha(2\pi)^D\delta(y)$, and integrating by parts in $y$,
$\hX=[(-i\partial_y)^\alpha e^{\frac i2y\hpi}c(\hx)e^{\frac i2y\hpi}]_{y=0}$. For $c=1$ the two exponentials merge,
$e^{iy\hpi}$, and $(-i\partial_y)^\alpha e^{iy\hpi}|_{y=0}$ is the coefficient of $y^\alpha$ in $(iy\hpi)^n/n!$
\corr{times $(-i)^n\alpha!$}, i.e.\ $\Sym(\hpi^\alpha)$: all orderings of the
$\hpi$'s appear with equal weight because $(y\hpi)^n$ is a symmetric function of the non-commuting factors. For
general $c$, apply the Leibniz rule to the product $e^{\frac i2y\hpi}\,c\,e^{\frac i2y\hpi}$: a subset $S$ of the
$n$ derivatives acts on the left exponential and its complement on the right one, each producing
$2^{-|S|}\Sym(\hpi_S)$, respectively $2^{-|S^c|}\Sym(\hpi_{S^c})$, which gives \eqref{eq:split}. (c) is the same
computation with $e^{ib\cdot p}$, for which the $p$-integral gives $\delta(y-b)$.
\end{proof}

The symbol of a product is the covariant star product, $(\hX\hY)_W=:X_W\star Y_W$. A degenerate case follows at
once from the exact symbols above: if one factor is $p$-independent and a gauge singlet, the covariant star product
reduces to the Moyal product, $c(x)\star Y_W=c(x)\star_MY_W$, because the Wilson loop in \eqref{eq:starWilson}
below has zero area.

\subsection{The star product and the triangular Wilson loop}

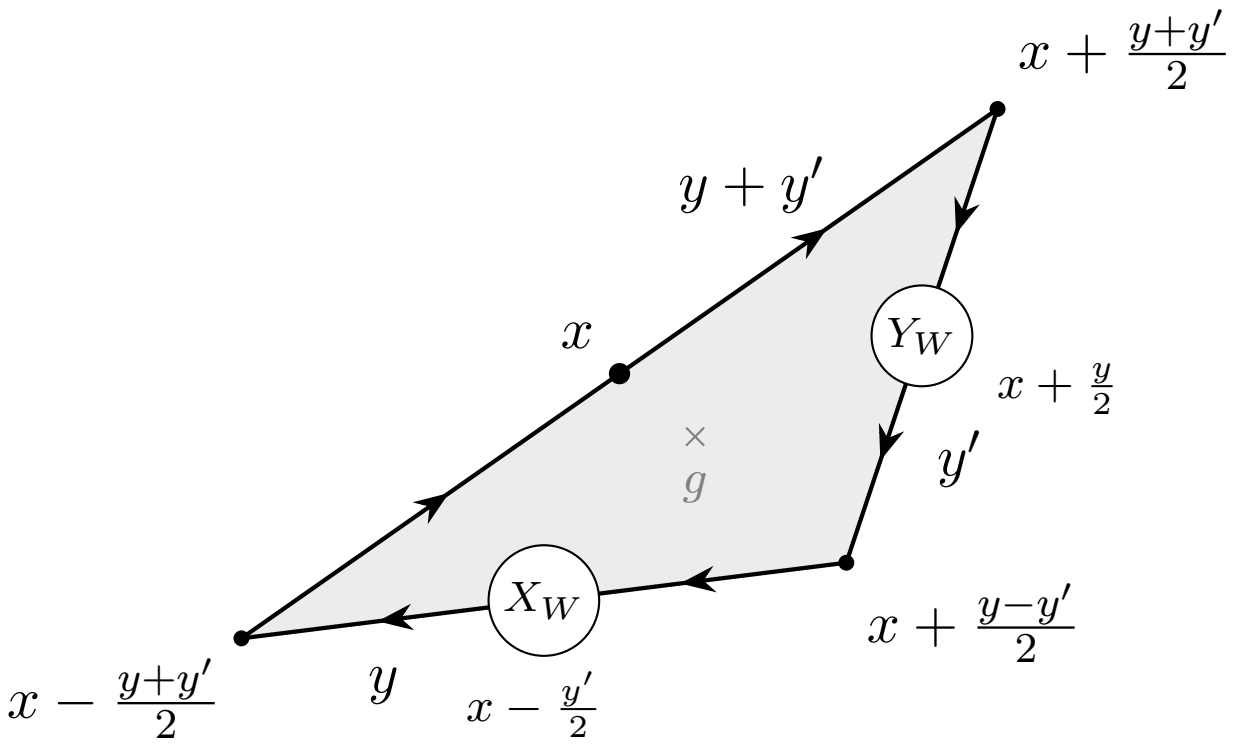
\begin{figure}[htb]
\centering
\begin{tikzpicture}[scale=1.25,>=Stealth,
  arr/.style={postaction={decorate},decoration={markings,mark=at position 0.55 with {\arrow{>}}}}]
\coordinate (x) at (0,0);
\coordinate (P1) at (-2.0,-1.4);
\coordinate (P2) at (1.2,-1.0);
\coordinate (P3) at (2.0,1.4);
\coordinate (XW) at (-0.4,-1.2);
\coordinate (YW) at (1.6,0.2);
\coordinate (g) at (0.4,-0.333);
\fill[gray!15] (P1)--(P2)--(P3)--cycle;
\draw[thick,arr] (x)--(P3);
\draw[thick,arr] (P3)--(YW);
\draw[thick,arr] (YW)--(P2);
\draw[thick,arr] (P2)--(XW);
\draw[thick,arr] (XW)--(P1);
\draw[thick,arr] (P1)--(x);
\fill (x) circle(1.6pt) node[above left=1pt]{$x$};
\fill (P1) circle(1.2pt) node[below left]{$x-\frac{y+y'}2$};
\fill (P2) circle(1.2pt) node[below right]{$x+\frac{y-y'}2$};
\fill (P3) circle(1.2pt) node[above right]{$x+\frac{y+y'}2$};
\node[draw,fill=white,inner sep=1.5pt,circle] at (XW) {\scriptsize$X_W$};
\node[draw,fill=white,inner sep=1.5pt,circle] at (YW) {\scriptsize$Y_W$};
\node at ($(XW)+(-0.05,-0.55)$) {\footnotesize$x-\frac{y'}2$};
\node at ($(YW)+(0.72,-0.28)$) {\footnotesize$x+\frac y2$};
\node[cross out,draw=gray,inner sep=1.5pt] at (g) {};
\node[gray,anchor=north] at ($(g)+(0,-0.08)$) {\footnotesize$g$};
\node at (-1.25,-1.65) {$y$};
\node at (1.8,-0.45) {$y'$};
\node at (0.7,1.0) {$y+y'$};
\end{tikzpicture}
\caption{The triangular Wilson loop of the covariant star product \eqref{eq:starWilson}. The parallel transporters run
along the closed path $x\to x+\frac{y+y'}2\to x+\frac{y-y'}2\to x-\frac{y+y'}2\to x$; the base point $x$ is the
midpoint of the side $y+y'$, and the symbols $X_W$, $Y_W$ are inserted at the midpoints $x-\frac{y'}2$ and $x+\frac y2$
of the sides $y$ and $y'$. The shaded triangle $\Delta$, with centroid $g=x+\frac{y-y'}6$, is the area entering the expansion
\eqref{eq:Wexp} of the loop.}
\label{fig:triangle}
\end{figure}

Applying \eqref{eq:inv} twice inside \eqref{eq:WT} and using \eqref{eq:braket} repeatedly one obtains
[Ref.~\cite{XZPRD}, Eq.~(63); Ref.~\cite{XZPLB}, Eq.~(24)]
\begin{align}
(X_W\star Y_W)(x,p)&=(2\pi)^{-2D}\int dy\,dk\,dy'dk'\,e^{-iy(k-p)-iy'(k'-p)}\notag\\
&\quad\times U\big(x,x-\tfrac{y+y'}2\big)U\big(x-\tfrac{y+y'}2,x-\tfrac{y'}2\big)X_W\big(x-\tfrac{y'}2,k\big)\notag\\
&\quad\times U\big(x-\tfrac{y'}2,x+\tfrac{y-y'}2\big)U\big(x+\tfrac{y-y'}2,x+\tfrac y2\big)Y_W\big(x+\tfrac y2,k'\big)\notag\\
&\quad\times U\big(x+\tfrac y2,x+\tfrac{y+y'}2\big)U\big(x+\tfrac{y+y'}2,x\big):
\label{eq:starWilson}
\end{align}
the integrand is the triangular Wilson loop $x\to x+\frac{y+y'}2\to x+\frac{y-y'}2\to x-\frac{y+y'}2\to x$ with
$X_W$ and $Y_W$ inserted at $x-\frac{y'}2$ and $x+\frac y2$ on its sides (Fig.~\ref{fig:triangle}). With $A=0$ it is
the Moyal product.

Using \eqref{eq:transl} to move the insertions to the point $x$ one obtains the differential representation
[Ref.~\cite{XZPRD}, Eqs.~(64), (70)--(71)]
\begin{align}
X_W\star Y_W&=(2\pi)^{-2D}\int dy\,dk\,dy'dk'\,e^{-iy(k-p)-iy'(k'-p)}\,
\mathcal W(x,y,y')\notag\\
&\quad\times\Big[e^{\frac{y+y'}2\cD}e^{-y'\cD}e^{-\frac y2\cD}X_W(x,k)\Big]\Big[e^{\frac{y+y'}2\cD}e^{-\frac{y'}2\cD}Y_W(x,k')\Big],
\label{eq:stardiff}
\end{align}
with the pure triangular loop
\begin{equation}
\mathcal W(x,y,y')=U\big(x,x-\tfrac{y+y'}2\big)U\big(x-\tfrac{y+y'}2,x+\tfrac{y-y'}2\big)
U\big(x+\tfrac{y-y'}2,x+\tfrac{y+y'}2\big)U\big(x+\tfrac{y+y'}2,x\big).
\end{equation}
Expanding \eqref{eq:stardiff} to first order in $y,y'$ (where $\mathcal W=1$) and converting powers of $y$ into
momentum derivatives by integration by parts gives the covariant Poisson bracket [Ref.~\cite{XZPRD}, Eq.~(72)]
\begin{equation}
X_W\star Y_W=X_WY_W-\frac i2\big(\partial_{p_\mu}X_W\,\cD_\mu Y_W-\cD_\mu X_W\,\partial_{p_\mu}Y_W\big)+\dots
\end{equation}

\begin{proposition}[Expansion of the Wilson loop, Ref.~\cite{XZPRD}, Eq.~(121); Ref.~\cite{XZPLB}, Eq.~(53)]
\begin{align}
\mathcal W(x,y,y')&=1-\frac i2y^\mu y'^\nu F_{\mu\nu}-\frac i{12}(y-y')^\alpha y^\mu y'^\nu\cD_\alpha F_{\mu\nu}
-\frac i{48}\big(y^\alpha y^\beta+y'^\alpha y'^\beta\big)y^\mu y'^\nu\cD_\alpha\cD_\beta F_{\mu\nu}\notag\\
&\quad-\frac18y^\mu y'^\nu y^\alpha y'^\beta F_{\mu\nu}F_{\alpha\beta}+O(\cD^5).
\label{eq:Wexp}
\end{align}
\end{proposition}
\begin{proof}
\reviewfix{Let $\Delta_+$ be the triangle with vertices
$x-(y+y')/2$, $x+(y-y')/2$, $x+(y+y')/2$, oriented with positive area
$\tfrac12y\wedge y'$. The Wilson path traverses its boundary in the opposite direction.
With $F$ parallel-transported to $x$, the surface-ordered expansion therefore starts as
$\mathcal W=1-i\int_{\Delta_+}F+\cdots$; the quadratic flux term through degree four is
$\tfrac12(-\tfrac i2y^\mu y'^\nu F_{\mu\nu})^2$.
The scalar moments below are those of this triangle and do not depend on its orientation.} Expanding $F(z)=F(x)+(z-x)^\alpha\cD_\alpha F+\frac12(z-x)^\alpha
(z-x)^\beta\cD_\alpha\cD_\beta F+\dots$ and using the moments of the uniform distribution on the triangle,
\[
\langle z-x\rangle_\Delta=\frac{y-y'}6,\qquad
\big\langle(z-x)^\alpha(z-x)^\beta\big\rangle_\Delta=\frac{y^\alpha y^\beta+y'^\alpha y'^\beta}{12},
\]
(the centroid of a triangle with vertices $a,b,c$ is $(a+b+c)/3$, and its second moments are
$\frac1{12}(a^\alpha a^\beta+b^\alpha b^\beta+c^\alpha c^\beta)+\corr{\frac34 g^\alpha g^\beta}$ with $g$ the centroid;
substituting the three vertices gives the values quoted) one obtains the second and third terms. The last term is
the second-order term of the exponential, $\frac12\big(-\frac i2y^\mu y'^\nu F_{\mu\nu}\big)^2$; at this order the
ordering ambiguity of the path-ordered exponential contributes only through $[F,F]$ terms multiplied by one more
power of $y$, i.e.\ at $O(\cD^5)$. The coefficients were checked independently by a normal-ordering computation of
\corr{$e^{-\frac i2(y+y')\hpi}e^{iy\hpi}e^{iy'\hpi}e^{-\frac i2(y+y')\hpi}$} [Ref.~\cite{XZPRD}, Eq.~(120)].
\end{proof}

\subsection{The class of operators whose symbol is gauge-field independent}
\label{sec:class}

The Dirac operators $\hQ=Q(\hx,\hpi)$ are built from gauge-singlet coefficient functions and covariant
derivatives, and the formalism rests on the statement that the covariant Weyl symbol of $\hQ$ does not depend on the
gauge field and equals the ordinary Wigner transform of $\hQ^{(A=0)}$.\footnote{Ref.~\cite{XZPRD}, Eqs.~(20), (74); Ref.~\cite{XZPLB}, Eq.~(34).} The
statement requires a specification of the ordering of the covariant derivatives, which do not commute:
$[\hpi_\mu,\hpi_\nu]=iF_{\mu\nu}$.

\begin{example}[The simplest example]
For the ordering $(-iD)^\alpha=\prod_\mu(-iD_\mu)^{\alpha_\mu}$,\footnote{This is the ordering written in Ref.~\cite{XZPLB}, Eq.~(12),
(30). Ref.~\cite{XZPLB}, Eq.~(34) therefore refers to the symmetrized class \eqref{eq:classQ}; the general ordering is covered by
Sec.~\ref{sec:homotopy}.} already at $m=2$
\begin{equation}
(-iD_\mu)(-iD_\nu)=\hpi_\mu\hpi_\nu=\tfrac12\{\hpi_\mu,\hpi_\nu\}+\tfrac i2F_{\mu\nu}(\hx)
\ \Longrightarrow\
\big[(-iD_\mu)(-iD_\nu)\big]_W=p_\mu p_\nu+\tfrac i2F_{\mu\nu}(x)\neq p_\mu p_\nu ,
\end{equation}
by Proposition~\ref{prop:quant}(a) and the exact symbol of $F(\hx)$. The symbol depends on $F$ and is not
proportional to the unit matrix in gauge space.
\end{example}

\begin{theorem}[Gauge-field independence of the symbol]\label{thm:class}
\corr{For finite-order differential operators built from $A$-independent gauge-singlet coefficients, the covariant Weyl symbol of $\hQ$ is independent of every gauge field if and only if $\hQ$}
can be written with all products of covariant derivatives completely symmetrized,
\begin{equation}
\hQ=\sum_\alpha c_\alpha(x)\,\Sym\big[(-iD)^\alpha\big],\qquad c_\alpha\ \text{independent of }A
\label{eq:classQ}
\end{equation}
\corr{(a finite sum). Then} $Q_W(x,p)=\sum_\alpha c_\alpha(x)\star_Mp^\alpha$ is the ordinary Wigner transform
of $\hQ^{(A=0)}=\sum_\alpha c_\alpha(\hx)\hp^\alpha$ and is proportional to unity in gauge space.
\end{theorem}
\begin{proof}
Sufficiency: by Proposition~\ref{prop:quant}(a) and the degenerate star product,
$[c_\alpha(\hx)\Sym(\hpi^\alpha)]_W=c_\alpha\star p^\alpha=c_\alpha\star_Mp^\alpha$, which is the ordinary Wigner
transform of $c_\alpha(\hx)\hp^\alpha$. Necessity: the transform and its inverse are bijective. If $Q_W$ is
$A$-independent, then $\hQ$ is the covariant quantization \eqref{eq:inv} of an $A$-independent function
$Q_W(x,p)$; \corr{for these polynomial symbols,} $Q_W=\sum_\alpha o_\alpha(x)p^\alpha$ and the
quantization is $\sum_\alpha\mathrm{Weyl}[o_\alpha p^\alpha]$, which by \eqref{eq:split} is of the form
\eqref{eq:classQ} (moving the coefficients to the left produces new $A$-independent coefficients built from
derivatives of $o_\alpha$).
\end{proof}

\begin{remark}
(i) The position of the coefficients is immaterial: $c\,\Sym$, $\Sym\,c$ and the split forms differ by class members
with $A$-independent coefficients, e.g.\ $c(\hx)\Sym(\hpi_1\hpi_2)=\mathrm{Weyl}[cp_1p_2]+\frac i2\mathrm{Weyl}
[\partial_1c\,p_2]+\frac i2\mathrm{Weyl}[\partial_2c\,p_1]-\frac14\partial_1\partial_2c$. (ii) The class contains all
first-order operators, all functions of a single combination $b\cdot(-iD)$ (such as $e^{ib\cdot(-iD)}$ or
$\cos(b\cdot(-iD))$, i.e.\ Peierls-substituted tight-binding operators: lattice hopping terms $e^{ib\cdot p}$ with $p\to-iD$, Appendix~\ref{app:ww}), and powers of $\hpi_1\pm i\hpi_2$; the
multi-Weyl example of Sec.~5 of Ref.~\cite{XZPLB} is in the class.
\end{remark}

\paragraph{The nested ordering of Ref.~\cite{XZPRD}.} That paper defines $Q(x,-iD)=\sum_{|\alpha|\le m}o_\alpha(x)\circ(-iD)^\alpha$ with
the nested anticommutator [Ref.~\cite{XZPRD}, Eq.~(18)]
\begin{equation}
o_\alpha\circ(-iD)^\alpha=2^{-|\alpha|}\{\dots\{\{o_\alpha,(-iD_1)\},\dots,(-iD_1)\},(-iD_2)\},\dots,(-iD_4)\}.
\label{eq:nested}
\end{equation}
Expanding the nested anticommutator, each term is a word in which the derivatives placed to the left of $o_\alpha$
appear in decreasing order of their position in the nesting sequence and those to the right in increasing order,
each word with weight $2^{-n}$. Averaging over the $n!$ orders of nesting therefore produces every ordering of the
left and right blocks with equal weight:

\begin{proposition}[Weyl ordering as an averaged nested anticommutator]\label{prop:nested}
\begin{align}
&\frac1{n!}\sum_{\sigma\in S_n}2^{-n}\Big\{\dots\big\{\{o_\alpha,(-iD_{\mu_{\sigma(1)}})\},(-iD_{\mu_{\sigma(2)}})\big\},\dots,
(-iD_{\mu_{\sigma(n)}})\Big\}\notag\\
&\qquad=\frac1{2^n}\sum_{S}\Sym[(-iD)_S]\,o_\alpha\,\Sym[(-iD)_{S^c}]=\mathrm{Weyl}[o_\alpha p^\alpha].
\end{align}
With this definition of $\circ$, $Q_W(x,p)=\sum_\alpha o_\alpha(x)p^\alpha$ holds exactly for arbitrary
non-Abelian fields.\footnote{Ref.~\cite{XZPRD}, Eq.~(18) uses the single nesting \eqref{eq:nested}; with it, Ref.~\cite{XZPRD}, Eq.~(20) and (74) hold up to
the $O(\cD F)$ terms listed below, which is beyond the $O(F)$ accuracy of the transport results of Ref.~\cite{XZPRD}. The averaged
form is the one to be used in exact statements.} The single nesting coincides with the Weyl ordering for
$|\alpha|\le2$ and for monomials in a single direction; for mixed $|\alpha|\ge3$ it differs by terms proportional to
covariant derivatives of $F$ (and no $F^2$ terms up to fourth order):
\begin{align}
2^{-2}\{\{o,\hpi_\mu\},\hpi_\nu\}-\mathrm{Weyl}[o\,p_\mu p_\nu]&=\tfrac18[o,[\hpi_\mu,\hpi_\nu]]=\tfrac i8[o,F_{\mu\nu}]=0,\\
2^{-3}\{\{\{o,\hpi_1\},\hpi_1\},\hpi_2\}-\mathrm{Weyl}[o\,p_1^2p_2]&=\tfrac16\,o\,\cD_1F_{12},\\
2^{-3}\{\{\{o,\hpi_1\},\hpi_2\},\hpi_3\}-\mathrm{Weyl}[o\,p_1p_2p_3]&=\tfrac1{12}\,o\,(\cD_2F_{13}+\cD_1F_{23}),
\end{align}
(for a gauge-singlet $o$).
\end{proposition}
\begin{proof}
The first statement is the counting above. For the differences, expand both sides into words and use
$\hpi_\mu\hpi_\nu=\hpi_\nu\hpi_\mu+iF_{\mu\nu}$ and $[\hpi_\rho,F_{\mu\nu}]=-i\cD_\rho F_{\mu\nu}$; the computations
were verified symbolically (Abelian background, generic $o(x)$, by realizing the operators as differential operators
acting on a test function).
\end{proof}

\paragraph{General ordering and the final results.} \corr{For the fixed ordering $(-iD)^\alpha=\prod_{\mu=1}^4\hpi_\mu^{\alpha_\mu}$ one has}
\begin{equation}
(-iD)^\alpha=\Sym[(-iD)^\alpha]+\frac i2\sum_{\mu<\nu}\alpha_\mu\alpha_\nu F_{\mu\nu}\,\Sym[(-iD)^{\alpha-e_\mu-e_\nu}]+\dots,
\end{equation}
\begin{GPTcorrection}
the dots containing covariant derivatives of $F$ and higher powers of $F$. The reordering terms have differential
order at most $|\alpha|-2$ and vanish wherever $F$ vanishes; \reviewfix{Hermiticity must be imposed separately on the two comparison operators. The displayed fixed-order
monomial is generally not Hermitian: $(\pi_\mu\pi_\nu)^\dagger=\pi_\nu\pi_\mu$ for Hermitian $\pi$.
Its difference $iF_{\mu\nu}/2$ from the symmetrized monomial is anti-Hermitian.} For chiral elliptic operators on a closed manifold the common
principal symbol fixes the integrated anomaly (Sec.~\ref{sec:homotopy}); the local currents and the finite-field
transport coefficients do depend on the ordering.
\end{GPTcorrection}

\paragraph{Infinite series.} Infinite series in the covariant derivatives (lattice operators, Peierls-substituted
dispersions) are defined as the covariant quantization \eqref{eq:inv} of an $A$-independent symbol, which keeps them
in the gauge-field-independent class (provided the series converge in the operator or symbol topology used).\footnote{A normal-ordered definition (Ref.~\cite{XZPRD}, footnote 2) is not compatible with the
gauge-field independence of the symbol. The reference ``[97]'' in the proof of Ref.~\cite{XZPRD}, Eq.~(74) should be read as a reference
to the inverse transform Ref.~\cite{XZPRD}, Eq.~(45).}

%======================================================================
\section{\reviewfix{Vector and axial currents: the Noether procedure and the response of $\log Z$}}
\label{sec:currents}

\reviewfix{The currents are defined in two ways. Section~\ref{sec:noether} follows the Noether procedure of
Refs.~\cite{XZPRD,XZPLB}. Section~\ref{sec:logZ} defines the vector current as the response of $\log Z$ to a variation
of the external gauge field, derives it within the covariant calculus (where the gauge field is not contained in the
symbol $Q_W$), and compares the two definitions.}

\subsection{\reviewfix{The Noether procedure}}
\label{sec:noether}

The action is $\corr{S=\int d^4x\,\bar\psi\,Q(x,-iD)\psi=-\trD\trG\trH(\hQ\hat\rho)}$ with $\langle x|\hat\rho|y\rangle=
\psi(x)\bar\psi(y)$ [Ref.~\cite{XZPRD}, Eq.~(73); Ref.~\cite{XZPLB}, Eq.~(31)].
\corr{We use the convention of Ref.~\cite{XZPLB}, $Z=\int D\bar\psi D\psi\,e^S$, so $\langle\hat\rho\rangle=-\hQ^{-1}$.
The minus sign in the trace form of $S$ comes from interchanging Grassmann fields.}
Under a local gauge rotation $\psi\to e^{i\alpha(x)}\psi$,
$\delta\hQ=[\hQ,i\alpha(\hx)]$, and by the trace formula [Ref.~\cite{XZPRD}, Eqs.~(77)--(81)]
\begin{equation}
\delta S=\trG\int d^4x\,\alpha(x)\Gamma(x),\qquad
\corr{\Gamma(x)=+i\trD\int\frac{d^4p}{(2\pi)^4}\big(Q_W\star\rho_W-\rho_W\star Q_W\big).}
\end{equation}
Using the star product \eqref{eq:starWilson} with one factor $p$-independent, integrating over $p$ and using
\eqref{eq:transl} to shift the arguments to $x$ [Ref.~\cite{XZPRD}, Eqs.~(82)--(84)],
\begin{align}
\corr{\Gamma(x)=+i\trD\int(2\pi)^{-8}}d^4y\,d^4k\,d^4k'\,e^{-iy(k-k')}\Big(&e^{-\frac i2\partial_k\cD}\big(Q_W(x,k)\rho_W(x,k')\big)\notag\\
&-e^{\frac i2\partial_{k'}\cD}\big(\rho_W(x,k)Q_W(x,k')\big)\Big),
\end{align}
and expanding in powers of $\cD$ [Ref.~\cite{XZPRD}, Eqs.~(85)--(87)],
\begin{equation}
\Gamma=\cD_\mu J_\mu,\qquad
\corr{J_\mu=+\trD\int\frac{d^4p}{(2\pi)^4}\sum_{k\ge0}}\frac{(-1)^k}{4^k(2k+1)!}
\big[\textstyle\prod_{i=1}^{2k}\cD_{\mu_i}\big]\Big(\rho_W\big[\prod_{i=1}^{2k}\partial_{p_{\mu_i}}\big]\partial_{p_\mu}Q_W\Big).
\label{eq:Jseries}
\end{equation}
(The coefficients follow from the Taylor expansion of $e^{\mp\frac i2\partial\cD}$: odd orders give the divergence
structure $\cD_\mu(\dots)$, with $(i/2)^{2k+1}/(2k+1)!$ producing $(-1)^k/4^k(2k+1)!$.) If
$\partial_{p_{\mu_1}}\cdots\partial_{p_{\mu_n}}Q_W=0$ for $n\ge3$, or if the higher terms are negligible because
the inhomogeneity is weak, the current reduces to [Ref.~\cite{XZPRD}, Eq.~(90)]
\begin{equation}
\corr{J_\mu=+\trD\int\frac{d^4p}{(2\pi)^4}\rho_W\,\partial_{p_\mu}Q_W},\qquad
\langle J_\mu\rangle=-\trD\int\frac{d^4p}{(2\pi)^4}G_W\,\partial_{p_\mu}Q_W,
\label{eq:Jvec}
\end{equation}
\corr{with $G_W=-\langle\rho_W\rangle=(\hQ^{-1})_W$}. The same formula follows directly by writing the variation as
$\corr{\delta S=+i\,\Tr(\hat\rho[\alpha(\hx),\hQ])}$, expanding $\alpha(x\pm y/2)$ to first order in the gradient and integrating by parts in
\begin{GPTcorrection}
momentum [Ref.~\cite{XZPRD}, Eqs.~(91)--(94)]. After a gauge trace, higher covariant divergences reduce to ordinary divergences
and integrate to zero with periodic boundary conditions or sufficient decay. This argument does not apply to
an untraced adjoint component: $\int\cD_\mu X$ need not vanish. For non-Abelian local or volume-averaged responses,
the truncation conditions above, or a separately justified covariant averaging prescription, are still required.
\end{GPTcorrection}

For the
axial current,
with $\{\gamma^5,\hQ\}=0$ and the chiral rotation $\psi\to e^{i\alpha\gamma^5}\psi$, $\bar\psi\to\bar\psi
e^{i\alpha\gamma^5}$, the same steps give [Ref.~\cite{XZPRD}, Eq.~(115)]
\begin{equation}
\langle J^{(5)}_\mu\rangle=-\frac12\trD\int\frac{d^4p}{(2\pi)^4}G_W\,\partial_{p_\mu}[Q_W,\gamma^5].
\label{eq:Jax}
\end{equation}
\reviewfix{Equation \eqref{eq:Jax} has the same truncation or averaging restrictions as
\eqref{eq:Jvec}; it is not the full local Noether current for an arbitrary higher-order operator.}
In Ref.~\cite{XZPLB} the same procedure is carried out with the point-split density $\hat\rho^\epsilon=e^{i\hpi\epsilon}\hat\rho
e^{i\hpi\epsilon}$, Sec.~\ref{sec:pointsplit}.

\subsection{\reviewfix{The current as the response of $\log Z$ to the gauge field}}
\label{sec:logZ}

\reviewfix{The Noether procedure determines the current through its covariant divergence $\Gamma=\cD_\mu J_\mu$. For
an external gauge field there is a direct definition: the current is the quantity to which the field couples,
\begin{equation}
\delta\log Z=\trG\int d^4x\,\delta A_\mu(x)\,\langle J_\mu(x)\rangle,\qquad
\langle J^a_\mu(x)\rangle\equiv\trG\big(t^a\langle J_\mu(x)\rangle\big)=\frac{\delta\log Z}{\delta A^a_\mu(x)} .
\label{eq:JlogZdef}
\end{equation}
This is the current that enters the response theory, and it is unique once the dependence of $\hQ$ on the gauge field
is specified. Its relation to the Noether procedure follows from the gauge invariance of the action,
\begin{equation}
S[\Omega\psi,\bar\psi\Omega^\dagger,A^\Omega]=S[\psi,\bar\psi,A],\qquad
A^\Omega_\mu=\Omega A_\mu\Omega^\dagger+i\,\Omega\,\partial_\mu\Omega^\dagger .
\label{eq:gaugeinv}
\end{equation}
For a non-Abelian field the finite transformation is not a shift of $A_\mu$. Only its first-order form is needed: for
$\Omega=e^{i\alpha}$ with infinitesimal $\alpha=\alpha^at^a$,
\begin{equation}
\delta_\alpha A_\mu=A^\Omega_\mu-A_\mu=\partial_\mu\alpha-i[A_\mu,\alpha]+O(\alpha^2)=\cD_\mu\alpha+O(\alpha^2),
\qquad \delta_\alpha\psi=i\alpha\psi,\quad\delta_\alpha\bar\psi=-i\bar\psi\alpha .
\label{eq:dAgauge}
\end{equation}
Below, $\delta A_\mu=\cD_\mu\alpha$ always denotes this infinitesimal transformation. To first order in $\alpha$,
\eqref{eq:gaugeinv} gives $\delta_\psi S+\delta_AS=0$, so that the variation of the fermion fields alone
is $\delta_\psi S=-\trG\int(\cD_\mu\alpha)\,\delta S/\delta A_\mu=\trG\int\alpha\,\cD_\mu(\delta S/\delta A_\mu)$, and $\delta S/\delta A_\mu$ has the covariant divergence $\Gamma$ of Sec.~\ref{sec:noether}. Whether the two currents
themselves coincide is not fixed by this argument, and is answered below.}

\reviewfix{With $Z=\int D\bar\psi D\psi\,e^S=\det\hQ$ and the trace formula \eqref{eq:trace},
\begin{equation}
\delta\log Z=\Tr\big(\hG\,\delta\hQ\big)=\int\frac{d^4x\,d^4p}{(2\pi)^4}\,\trD\trG\big[G_W\,(\delta\hQ)_W\big],
\label{eq:dlogZ}
\end{equation}
where $\Tr=\trD\trG\trH$. The object $(\delta\hQ)_W$ in \eqref{eq:dlogZ} is the symbol of the variation of the operator.
It is not the variation of the symbol, and the difference is essential here. Denote by $\mathsf W_A$ the covariant
Wigner transform \eqref{eq:WT} in the field $A$ and by $\mathsf W_A^{-1}$ the quantization map \eqref{eq:inv}, so
that $X_W=\mathsf W_A[\hX]$; both maps depend on $A$ through the transporters. An operator of the class
\eqref{eq:classQ} is $\hQ[A]=\mathsf W_A^{-1}[Q_W]$ with a symbol $Q_W(x,p)$ that does not contain the gauge field.
Three objects are to be distinguished:
\begin{align}
\delta(Q_W)&:=\mathsf W_{A+\delta A}\big[\hQ[A+\delta A]\big]-\mathsf W_A\big[\hQ[A]\big]=0,\notag\\
\delta\hQ&:=\hQ[A+\delta A]-\hQ[A]=\mathsf W^{-1}_{A+\delta A}[Q_W]-\mathsf W^{-1}_A[Q_W]\neq0,\notag\\
(\delta\hQ)_W&:=\mathsf W_A\big[\delta\hQ\big]=-\big(\delta\mathsf W_A\big)\big[\hQ\big]\neq0 .
\label{eq:dQWdef}
\end{align}
The variation of the symbol vanishes and does not contain $\delta A$. The operator itself changes, because the
quantization map changes. The symbol of this change, taken with the transform in the unvaried field $A$, is linear in
$\delta A$; the last equality in \eqref{eq:dQWdef} follows from $0=\delta(Q_W)=(\delta\mathsf W_A)[\hQ]+\mathsf
W_A[\delta\hQ]$ and shows that $(\delta\hQ)_W$ is produced entirely by the variation of the transporters. The simplest
examples are
\begin{align}
\hQ=\hpi_\mu=\hp_\mu-A_\mu(\hx):&\quad Q_W=p_\mu,\quad \delta\hQ=-\delta A_\mu(\hx),\quad (\delta\hQ)_W=-\delta A_\mu(x);\notag\\
\hQ=\hpi_\mu\hpi_\mu:&\quad Q_W=p^2,\quad \delta\hQ=-\{\delta A_\mu(\hx),\hpi_\mu\},\notag\\
&\quad(\delta\hQ)_W=-\big(\delta A_\mu\star p_\mu+p_\mu\star\delta A_\mu\big)=-2\,\delta A_\mu\,p_\mu .
\label{eq:dQWex}
\end{align}
In both cases $\delta(Q_W)=0$. The trace formula \eqref{eq:trace} holds for any pair of operators, with both symbols
taken in the same field $A$; applied to $\hG$ and $\delta\hQ$ it gives the second equality in \eqref{eq:dlogZ}, with
$G_W=\mathsf W_A[\hG]$ and $(\delta\hQ)_W=\mathsf W_A[\delta\hQ]$.}

\reviewfix{To find $(\delta\hQ)_W$ for a general operator of the class we use the kernel of the operator. Inverting
\eqref{eq:WT},
\begin{equation}
\langle x-\tfrac y2|\hQ|x+\tfrac y2\rangle=U(x-\tfrac y2,x)\,\tilde Q(x,y)\,U(x,x+\tfrac y2),\qquad
\tilde Q(x,y)=\int\frac{d^4p}{(2\pi)^4}\,e^{-ipy}\,Q_W(x,p).
\label{eq:kernelQ}
\end{equation}
The gauge field enters the kernel only through the two straight transporters; $\tilde Q$ is not varied. The kernel of
$\delta\hQ$ is therefore obtained by varying the two transporters in \eqref{eq:kernelQ}, and $(\delta\hQ)_W$ by
applying to it the transform \eqref{eq:WT} with the unvaried transporters.}

\begin{lemma}[\reviewfix{Variation of the straight transporters}]\label{lem:dU}
\reviewfix{To first order in $\delta A$,
\begin{align}
\delta U(x-\tfrac y2,x)&=U(x-\tfrac y2,x)\,\Big(-\frac i2\Big)y^\mu\sum_{n\ge0}\frac1{(n+1)!}\Big(-\frac{y\cdot\cD}2\Big)^n\delta A_\mu(x),\notag\\
\delta U(x,x+\tfrac y2)&=\Big(-\frac i2\Big)y^\mu\sum_{n\ge0}\frac1{(n+1)!}\Big(\frac{y\cdot\cD}2\Big)^n\delta A_\mu(x)\;U(x,x+\tfrac y2).
\label{eq:dU}
\end{align}}
\end{lemma}
\begin{proof}
\reviewfix{The first-order variation of the path-ordered exponential is
\begin{align*}
\delta U(w,x)&=i\int_{x\to w}U(w,z)\,\delta A_\mu(z)\,dz^\mu\,U(z,x),\\
U(w,z)\,\delta A_\mu(z)\,U(z,x)&=U(w,x)\,\big[U(x,z)\,\delta A_\mu(z)\,U(z,x)\big].
\end{align*}
For $w=x-\frac y2$ put $z=x-s\frac y2$, $0\le s\le1$; the variation $\delta A_\mu$ transforms in the adjoint representation, and
by \eqref{eq:transl} the bracket equals $e^{-s\frac y2\cD}\delta A_\mu(x)$. The integral over $s$ gives
$\int_0^1ds\,s^n=1/(n+1)$. The second line is obtained in the same way with $z=x+s\frac y2$.}
\end{proof}

\begin{proposition}[\reviewfix{Symbol of the variation}]\label{prop:dQW}
\reviewfix{Let $\hQ$ be the covariant quantization \eqref{eq:inv} of a symbol $Q_W(x,p)$ that does not depend on the
gauge field. To first order in $\delta A$ and to all orders in the derivatives,
\begin{align}
(\delta\hQ)_W=-\frac12\sum_{n\ge0}\frac{(i/2)^n}{(n+1)!}\Big[&\big(\cD_{\nu_1}\cdots\cD_{\nu_n}\delta A_\mu\big)\,
\partial_{p_\mu}\partial_{p_{\nu_1}}\cdots\partial_{p_{\nu_n}}Q_W\notag\\
&+(-1)^n\,\partial_{p_\mu}\partial_{p_{\nu_1}}\cdots\partial_{p_{\nu_n}}Q_W\,\big(\cD_{\nu_1}\cdots\cD_{\nu_n}\delta A_\mu\big)\Big].
\label{eq:dQW}
\end{align}
If $Q_W$ is a gauge singlet, the terms with odd $n$ cancel:
\begin{equation}
(\delta\hQ)_W=-\sum_{k\ge0}\frac{(-1)^k}{4^k(2k+1)!}\big(\cD_{\nu_1}\cdots\cD_{\nu_{2k}}\delta A_\mu\big)\,
\partial_{p_\mu}\partial_{p_{\nu_1}}\cdots\partial_{p_{\nu_{2k}}}Q_W
=-\delta A_\mu\,\partial_{p_\mu}Q_W+O(\cD^2\delta A).
\label{eq:dQWsinglet}
\end{equation}}
\end{proposition}
\begin{proof}
\reviewfix{Insert \eqref{eq:dU} into \eqref{eq:kernelQ} and apply the transform \eqref{eq:WT}; the unvaried
transporters cancel, and
\begin{align*}
(\delta\hQ)_W&=\int d^4y\,e^{ipy}\Big[Y_-(y)\,\tilde Q(x,y)+\tilde Q(x,y)\,Y_+(y)\Big],\\
Y_\mp(y)&=-\frac i2\,y^\mu\sum_{n\ge0}\frac1{(n+1)!}\Big(\mp\frac{y\cdot\cD}2\Big)^n\delta A_\mu(x).
\end{align*}
Since $y^\nu\tilde Q(x,y)=\int\frac{d^4p}{(2\pi)^4}e^{-ipy}(-i\partial_{p_\nu})Q_W(x,p)$, each power of $y$ becomes
$-i\partial_p$ acting on $Q_W$, which gives \eqref{eq:dQW}. The covariant derivatives are contracted with a symmetric
tensor, so only their symmetrized product enters.}
\end{proof}

\reviewfix{The leading term of \eqref{eq:dQWsinglet} is the minimal substitution $p\to p-\delta A$ at the level of
symbols, although $A$ itself does not appear in $Q_W$. Two independent checks of \eqref{eq:dQWsinglet}: (i) for
$Q_W=e^{ib\cdot p}$, $\hQ=e^{ib\hpi}$ (Proposition~\ref{prop:quant}(c)) the Duhamel formula gives
$\delta\hQ=-i\int_0^1ds\,e^{isb\hpi}\,b^\mu\delta A_\mu(\hx)\,e^{i(1-s)b\hpi}$; with
$e^{isb\hpi}c(\hx)e^{-isb\hpi}=(e^{sb\cD}c)(\hx)$ and $[c(\hx)e^{ib\hpi}]_W=e^{ib\cdot p}\,e^{-\frac12b\cD}c(x)$ one
obtains $(\delta\hQ)_W=-ie^{ib\cdot p}\,b^\mu\int_{-1/2}^{1/2}dt\,e^{tb\cD}\delta A_\mu$, which is
\eqref{eq:dQWsinglet}; this holds for an arbitrary non-Abelian field, and the exponentials span the class. (ii) For
$\hQ=\Sym(\hpi^\alpha)$ with $|\alpha|\le4$ in two dimensions, in an Abelian background with arbitrary $A(x)$
($F\ne0$) and arbitrary $\delta A(x)$, the operator $\frac d{d\varepsilon}\Sym[(\hpi-\varepsilon\,\delta
A)^\alpha]_{\varepsilon=0}$ was compared symbolically with the quantization \eqref{eq:split} of
\eqref{eq:dQWsinglet}, acting on a test function; the two agree.}

\begin{theorem}[\reviewfix{The current defined through $\log Z$}]\label{thm:JlogZ}
\reviewfix{For $\hQ$ of the class \eqref{eq:classQ} (gauge-singlet symbol independent of the gauge field) the
current \eqref{eq:JlogZdef} is
\begin{equation}
\langle J_\mu\rangle=-\trD\int\frac{d^4p}{(2\pi)^4}\sum_{k\ge0}\frac{(-1)^k}{4^k(2k+1)!}
\big[\textstyle\prod_{i=1}^{2k}\cD_{\mu_i}\big]\Big(G_W\big[\prod_{i=1}^{2k}\partial_{p_{\mu_i}}\big]\partial_{p_\mu}Q_W\Big).
\label{eq:JlogZ}
\end{equation}
This is the expectation value of the Noether series \eqref{eq:Jseries} with $\langle\rho_W\rangle=-G_W$. The two
definitions of the vector current coincide identically, to all orders of the derivative expansion and for an arbitrary
non-Abelian field; the leading term is \eqref{eq:Jvec}.}
\end{theorem}
\begin{proof}
\reviewfix{Insert \eqref{eq:dQWsinglet} into \eqref{eq:dlogZ}. Since $\partial_p^{2k+1}Q_W$ is a gauge singlet,
$\trD\trG[G_W(\cD^{2k}\delta A_\mu)\partial_p^{2k+1}Q_W]=\trG[(\cD^{2k}\delta A_\mu)\,\trD(G_W\partial_p^{2k+1}Q_W)]$.
The covariant derivatives are moved from $\delta A$ to the second factor with
$\trG[(\cD_\nu X)Y]=\partial_\nu\trG(XY)-\trG[X\cD_\nu Y]$; the boundary terms vanish for a localized $\delta A$, and
the $2k$ sign factors compensate.}
\end{proof}

\reviewfix{Three remarks. (a) Nothing has been truncated. The derivation uses the straight transporters only; the
triangular Wilson loop of the star product does not appear, because the trace formula \eqref{eq:trace} contains the
ordinary product of the symbols. The dependence of the current on the field strength is contained in $G_W$. (b) The
current \eqref{eq:JlogZ} is covariantly conserved, and this can be seen directly. For a $p$-independent adjoint
function $\alpha(x)$ the star product is $\alpha\star Q_W=\sum_n\frac{(i/2)^n}{n!}(\cD_{\nu_1}\cdots\cD_{\nu_n}\alpha)
\partial_{p_{\nu_1}}\cdots\partial_{p_{\nu_n}}Q_W$ and $Q_W\star\alpha=\sum_n\frac{(-i/2)^n}{n!}\partial_{p_{\nu_1}}
\cdots\partial_{p_{\nu_n}}Q_W(\cD_{\nu_1}\cdots\cD_{\nu_n}\alpha)$. For the infinitesimal gauge transformation
\eqref{eq:dAgauge}, $\delta A_\mu=\cD_\mu\alpha$, the series \eqref{eq:dQW} is therefore
\begin{equation}
(\delta\hQ)_W\big|_{\delta A=\cD\alpha}=i\big(\alpha\star Q_W-Q_W\star\alpha\big)-i[\alpha,Q_W],
\label{eq:dQgauge}
\end{equation}
which expresses the gauge covariance of the quantization map, $\Omega Q_W\Omega^\dagger$ quantized in the field
$A^\Omega$ being $\Omega(\hx)\hQ\,\Omega^\dagger(\hx)$: to first order in $\alpha$, the quantization of
$Q_W+i[\alpha,Q_W]$ in the field $A+\cD\alpha$ is $\hQ+i[\alpha(\hx),\hQ]$. For a singlet symbol the last term of \eqref{eq:dQgauge} is absent,
$\delta\log Z=i\,\Tr(\hG[\alpha(\hx),\hQ])=0$, and $\cD_\mu\langle J_\mu\rangle=0$ exactly. (c) The definition
\eqref{eq:JlogZdef} requires a gauge field that couples to the current. It is not available for the axial current
unless an axial background field is introduced, and the way in which such a field enters a higher-order $\hQ$ is a
part of the regularization (consistent versus covariant current); for the axial current we use the Noether form
\eqref{eq:Jax} and the covariant point splitting of Sec.~\ref{sec:pointsplit}.}

\paragraph{\reviewfix{When the leading term is sufficient.}}
\reviewfix{The series \eqref{eq:JlogZ} can be summed. Since $\sum_{k\ge0}\frac{(-1)^kz^{2k}}{4^k(2k+1)!}=
\int_{-1/2}^{1/2}dt\,e^{itz}$,
\begin{align}
\langle J_\mu(x)\rangle&=-\int_{-1/2}^{1/2}dt\;\trD\int\frac{d^4p}{(2\pi)^4}\,e^{it\,\cD\cdot\partial^{(Q)}_p}\Big(G_W\,\partial_{p_\mu}Q_W\Big)\notag\\
&=-i\int d^4y\,y_\mu\int_{-1/2}^{1/2}dt\;U(x,x+ty)\,\trD\Big[\tilde G(x+ty,-y)\,\tilde Q(x+ty,y)\Big]\,U(x+ty,x),
\label{eq:Jsegment}
\end{align}
where $\partial^{(Q)}_p$ acts on $Q_W$ only, $\cD$ acts on the product, and $\tilde G(x,y)=\int\frac{d^4p}{(2\pi)^4}
e^{-ipy}G_W(x,p)$. The second form follows directly from Lemma~\ref{lem:dU}: for a singlet symbol the two terms of
the proof of Proposition~\ref{prop:dQW} combine into $Y_-+Y_+=-iy^\mu\int_{-1/2}^{1/2}dt\,e^{ty\cD}\delta A_\mu(x)$.
Its meaning is the following. The kernel $\tilde Q(x',y)$ connects the points $x'\mp\frac y2$, and this pair
contributes to the current uniformly along the straight segment that joins them, parallel-transported to the point of
observation. For a tight-binding operator $\tilde Q\propto\delta(y-b)$, and the current of the hop $b$ is distributed
along the link. The leading term \eqref{eq:Jvec} is the value $t=0$ of the integrand: it places the whole contribution
at the midpoint. In terms of the moments
\begin{align}
C^{(k)}_{\mu\nu_1\dots\nu_{2k}}(x)&=\trD\int\frac{d^4p}{(2\pi)^4}\,G_W\,\partial_{p_\mu}\partial_{p_{\nu_1}}\cdots\partial_{p_{\nu_{2k}}}Q_W,\notag\\
\langle J_\mu\rangle&=-C^{(0)}_\mu+\frac1{24}\,\cD_\nu\cD_\lambda C^{(1)}_{\mu\nu\lambda}-\frac1{1920}\,\cD^4C^{(2)}_\mu+\dots
\label{eq:Jcorr}
\end{align}
(in the last term the four covariant derivatives are contracted with the four additional indices of $C^{(2)}$).}

\reviewfix{The expression \eqref{eq:Jvec}, $\langle J_\mu\rangle=-C^{(0)}_\mu$, is therefore exact in three cases, without
any condition on the strength of the field or on the gauge potential. (i) $\partial^3_pQ_W=0$: operators at most
quadratic in the momenta (Dirac and Wilson-type operators, non-relativistic fermions); for a polynomial $Q_W$ of
degree $m$ the series terminates at $k\le(m-1)/2$. (ii) Covariantly constant moments, $\cD C^{(k)}=0$. An example is
a homogeneous $Q_W$ in a covariantly constant field, $\cD F=0$, of arbitrary strength: then $[F_{\mu\nu},
F_{\rho\sigma}]=0$, in the gauge $A_\mu=-\frac12F_{\mu\nu}x_\nu$ the operator $\hQ$ commutes with the magnetic
translations, and $\cD G_W=0$. This is the case of the Landau levels. (iii) Volume-averaged currents after the gauge
trace (or after the contraction with a covariantly constant weight), with periodic boundary conditions or sufficient
decay: the terms with $k\ge1$ are total derivatives. This is the case used for the coefficients of the CSE, of the
QHE and of the chiral magnetic like effect.}

\reviewfix{Case (iii) is seen directly from the first line of \eqref{eq:Jsegment}. Let $n(x)$ be a covariantly
constant weight, $\cD_\mu n=0$ ($n=1$ for the gauge trace; a generator $t^a$ with $[A_\mu,t^a]=0$, e.g.\ for a field
directed along one generator or lying in the Cartan subalgebra, in the gauge where $A$ lies there as well). In the
expansion of the exponential every term with $m\ge1$ is of the form $\cD_{\nu_1}\cdots\cD_{\nu_m}X$, and
\[
\int d^4x\,\trG\big[n\,\cD_\nu X\big]=\int d^4x\,\partial_\nu\trG(nX)-\int d^4x\,\trG\big[(\cD_\nu n)X\big]=0
\]
with periodic boundary conditions or sufficient decay. The exponential is replaced by unity, the integral over $t$
gives $1$, and
\begin{equation}
\bar J_\mu:=\frac1V\int d^4x\,\trG\big[n\,\langle J_\mu\rangle\big]
=-\frac1V\int d^4x\,\trG\Big[n\,\trD\int\frac{d^4p}{(2\pi)^4}\,G_W\,\partial_{p_\mu}Q_W\Big]
\label{eq:Javexact}
\end{equation}
exactly, to all orders of the derivative expansion and for any strength of the field. The same holds for the axial
current. Its Noether series has the same structure: the variation of the action under the chiral rotation is
$\delta S=i\bar\psi[\hQ,\alpha]\gamma^5\psi$, the vector expression with $\gamma^5$ inserted, so that for a singlet
symbol the even orders cancel and the odd ones give \eqref{eq:Jsegment} with $\partial_{p_\mu}Q_W\to
\partial_{p_\mu}Q_W\gamma^5=\frac12\partial_{p_\mu}[Q_W,\gamma^5]$; hence
\begin{equation}
\bar J^{(5)}_\mu=-\frac1{2V}\int d^4x\,\trG\Big[n\,\trD\int\frac{d^4p}{(2\pi)^4}\,G_W\,\partial_{p_\mu}[Q_W,\gamma^5]\Big]
\label{eq:Jaxavexact}
\end{equation}
exactly. For an adjoint component with a weight that is not covariantly constant the statement fails:
$\int\trG(t^a\cD X)=-i\int\trG([t^a,A]X)\ne0$ in general. Equations \eqref{eq:Javexact}, \eqref{eq:Jaxavexact} remove
the truncation of the current. They do not remove the expansion of $G_W$ in the field: the exact $G_W$ enters them.}

\reviewfix{For the local current in an inhomogeneous situation the condition is
\begin{equation}
\frac1{24}\Big|\cD_\nu\cD_\lambda C^{(1)}_{\mu\nu\lambda}\Big|\ll\Big|C^{(0)}_\mu\Big|,
\qquad\text{i.e.}\qquad \frac1{24}\Big(\frac\ell L\Big)^2\ll1 ,
\label{eq:slowcond}
\end{equation}
and similarly for the higher terms. Here $L$ is the length over which the gauge-covariant quantities vary: the field
strength (through $\cD F$) and the background (through $\partial_xQ_W$). The length $\ell$ is the range of the
kernel of $\hQ$, $\ell^2\sim|\partial_p^3Q_W|/|\partial_pQ_W|$ at the momenta that dominate the integral: the hopping
range (the lattice spacing $a$) for a tight-binding operator, and of the order of the inverse momentum at the Fermi
surface or at the Fermi point for a polynomial $Q_W$. The condition \eqref{eq:slowcond} states that the covariant
quantities change little over the segment of \eqref{eq:Jsegment}. It restricts the gradients $\cD F$ and
$\partial_xQ_W$; it does not restrict the magnitude of $F$, and it does not involve the potential $A$. For lattice
models it is the requirement that the fields vary slowly on the scale of the lattice spacing, and the corrections
vanish as $(a/L)^2$ in the continuum limit. In the linear response of a homogeneous system the correction is
proportional to $\cD\cD F$; for a field with the wave vector $q$ its relative size is of the order of
$(q\ell)^2/24$.}

\reviewfix{This truncation is to be distinguished from the expansion \eqref{eq:Gexp} of $G_W$ itself in the field,
which has its own condition (a weak and slowly varying field strength) and is needed for the transport coefficients
also when \eqref{eq:Jvec} is exact. At the order $\cD^2F$ the higher terms of \eqref{eq:Gexp} contribute to the local
current together with the $k=1$ term of \eqref{eq:Jcorr}.}

\paragraph{\reviewfix{Comparison of the two currents for operators outside the class.}}
\reviewfix{For other orderings of the covariant derivatives, or in the presence of Pauli terms, the symbol depends
on the gauge field through $F_{\mu\nu}$ and its covariant derivatives and transforms in the adjoint representation,
$Q_W\to\Omega Q_W\Omega^\dagger$ (an example is $[\hpi_\mu\hpi_\nu]_W=p_\mu p_\nu+\frac i2F_{\mu\nu}$,
Sec.~\ref{sec:class}). Then \eqref{eq:dlogZ} contains the variation of the quantization map at fixed symbol and the
variation of the symbol,
\begin{equation}
\langle J_\mu\rangle=J^{(1)}_\mu+J^{(2)}_\mu,\qquad
\trG\int d^4x\,\delta A_\mu J^{(2)}_\mu=\int\frac{d^4x\,d^4p}{(2\pi)^4}\,\trD\trG\big[G_W\,\delta Q_W\big],
\label{eq:J12}
\end{equation}
where, from \eqref{eq:dQW},
\begin{equation}
J^{(1)}_\mu=-\frac12\trD\int\frac{d^4p}{(2\pi)^4}\sum_{n\ge0}\frac{(-i/2)^n}{(n+1)!}\,\cD_{\nu_1}\cdots\cD_{\nu_n}
\Big[\partial_{p_\mu}\partial^n_{p_\nu}Q_W\,G_W+(-1)^nG_W\,\partial_{p_\mu}\partial^n_{p_\nu}Q_W\Big],
\label{eq:J1}
\end{equation}
with $\partial^n_{p_\nu}=\partial_{p_{\nu_1}}\cdots\partial_{p_{\nu_n}}$. The term $J^{(1)}$ is the current that the
Noether procedure of Sec.~\ref{sec:noether} produces: keeping the order of the factors in the expansion of
$\Gamma=i\,\trD\int\frac{d^4p}{(2\pi)^4}(Q_W\star\rho_W-\rho_W\star Q_W)$ one finds
\begin{equation}
\Gamma=\cD_\mu J^{(1)}_\mu[\rho]+i\,\trD\int\frac{d^4p}{(2\pi)^4}\,[Q_W,\rho_W],
\label{eq:GammaGeneral}
\end{equation}
where $J^{(1)}_\mu[\rho]$ is \eqref{eq:J1} with $G_W\to-\rho_W$; for a singlet symbol the commutator vanishes under
$\trD$ and \eqref{eq:GammaGeneral} reduces to \eqref{eq:Jseries}. Since $\langle\Gamma\rangle=-i\,\trD\int
\frac{d^4p}{(2\pi)^4}(Q_W\star G_W-G_W\star Q_W)=0$ identically,
\begin{equation}
\cD_\mu J^{(1)}_\mu=i\,\trD\int\frac{d^4p}{(2\pi)^4}\,[Q_W,G_W].
\label{eq:divJ1}
\end{equation}
The same follows from \eqref{eq:dQgauge}. On the other hand, for $\delta A_\mu=\cD_\mu\alpha$ the covariance of the
symbol gives $\delta Q_W=i[\alpha,Q_W]$, and \eqref{eq:J12} yields
\begin{equation}
\cD_\mu\big(\langle J_\mu\rangle-J^{(1)}_\mu\big)=\cD_\mu J^{(2)}_\mu=-i\,\trD\int\frac{d^4p}{(2\pi)^4}\,[Q_W,G_W].
\label{eq:divJ2}
\end{equation}
Equations \eqref{eq:divJ1} and \eqref{eq:divJ2} give $\cD_\mu\langle J_\mu\rangle=0$: the current defined through
$\log Z$ is covariantly conserved for any gauge-covariant $\hQ$. The covariant divergence of the difference of the two
currents is the commutator term \eqref{eq:divJ2}, and three cases are to be distinguished.}

\reviewfix{(i) Operators of the class \eqref{eq:classQ}: $J^{(2)}=0$ and the two currents coincide
(Theorem~\ref{thm:JlogZ}). (ii) Gauge-singlet symbols that depend on the field strength; this includes every ordering
and every Pauli term for an Abelian field. The commutator in \eqref{eq:divJ2} vanishes, both currents are conserved,
and they differ by the identically conserved current $J^{(2)}$. For a linear dependence,
$Q_W=Q^0_W+\frac12S_{\mu\nu}(x,p)F_{\mu\nu}$ with $S_{\mu\nu}=-S_{\nu\mu}$, one has $\delta Q_W=S_{\mu\nu}\cD_\mu\delta
A_\nu$ and
\begin{equation}
J^{(2)}_\mu=\cD_\nu M_{\mu\nu},\qquad M_{\mu\nu}=\trD\int\frac{d^4p}{(2\pi)^4}\,S_{\mu\nu}\,G_W=-M_{\nu\mu},
\label{eq:Jmagn}
\end{equation}
a magnetization current, with $\partial_\mu\partial_\nu M_{\mu\nu}=0$ in the Abelian case. (iii) Non-Abelian field
and a symbol that is not a gauge singlet. For \eqref{eq:Jmagn} with a singlet $S_{\mu\nu}$,
$\cD_\mu\cD_\nu M_{\mu\nu}=\frac12[\cD_\mu,\cD_\nu]M_{\mu\nu}=-\frac i2[F_{\mu\nu},M_{\mu\nu}]$, in agreement with
\eqref{eq:divJ2}; this does not vanish in general. The Noether expression $J^{(1)}$ alone is then not covariantly
conserved, and only the sum $J^{(1)}+J^{(2)}$, the current defined through $\log Z$, is.}

\reviewfix{In all cases $J^{(2)}$ is a total covariant derivative of a local expression. After the gauge trace with
a covariantly constant weight it does not contribute to the volume-averaged responses, so that the results of
Secs.~\ref{sec:CSE}--\ref{sec:QHE} do not depend on the choice between the two definitions; the local current density,
the magnetization and the edge currents do. The restriction on untraced adjoint components stated in
Sec.~\ref{sec:noether} applies to both.}

%======================================================================
\section{The generalized Groenewold equation and its iterative solution}
\label{sec:groenewold}

The symbol of $\hG=\hQ^{-1}$ satisfies the generalized Groenewold equation $Q_W\star G_W=1$ (Appendix~\ref{app:ww}). Writing the star product in the form \eqref{eq:stardiff},
separating the $x$-dependence of $Q_W$ (background) from the dependence on the gauge field (introduced through an
auxiliary argument $z$ on which the covariant derivatives act) and expanding in powers of $\cD_z$ with
\eqref{eq:Wexp} [Ref.~\cite{XZPRD}, Eqs.~(116)--(121)], one finds order by order [Ref.~\cite{XZPRD}, Eqs.~(122)--(126)]:
\begin{align}
&\text{order }0:\quad Q_W\star_MG^{(0)}=1,\qquad \star_M=e^{\frac i2(\overleftarrow\partial_x\overrightarrow\partial_p-\overleftarrow\partial_p\overrightarrow\partial_x)};\\
&\text{order }1:\quad G^{(1)}=0;\\
&\text{order }2:\quad 0=Q_W\star_MG^{(2)}+\frac i2\,\partial_{p_\mu}Q_W\star_M\partial_{p_\nu}G^{(0)}\,F_{\mu\nu},\\
&\phantom{\text{order }2:}\quad G^{(2)}=-\frac i2\,G^{(0)}\star_M\partial_{p_\mu}Q_W\star_M\partial_{p_\nu}G^{(0)}\,F_{\mu\nu}.
\end{align}
Hence
\begin{equation}
G_W=G^{(0)}+\Gamma_{\mu\nu}F_{\mu\nu}+O(\cD^3),\qquad
\Gamma_{\mu\nu}=-\frac i4\,G^{(0)}\star_M\partial_{p_{[\mu}}Q_W\star_M\partial_{p_{\nu]}}G^{(0)} .
\label{eq:Gexp}
\end{equation}
\begin{GPTcorrection}
The order-one term in the external-field expansion vanishes because a single external covariant derivative
produces no gauge-covariant field strength for singlet zeroth-order symbols. Ordinary background $x$-derivatives
are already included in $\star_M$ and need not vanish. The order-two correction comes from the first curvature
term in \eqref{eq:Wexp}. This is an asymptotic derivative expansion: small external-field gradients and
appropriate bounds on the symbol derivatives and inverse are needed to control its remainder.
\end{GPTcorrection}

Equation \eqref{eq:Gexp} is the input of all transport results; it is where the derivative expansion in the field
strength is made (background inhomogeneity is treated exactly, through $\star_M$). For $Q_W=\tilde Q(p)$ independent
of $x$ one has $\tilde G_W=\tilde G^{(0)}+\frac i2\tilde G^{(0)}\partial_{p_\alpha}\tilde Q\tilde G^{(0)}\partial_{p_\beta}
\tilde Q\tilde G^{(0)}F_{\alpha\beta}+\reviewfix{O(\cD F,F^2)}$ [Ref.~\cite{XZPLB}, Eq.~(65)].

%======================================================================
\section{The non-Abelian chiral separation effect}
\label{sec:CSE}

\subsection{Derivation}
\label{sec:CSEder}

The Pauli--Villars regularization replaces $\Omega=-T\log Z$ by $\Omega=-T\sum_iC_i\log Z(M_i)$ with regulator
operators $Q_{(M_i)}$ of mass $M_i$ and $\sum_iC_i=0$, $C_0=1$ (and the further conditions needed for the cancellation
of all divergences) [Ref.~\cite{XZPRD}, Eqs.~(35), (127)]. \reviewfix{For the Hermitian massless Dirac symbol $Q=\gamma\cdot p$ of our conventions a gapped regulator is
$Q_{(M_i)}=Q+iM_i$ (the usual $i\gamma\cdot p+M_i$ up to an overall phase); a real mass added to $\gamma\cdot p$
would not gap it, since $\gamma\cdot p$ has the eigenvalues $\pm|p|$.} The regularized axial current is [Ref.~\cite{XZPRD}, Eq.~(128)]
\begin{equation}
J^{(5)a}_\mu=-\frac12\sum_iC_i\int\frac{d^4p}{(2\pi)^4}\trD\trG\big(G_{(M_i),W}\,\partial_{p_\mu}[Q_{(M_i)},t^a\gamma^5]\big).
\end{equation}
Inserting \eqref{eq:Gexp} [Ref.~\cite{XZPRD}, Eqs.~(129)--(132)],
\begin{align}
J^{(5)a}_\mu&=S_\mu\,\trG t^a+M_{\mu\alpha\beta}\,\trG(t^aF_{\alpha\beta}),\\
M_{\mu\alpha\beta}&=-\frac i8\sum_iC_i\int\frac{d^4p}{(2\pi)^4}\trD\big([\gamma^5,\partial_{p_\mu}Q]\,G^{(0)}\star_M\partial_{p_{[\alpha}}Q\star_M\partial_{p_{\beta]}}G^{(0)}\big).
\end{align}
The CSE conductivity is the response to both the colour-magnetic field and the chemical potential,
\corr{$\partial J^{\rm CSE}_i/\partial\mu=(\partial M_{ijk}/\partial\mu)\epsilon_{jkl}B_l$} [Ref.~\cite{XZPRD}, Eq.~(133)]. The chemical potential enters as the
shift $ip_4\to ip_4+\mu$, so \corr{$\partial/\partial\mu$ acts as $-i\partial/\partial p_4$} under the integral, and the
$p_4$-integral of the total derivative reduces to the hyperplanes $p_4=\pm\epsilon$ and $p_4=\pm\infty$
[Ref.~\cite{XZPRD}, Eqs.~(134)--(135)]:
\begin{equation}
\frac{\partial}{\partial\mu}M_{ijk}=\corrB{+\frac18}\sum_iC_i\sum_{p_4=\pm\epsilon,\pm\infty}\int_{\Sigma_{p_4}}\frac{d^3p}{(2\pi)^4}
\trD\big([\gamma^5,\partial_{p_i}Q]\,G^{(0)}\star_M\partial_{p_{[j}}Q\star_M\partial_{p_{k]}}G^{(0)}\big).
\end{equation}
\corrB{The four hyperplanes are oriented as in the Conventions: $+dp_1\wedge dp_2\wedge dp_3$ on $p_4=+\epsilon$ and on
$p_4=-\infty$, its negative on $p_4=-\epsilon$ and on $p_4=+\infty$; the integral of $\partial_{p_4}$ over $|p_4|>\epsilon$
produces the lower end of each interval with a minus sign, which together with $(-i)(-\tfrac i8)=-\tfrac18$ gives the
sign $+\tfrac18$ of the oriented sum.\footnote{Ref.~\cite{XZPRD}, Eq.~(135) carries the factor $-\frac18$, the
orientation of the hyperplanes being left implicit there.}}
\corr{With a common chirally symmetric ultraviolet asymptotic symbol for the physical field and regulators, their boundary contributions at $p_4\to\pm\infty$ cancel by $\sum_iC_i=0$;}
\reviewfix{the regulators do not contribute to the $\epsilon\to0$ slice difference provided they remain
gapped at the chosen $\mu$ (in particular $|\mu|<|M_i|$ in the elementary relativistic example);} what remains is the massless field at
$p_4=\pm\epsilon$. Using chirality near the Fermi surface ($\gamma^5$ anticommutes with $Q$ and $G^{(0)}$ there),
$\trD([\gamma^5,\partial_iQ]G\star\partial_{[j}Q\star\partial_{k]}G)=2\trD(\gamma^5\partial_iQ\,G\star\partial_{[j}Q\star
\partial_{k]}G)$ and the antisymmetrization in $i,j,k$ turns the integrand into the 3-form of \eqref{eq:N3hom} with
Moyal products. The result is [Ref.~\cite{XZPRD}, Eqs.~(136)--(138)]
\begin{equation}
\frac{\partial}{\partial\mu}\langle J^{(5)}_i\rangle=\frac{N_3}{2\pi^2}B_i,\qquad
N_3=-\frac{1}{48\pi^2V}\sum_{p_4=\pm\epsilon}\int d^3x\int_{\Sigma_{p_4}}\trD\big(G^{(0)}\star_MdQ\star_M\wedge dG^{(0)}\star_M\wedge dQ\,\gamma^5\big),
\label{eq:CSE}
\end{equation}
identical in form to the Abelian result \cite{SuleymanovZubkov,ZA}, with $F\to F^a$ and the colour trace
$\trG(t^aF)$ supplying the index $a$ of the adjoint current. \reviewfix{The reduction assumes a static background, $T=0$ (continuous $p_4$; at finite $T$ the Matsubara sum is not
a total derivative) and linear response to the slowly varying field.}
\reviewfix{The precise statement, with the conditions of Sec.~\ref{sec:logZ}, is given at the end of this subsection in
two versions (paragraph ``The two versions of the non-Abelian CSE'').} Here $\frac1V\int d^3x$ is the \emph{trace per
unit volume}, and $\langle J^{(5)}_i\rangle$ on the left is the same trace per unit volume of the current. It is defined
by the periodization construction of \cite{KMMZZ}: space is divided into supercells of volume $V$, the symbol $Q_W$
(and with it $G_W$ and the background fields) is taken periodic under translations from one supercell to another, the
$x$-integration is restricted to one supercell and divided by its volume, and the limit of infinite supercell volume
is taken at the end. For an $x$-independent symbol this is the integral over the whole volume divided by the volume,
and \eqref{eq:CSE} reduces to the homogeneous result; for an $x$-dependent symbol the $x$-integral of the Moyal
expression is what makes $N_3$ a topological invariant (Sec.~\ref{sec:stokes}). The same trace per unit volume is
understood in \eqref{eq:N3T}, \eqref{eq:QHE2}, \eqref{eq:PLB13} and in the lattice form of the invariants
(Sec.~\ref{sec:NCG}, Appendix~\ref{app:ncg}). The adjoint components $J^a$, $F^a$ are not gauge covariant when
integrated; the relation is written for a colour-magnetic field $F^a$ that is constant over the supercell in the
chosen gauge, while $N_3$ itself is gauge invariant.

\paragraph{\reviewfix{The two versions of the non-Abelian CSE.}}
\reviewfix{The result \eqref{eq:CSE} admits two formulations. They differ in the component of the current that is
considered and in the approximations that are made; the conditions are those of Sec.~\ref{sec:logZ} (paragraph ``When
the leading term is sufficient'').}

\reviewfix{\emph{Version I: truncated current, arbitrary generator.} The current is represented by the leading term
\eqref{eq:Jax} of its series, and $G_W$ by \eqref{eq:Gexp}. This gives a local relation between colour matrices,
$J^{(5)}_\mu(x)=S_\mu+M_{\mu\alpha\beta}(x)F_{\alpha\beta}(x)$, and for every generator $t^a$
\begin{equation}
J^{(5)a}_i(x)=\trG\big(t^aJ^{(5)}_i(x)\big),\qquad
\frac{\partial}{\partial\mu}\bar J^{(5)a}_i=\frac{N_3}{2\pi^2}\,B^a_i,\qquad
B^a_i=\frac12\epsilon_{ijk}\trG\big(t^aF_{jk}\big),
\label{eq:CSEI}
\end{equation}
the bar denoting the trace per unit volume. No condition is imposed on $t^a$, because no integration by parts in $x$
is used. The conditions are: (a) linear response in $F$; (b) for the local relation, the slow variation
\eqref{eq:slowcond}, $(\ell/L)^2/24\ll1$, or one of the cases (i), (ii) of Sec.~\ref{sec:logZ}, in which the leading
term is exact locally; (c) for the averaged relation \eqref{eq:CSEI}, a covariantly constant field, $\cD F=0$, with
$B^a_i$ constant over the supercell in the chosen gauge. Under (c) the omitted terms of the series of the current are,
at the linear order in $F$, of the form $(\partial^{2k}c(x))\,\trG(t^aF)$ with a gauge-singlet coefficient $c(x)$, and
average to zero; the terms of $G_W$ with $\cD F$ are absent. The components $\bar J^{(5)a}$ and $B^a$ refer to the
chosen gauge and are not gauge invariant.}

\reviewfix{\emph{Version II: exact current, covariantly constant weight.} Let $n(x)$ be a covariantly constant colour
matrix, $\cD_\mu n=0$. Such a weight is $n=1$ for an Abelian field or for the gauge-singlet current; $n=t^a$ with
$[A_\mu,t^a]=0$ for a field directed along the generator $t^a$ (or lying in the Cartan subalgebra) in the gauge where
$A_\mu$ lies in the same subalgebra, and $n(x)=\Omega(x)t^a\Omega^\dagger(x)$ in a general gauge; for a covariantly
constant field the colour direction of $F$ itself is such a weight. By \eqref{eq:Jaxavexact} the averaged axial
current is given by the leading term exactly,
\begin{equation}
\bar J^{(5)}_i[n]=-\frac1{2V}\int d^4x\,\trG\Big[n\,\trD\int\frac{d^4p}{(2\pi)^4}\,G_W\,\partial_{p_i}[Q_W,\gamma^5]\Big],
\qquad\cD n=0,
\label{eq:CSEexact}
\end{equation}
with the regulator contributions understood, to all orders of the derivative expansion and for any strength of the
field: no truncation of the current is made. Here $G_W$ is the exact symbol. When its expansion in the field is
inserted, $\trG(n\,\cD^mF)=\partial^m\trG(nF)$, so that only the gauge-invariant field $\trG(nF)$ enters, and
\begin{equation}
\frac{\partial}{\partial\mu}\bar J^{(5)}_i[n]=\frac{N_3}{2\pi^2}\,B^{(n)}_i+O(F^2),\qquad
B^{(n)}_i=\frac12\epsilon_{ijk}\trG\big(nF_{jk}\big),
\label{eq:CSEII}
\end{equation}
for $B^{(n)}_i$ constant over the supercell. Both sides are gauge invariant. For a varying $B^{(n)}$ the terms with
its derivatives enter the average only through $\int c(x)\,\partial^mB^{(n)}=(-1)^m\int(\partial^mc)\,B^{(n)}$, that
is, only when the background is inhomogeneous as well; for a homogeneous background $B^{(n)}_i$ in \eqref{eq:CSEII} is
replaced by its average. The conductivity $\sigma_{\rm CSE}=N_3/2\pi^2$ is therefore exact as the coefficient of the
first power of the field, without gradient corrections. What is not covered is the nonlinear response, the terms of
$G_W$ of the order $F^2$ and higher. For free massless Dirac fermions they are known to be absent (only the lowest
Landau level contributes to the axial current), but the present derivation does not show this for a general $\hQ$.}

\reviewfix{\emph{Relation of the two versions.} For a covariantly constant field the linear response
$M_{\mu\alpha\beta}F_{\alpha\beta}$ is directed along $F$ in colour space, and this direction is covariantly constant.
The component with a nonzero response in Version~I is then one for which $n$ can be chosen, and the two versions
coincide at the linear order in the field. Version~II in addition is gauge invariant and gives, in
\eqref{eq:CSEexact}, the exact expression from which the nonlinear response is to be computed. If the field admits no
covariantly constant weight other than $n=1$, only Version~I is available for the adjoint components, as a local
relation under the condition \eqref{eq:slowcond}.}

\begin{figure}[htb]
\centering
\begin{tikzpicture}[scale=1.0,>=Stealth]
\fill[gray!12] (-2.75,-0.35) rectangle (2.75,0.35);
\draw[->] (-3.3,0)--(3.5,0) node[right]{$\vec p$};
\draw[->] (0,-3.3)--(0,3.4) node[above]{$p_4$};
\draw[thick] (0,0) circle (2.8);
\node at (2.25,2.25) {$S^3_\infty$};
\foreach \a in {30,110,200,290} {\draw[->,thick] (\a:2.8)--(\a:3.25);}
\draw[thick,dashed] (-2.78,0.35)--(2.78,0.35) node[right]{$\Sigma_{0^+}$};
\draw[thick,dashed] (-2.78,-0.35)--(2.78,-0.35) node[right]{$\Sigma_{0^-}$};
\foreach \u in {-1.9,-0.7,0.5,1.7} {\draw[->,thick] (\u,0.35)--(\u,-0.05);}
\foreach \u in {-1.3,-0.1,1.1,2.3} {\draw[->,thick] (\u,-0.35)--(\u,0.05);}
\fill (0,0) circle (2pt);
\fill (1.4,0) circle (1.6pt); \fill (-1.4,0) circle (1.6pt);
\node at (0,-0.95) {\footnotesize singularities of $G_W$ at $p_4=0$};
\node[cross out,draw,thick,inner sep=2pt] at (0,1.6) {};
\draw[gray,dotted,thick] (0.15,1.6)--(3.45,1.6);
\node[anchor=west] at (3.5,1.6) {\footnotesize $p_4=b$: singularity of $O=p_4-b+i\vec\sigma\!\cdot\!\vec p$};
\end{tikzpicture}
\caption{Momentum-space surfaces entering the invariant $N_3$. \reviewfix{For a Hamiltonian-form operator satisfying the smooth-resolvent symbol assumptions,
the singularities relevant to this construction lie in the hyperplane $p_4=0$} (dots: a Fermi point or a Fermi surface), \reviewfix{so the integral of a closed three-form over the sphere at infinity $S^3_\infty$ (outward normals)
is the \emph{negative} of the inward-oriented slice difference} $\Sigma(0^+)-\Sigma(0^-)$ of the integrals over the
hyperplanes $\Sigma_{0^\pm}=\{p_4=\pm\epsilon\}$ taken with the inward orientation of the slab between them (arrows);
\reviewfix{Sec.~\ref{sec:stokes}. The minus sign in the outward-sphere definition of $N_3$ compensates this
orientation reversal. For an operator whose singularity is displaced} from $p_4=0$ (cross) the hyperplane expression
misses it and only the sphere at infinity, or a surface enclosing all singularities, gives the invariant.}
\label{fig:slab}
\end{figure}
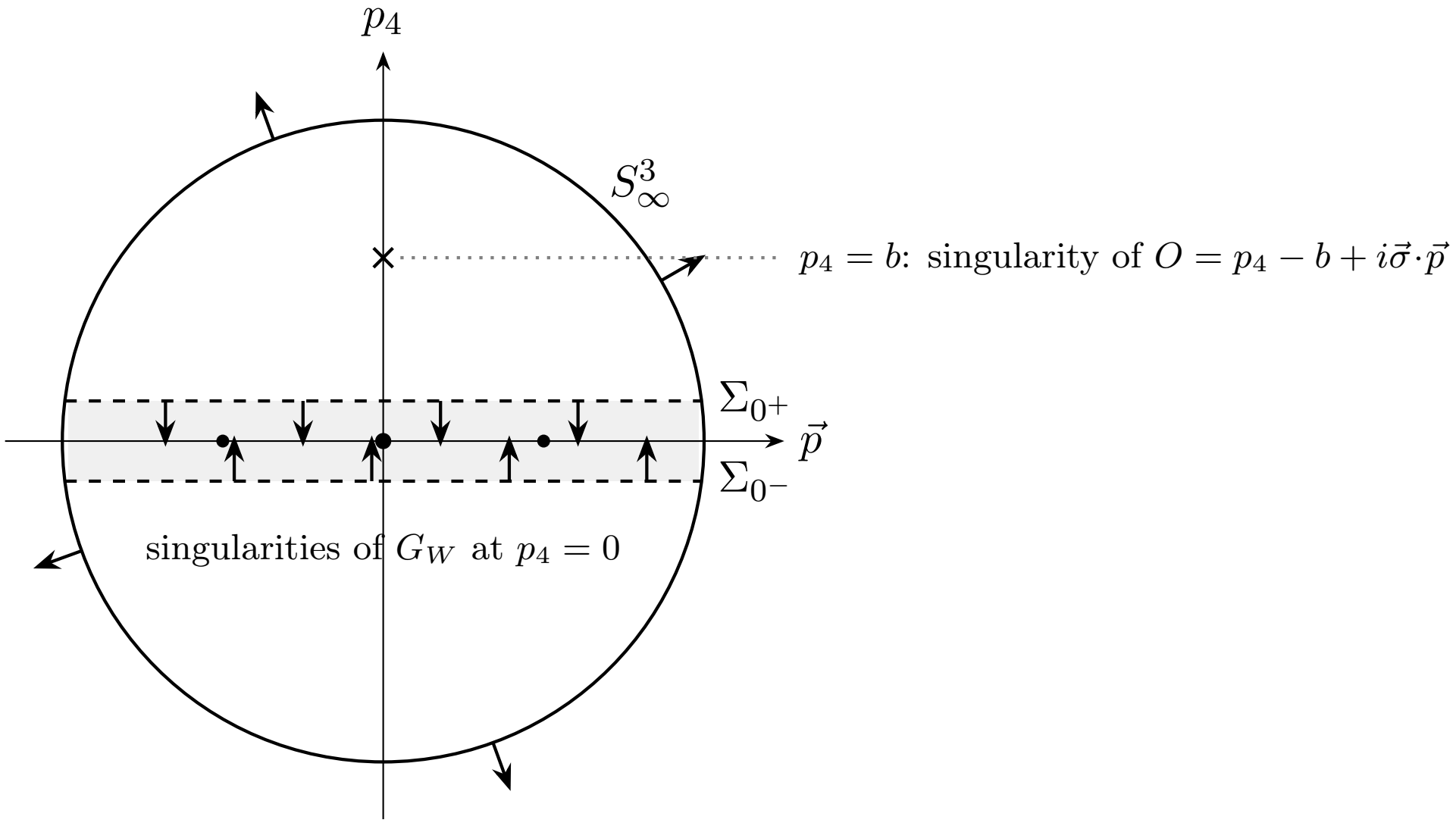

\begin{remark}[Hyperplanes versus the sphere at infinity]
The reduction to $p_4=\pm\epsilon$ assumes that all singularities of the integrand lie at $p_4=0$. \reviewfix{This is the case for Hamiltonian-form operators $Q=i\omega-H$, $H=H^\dagger$ (no pole at real
$\omega\ne0$), and a normalized positive spectral representation also excludes zeros of the canonical Green function
there (Appendix~\ref{app:manybody}). For an inhomogeneous system these are operator statements;
\corrB{what enters the slice integrals is the Moyal inverse $G_W=Q_W^{\star-1}$, the symbol of the operator inverse,
which exists for $\omega\ne0$ whenever the operator is invertible (pointwise invertibility of $Q_W$ is neither needed
nor implied); its smoothness and the trace estimates on the slices are part of the hypotheses.}
Spectral positivity is not automatic for a gauge-fixed quark propagator.} It need not hold for a general differential symbol: for $O=-iD_4-b+i\vec\sigma\cdot(-i\vec D)$ the singularity sits at
$p_4=b$. In general one should define $N_3$ by the surface enclosing all singularities, or by the sphere at infinity
(Sec.~\ref{sec:stokes}), which is what Theorem~1 of Ref.~\cite{XZPLB} produces (Fig.~\ref{fig:slab}).
\end{remark}

\subsection{Examples and prior work}

\begin{example}[Dirac fermions with background $U(1)$ field and vierbein, Ref.~\cite{XZPRD}, Eqs.~(139)--(141)]
$\hQ=\gamma^a\{|e|e^\mu_a,(\hp_\mu-A_\mu-C_\mu)\}/2-m|e|$. For constant background $Q(p)=\gamma^a|e|e^\mu_a(p_\mu-C_\mu)-m|e|$,\footnote{Ref.~\cite{XZPRD}, Eqs.~(140)--(141), writes the inverse with the denominator $\tilde m^2+l^2$, i.e.\
uses anti-Hermitian Euclidean $\gamma$-matrices there. In the conventions of this review ($\gamma$ Hermitian,
$(\gamma\cdot p)^2=p^2$) the gapped operator reads $\gamma^a|e|e^\mu_a(p_\mu-C_\mu)-im|e|$; with a real mass
$\gamma\cdot p-m$ would vanish at $|p|=|m|$.} $G^{(0)}=Q^{-1}$, and the integral \eqref{eq:CSE} is reduced to the standard one by deforming $e^\mu_a$ to the
\corr{unit matrix within its orientation-preserving component. In the massless case $m=0$, with no singularity crossing the integration surface, $N_3=1$ per Dirac fermion in this component and $N_3=n$ for $n$ such fermions. Reversing the vierbein orientation reverses the signed charge. A massive field does not satisfy the chiral assumptions of the CSE argument; a smooth background deformation preserves the invariant only while those assumptions and surface invertibility remain valid.}
\end{example}

\begin{example}[$N_f$ flavours with a dynamical $SU(N_c)$ field, Sec.~VI\,B\,2 of Ref.~\cite{XZPRD}]
\corr{With the interacting Green function and its inverse inserted into \eqref{eq:CSE}, $N_3=N_cN_f$ provided they are connected to the massless free theory without singularities crossing the defining surface.} The justification of
this replacement is the subject of \cite{ZA}, reviewed in Sec.~\ref{sec:CSEint}; it requires exact chiral symmetry
near the singularities and the renormalized axial current. The condition $NN_f>11N_c/2$ (\corrB{$N$ being the dimension of the external gauge multiplet of Sec.~\ref{sec:WW},
$G\subset U(N)$, so that} the dynamical $SU(N_c)$ field sees $NN_f$ Dirac fermions in its fundamental representation) \reviewfix{makes the one-loop coefficient $\beta_0=(11N_c-2NN_f)/3$ negative, so that
$\beta(g)=-\beta_0g^3/16\pi^2+O(g^5)$ drives a small coupling to zero in the infrared (a perturbative statement,
which is what makes the weak-coupling assumption consistent).}\footnote{Ref.~\cite{XZPRD} refers at this point to the
conformal-window analysis of its Ref.~[135]; the relevant property is the loss of asymptotic freedom.}
\end{example}

\paragraph{Prior work.} The non-Abelian counterparts of the chiral magnetic and chiral vortical effects for free
fermions were obtained in the non-Abelian chiral kinetic theory of Luo and Gao \cite{LuoGao} (see also
\cite{LuoMaGao}), based on the gauge-covariant Wigner function; the CSE counterpart follows in the same framework
although it is not stated there. Chiral conductivities of $SU(2)$ currents in multi-Weyl semimetals were derived from
the consistent anomaly by Dantas, Pe\~na-Benitez, Roy and Sur\'owka \cite{Dantas}. Lattice studies of the CSE in an
Abelian field with dynamical gluons \cite{PuhrBuividovich,BuividovichSU2,Brandt} find the free value at high density
and temperature and deviations in the confining phase. What is new in Ref.~\cite{XZPRD} is the expression of the conductivity as a
topological invariant of the (inhomogeneous, interacting) Green function for a general operator $\hQ$.

\subsection{Non-renormalization of $N_3$ by interactions}
\label{sec:CSEint}

Ref.~\cite{ZA} considers lattice-regularized fermions with a dynamical $SU(3)$ field (or Coulomb interactions),
external Abelian field $\mathcal A$, complete two-point function $\bG$ and $\bQ=\bG^{-1}$, and uses the approximate
lattice Wigner--Weyl calculus; the derivation below is its continuum form, which is exact, the compactness of the
Brillouin zone being replaced by the conditions at $|p|\to\infty$ of Sec.~\ref{sec:latcont}. The renormalized axial
current is taken as the definition,
\begin{equation}
j^5_k(x)=-\int\frac{d^Dp}{(2\pi)^D}\tr\big[\gamma^5\bG_W(x,p)\,\partial_{p_k}\bQ_W(x,p)\big],
\end{equation}
i.e.\ the bare velocity $\gamma_k$ is replaced by the renormalized velocity $\partial_{p_k}\bQ_W$ (at $T=\mu=0$,
$\partial_p\bQ\simeq Z_F\gamma$). \corr{The replacement of a chemical-potential derivative by a frequency derivative also requires that the relevant $\mu$ dependence be exhausted by the frequency shift, or that additional medium-dependent vertex terms be shown to cancel. A self-energy with explicit state dependence on $\mu$ need not satisfy this automatically.}
The Groenewold equation $\bQ_W\star\bG_W=1$ and the expansion
$\bQ_W=\bQ^{(0)}(p-\mathcal A)+\bQ^{(1)}_{ij}F_{ij}+\dots$, with $\bQ^{(1)}$ the field-dependent part of the
self-energy (vertex corrections), give
\begin{equation}
\corr{\bG^{(1)}=\left(\frac i2\bG^{(0)}\star\partial_{p_i}\bQ^{(0)}\star\bG^{(0)}\star\partial_{p_j}\bQ^{(0)}\star\bG^{(0)}-\bG^{(0)}\star\bQ^{(1)}_{ij}\star\bG^{(0)}\right)F_{ij}.}
\label{eq:G1}
\end{equation}
\reviewfix{After the cyclic spatial phase-space trace the two terms containing $\bQ^{(1)}$ reduce to a total
momentum derivative wherever $\gamma^5$ anticommutes with $\bQ$ and $\bG$ (exact chiral symmetry); for the surface
form of the invariant it suffices that this hold on the slices $\Sigma_{0^\pm}$ and that the remaining boundary terms
vanish (Sec.~\ref{sec:latcont}). The $\bQ^{(1)}$ terms then cancel,} and one arrives at \eqref{eq:CSE} with the complete Green function:
$\sigma_{\rm CSE}=\cN/2\pi^2$ with $\cN=N$ for $N$ chiral Dirac fermions and $\cN=N_cN_f$ for cold quark matter
\corr{for a nonsingular deformation of the massless free theory. Treating quarks with nonzero masses as chiral is an approximation, even if their masses are smaller than $\mu$.} \reviewfix{Assumptions: equilibrium; a controlled weak-field limit ($\lambda^2|F|\ll1$ with $\lambda$ the
correlation length where it is finite; at a gapless Fermi surface the infrared regulator and the order of limits
take its place);}
\corrB{} exact chiral symmetry near the singularities, no colour-superconducting gap, infrared finiteness. Not protected: the
\corr{current defined independently by a specified local axial operator (its vertex and medium corrections cannot generally be reduced to one wave-function factor; cf.\ \cite{Gorbar})}, \reviewfix{the broken-symmetry phase, and finite temperature}\corrB{, where the field-correlator estimate of \cite{ZA}
gives a suppression by $\sim2$ at $\mu,T\sim T_c$ and the topological value only for $\mu\gg m_D$}\reviewfix{ (a
model estimate, not a consequence of the topological argument).}

\begin{GPTcorrection}
\paragraph{When the interacting invariant equals the free one.}
The equality follows from a smooth deformation to the free Green function that keeps both the Green function
and its inverse nonsingular on the integration surface and preserves chirality and the ultraviolet boundary
conditions. In a continuum elliptic model one may instead use Sec.~\ref{sec:stokes}, provided its uniform
symbol estimates hold and the ultraviolet symbol is homotopic to the free one. A subleading self-energy, or
$\bQ\sim Z o$ with an invertible homotopically trivial $Z$, is an additional assumption, not a consequence of
interactions in general. Positive spectral representations exclude off-real-energy singularities for physical
fermion Green functions; positivity need not hold for gauge-fixed quarks. On a lattice the appropriate argument
uses the compact Brillouin zone and a fixed frequency regularization. These observations protect the stated
Green-function invariant under the specified deformation; they are not an Adler--Bardeen theorem for an
arbitrarily defined physical CSE current.
\end{GPTcorrection}

%======================================================================
\section{\reviewfix{Non-elliptic Fermi-surface models: the invariant, the chiral magnetic like effect and the chiral separation effect}}
\label{sec:CME}

\reviewfix{This section collects the non-elliptic Fermi-surface models of Ref.~\cite{ZubJETP}, the operators
$Q=ip_4X-Y$ whose frequency coefficient $X$ changes sign. Section~\ref{sec:FS} treats the basic model with plane
Fermi surfaces: its invariant $N_3^{(T)}$, computed by residues, its lattice regularization, and the equilibrium current
of the conserved charge along $B$ that the invariant implies -- a chiral magnetic like effect, which differs from the chiral magnetic
effect proper as explained at the end of Sec.~\ref{sec:FS}. Section~\ref{sec:CMEsphere}
treats the model with spherical Fermi surfaces, one of whose Fermi spheres shrinks to a Weyl-like Fermi point: its
chiral separation effect, the absence of the chiral magnetic like effect, and the static-gauge divergence of its
$\gamma^5$ current. Section~\ref{sec:CMEsphere2}
modifies that model so that the two effects exchange their roles -- a chiral magnetic like effect with spherical Fermi surfaces.
The static responses of these models in an electric field, and the reasons why they do not by themselves establish
an anomaly, are the subject of Sec.~\ref{sec:nonell}.}

\subsection{\reviewfix{The basic model with plane Fermi surfaces: general consideration and the chiral magnetic like effect}}
\label{sec:FS}

\reviewfix{Ref.~\cite{ZubJETP} replaces $\gamma^5$ by a symmetry generator $\hat T$ and normalizes the invariant with
the Hall-type prefactor,
\begin{equation}
N_3^{(T)}=-\frac{1}{24\pi^2V}\sum_{p_4=\pm\epsilon}\int d^3x\int_{\Sigma_{p_4}}\trD\big(\hat T\,G^{(0)}\star dQ\star\wedge
dG^{(0)}\star\wedge dQ\big),\qquad \frac{\partial\langle J^{(T)}\rangle_i}{\partial\mu}=\frac{N_3^{(T)}}{4\pi^2}B_i
\label{eq:N3T}
\end{equation}
[its Eqs.~(53)--(55)], the derivation being that of Sec.~\ref{sec:CSEder} for a generator that commutes with $Q$
and $G^{(0)}$ near the singularities, $\{\hat T,\partial_pQ\}=2\hat T\partial_pQ$ -- the case of $\hat T=1$ and of the
$\gamma^5$ of the $M=4$ models of \cite{ZubJETP}, where $\gamma^5$ commutes with the Hamiltonian; for the anticommuting
$\gamma^5$ of the Dirac convention of Sec.~\ref{sec:CSEder} the axial derivation given there applies instead. The two
normalizations are related by $N_3^{(\gamma^5)}=2N_3$ with $N_3$ of \eqref{eq:CSE}, so that both give
$\sigma_{\rm CSE}=N_3/2\pi^2$. The invariant controls the stability of
coordinate-dependent Fermi surfaces, and the paper constructs models in which it is nontrivial and computable from
Landau levels (the levels $E_n$ of the transverse kinetic energy $\pi_\perp^2/2m$ in the field $B>0$, each with
degeneracy $B/2\pi$ per unit area; Appendix~\ref{app:manybody}).} The basic model is
\begin{equation}
\hQ=i\kappa\hat p_4\hat p_3-\hat p_3-\frac{\hat\pi_\perp^2}{2m}+\mu_0,\qquad E_n=\frac Bm\Big(n+\frac12\Big),
\label{eq:FSmodel}
\end{equation}
\begin{GPTcorrection}
Here $B>0$, $m>0$, $\kappa>0$, the charge is one, and $\mu_0\ne E_n$. \reviewfix{The parameter $\mu_0$ fixes the positions of the Fermi sheets; it is a coupling of the model and must be distinguished from the chemical potential $\mu$ of \eqref{eq:N3T}, which enters through $ip_4\to ip_4+\mu$ (in the model: through the term $\kappa p_3\mu$) and is set to zero in \eqref{eq:FSmodel}. In Ref.~\cite{ZubJETP} this parameter is denoted $\mu$.} The singularities are at $p_4=0$,
$p_3=\mu_0-E_n$, with one sheet for every Landau level. The contour-selected count is
\[
n_{\rm filled}=\#\{n\ge0:E_n<\mu_0\}
=\max\!\left(0,\left\lfloor\frac{m\mu_0}{B}+\frac12\right\rfloor\right),\qquad \corrB{N_3^{(T)}=2n_{\rm filled}}.
\]
At a Landau threshold the slice is singular and this invariant is not defined. The selected count is
not the total number of sheets of this continuum model. \corrB{In the normalization \eqref{eq:N3T} of \cite{ZubJETP} (prefactor $1/24\pi^2$, $\hat T=1$) each slice
contributes $\pm n_{\rm filled}$ (Proposition~\ref{prop:FS}); in the normalization of \eqref{eq:CSE} (prefactor
$1/48\pi^2$, no $\gamma^5$ trace over two chiral blocks for this scalar symbol) the slices give $\pm\frac12n_{\rm
filled}$ and the invariant is $n_{\rm filled}$.}\footnote{In \cite{ZubJETP} the result is written with
the integer part of $(\mu_0-B/2m)/(B/m)$; read literally, this misses the lowest occupied level. The number of
Landau levels with $E_n<\mu_0$ is $n_{\rm filled}=\max(0,\lfloor m\mu_0/B+\frac12\rfloor)$ away from the thresholds,
equal to one just above $\mu_0=B/2m$.}
\end{GPTcorrection}

\corrB{Variants with $M=4$ ($\hat T=\gamma^5$, Pauli--Villars, $N_3^{(T)}=(4\lfloor\mu_0^2/2B\rfloor+2)\sgn\mu_0$
[Ref.~\cite{ZubJETP}, Eq.~(78)]: the lowest Dirac level counts once and every higher level twice, and the $\gamma^5$
trace over the two chiral blocks doubles the count relative to the scalar model) and with cylindrical or spherical
Fermi surfaces have the same structure $Q=ip_4X-Y$; the value for $M=4$ is taken from \cite{ZubJETP}, the spherical
model is taken up in Sec.~\ref{sec:CMEsphere}.}

\begin{figure}[!htb]
\centering
\begin{tikzpicture}[scale=1.0,>=Stealth]
% panel (a): Fermi sheets
\begin{scope}[xshift=-4.2cm]
\draw[->] (-2.6,0)--(2.6,0) node[right]{$p_3$};
\draw[->] (0,-1.3)--(0,1.2) node[above]{$\omega$};
\foreach \x/\n in {2.2/0,1.5/1,0.8/2} {
  \fill (\x,0) circle (1.6pt);
  \draw[thick] (\x-0.35,-0.5)--(\x+0.35,0.5);
  \draw[->,thick] (\x+0.2,0.28)--(\x+0.35,0.5);
  \node[below] at (\x,-0.1) {\scriptsize$\n$};
}
\foreach \x/\n in {-0.6/3,-1.3/4} {
  \draw (\x,0) circle (1.6pt);
  \draw[thick] (\x+0.35,-0.5)--(\x-0.35,0.5);
  \node[below] at (\x,-0.1) {\scriptsize$\n$};
}
\draw[dotted,thick] (0,-1.1)--(0,1.1);
\node[anchor=north] at (0,-1.3) {\scriptsize open sea, $\kappa p_3\to0$};
\node at (0,1.95) {(a) sheets $p_3=\mu_0-E_n$ (labels: $n$)};
\end{scope}
% panel (b): contour
\begin{scope}[xshift=3.2cm]
\draw[->] (-2.9,0)--(2.9,0) node[right]{$\mathrm{Re}\,z$};
\draw[->] (0,-1.9)--(0,1.9) node[right]{$\mathrm{Im}\,z$};
\foreach \x/\n in {-2.2/0,-1.4/1,-0.6/2,0.9/3,1.7/4} {
  \node[cross out,draw,inner sep=1.5pt] at (\x,0) {};
  \node[below] at (\x,-0.1) {\scriptsize$E_\n$};
}
\fill (0,0) circle (1.6pt); \node[below left] at (0,-0.1) {\scriptsize$\mu_0$};
\draw[thick,->] (1.0,-1.7) node[below right]{\scriptsize$z=\mu_0+(i\kappa p_4-1)p_3$} --(-1.0,1.7);
\draw[thick,dashed] (0,-1.8)--(0,1.8);
\draw[->,thick,gray] (-0.75,1.3) arc (120:92:1.5);
\draw[->,thick,gray] (0.75,-1.3) arc (300:272:1.5);
\draw[thick,dotted] (0,1.8) arc (90:270:1.8);
\node at (-2.15,1.25) {\scriptsize$n_{\rm filled}=3$};
\node at (0,2.65) {(b) slice integral as a contour integral};
\end{scope}
\end{tikzpicture}
\caption{The Fermi-surface model \eqref{eq:FSmodel}. (a) Each Landau level $n$ produces one sheet $p_3=\mu_0-E_n$ of
the Fermi surface; near it the spectrum is a chiral one-dimensional branch of slope \corrB{$+1/\kappa p_3$} (filled dots:
$E_n<\mu_0$, $p_3>0$; open dots: $E_n>\mu_0$). The sea is open at $p_3\to0$, where the coefficient $\kappa p_3$ of the
frequency vanishes. (b) On a slice $p_4={\rm const}$ the dependence on $p_3$ enters only through
$z=\mu_0+(i\kappa p_4-1)p_3$, so the slice integral is the integral of the two-dimensional Hall function $F(z)$ along
the straight line through $\mu_0$ (solid); rotating it about $\mu_0$ to the line $\mathrm{Re}\,z=\mu_0$ (dashed) without
crossing the poles $E_n$ (crosses) and closing the contour gives $\pm n_{\rm filled}$ for $p_4\gtrless0$; equivalently, the
$p_3$-integral done level by level by residues gives the Hall sum over occupied and empty Landau levels \corrB{in
units of the two-dimensional Hall number, i.e.\ $\Sigma^{(T)}(p_4)=\pm n_{\rm filled}$ in the normalization
\eqref{eq:N3T} and $\pm\frac12n_{\rm filled}$ in that of \eqref{eq:CSE}}
(Proposition~\ref{prop:FS}).}
\label{fig:contour}
\end{figure}
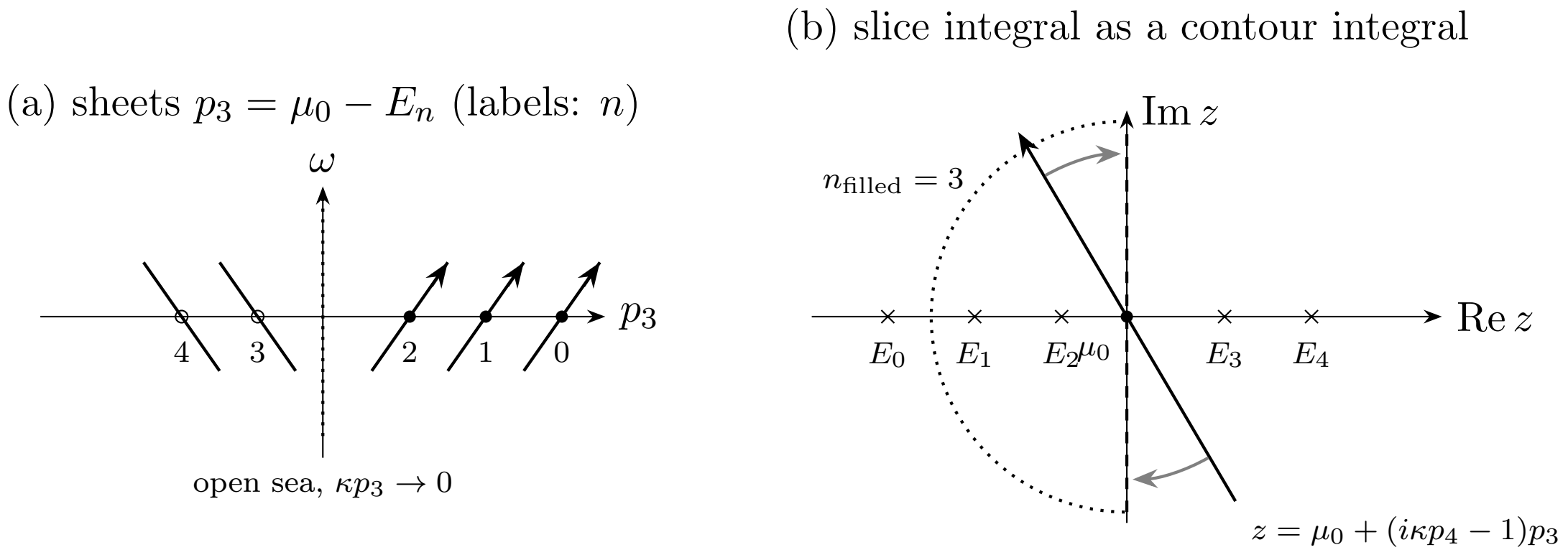

\begin{proposition}\label{prop:FS}
For \eqref{eq:FSmodel} the slice integrals of \eqref{eq:N3T} are $\Sigma^{(T)}(p_4)=\pm n_{\rm filled}$ for
$p_4\gtrless0$, so that $N_3^{(T)}=\Sigma^{(T)}(0^+)-\Sigma^{(T)}(0^-)=2n_{\rm filled}$ \corrB{(in the normalization of
\eqref{eq:CSE}: $\pm\tfrac12n_{\rm filled}$ and $n_{\rm filled}$)}, while the principal symbol, being non-invertible on
large spheres, has no degree.
\end{proposition}
\begin{proof}
\reviewfix{We follow Sec.~I of the Supplementary Material of Ref.~\cite{ZubJETP}, Eqs.~(S2)--(S7) there, and
compute the two slice integrals of \eqref{eq:N3T} directly, for $\hat T=1$. The three-form is integrated over the
hyperplane $p_4=\epsilon$ with the orientation $dp_1\wedge dp_2\wedge dp_3$,
\[
\Sigma(\epsilon)=-\frac{1}{24\pi^2V}\int d^3x\int d^3p\;\big[\trD\big(G\star dQ\star\wedge dG\star\wedge dQ\big)\big]_{123},
\]
and the sum over the two faces $p_4=\pm\epsilon$ in \eqref{eq:N3T}, taken with the orientations of the Conventions
(upper slice positive, lower slice negative), is
$N_3^{(T)}=\Sigma(\epsilon)-\Sigma(-\epsilon)$, as in \eqref{eq:slices}.

\emph{Step 1: the three-form on a slice.} The symbol of \eqref{eq:FSmodel} is gauge-field independent
(Sec.~\ref{sec:class}; $\pi_\perp^2$ is Weyl ordered), $Q=i\kappa p_4p_3-p_3-H+\mu_0$ with $H=p_\perp^2/2m$, and for
the constant field $B$ along $x_3$ the covariant star product acts only through $F_{12}$, i.e.\ only on the dependence
on $p_1,p_2$: $p_3$ and $p_4$ are c-numbers in all star products. On the slice put
\[
a=i\kappa\epsilon-1,\qquad Q=a\,p_3+\mu_0-H,\qquad G=(a\,p_3+\mu_0-H)^{-1}_\star ,
\]
the star inverse being taken in the transverse variables. With $dG=-G\star dQ\star G$ the three-form is
$-\epsilon_{ijk}\tr(G\,\partial_iQ\,G\,\partial_jQ\,G\,\partial_kQ)\,d^3p$ (stars understood), where $\partial_3Q=a$ is
a c-number and $\partial_bQ=-\partial_bH$ for $b=1,2$. Of the six terms of $\epsilon_{ijk}$ exactly one index is $3$;
pulling $a$ out and using the cyclicity of the transverse trace with the antisymmetry of $\epsilon_{bc}$, the three
positions of $a$ contribute equally (the identity holds under the transverse trace, by cyclicity, not pointwise
for the symbols), and
\begin{equation}
\big[\tr(G\star dQ\star\wedge dG\star\wedge dQ)\big]_{123}=-3a\,\epsilon^{bc}\,\tr\big(G\star\partial_bH\star G\star\partial_cH\star G\big).
\label{eq:FS3form}
\end{equation}

\emph{Step 2: Landau levels.} For a constant field the covariant Moyal algebra of the symbols depending on
$p_\perp$ is the operator algebra generated by $\hat\pi_1,\hat\pi_2$ with $[\hat\pi_1,\hat\pi_2]=iB$; the symbol
$\partial_bH=p_b/m$ corresponds to the velocity $v_b=\hat\pi_b/m=i[\hat H,\hat x_b]$ and $G$ to
$(a\,p_3+\mu_0-\hat H)^{-1}$, $\hat H=\hat\pi_\perp^2/2m$, $\hat H|n\rangle=E_n|n\rangle$, $E_n=\omega(n+\frac12)$,
$\omega=B/m$. The transverse trace per unit area of the symbols is the operator trace per unit area,
$\frac{1}{V_\perp}\int d^2x_\perp d^2p_\perp\,A_W=(2\pi)^2\,{\rm Tr}\,\hat A/V_\perp$, which for an operator that is a
function of $\hat\pi_\perp$ only -- degenerate in the guiding centre, $B/2\pi$ states per unit area and level -- is
$(2\pi)^2\frac{B}{2\pi}\sum_n\langle n|\hat A|n\rangle$. Inserting the Landau basis twice in \eqref{eq:FS3form} and
cancelling $\int dx_3$ against $V=V_\perp L_3$,
\begin{equation}
\Sigma(\epsilon)=\frac{B}{4\pi}\sum_{n,k\ge0}M_{nk}\int_{-\infty}^{\infty}dp_3\,
\frac{a}{(a\,p_3+\mu_0-E_n)^2\,(a\,p_3+\mu_0-E_k)},\qquad
M_{nk}=\epsilon^{bc}\langle n|v_b|k\rangle\langle k|v_c|n\rangle ,
\label{eq:FSsum}
\end{equation}
In the notation of Sec.~I of the Supplementary Material of \cite{ZubJETP} -- velocities written through
$[\hat H,\hat x_i]=-iv_i$ (for $\hat p=-i\partial$), levels $E^{\rm J}_n=E_n-\mu_0$ measured from $\mu_0$ (the chemical
potential of \cite{ZubJETP} is our $\mu_0$), and $\sum_{n,k}$ running over the full transverse basis per area $A$,
so that $\frac1A\sum_{\rm orbitals}=\frac B{2\pi}$ -- Eq.~\eqref{eq:FSsum} reads
\begin{equation}
\Sigma(\epsilon)=-(1-i\kappa\epsilon)\,\frac{\epsilon_{ij}}{2A}\sum_{n,k}\int dp_3\,
\frac{\langle n|[\hat H,\hat x_i]|k\rangle\langle k|[\hat H,\hat x_j]|n\rangle}
{(p_3+E^{\rm J}_n-i\epsilon\kappa p_3)^2\,(p_3+E^{\rm J}_k-i\epsilon\kappa p_3)}\,,
\label{eq:S3}
\end{equation}
which is Eq.~(S3) of \cite{ZubJETP} with the prefactor $-(1-i\kappa\epsilon)$ made explicit: it follows from
\eqref{eq:FSsum} with $a=-(1-i\kappa\epsilon)$ and $[\hat H,\hat x_i]=-iv_i$, the sign being that of $(-i)^2$ in
$\langle n|[\hat H,\hat x_i]|k\rangle\langle k|[\hat H,\hat x_j]|n\rangle=-\langle n|v_i|k\rangle\langle k|v_j|n\rangle$.
As printed, (S3) of \cite{ZubJETP} lacks this prefactor [the factor $1-i\kappa\epsilon\to1$ is kept there only for the
surfaces $p_4=\omega(p_3)$, Eq.~(S10)], and the passage from its (S5) to (S6) carries $+2\pi i\,{\rm sign}\,\delta$
where the residue theorem gives $-2\pi i\,{\rm sign}\,\delta$ (Step 3 below); the two signs compensate and (S7) is
correct. With $\hat\pi_1=\sqrt{B/2}\,(\hat a+\hat a^\dagger)$,
$\hat\pi_2=-i\sqrt{B/2}\,(\hat a-\hat a^\dagger)$ the only nonzero velocity matrix elements are
$\langle n|v_1|n{\pm}1\rangle$ and $\langle n|v_2|n{\pm}1\rangle$, and
\[
M_{n,n+1}=\frac{iB}{m^2}\,(n+1),\qquad M_{n,n-1}=-\frac{iB}{m^2}\,n,\qquad M_{kn}=-M_{nk},\qquad M_{nk}=0\ \text{otherwise}.
\]

\emph{Step 3: the $p_3$-integral.} The integrand of \eqref{eq:FSsum} is a rational function of $p_3$ decaying as
$p_3^{-3}$, with a double pole at $p_3^{(n)}=(E_n-\mu_0)/a$ and a simple pole at $p_3^{(k)}=(E_k-\mu_0)/a$. Since
$1/a=-(1+i\kappa\epsilon)/(1+\kappa^2\epsilon^2)$, ${\rm Im}\,p_3^{(n)}=-\kappa\epsilon(E_n-\mu_0)/(1+\kappa^2\epsilon^2)$:
for $\epsilon>0$ the pole of an occupied level, $E_n<\mu_0$, lies in the upper half plane and that of an empty level
in the lower one. For $\epsilon\to0^+$ the only trace of $\epsilon$ is the position of the poles,
$-i\epsilon\kappa p_3\to+i\epsilon\kappa E^{\rm J}_n$ at the pole of the level $n$, and \eqref{eq:S3} becomes
\begin{equation}
\Sigma(0^+)=-\frac{\epsilon_{ij}}{2A}\sum_{n,k}\int dp_3\,
\frac{\langle n|[\hat H,\hat x_i]|k\rangle\langle k|[\hat H,\hat x_j]|n\rangle}
{(p_3+E^{\rm J}_n+i\delta\,{\rm sign}\,E^{\rm J}_n)^2\,(p_3+E^{\rm J}_k+i\delta\,{\rm sign}\,E^{\rm J}_k)}\,,\qquad\delta\to0^+,
\label{eq:S5}
\end{equation}
Eq.~(S5) of \cite{ZubJETP} with the prefactor of \eqref{eq:S3}, and with $\delta\to0^-$ for $\Sigma(0^-)$. If both
levels are
occupied, or both empty, all poles lie in one half plane and the integral vanishes (close the contour in the other
half plane). If $n$ is occupied and $k$ empty, closing in the lower half plane picks up the simple pole,
with residue $1/(a\,p_3^{(k)}+\mu_0-E_n)^2=1/(E_k-E_n)^2$, so the integral is $-2\pi i/(E_k-E_n)^2$; if $n$ is empty and
$k$ occupied it is $+2\pi i/(E_k-E_n)^2$. Hence, using $M_{kn}=-M_{nk}$ to combine the two cases,
\begin{equation}
\Sigma(\epsilon)=-iB\,{\rm sgn}(\epsilon)\sum_{E_n<\mu_0}\ \sum_{E_k>\mu_0}\frac{M_{nk}}{(E_k-E_n)^2},
\label{eq:FStknn}
\end{equation}
the two-dimensional Hall (TKNN) sum over occupied and empty Landau levels [Eq.~(S6) of \cite{ZubJETP}, up to the
sign discussed after \eqref{eq:S3}]: the dimensionless quantity $-iB\sum_{\rm occ,emp}M_{nk}/(E_k-E_n)^2$ is the Hall
number $C_H=2\pi\sigma_{12}$ of the transverse system (unit charge), so that $\Sigma(\epsilon)={\rm sgn}(\epsilon)\,C_H$.
For $\epsilon<0$ the half planes are exchanged and the sign reverses, which is the factor ${\rm sgn}\,\epsilon$.

\emph{Step 4: the Hall sum.} $M_{nk}$ connects adjacent levels only, so the only term of \eqref{eq:FStknn} is the
pair across the Fermi level, $n=n_{\rm filled}-1$ occupied and $k=n_{\rm filled}$ empty, with
$M_{nk}=iB\,n_{\rm filled}/m^2$ and $(E_k-E_n)^2=\omega^2=B^2/m^2$:
\[
\Sigma(\epsilon)=-iB\,{\rm sgn}(\epsilon)\,\frac{iB\,n_{\rm filled}/m^2}{B^2/m^2}=n_{\rm filled}\,{\rm sgn}(\epsilon),
\]
Eq.~(S7) of \cite{ZubJETP} (for $\mu_0<E_0$ the sum is empty and $\Sigma=0$). The result is independent of
$|\epsilon|$, as $d\Sigma/dp_4=0$ for $p_4\ne0$ requires. Each $p_3$-integral converges absolutely, and the interchange
of the level sum with the $p_3$-integral is justified for the antisymmetrized pairs $(n,n+1)+(n+1,n)$, whose combined
integrand is $-\frac{iB^2}{m^3}\,\frac{(n+1)\,a}{(ap_3+\mu_0-E_n)^2(ap_3+\mu_0-E_{n+1})^2}$, with absolute $p_3$-integral
$O(n^{-2})$ at fixed $\epsilon\ne0$ (the two unpaired terms are $O(n^{-1})$ each); a finite-level cutoff above the
occupied levels gives the same value. No ultraviolet prescription enters the slice invariant. Finally
\[
N_3^{(T)}=\Sigma(0^+)-\Sigma(0^-)=2n_{\rm filled}.
\]
\corrB{In the normalization of \eqref{eq:CSE} (prefactor $1/48\pi^2$) the same computation gives
$\Sigma(\epsilon)=\pm\frac12n_{\rm filled}$ and $N_3=n_{\rm filled}$; for a Dirac operator the missing factor two is
supplied by the trace with $\gamma^5$ over the two chiral blocks, which the scalar symbol \eqref{eq:FSmodel} does not
have.} At a Landau threshold, $\mu_0=E_n$, that level has its pole at $p_3=0$ for every $\epsilon$, and
$\Sigma$ is undefined, which is the statement that the invariant jumps there. The principal symbol
$i\kappa p_4p_3-\pi_\perp^2/2m$ is not invertible on any large sphere, so that no degree is defined for it (a scalar
symbol, if invertible, would have zero winding).}
\end{proof}
\reviewfix{\emph{Remark (the same computation as a contour integral).} Summing the antisymmetrized pairs of Step 4 over
$n$ before integrating over $p_3$ writes \eqref{eq:FSsum} as $\Sigma(\epsilon)=\frac{B}{4\pi}\int_{\mathcal L}F(z)\,dz$ along the line
$z=\mu_0+a\,p_3$, with the transverse Hall function
$F(z)=\epsilon^{bc}\sum_n\langle n|G_zv_bG_zv_cG_z|n\rangle=-\frac{iB^2}{m^3}\sum_{n\ge0}\frac{n+1}{(z-E_n)^2(z-E_{n+1})^2}$,
$G_z=(z-\hat H)^{-1}$, which is analytic off the Landau levels, decays as $|z|^{-2}$ in every sector off the real axis
and has the residue $-2i/B$ at every level. Rotating $\mathcal L$ about $\mu_0$ to the vertical line
${\rm Re}\,z=\mu_0$ (upward for $\epsilon>0$) and closing it to the left encloses the levels below $\mu_0$ and gives
$\frac{B}{4\pi}\cdot2\pi i\cdot(-2i/B)\,n_{\rm filled}=n_{\rm filled}$ again; this is the picture of
Fig.~\ref{fig:contour}b, and the form in which the lattice regularization below is treated.}

The operator is not elliptic: its principal symbol vanishes on non-compact subsets of the fibre ($p_4=\pi_\perp=0$,
$p_3=\pi_\perp=0$) and \corr{$Q(p_4,\vec p=0)=\mu_0$, so its inverse has no uniform ultraviolet decay along that axis when $\mu_0\ne0$.} On the far plane $p_4\to\infty$
the window $p_3\sim1/\kappa p_4$ keeps $\omega'=\kappa p_4p_3$ finite, $Q\to i\omega'-H_\perp+\mu_0$, and the slice
integral collapses to the Hall invariant: $N_3$ is an infrared quantity here, not a degree. The consequences for
the anomaly argument are discussed in Sec.~\ref{sec:nonell}.

\begin{figure}[!htb]
\centering
\begin{tikzpicture}[>=Stealth]
% panel (a): c(p3), s(p3) and the roots
\begin{scope}[xshift=-4.6cm]
\draw[->] (-2.3,0)--(2.4,0) node[right]{\scriptsize$p_3$};
\draw[->] (0,-1.15)--(0,1.95);
\draw[dotted,thick] (-2.042,-1.0)--(-2.042,1.8);
\draw[dotted,thick] (2.042,-1.0)--(2.042,1.8);
\node[below] at (-2.042,-1.0) {\scriptsize$-\pi$};
\node[below] at (2.042,-1.0) {\scriptsize$\pi$};
\node[below left] at (0,0) {\scriptsize$0$};
\draw[thick] plot[domain=-3.1416:3.1416,samples=140,smooth] (\x*0.65,{0.9*(sin(\x r)+0.7*(1-cos(\x r)))});
\draw[thick,dashed] plot[domain=-3.1416:3.1416,samples=140,smooth] (\x*0.65,{0.9*sin(\x r)});
\node[right] at (1.1,1.85) {\scriptsize$c$};
\node[right] at (1.2,0.72) {\scriptsize$s$};
\foreach \y/\n in {1.53/0,0.63/1,-0.27/2} {
  \draw[dotted] (-2.1,\y)--(2.1,\y);
  \node[right] at (2.1,\y) {\scriptsize$p_{3,\n}$};
}
% roots: filled = upper half plane for eps>0, open = lower
\fill (1.021,1.53) circle (1.7pt); \draw (1.815,1.53) circle (1.7pt);
\fill (0.397,0.63) circle (1.7pt); \fill (-1.645,0.63) circle (1.7pt);
\draw (-0.227,-0.27) circle (1.7pt); \fill (-1.021,-0.27) circle (1.7pt);
\node at (0,2.45) {(a) $c(p_3)$, $s(p_3)$ and the roots of $c=p_{3,n}$};
\end{scope}
% panel (b): the curve w(p3)
\begin{scope}[xshift=3.3cm]
\draw[->] (-2.5,0)--(3.5,0) node[right]{\scriptsize$\mathrm{Re}\,w$};
\draw[->] (0,-1.2)--(0,1.35) node[above]{\scriptsize$\mathrm{Im}\,w$};
\draw[thick,postaction={decorate},decoration={markings,
   mark=at position 0.18 with {\arrow{Stealth}},
   mark=at position 0.42 with {\arrow{Stealth}},
   mark=at position 0.66 with {\arrow{Stealth}},
   mark=at position 0.90 with {\arrow{Stealth}}}]
   plot[domain=-3.1416:3.1416,samples=200] ({1.6*(sin(\x r)+0.7*(1-cos(\x r)))},{-0.8*sin(\x r)});
\foreach \x/\n/\w in {2.72/0/0,1.12/1/1,-0.48/2/0,-2.08/3/0} {
  \node[cross out,draw,inner sep=1.5pt] at (\x,0) {};
  \node at (\x,-1.05) {\scriptsize$p_{3,\n}$};
  \node at (\x,1.05) {\scriptsize$N=\w$};
}
\node[above right,inner sep=1pt] at (0.02,0.02) {\scriptsize$c_0$};
\node[above right,inner sep=1pt] at (2.26,0.02) {\scriptsize$c_\pi$};
\node at (0.5,1.95) {(b) the closed curve $w(p_3)$, $\epsilon>0$};
\end{scope}
\end{tikzpicture}
\caption{The poles of the lattice integrand \eqref{eq:S3lat} for the Wilson-type example, $s=\sin p_3$,
$c=\sin p_3+\alpha(1-\cos p_3)$, with $\alpha=0.7$, $\mu_0=2.2\,\omega$ ($\omega=B/m=1$ in the plotted units) and
$\kappa\epsilon=0.5$ (exaggerated for
visibility). (a) $c$ (solid) and $s$ (dashed) over the zone; the dotted horizontals are the levels $p_{3,n}=\mu_0-E_n$,
$n=0,1,2$, and the real roots of $c(p_3)=p_{3,n}$ -- the $\epsilon=0$ positions of the poles \eqref{eq:latpoles},
which move off the axis for $\epsilon\ne0$ -- are their intersections with $c$, two for each of these levels.
By \eqref{eq:lathalf} a root moves into the upper half plane for $\epsilon>0$ if $s\,c'>0$ there (filled) and into the
lower one otherwise (open): the level $n=1$, inside the window $(\mu_0-2\alpha,\mu_0)$, has both roots above the axis;
$n=0$ and $n=2$ have one root on each side, with the opposite $i0$ for the two roots. (b) The same in the $w$-plane:
$w(p_3)=c-i\kappa\epsilon s$ traces a closed curve as $p_3$ runs once around the zone, crossing the real axis only at
the junctions, downward at $w=c_0=0$ ($s'_0=1$) and upward at $w=c_\pi=2\alpha$ ($s'_\pi=-1$). The pole of the level
$n$ is the real point $w=-E^{\rm J}_n=p_{3,n}$ (crosses): it lies inside the curve for the level in the window, $N_n=1$,
and outside for the others, $N_n=0$, Eq.~\eqref{eq:latres}; a level with the pole inside has both roots of panel (a)
in the upper half plane, a level with the pole outside has one root on each side. $p_{3,3}$ lies outside the range
of $c$, where the two roots are a complex pair.}
\label{fig:wcurve}
\end{figure}
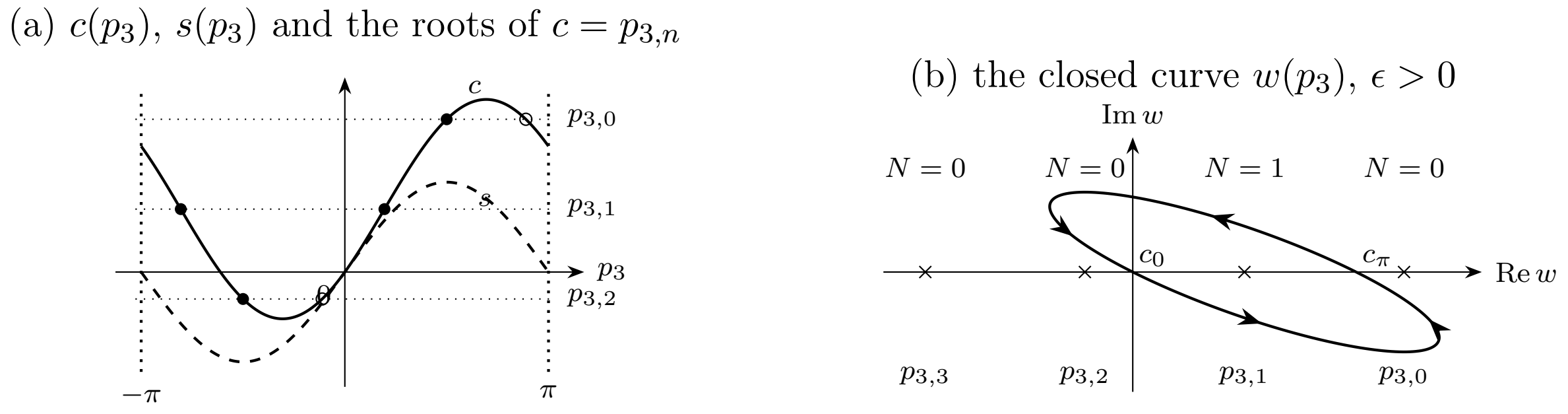

\paragraph{\reviewfix{Lattice regularization.}}
\reviewfix{Regularize the model on a lattice in the $x_3$ direction (spacing one), keeping the transverse motion
continuous: $p_3$ becomes a coordinate on the circle and
\begin{equation}
Q_{\rm lat}=i\kappa p_4\,s(p_3)-c(p_3)-H_\perp+\mu_0 ,
\label{eq:FSlat}
\end{equation}
with $s$, $c$ real, smooth and $2\pi$-periodic, $s\simeq c\simeq p_3$ near $p_3=0$, the junctions simple,
$s'(k_j)\ne0$, and $\mu_0-c_j\ne E_n$ for all $j$, $n$ (for the complex roots below $s$ and $c$ are trigonometric
polynomials); for instance $s=\sin p_3$,
$c=\sin p_3+\alpha(1-\cos p_3)$, with a Wilson-type term. The symbol remains gauge-field independent and
non-elliptic: $X=\kappa s(p_3)$ vanishes at the junctions, the zeros $k_j$ of $s$ -- an even number, with slopes
$s'_j$ of alternating sign -- where $c$ takes the values $c_j$; for $s=\sin p_3$ they are $k_j=0,\pi$, with $c_0=0$ and
$c_\pi=2\alpha$. The invariant is computed exactly as in the proof of Proposition~\ref{prop:FS}. Steps 1 and 2 are
unchanged, the c-number $\partial_3Q=a(p_3)=i\kappa\epsilon\,s'(p_3)-c'(p_3)$ now depending on $p_3$:
\[
\Sigma(\epsilon)=\frac{B}{4\pi}\sum_{n,k}M_{nk}\oint dp_3\,\frac{a(p_3)}{D_n(p_3)^2\,D_k(p_3)},\qquad
D_n(p_3)=i\kappa\epsilon\,s(p_3)-c(p_3)+\mu_0-E_n ,
\]
the integral running once around the Brillouin zone. With $w(p_3)=c(p_3)-i\kappa\epsilon\,s(p_3)$ this is, in the form
\eqref{eq:S3},
\begin{equation}
\Sigma(\epsilon)=-\frac{\epsilon_{ij}}{2A}\sum_{n,k}\oint dp_3\,\partial_{p_3}w(p_3)\,
\frac{\langle n|[\hat H,\hat x_i]|k\rangle\langle k|[\hat H,\hat x_j]|n\rangle}{(w(p_3)+E^{\rm J}_n)^2\,(w(p_3)+E^{\rm J}_k)}\,,
\label{eq:S3lat}
\end{equation}
the lattice analogue of Eq.~(S3) of \cite{ZubJETP} [and of its Eq.~(S10), with $p_3(1-i\kappa\omega(p_3))\to w(p_3)$].
Step 3 is where the lattice differs. The poles of \eqref{eq:S3lat} are the solutions of $w(p_3)=-E^{\rm J}_n$,
\begin{equation}
c(p_3)-i\kappa\epsilon\,s(p_3)=p_{3,n},\qquad p_{3,n}=\mu_0-E_n ,
\label{eq:latpoles}
\end{equation}
and their number is doubled with respect to \eqref{eq:S5}: for $s$, $c$ trigonometric polynomials of degree $d$
(hopping of range $d$) this is an equation of degree at most $2d$ in $e^{ip_3}$, exactly $2d$ when the extreme Laurent
coefficients do not vanish -- two roots per level in the nearest-neighbour example below. For $\epsilon\to0$ the
solutions are the simple real roots $p^*_{n,r}$ of $c(p^*)=p_{3,n}$ (with $c'(p^*)\ne0$), displaced into the complex
plane,
\begin{equation}
p_3=p^*_{n,r}+i\eta_{n,r}+O(\epsilon^2),\qquad \eta_{n,r}=\kappa\epsilon\,\frac{s(p^*_{n,r})}{c'(p^*_{n,r})}
\label{eq:lathalf}
\end{equation}
[expand $c(p^*+i\eta)=p_{3,n}+i\eta\,c'(p^*)$ in \eqref{eq:latpoles}], together with the nonreal roots of
$c(p_3)=p_{3,n}$, if any, which also move under the perturbation. This is the analogue of \eqref{eq:S5}: $w+E^{\rm J}_n\to c(p_3)+E^{\rm J}_n-i\delta\,\sgn s(p_3)$,
$\delta\to0^+$, with $p_3$ at the pole. In the continuum $s=c=p_3$, the single root $p^*=p_{3,n}$ has
$\eta=\kappa\epsilon\,p_{3,n}$, in the upper half plane for an occupied level at $\epsilon>0$, which is the
$i\delta\,\sgn E^{\rm J}_n$ of \eqref{eq:S5}. On the lattice the half plane of a root is fixed by the sign of
$\epsilon\,s\,c'$ \emph{at that root}, not by the sign of $E^{\rm J}_n$: the roots of one and the same level may lie in
opposite half planes, and the $i0$ prescription differs from root to root. The integral itself is best done in
the $w$-plane: since $dp_3\,\partial_{p_3}w=dw$,
\[
\oint dp_3\,\frac{\partial_{p_3}w}{(w+E^{\rm J}_n)^2(w+E^{\rm J}_k)}=\oint_\Gamma\frac{dw}{(w+E^{\rm J}_n)^2(w+E^{\rm J}_k)},
\]
$\Gamma$ being the closed curve traced by $w(p_3)$ as $p_3$ runs once around the zone (Fig.~\ref{fig:wcurve}b). There
each term $n\ne k$ has a double pole at $w=-E^{\rm J}_n$ and a simple pole at $w=-E^{\rm J}_k=p_{3,k}$, both real -- one
pole per level; the residues are those of \eqref{eq:S5},
$-(E_k-E_n)^{-2}$ at the double pole $w=-E^{\rm J}_n$ and $+(E_k-E_n)^{-2}$ at the simple pole $w=-E^{\rm J}_k$, and
\begin{equation}
\oint dp_3\,\frac{\partial_{p_3}w}{(w+E^{\rm J}_n)^2(w+E^{\rm J}_k)}=2\pi i\,\frac{N_k-N_n}{(E_k-E_n)^2},\qquad
N_k={\rm wind}(\Gamma,-E^{\rm J}_k)=\frac1{2\pi i}\oint_\Gamma\frac{dw}{w+E^{\rm J}_k}\in\mathbb Z ,
\label{eq:latres}
\end{equation}
the winding number of $\Gamma$ about the pole of the level $k$. For the simple loop of the example
(Fig.~\ref{fig:wcurve}b) at $\epsilon>0$ it is $1$ if $-E^{\rm J}_k$ lies inside $\Gamma$ and $0$ if outside;
$\epsilon\to-\epsilon$ conjugates the curve and reverses its windings about the real poles ($-1$ inside). Which real
points are
inside is read off from the crossings: ${\rm Im}\,w=-\kappa\epsilon\,s(p_3)$ vanishes only at the junctions $k_j$, the
zeros of $s$, so $\Gamma$ meets the real axis only there, at $w=c_j=c(k_j)$, and it does so downward where
$\epsilon s'_j>0$ and upward otherwise, $s'_j=s'(k_j)$ being the slope of $s$ at the junction ($s'_0=1$, $s'_\pi=-1$ for
$s=\sin p_3$). A real point $x$ is therefore encircled $\sgn(\epsilon)\sum_j\sgn(s'_j)\,\theta(x-c_j)$ times and
\[
N_n=\sgn(\epsilon)\sum_j\sgn(s'_j)\,\theta(\mu_0-c_j-E_n) .
\]
For the Wilson-type example, with its two junctions, $-E^{\rm J}_n=p_{3,n}$ is inside $\Gamma$ iff $c_0<p_{3,n}<c_\pi$,
i.e.\ for $\epsilon>0$
\[
N_n=\begin{cases}1,&\mu_0-c_\pi<E_n<\mu_0-c_0,\ \text{i.e.}\ \mu_0-2\alpha<E_n<\mu_0,\\0,&\text{otherwise:}\end{cases}
\]
the levels that count are those in the energy window between the effective chemical potentials $\mu_0-c_j$ of the two
junctions [for the symmetric choice $c\to c-\alpha$, with $c_0=-\alpha$, $c_\pi=\alpha$, the window is
$\mu_0-\alpha<E_n<\mu_0+\alpha$]. It is not the condition that $p_{3,n}$ lie in the range $(\min c,\max c)$ of $c$: a
point of the range beyond
$c_0$ or $c_\pi$ lies under a lobe of the curve, which passes it twice in the same half plane ($p_{3,0}$ and $p_{3,2}$
in Fig.~\ref{fig:wcurve}b). For the continuum model, one junction with $s'=1$, $N_n=\sgn(\epsilon)\,\theta(\mu_0-E_n)$,
the pole bookkeeping of Step 3 of the proof. In the $p_3$-plane the same count is the argument principle in $z=e^{ip_3}$: $w(z)$ is a Laurent polynomial with
a pole of order $d_-$ at $z=0$, and if $Z_n$ is the number of zeros of $w(z)+E^{\rm J}_n$ in $|z|<1$ (with multiplicity;
none lies on $|z|=1$ because $\mu_0-c_j\ne E_n$), then $N_n=Z_n-d_-$ and $N_k-N_n=Z_k-Z_n$. The zeros inside the disk
are the roots of \eqref{eq:latpoles} with ${\rm Im}\,p_3>0$, so $N^+_n\equiv Z_n$ counts them: each root contributes
$0$ or $1$, $N^+_n$ ranges from $0$ to $2d$, and $N_n=N^+_n-N^+_\infty$ with $N^+_\infty=d_-$ the common value for the
levels far above $\mu_0$. Indeed, by \eqref{eq:lathalf} the root passing through
$k_j$ when $p_{3,n}$ passes $c_j$ has $\eta=\kappa\epsilon\,s'_j\,(p_{3,n}-c_j)/c_j'^2$ and crosses the real axis upward
if $\epsilon s'_j>0$, which is the crossing of $\Gamma$ just described, whereas at an extremum of $c$ two real roots of
opposite $c'$ merge into a complex pair with one member in each half plane and $N^+_n$ does not change.}

\reviewfix{\emph{The roots for the Wilson-type example.} For $s=\sin p_3$, $c=\sin p_3+\alpha(1-\cos p_3)=\sqrt{1+\alpha^2}\,
\sin(p_3-\varphi)+\alpha$, $\tan\varphi=\alpha$, Eq.~\eqref{eq:latpoles} is the quadratic
$(1-i\kappa\epsilon-i\alpha)z^2+2i(\alpha-p_{3,n})z-(1-i\kappa\epsilon+i\alpha)=0$ in $z=e^{ip_3}$: two roots for
every level, with $|z_1z_2|=[1+(\alpha-\kappa\epsilon)^2]^{1/2}[1+(\alpha+\kappa\epsilon)^2]^{-1/2}$, equal to one
at $\epsilon=0$. For $p_{3,n}$ outside the range $[\alpha-\sqrt{1+\alpha^2},\alpha+\sqrt{1+\alpha^2}]$ of $c$ the
roots are a conjugate pair at $\epsilon=0$ and, no longer conjugate but still one in each half plane, for
$\epsilon\ne0$: $N^+_n=1$, $N_n=0$. Strictly inside the range the two real roots at $\epsilon=0$ are
(Fig.~\ref{fig:wcurve})
\[
p^{(1)}_n=\varphi+\theta_n,\qquad p^{(2)}_n=\varphi+\pi-\theta_n,\qquad
\sin\theta_n=\frac{p_{3,n}-\alpha}{\sqrt{1+\alpha^2}},\quad|\theta_n|<\frac\pi2 ,
\]
with $c'(p^{(1)}_n)=\sqrt{1+\alpha^2}\cos\theta_n>0$ and $c'(p^{(2)}_n)<0$; as $p_{3,n}$ grows from $0$ to
$2\alpha$, $p^{(1)}_n$ runs from the junction $0$ to $2\varphi$ and $p^{(2)}_n$, taken in $[-\pi,\pi]$, decreases from
$2\varphi-\pi$ to the junction $-\pi\equiv\pi$. Since $s(p^{(1)}_n)=\sin(\varphi+\theta_n)$ has the sign of $p_{3,n}$ and $s(p^{(2)}_n)=\sin(\theta_n-\varphi)$
the sign of $p_{3,n}-2\alpha$, \eqref{eq:lathalf} gives
\[
{\rm Im}\,p^{(1)}_n>0\iff\epsilon\,(\mu_0-E_n)>0,\qquad {\rm Im}\,p^{(2)}_n>0\iff\epsilon\,(E_n-\mu_0+2\alpha)>0 .
\]
The root associated with the junction $0$ (the $\epsilon=0$ root and its continuation) follows the rule of
\eqref{eq:S5} -- upper half plane for the occupied levels at
$\epsilon>0$. The root near the junction $\pi$ follows the opposite rule -- upper half plane for the levels
\emph{above} $\mu_0-2\alpha$: its $i0$ is that of $-(E^{\rm J}_n+2\alpha)$, the pole of a level counted from the
effective chemical potential $\mu_0-c_\pi$ of the second junction with the sign of $\epsilon$ reversed, as
$s'_\pi=-1$ requires. Hence, for $\epsilon>0$,
\[
N^+_n=\theta(\mu_0-E_n)+\theta(E_n-\mu_0+2\alpha)=\begin{cases}2,&\mu_0-2\alpha<E_n<\mu_0,\\1,&\text{otherwise,}\end{cases}
\]
and $0$ resp.\ $1$ for $\epsilon<0$. A level with its pole outside $\Gamma$ has one root in each half plane,
$N_n=0$: it contributes nothing to \eqref{eq:latres} when paired with another such level, exactly as a complex pair
does, but it does contribute when paired with a level of different winding. A level with its pole inside $\Gamma$
has, for $\epsilon>0$, both roots in the upper half plane: $N_n=N^+_n-1=\sgn(\epsilon)[\theta(\mu_0-E_n)-\theta(\mu_0-2\alpha-E_n)]$, one for the levels
in the window $\mu_0-2\alpha<E_n<\mu_0$ and zero otherwise, and \eqref{eq:latres} is nonzero only for
the pairs $(n,k)$ straddling one of the two effective Fermi levels $\mu_0$ and $\mu_0-2\alpha$ -- the Fermi-level pair
of the continuum, Step 4 of the proof, and a second one at $\mu_0-2\alpha$ with the opposite sign. For $\alpha=0$ the
roots are $\theta_n$ and $\pi-\theta_n$, with equal $s$ and opposite $c'$: they lie in opposite half planes for every
level, $N^+_n=1$ and $N_n=0$ for all $n$ (the curve $\Gamma$ encloses no real point), and the invariant vanishes.
Step 4 is the same telescoping sum: with $M_{kn}=-M_{nk}$ and
$M_{n,n+1}+M_{n,n-1}=iB/m^2$ for every $n$,
\[
\Sigma(\epsilon)=-\frac{iB}{2}\sum_{n\ne k}M_{nk}\,\frac{N_n-N_k}{(E_k-E_n)^2}=-iB\sum_nN_n\,\frac{iB/m^2}{B^2/m^2}=\sum_nN_n ,
\]
and therefore
\begin{equation}
\Sigma(\pm\epsilon)=\pm\sum_j\sgn(s'_j)\,n_{\rm filled}(\mu_0-c_j),\qquad
N_3^{(T)}=2\sum_j\sgn(s'_j)\,n_{\rm filled}(\mu_0-c_j),
\label{eq:N3lat}
\end{equation}
where $n_{\rm filled}(\mu)$ is the number of Landau levels below $\mu$. The sum over levels is finite because
$\sum_j\sgn s'_j=0$, each slice is an integer by the residue count (the model is discretized in $x_3$ only, so the
lattice index theorem of Sec.~\ref{sec:NCG} does not apply to it directly), and every
junction contributes the Hall number of the transverse system at its own effective level $\mu_0-c_j$, with the sign
of the crossing. For the symmetric choice $\alpha=0$ the two junctions sit at the same level and $N_3^{(T)}=0$; for
$\alpha\ne0$, $N_3^{(T)}=2[n_{\rm filled}(\mu_0)-n_{\rm filled}(\mu_0-2\alpha)]$, nonzero whenever a Landau level lies
between $\mu_0-2\alpha$ and $\mu_0$.}

\paragraph{\reviewfix{The chiral magnetic like effect.}}\label{sec:CMEplane}
\reviewfix{For $\hat T=1$ the current in \eqref{eq:N3T} is the conserved vector current, and a nonzero $N_3^{(1)}$ means
an equilibrium current of the conserved charge along $B$ -- the vector analogue of the CSE, which we call a
chiral magnetic like effect (the difference from the chiral magnetic effect proper is explained in the symmetry paragraph below).
For a system of Hamiltonian form, $Q=ip_4-H$ with $H$ bounded from below, such a current is excluded by the
generalized Bloch theorem \cite{Bohm,Yamamoto}; in one dimension this is the statement that the $\mu$-derivative of
the current of a band is supported at its Fermi points, $\partial j/\partial\mu=\frac1{2\pi}\sum_{\rm FP}\sgn v$, and
that a band bounded from below crosses $\mu$ an even number of times with alternating slopes. The non-elliptic models
of \cite{ZubJETP} are not of this form, and the question is decided model by model: the basic model \eqref{eq:FSmodel}
with plane Fermi surfaces, together with its lattice regularization above, carries the current; the model with
spherical Fermi surfaces of \cite{ZubJETP}, which has Weyl-like Fermi points, does not (Sec.~\ref{sec:CMEsphere}).}

\reviewfix{For \eqref{eq:FSmodel}, Eq.~\eqref{eq:N3T} with $\hat T=1$ states that the model carries in equilibrium a
current along $B$ with
$\partial J_3/\partial\mu=N_3^{(T)}B/4\pi^2=2n_{\rm filled}B/4\pi^2\neq0$. Writing $Q=\kappa p_3(ip_4-\varepsilon_n)$ in
the level $n$, the effective dispersion $\varepsilon_n(p_3)=(p_3-p_{3,n})/\kappa p_3$ has its Fermi point at the sheet
$p_{3,n}$, with slope $v_n=1/\kappa p_{3,n}$, and passes through infinity at the junction $p_3=0$, where the coefficient
$X=\kappa p_3$ of $ip_4$ changes sign: all sheets of the levels below $\mu_0$ have the same chirality, their partner
being the junction, which is not a Fermi point, and Bloch's theorem does not apply. The Fermi-point count of the
model is $\frac{B}{4\pi^2}\sum_{E_n<\mu_0}\sgn v_n=n_{\rm filled}B/4\pi^2$ -- the $\mu$-derivative of
$-\Tr\,G\partial_{p_3}Q$ per mode is $-\partial_{k_3}\varepsilon_n(ip_4+\mu-\varepsilon_n)^{-2}$, whose frequency integral
$\partial_{k_3}\varepsilon_n\,\delta(\mu-\varepsilon_n)$ is supported at the Fermi points; the levels above $\mu_0$, whose
sheets lie at $p_3<0$ with $v_n<0$, are infinitely many and are not counted -- which is half of \eqref{eq:N3T}: the two half-lines $p_3\gtrless0$ of the
slice integral of Proposition~\ref{prop:FS} have equal real parts, $\frac12n_{\rm filled}$ each, and opposite imaginary
parts [$F(\bar z)=-\overline{F(z)}$ for Hermitian $H_\perp$], and the half $p_3<0$ has no Fermi point among the
selected levels.

On the lattice (the regularization above) the relation \eqref{eq:N3T} between the current and the invariant holds as in the
continuum, so that the equilibrium current along $B$ is, for
the Wilson-type example,
\[
\frac{\partial J_3}{\partial\mu}=\frac{N_3^{(T)}B}{4\pi^2}=\frac{B}{2\pi^2}\big[n_{\rm filled}(\mu_0)-n_{\rm filled}(\mu_0-2\alpha)\big]:
\]
the chiral magnetic like effect of the lattice model is carried by the Landau levels in the window $\mu_0-2\alpha<E_n<\mu_0$
between the effective chemical potentials of the two junctions, it is present whenever that window contains a level,
and it vanishes for $\alpha=0$. Bloch's theorem does not apply to the lattice model either, which remains
non-Hamiltonian at its junctions.}

\reviewfix{\paragraph{Symmetries, and the difference from the chiral magnetic effect.}
The chiral magnetic effect proper is the vector current $\mathbf J=\mu_5\mathbf B/2\pi^2$ of chiral fermions with a
chirality imbalance $\mu_5=(\mu_R-\mu_L)/2$: under parity $\mathbf J$ is odd, $\mathbf B$ is even and $\mu_5$ is odd.
In \eqref{eq:N3T} with $\hat T=1$ the current is driven instead by the ordinary chemical potential, which is parity
even, so that the coefficient $N_3^{(1)}$ must be a pseudoscalar: a parity-invariant system has $N_3^{(1)}=0$, and a
nonzero value requires that parity be broken by the system itself. For \eqref{eq:FSmodel} parity, $\vec p\to-\vec p$
at fixed $p_4$, gives $Q(p_4,-\vec p)=-i\kappa p_4p_3+p_3-H_\perp+\mu_0\ne Q(p_4,\vec p)$: the two terms odd in $p_3$
break it and select an orientation $\mathbf n$ of the $x_3$ axis, a polar vector built into the model. Its parity
image is the mirror model, with the sheets at $p_3=-p_{3,n}$ and velocities of the opposite sign, and at the same $B$
it has the opposite invariant. The sign of $B$ enters in the same way. The formulas above are written for $B>0$; in
general the Landau degeneracy is $|B|/2\pi$ while the residue of the Hall function is $-2i/B$, so that
$\Sigma^{(T)}(\pm\epsilon)=\pm\sgn(B)\,n_{\rm filled}$ and
\[
N_3^{(T)}=2\,n_{\rm filled}\,\sgn(B)=2\,n_{\rm filled}\,(\mathbf n\cdot\hat{\mathbf B}),\qquad
\frac{\partial\mathbf J}{\partial\mu}=\frac{n_{\rm filled}\,|B|}{2\pi^2}\,\mathbf n ,
\]
a pseudoscalar, as required, and a current that is even in $B$: its direction is fixed by $\mathbf n$, not by the
field. Time reversal, $Q(p_4,\vec p)\to Q^*(-p_4,-\vec p)$, also maps the model onto its mirror image. Under the
physical transformation, with $\mathbf J$ and $\mathbf B$ odd and $\mu$ even, $N_3^{(1)}$ is invariant when the model
and the field are reversed together, and it changes sign when the field is held fixed; the model breaks parity and
time reversal separately and keeps their product. The lattice regularization illustrates the selection rule. For
$\alpha=0$, $s=c=\sin p_3$, and the model is invariant under $p_3\to\pi-p_3$, a mirror $x_3\to-x_3$ combined with
the staggered phase $e^{i\pi x_3}$. This symmetry reverses $J_3$ and leaves $B$ unchanged, so it forces
$N_3^{(T)}=0$, in agreement with \eqref{eq:N3lat}; the Wilson term $\alpha(1-\cos p_3)$, which becomes
$\alpha(1+\cos p_3)$, breaks it.

The effect is therefore not the chiral magnetic effect. There is no chirality imbalance, the current responds to
$\mu$ rather than to $\mu_5$, it is even in $B$, and its parity-odd input is the structural mirror asymmetry of the
model: all selected sheets are one-dimensional branches of one handedness along $\mathbf n$, their partner being the
junction rather than a Fermi point of the opposite velocity. Its kinship with the chiral magnetic effect is the
equilibrium vector current along the magnetic field carried by unpaired chiral branches; we call it a
\emph{chiral magnetic like effect}. In the spherical model of Sec.~\ref{sec:CMEsphere2} the parity-odd input is instead the
parameter $\mu_0$, which multiplies $\gamma^5$ and is a pseudoscalar if parity exchanges the chiral blocks: it plays
the role of $\mu_5$, and that effect is closer to the chiral magnetic effect, although its response $\partial J/\partial\mu$
is again driven by $\mu$ and depends on $\mu_0$ only through the count $C(\mu_0)$.}

\subsection{\reviewfix{The chiral separation effect in the model with spherical Fermi surfaces}}
\label{sec:CMEsphere}

\reviewfix{The last model of the Supplementary Material of \cite{ZubJETP} (its Sec.~C.2, Eqs.~(S67)--(S70)) replaces the
planes by concentric spheres. With $p_r=|\vec p|$ and $(p_\theta,p_\varphi)=(\theta/b,\varphi/a)$ the rescaled angular
coordinates of momentum space, the model is, in the flat angular limit \reviewfix{$a,b\to0$} in which \cite{ZubJETP} works
(the periodic function $f_a$ of \cite{ZubJETP} replaced by its argument; \reviewfix{\cite{ZubJETP} writes
$a,b\to\infty$, but with $p_\varphi=\varphi/a$ and $p_\theta=\theta/b$ it is $a,b\to0$ that makes the angular domain
infinite; the periodic implementation, the poles and finite $a,b$ are discussed at the end of this subsection}),
\begin{equation}
\hQ=i\kappa\lambda\hat p_4(\hat H-\mu_0\gamma^5)-\hat p_r-(\hat H-\mu_0\gamma^5)\lambda,\qquad
\hat H=(p_\varphi+B\hat x_\theta)\gamma^0\gamma^1+p_\theta\gamma^0\gamma^2 ,\qquad \hat x_\theta=i\partial_{p_\theta},
\label{eq:sphQ}
\end{equation}
with $B,\kappa,\lambda>0$. Here $\hat H$ is a two-dimensional Dirac Hamiltonian on the angular variables, $B$ being a
parameter of this angular operator and not a magnetic field in coordinate space; it commutes with $\gamma^5$ and with
$p_r$, its levels are the relativistic Landau levels $E_n=\sgn(n)\sqrt{2B|n|}$, $n\in\mathbb Z$, with degeneracy
$B/2\pi$ per unit area of $(x_\theta,x_\varphi)$, and in the block of chirality $s=\pm1$ it acts as $s\hat H_2$,
$\hat H_2=(p_\varphi+B\hat x_\theta)\sigma_1+p_\theta\sigma_2$. The fixed parameter of the model is denoted $\mu_0$, as
in the basic model; the chemical potential $\mu$ of \eqref{eq:N3T} enters through $ip_4\to ip_4+\mu$ and is set to
zero. On the slice $p_4=\epsilon$ the sector of chirality $s$ factorizes exactly,
\[
Q_s=s\lambda(1-i\kappa\epsilon)\,(z_s-\hat H_2),\qquad z_s=\mu_0-\frac{s\,p_r}{\lambda(1-i\kappa\epsilon)} ,
\]
and in the sector of the level $n$ the operator is a c-number function of $p_r$,
\[
Q_{n,s}=s\kappa\lambda(E_n-\mu_0)\,(ip_4-\varepsilon_{n,s}),\qquad
\varepsilon_{n,s}(p_r)=\frac1\kappa+\frac{p_r}{s\kappa\lambda(E_n-\mu_0)} ,
\]
with the Fermi surface $\varepsilon_{n,s}=0$ at $p_r=s\lambda(\mu_0-E_n)$ [Eq.~(S70) of \cite{ZubJETP}]: a sphere for
every level below $\mu_0$ in the sector $s=+1$ and for every level above $\mu_0$ in the sector $s=-1$, all with
$\partial\varepsilon/\partial p_r=-1/[\kappa\lambda|E_n-\mu_0|]<0$. The coefficient $X_{n,s}=s\kappa\lambda(E_n-\mu_0)$
of $ip_4$ is constant within a sector and changes sign from sector to sector. The junction of the plane model, $X=0$
inside a sector, has no analogue here ($X$ vanishes only at the thresholds $\mu_0=E_n$); the model as a whole is
nevertheless not of Hamiltonian form, and in every sector that carries a sphere $X<0$, i.e.\ the spectral weight of
$G_{n,s}=X_{n,s}^{-1}(ip_4-\varepsilon_{n,s})^{-1}$ is negative. For $\mu_0=\delta\to0$ the sphere of the level $n=0$,
of chirality $s=\sgn\delta$ and radius $\lambda|\delta|$, shrinks to the origin -- a Fermi point in the limit, and in
this sense the model is a semi-realistic, Weyl-like one -- while the spheres of the levels $n\ne0$ remain, with radii
$\lambda\sqrt{2B|n|}$; the small sphere is isolated from them inside a ball of radius $R$,
$\lambda|\delta|<R<\lambda(\sqrt{2B}-|\delta|)$. At $\delta=0$ itself the zero-level sector is $Q_{0,s}=-p_r$,
without frequency dependence: this is the junction of the model, and the invariant is defined for $\delta\ne0$ only.
The sphere does not disappear there: for $\delta<0$ it is the sphere of the level $n=0$ in the sector $s=-1$, at
$p_r=\lambda|\delta|$ -- it contracts to a point at $\delta=0$ and re-expands with the opposite chirality, its local
invariant passing from $+1$ to $-1$, the jump of $N_3^{(\gamma^5)}$ below. This is the sense in which the invariant
protects it: the Fermi surface can be passed through a Fermi point, where its chirality flips, but not removed.

\emph{Derivation of the invariant.} The slices of \eqref{eq:N3T} are computed directly, along the lines of the
proof of Proposition~\ref{prop:FS} and of Eqs.~(S54)--(S65) of \cite{ZubJETP}, for $\hat T=1$ and $\hat T=\gamma^5$ at
once; $b=1-i\kappa\epsilon$ throughout.

\emph{Step 1: spherical parametrization of the slice.} Following Eqs.~(S55)--(S57) of \cite{ZubJETP}, the slice
$p_4=\epsilon$ is parametrized by $k=(p_r,p_\theta,p_\varphi)$ and coordinate space by the conjugate variables
$\xi_i=(\partial p_j/\partial k_i)\,x_j$. The change of variables is canonical: $d^3k\,d^3\xi=d^3p\,d^3x$, the Moyal
product keeps its form in the variables $(k,\xi)$ [Eq.~(S56) there; we write it $\circ$], and the integrand is a
three-form, so that
\[
\Sigma^{(T)}(\epsilon)=-\frac1{24\pi^2V}\int d^3\xi\int d^3k\,
\big[\trD\big(\hat T\,G\circ dQ\circ\wedge dG\circ\wedge dQ\big)\big]_{r\theta\varphi},\qquad V=\int d^3\xi ,
\]
with $(p_r,p_\theta,p_\varphi)$ right-handed, i.e.\ with the orientation $dp_1\wedge dp_2\wedge dp_3$ of
\eqref{eq:N3T}. In the flat angular limit the symbol depends on the coordinates only through $x_\theta$, inside
$\hat H$; $p_r$ is a c-number in all products, and $\partial_{p_r}Q=-1$. The algebra of Step~1 of the proof of
Proposition~\ref{prop:FS} -- one index radial, the radial derivative a c-number, its three positions equal by the
cyclicity of the angular trace -- gives
\begin{equation}
\big[\tr\big(G\circ dQ\circ\wedge dG\circ\wedge dQ\big)\big]_{r\theta\varphi}
=-3\,a_r\,\epsilon^{bc}\,\tr\big(G\circ\partial_bQ\circ G\circ\partial_cQ\circ G\big),\qquad a_r=\partial_{p_r}Q=-1 ,
\label{eq:sph3form}
\end{equation}
with $b,c\in\{\theta,\varphi\}$ and $\epsilon^{\theta\varphi}=1$; \eqref{eq:FS3form} is the case $a_r=a$,
$\partial_bQ=-\partial_bH$.

\emph{Step 2: the chiral blocks and the Landau levels.} In the block of chirality $s$,
$Q_s=s\lambda b\,(z_s-\hat H_2)$ with $z_s=\mu_0-s\,p_r/\lambda b$, so that
\[
G_s=(s\lambda b)^{-1}g(z_s),\qquad g(z)=(z-\hat H_2)^{-1},\qquad \partial_bQ_s=-s\lambda b\,\partial_b\hat H_2 ,
\]
and the factors $s\lambda b$ cancel up to one:
$G_s\circ\partial_bQ_s\circ G_s\circ\partial_cQ_s\circ G_s=(s\lambda b)^{-1}\,g\,\partial_b\hat H_2\,g\,\partial_c\hat H_2\,g$.
Here $\partial_{p_\varphi}\hat H_2=\sigma_1$ and $\partial_{p_\theta}\hat H_2=\sigma_2$ are the velocities $v_1$, $v_2$
along the kinetic momenta $\pi_1=p_\varphi+B\hat x_\theta$, $\pi_2=p_\theta$, $[\pi_1,\pi_2]=iB$; the ordered pair
$(\theta,\varphi)$ is the pair $(2,1)$, so that $\epsilon^{\theta\varphi}$ becomes $-\epsilon^{ij}$ with
$\epsilon^{12}=1$. The trace over the angular phase space per unit area is $(2\pi)^2\frac B{2\pi}\sum_n\langle
n|\cdot|n\rangle$, as in Step~2 of the proof of Proposition~\ref{prop:FS}, and $\hat T$ acts in the block $s$ as the
number $T_s$: $T_s=1$ for $\hat T=1$ and $T_s=s$ for $\hat T=\gamma^5$. With the Hall function of the relativistic
Landau levels,
\[
F(z)=\epsilon^{ij}\sum_n\langle n|g(z)\,v_i\,g(z)\,v_j\,g(z)|n\rangle=\sum_{n,k}\frac{M_{nk}}{(z-E_n)^2(z-E_k)},\qquad
M_{nk}=\epsilon^{ij}\langle n|v_i|k\rangle\langle k|v_j|n\rangle ,
\]
Eq.~\eqref{eq:sph3form} and $dz_s=-s\,dp_r/\lambda b$ give
\begin{equation}
\Sigma^{(T)}(\epsilon)=\frac B{4\pi}\sum_{s=\pm1}T_s\int_0^\infty\frac{dp_r}{s\lambda b}\,F(z_s)
=-\frac B{4\pi}\sum_{s=\pm1}T_s\int_{\ell_s}F(z)\,dz ,
\label{eq:sphrays}
\end{equation}
with $\ell_s$ the ray from $\mu_0$ ($p_r=0$) to infinity in the direction $-s/b$. This is Eq.~(S62) of
\cite{ZubJETP}. The difference with Proposition~\ref{prop:FS} is the domain $p_r>0$: each chirality supplies a ray,
not a line. Listing the angular coordinates in the order $(p_\varphi,p_\theta)$, as in (S55) of \cite{ZubJETP},
reverses both the orientation of the parametrization and the sign of $\epsilon^{bc}$, and leaves \eqref{eq:sphrays}
unchanged.

\emph{Step 3: the residues.} The poles of $F$ are the real levels $E_m$. With $M_{km}=-M_{mk}$, and with
$v_i=i[\hat H_2,\hat x_i]$, $(\hat x_1,\hat x_2)=(\hat x_\varphi,\hat x_\theta)$, which gives
$\langle m|v_i|k\rangle\langle k|v_j|m\rangle=(E_m-E_k)^2\langle m|\hat x_i|k\rangle\langle k|\hat x_j|m\rangle$,
\[
{\rm Res}_{z=E_m}F=-2\sum_k\frac{M_{mk}}{(E_m-E_k)^2}
=-2\,\epsilon^{ij}\sum_{k:\,E_k\ne E_m}\langle m|\hat x_i|k\rangle\langle k|\hat x_j|m\rangle .
\]
Write $\hat x_i=X_i+\eta_i$ with $\eta_1=-\pi_2/B$, $\eta_2=\pi_1/B$. The guiding centre $X_i$ commutes with $\pi_1$,
$\pi_2$, hence with $\hat H_2$, and does not connect different levels. The cyclotron part $\eta_i$ has no matrix
elements within a level: in the oscillator basis of $a=(\pi_1+i\pi_2)/\sqrt{2B}$ the eigenstates of
$\hat H_2=\sqrt{2B}\left(\begin{smallmatrix}0&a^\dagger\\ a&0\end{smallmatrix}\right)$ are $(|0\rangle,0)$ and
$|\pm n\rangle=(|n\rangle,\pm|n-1\rangle)/\sqrt2$, and $a$, $a^\dagger$ change the oscillator index by one. Hence
\begin{equation}
{\rm Res}_{z=E_m}F=-2\,\epsilon^{ij}\langle m|\eta_i\eta_j|m\rangle=-2\,\langle m|[\eta_1,\eta_2]|m\rangle=-\frac{2i}B
\label{eq:sphres}
\end{equation}
at every level, independently of the dispersion -- the residue of the nonrelativistic levels in the Remark after
Proposition~\ref{prop:FS}. The sum over the infinitely many levels below $\mu_0$ requires a prescription. Truncating
symmetrically to the levels $-L\le n\le L$ keeps \eqref{eq:sphres} for the interior levels, whose neighbours are
kept. The sum of all residues of the truncated $F$ vanishes identically ($M$ antisymmetric, the denominators
symmetric), so the two endpoint levels carry $+(2L-1)i/B$ each, equally by $\sigma_3\hat H_2\sigma_3=-\hat H_2$; and
the truncated $F$ is $O(|z|^{-4})$ at infinity, since $\sum_{n,k}M_{nk}=0$.

\emph{Step 4: the chiral separation invariant.} For $\hat T=\gamma^5$, $T_s=s$, and the two rays of
\eqref{eq:sphrays} join into the full line $\ell$ through $\mu_0$, traversed in the direction $-1/b$ -- the step from
(S62) to (S63) of \cite{ZubJETP}:
\[
\Sigma^{(\gamma^5)}(\epsilon)=-\frac B{4\pi}\Big[\int_{\ell_+}-\int_{\ell_-}\Big]F\,dz=-\frac B{4\pi}\int_\ell F\,dz .
\]
For $\epsilon>0$ the direction $-1/b=-(1+i\kappa\epsilon)/|b|^2$ points into the lower left. The line is rotated about
$\mu_0$ onto ${\rm Re}\,z=\mu_0$, traversed downward, without crossing a pole, and closed on the left, which encircles
the levels below $\mu_0$ clockwise:
\[
\int_\ell F\,dz=-2\pi i\sum_{E_m<\mu_0}{\rm Res}_{z=E_m}F=-2\pi i\Big(-\frac{2i}B\Big)\Big[N_<-\frac{2L+1}2\Big],
\]
$N_<=\#\{-L\le n\le L:\,E_n<\mu_0\}$,
for $\mu_0$ between the endpoint levels, the endpoint $E_{-L}$ contributing $+(2L-1)i/B$. Hence
$\Sigma^{(\gamma^5)}(\epsilon)=N_<-\frac{2L+1}2$, which no longer depends on $L$ once $L$ is large. For $\epsilon<0$ the
line points into the upper left, the rotation gives the upward line, the contour encircles the levels below $\mu_0$
counterclockwise, and the sign is reversed. Thus
\[
\Sigma^{(\gamma^5)}(0^\pm)=\pm C(\mu_0),\qquad
\begin{aligned}
C(\mu_0)&=\lim_{L\to\infty}\Big[\sum_{n=-L}^{L}\theta(\mu_0-E_n)-\frac{2L+1}2\Big]\\
&=\sgn\mu_0\Big(\frac12+\sum_{n>0}\theta(|\mu_0|-\sqrt{2Bn})\Big),\qquad \mu_0\notin\{E_n\},
\end{aligned}
\]
and
\begin{equation}
N_3^{(\gamma^5)}=\Sigma^{(\gamma^5)}(0^+)-\Sigma^{(\gamma^5)}(0^-)=2C(\mu_0)
=2\sgn\mu_0\Big(\frac12+\sum_{n>0}\theta(|\mu_0|-\sqrt{2Bn})\Big)
\label{eq:N3sph}
\end{equation}
[\cite{ZubJETP}, written there for $\mu_0>0$; in the orientation of \eqref{eq:N3T} the sign agrees with
\cite{ZubJETP}] -- the half-integer Hall sequence of the relativistic Landau problem. Without the truncation the
residue count gives $\Sigma(\epsilon)=\sgn(\epsilon)\sum_n\theta(\mu_0-E_n)$, Eq.~(S65) of \cite{ZubJETP}, divergent
because infinitely many negative levels lie below $\mu_0$. The symmetric truncation is the symmetric Dirac-sea
prescription, which the Pauli--Villars regularization of \cite{ZubJETP} implements (the negative levels cancelled one
by one, the zero level by one half). The jumps at the thresholds are fixed by the residues \eqref{eq:sphres}; the
absolute value depends on the subtraction, i.e.\ on the ultraviolet completion, as for the $M=4$ model of
Sec.~\ref{sec:FS}. For $0<|\delta|<\sqrt{2B}$, $N_3^{(\gamma^5)}=\sgn\delta$: the Fermi surface is protected like a
Weyl point, and the CSE response \eqref{eq:N3T}, $\partial J^{(5)}_i/\partial\mu=\sgn\delta\,\mathcal B_i/4\pi^2$ for a
probe field $\mathcal B$ (under the assumptions of \eqref{eq:N3T}; the $B$ of \eqref{eq:sphQ} is not this field), has
the coefficient of a single Weyl fermion of chirality $\sgn\delta$, half that of a Dirac fermion.

\emph{Step 5: the conserved current.} For $\hat T=1$, $T_s=1$ for both chiralities, and the rays add as rays, not as
a line. For $\epsilon>0$, $\ell_+$ points into the lower left and $\ell_-$ into the upper right. Each is rotated about
$\mu_0$ within its own quadrant, without crossing a pole and with vanishing arcs, onto the corresponding half of the
line ${\rm Re}\,z=\mu_0$:
\[
\Sigma^{(1)}(\epsilon)=-\frac B{4\pi}\Big[\int_{\mu_0}^{\mu_0-i\infty}+\int_{\mu_0}^{\mu_0+i\infty}\Big]F\,dz
=-\frac{iB}{2\pi}\int_0^\infty{\rm Re}\,F(\mu_0+iy)\,dy ,
\]
the last form by the reality identity $F(\bar z)=-\overline{F(z)}$, the $M_{nk}$ being purely imaginary. For
$\epsilon<0$, $\ell_+$ points into the upper left and $\ell_-$ into the lower right, and the rotations give the same
two outward half-lines. The slice value is therefore the same on both sides, and
\[
N_3^{(1)}=\Sigma^{(1)}(0^+)-\Sigma^{(1)}(0^-)=0
\]
for every $\mu_0$ away from the thresholds; the truncation only makes each half-line integral finite. In terms of
the contributions of the two chiralities, $A(\epsilon)=-\frac B{4\pi}\int_{\ell_+}F\,dz$ and
$-D(\epsilon)=-\frac B{4\pi}\int_{\ell_-}F\,dz$, the same rotations give $D(\epsilon)=\overline{A(\epsilon)}$ and
$A(-\epsilon)=-\overline{A(\epsilon)}$, so that $\Sigma^{(\gamma^5)}=A+D=2\,{\rm Re}\,A$ and
$\Sigma^{(1)}=A-D=2i\,{\rm Im}\,A$, with ${\rm Re}\,A=\frac12\sgn(\epsilon)\,C(\mu_0)$ -- the statement of
Sec.~\ref{sec:CMEplane} that the two half-lines of the plane model have equal real parts. We have checked Steps 3--5
numerically for the truncation $L=8$: the residues \eqref{eq:sphres} and $+15i/B$ at the endpoints, the
$\gamma^5$ slices $\pm C(\mu_0)$ for $\mu_0=0.3$, $1.6$, $-0.3$, $-2.3$ (with $B=1$), and the $\hat T=1$ slices equal on
both sides (for instance $0.342\,i$ at $\mu_0=0.3$).

The response coefficient \eqref{eq:N3T} of the conserved charge thus vanishes: the spherical model carries no
equilibrium current of the conserved charge in the sense in which the plane model does. The comparison with
Sec.~\ref{sec:CMEplane} locates the difference: in the plane model the second half-line is the sector $p_3<0$ of the
same scalar field and adds to the first for $\hat T=1$ -- the factor two of \eqref{eq:N3T}; in the spherical model it
is the opposite chirality, which adds for $\gamma^5$ and subtracts for $1$. This is the pattern of a Dirac fermion,
CSE present and no equilibrium current of the conserved charge, reproduced by a model that is not of Hamiltonian form.
The individual $\hat T=1$ slice values are not real, the rays not being closed contours; only their difference is the
invariant. The cylindrical model of \cite{ZubJETP} (Sec.~C.1 of its Supplement) has the same \reviewfix{local}
structure\reviewfix{; its periodic functions in both angular directions, however, bring doublers, see the end of this
subsection}.

\emph{Static-gauge divergence of the $\gamma^5$ current.} The operator \eqref{eq:sphQ} commutes with $\gamma^5$, and the $\gamma^5$ current is the
one whose response \eqref{eq:N3T} is the CSE above. In a static electric field, in the static gauge of
Sec.~\ref{sec:nonell} (local chemical potential $\mu(\mathbf x)=\mu+\mathbf E\cdot\mathbf x$, $\mu_0$ fixed, constant
probe $\mathcal B$), the spatial divergence of that current is, to the indicated order in the probes,
\[
\partial_iJ^{(5)}_i=\frac{\partial J^{(5)}_i}{\partial\mu}E_i=\frac{N_3^{(\gamma^5)}}{4\pi^2}\,
\mathbf E\cdot\boldsymbol{\mathcal B},
\]
and for the Fermi point, $0<|\delta|<\sqrt{2B}$, $\partial_iJ^{(5)}_i=\sgn\delta\,\mathbf E\cdot\boldsymbol{\mathcal B}/4\pi^2$,
the coefficient of the anomaly of a single Weyl fermion of chirality $\sgn\delta$. As in Sec.~\ref{sec:nonell} this is
the static-gauge identity for the divergence in a static field, not a derivation of the anomalous Ward identity
$\partial_t\rho_5+\nabla\cdot\mathbf J_5$ in real time, which would require the regulated current and state and
control of the ultraviolet and of the density; there is no ultraviolet degree and no index theorem behind it. A
radial branch count shows where the spherical model differs from the plane one. There the window current of the
selected sheets carried half of \eqref{eq:N3T}, Eq.~\eqref{eq:FSanomaly} with $1/8\pi^2$: the selected sheets count
$n_{\rm filled}$, whereas \eqref{eq:N3T} is the reference-subtracted signed count of all sheets, $2n_{\rm filled}$
[Sec.~\ref{sec:nonell}(e)]. Here every sphere of either chiral sector is a Fermi point with inward radial velocity,
$\partial\varepsilon/\partial p_r<0$, and the chirality-weighted count is
\[
\lim_{L\to\infty}\big[\#\{-L\le n\le L:\,E_n>\mu_0\}-\#\{-L\le n\le L:\,E_n<\mu_0\}\big]=-2C(\mu_0)=-N_3^{(\gamma^5)}:
\]
the full invariant in magnitude, both half-lines carrying Fermi points, with the minus sign of the inward velocity.
For $\hat T=1$ the count is $-(2L+1)$, independent of $\mu_0$ away from the thresholds: no change of the vector
current, in line with $N_3^{(1)}=0$. These are counts of radial branches in momentum space, not yet currents along a
direction of coordinate space: the step to a Cartesian current, and with it the relative sign between the branch
count and \eqref{eq:N3T}, requires the angular measure, the velocity projections and the probe coupling of the model,
which \cite{ZubJETP} does not provide beyond the invariant.}

\reviewfix{\emph{Periodicity in $\varphi$, the poles, and finite $a,b$.} The model \eqref{eq:sphQ} is the limit
$a,b\to0$ of a family in which $\varphi=ap_\varphi$ is an angle and the poles $\theta=0,\pi$ are points at which
$\varphi$ is undefined. The invariant is topological: it does not change under a deformation of $\hQ$ that keeps the
integrands of the slices regular. The value obtained for $a,b\to0$ therefore persists at finite $a,b$ until a phase
transition is reached -- a level $E_n(a,b)$ crossing $\mu_0$ (a Lifshitz transition, in which a Fermi sphere appears
or disappears), or a broadening of the levels into bands that closes the gap at $\mu_0$. As long as the angular
spectrum stays discrete and gapped at $\mu_0$, the description of the CSE by $N_3^{(\gamma^5)}$, and in
Sec.~\ref{sec:CMEsphere2} of the chiral magnetic like effect by $N_3^{(1)}$, extends to finite $a,b$. The argument requires that the limit
$a,b\to0$ of the periodic family be the model \eqref{eq:sphQ} itself, and not several copies of it.

The periodic function of \cite{ZubJETP}, $\hat H=f_a(p_\varphi+B\hat x_\theta)\gamma^0\gamma^1+p_\theta\gamma^0\gamma^2$
with $f_a(u)=[\sin(au)+1-\cos(au)]/a$ [Eq.~(S68) there], does not meet this requirement. A real single-valued periodic
function changes sign an even number of times per period: $f_a$ vanishes at $au=0$ with slope $+1$ and at
$au=3\pi/2$ with slope $-1$, for every $a$. Each zero carries a complete ladder of Landau levels, and the second one
does not leave the sphere as $a\to0$. Its residues are again $-2i/B$ per level, Eq.~\eqref{eq:sphres}, since they
follow from $[\eta_1,\eta_2]=i/B$ alone and not from the sign of the slope -- as the two valleys of graphene add in the
quantum Hall effect. The limit $a\to0$ of this family is thus two copies of \eqref{eq:sphQ}, and the deformation
argument gives $N_3^{(\gamma^5)}=4C(\mu_0)$ and a doubly degenerate Fermi point with invariant $2\sgn\delta$
(and $N_3^{(1)}=4C(\mu_0)$ in Sec.~\ref{sec:CMEsphere2}). The term $(1-\cos au)/a$ does not remove the doubler: it
lies in the channel of the kinetic term, not of a mass term, and only moves the second zero from $au=\pi$ to
$au=3\pi/2$. The two ladders, moreover, hybridize already at moderate $a$ (Table~\ref{tab:sphperiodic}). The
cylindrical model of Sec.~C.1 of \cite{ZubJETP}, with periodic functions in both angular directions, has four zeros.

The doubler is removed if the spinor frame, rather than the symbol, is single-valued only up to a rotation -- as the
tangent frame of a sphere is around its poles:
\begin{equation}
\hat H_a=f(p_\varphi+B\hat x_\theta)\gamma^0\gamma^1+p_\theta\gamma^0\gamma^2,\qquad f(u)=\frac2a\sin\frac{au}2,\qquad
\psi\big|_{\varphi+2\pi}=W\psi\big|_{\varphi},\quad W={\rm diag}(\sigma_2,\sigma_2).
\label{eq:sphperiodic}
\end{equation}
Here $f(u+2\pi/a)=-f(u)$, and the twist $W$, applied under the shift $p_\varphi\to p_\varphi+2\pi/a$
($\varphi\to\varphi+2\pi$), compensates the sign: in the chiral basis, where $\gamma^0\gamma^{1,2}={\rm
diag}(\sigma_{1,2},-\sigma_{1,2})$ and $\gamma^5={\rm diag}(1,-1)$, $W$ anticommutes with $\gamma^0\gamma^1$ and
commutes with $\gamma^0\gamma^2$, with $\gamma^5$, and with the terms $\hat p_r$, $\hat p_r\gamma^5$ and
$\mu_0\gamma^5$ of \eqref{eq:sphQ} and \eqref{eq:sphQ2}. The operator $\hat H_a$ thus returns to itself up to the frame
rotation $W$ after one turn, $f$ has a single zero per turn, and $f(u)\to u$ as $a\to0$: the limit is \eqref{eq:sphQ}
itself. Table~\ref{tab:sphperiodic} compares the two periodic implementations ($B=1$; flat values
$0,\pm1.414,\pm2,\pm2.449$). In the twisted model the ladder is single, the zero mode stays exactly at zero -- one zero
of $f$ per turn, protected by the index -- and the levels deform smoothly without crossing. The results
$N_3^{(\gamma^5)}=2C(\mu_0)$, $N_3^{(1)}=0$, the single Weyl-like Fermi point with $\sgn\delta$, and in
Sec.~\ref{sec:CMEsphere2} $N_3^{(1)}=2C(\mu_0)$, $N_3^{(\gamma^5)}=0$, therefore hold for the family
\eqref{eq:sphperiodic} at finite $a$, up to the first phase transition.
\begin{table}[h]
\centering\small
\begin{tabular}{c|l|l}
$a$ & $f_a$ of \cite{ZubJETP}, Eq.~(S68) & twisted model \eqref{eq:sphperiodic}\\ \hline
$0.05$ & $0,0,\ \pm1.412\,(\times2),\ \pm1.995\,(\times2)$ & $0,\ \pm1.414,\ \pm2.000,\ \pm2.449$\\
$0.3$ & $\pm0.005,\ \pm1.276,\ \pm1.381,\ \pm1.720$ & $0,\ \pm1.410,\ \pm1.989,\ \pm2.428$\\
$0.6$ & $\pm0.163,\ \pm1.238,\ \pm1.693$ & $0,\ \pm1.398,\ \pm1.952,\ \pm2.358$
\end{tabular}
\caption{Lowest levels of the angular operator with $B=1$ for the two periodic implementations, from exact
diagonalization in a plane-wave basis on one period (both sectors of the twist give the same spectrum).}
\label{tab:sphperiodic}
\end{table}

The poles are treated in the same spirit. The Mercator variable $p_\theta=b^{-1}\ln\tan(\theta/2)$ maps the sphere
without its poles onto the cylinder $p_\theta\in\mathbb R$, $\varphi\in[0,2\pi)$, and sends the poles to
$p_\theta\to\pm\infty$, where $\varphi$ is undefined. No boundary condition is needed there, $\hat x_\theta$ is
self-adjoint, and the position operators used in Step~3 are defined globally, so that the residue \eqref{eq:sphres}
holds as derived. In this form $b$ enters only through the map. The round metric would add a conformal factor
$1/\cosh(bp_\theta)$, which the model does not contain. If it is included with a uniform field, the angular problem
becomes the Dirac operator on a sphere of radius $1/b$ with $N=2B/b^2$ flux quanta, $N$ an integer, with the levels
$E_n=\sgn(n)\sqrt{2B|n|+b^2n^2}$ of degeneracy $N+2|n|$ and $N$ zero modes of one chirality by the index theorem: the
levels never cross and the zero modes are protected, but at fixed $B$ the radius takes discrete values. The invariant
on the round sphere itself, where the position operator is not global, has not been recomputed here.}

\subsection{\reviewfix{A spherical model with the chiral magnetic like effect}}
\label{sec:CMEsphere2}

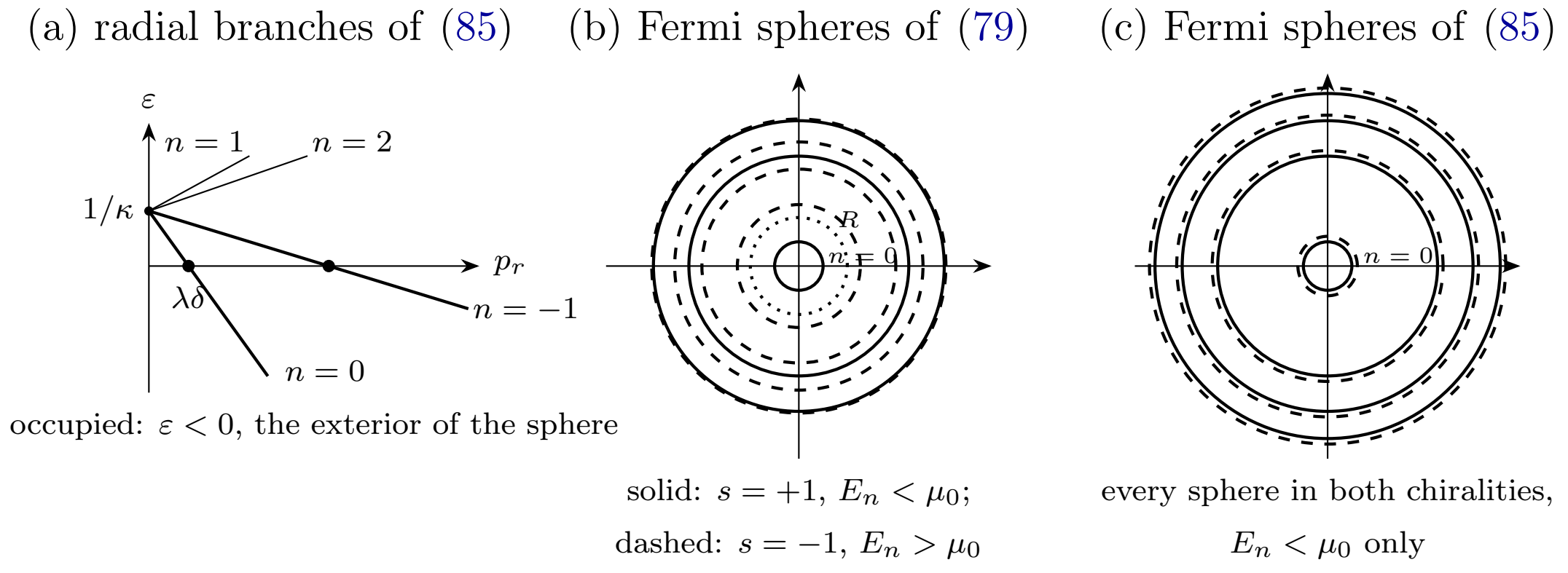
\begin{figure}[!htb]
\centering
\begin{tikzpicture}[>=Stealth]
% (a) radial dispersions of (55)
\begin{scope}[xshift=-5.6cm]
\draw[->] (0,0)--(3.0,0) node[right]{\scriptsize$p_r$};
\draw[->] (0,-1.15)--(0,1.3) node[above]{\scriptsize$\varepsilon$};
\node[left] at (0,0.5) {\scriptsize$1/\kappa$};
\fill (0,0.5) circle (1.2pt);
% n=0: eps = 1 - p/0.4  (y scale 0.5, x scale 0.9)
\draw[thick] (0,0.5)--(1.08,-1.0);
\node[right] at (1.1,-0.95) {\scriptsize$n=0$};
% n=-1: eps = 1 - p/1.814
\draw[thick] (0,0.5)--(2.9,-0.388);
\node[right,inner sep=1pt] at (2.9,-0.4) {\scriptsize$n=-1$};
% n=1: eps = 1 + p/1.014 ; n=2: 1 + p/1.6 (empty sectors, thin)
\draw[thin] (0,0.5)--(0.913,1.0); \node[above left,inner sep=1pt] at (0.913,1.0) {\scriptsize$n=1$};
\draw[thin] (0,0.5)--(1.44,1.0); \node[above right,inner sep=1pt] at (1.44,1.0) {\scriptsize$n=2$};
% Fermi spheres (zeros)
\foreach \x in {0.36,1.633} {\fill (\x,0) circle (1.6pt);}
\node[below] at (0.36,-0.05) {\scriptsize$\lambda\delta$};
\node[anchor=north,align=center] at (1.5,-1.2) {\scriptsize occupied: $\varepsilon<0$, the exterior of the sphere};
\node at (1.1,2.15) {(a) radial branches of \eqref{eq:sphQ2}};
\end{scope}
% (b) spheres of (53)
\begin{scope}[xshift=0.3cm]
\draw[->] (-1.75,0)--(1.75,0); \draw[->] (0,-1.75)--(0,1.75);
\foreach \r in {0.22,0.998,1.32} {\draw[thick] (0,0) circle (\r);}
\foreach \r in {0.558,0.88,1.127,1.335} {\draw[thick,dashed] (0,0) circle (\r);}
\draw[dotted,thick] (0,0) circle (0.44);
\node[right,inner sep=1pt] at (0.22,0.1) {\tiny$n=0$};
\node[anchor=south west,inner sep=1pt] at (0.31,0.31) {\tiny$R$};
\node at (0,2.15) {(b) Fermi spheres of \eqref{eq:sphQ}};
\node[anchor=north,align=center] at (0,-1.8) {\scriptsize solid: $s=+1$, $E_n<\mu_0$;\\ \scriptsize dashed: $s=-1$, $E_n>\mu_0$};
\end{scope}
% (c) spheres of (55)
\begin{scope}[xshift=5.1cm]
\draw[->] (-1.75,0)--(1.75,0); \draw[->] (0,-1.75)--(0,1.75);
\foreach \r in {0.22,0.998,1.32,1.567} {\draw[thick] (0,0) circle (\r); \draw[thick,dashed] (0,0) circle (\r+0.05);}
\node[right,inner sep=1pt] at (0.28,0.1) {\tiny$n=0$};
\node at (0,2.15) {(c) Fermi spheres of \eqref{eq:sphQ2}};
\node[anchor=north,align=center] at (0,-1.8) {\scriptsize every sphere in both chiralities,\\ \scriptsize $E_n<\mu_0$ only};
\end{scope}
\end{tikzpicture}
\caption{The spherical models, for $\mu_0=\delta=0.4$ and $B=\kappa=\lambda=1$ (levels $E_n=\sgn(n)\sqrt{2|n|}$).
(a) Radial dispersions $\varepsilon_n(p_r)=1/\kappa+p_r/X_n$ of the model \eqref{eq:sphQ2}, the same for both
chiralities. All branches start at $1/\kappa$; the levels below $\mu_0$ ($X_n<0$; $n=0,-1$ drawn) fall through
zero at their Fermi spheres (dots) and continue to $-\infty$, so that the occupied set of a sector is the exterior of
its sphere, with a single boundary and no junction; the levels above $\mu_0$ ($X_n>0$; $n=1,2$, thin) are empty.
(b) Section through the origin of momentum space for the model \eqref{eq:sphQ}: spheres of chirality $s=+1$ (solid)
for the levels below $\mu_0$, radii $\lambda(\mu_0-E_n)$, and of chirality $s=-1$ (dashed) for the levels above,
radii $\lambda(E_n-\mu_0)$; the dotted circle is a ball of radius $R$, $\lambda\delta<R<\lambda(\sqrt{2B}-\delta)$,
that isolates the small sphere of the level $n=0$. (c) The same for the model \eqref{eq:sphQ2}: every
level below $\mu_0$ has its sphere in both chiralities (coincident; the dashed circle is drawn slightly offset for
visibility), the levels above have none. The two copies carry the same inward radial velocity: they add in the conserved current and cancel in the
$\gamma^5$ current, Secs.~\ref{sec:CMEsphere2} and~\ref{sec:nonell}(g).}
\label{fig:sph}
\end{figure}

\reviewfix{The two results of Sec.~\ref{sec:CMEsphere} have a common origin: in the sector of chirality $s$ the slice
variable is $z_s=\mu_0-\frac{s\,p_r}{\lambda(1-i\kappa\epsilon)}$, so that the two sectors run along opposite rays from
$\mu_0$. The weight $\gamma^5=s$ compensates the orientation $s$ and the rays add; the weight $1$ of the conserved
current does not, and they subtract. Two blocks that coincide up to a constant sign, with the same domains and the
same regulator, add for $\hat T=1$ and cancel for $\hat T=\gamma^5$; any modification that gives the two sectors the
same orientation thus exchanges the roles of the two currents. The simplest one correlates the radial term with the
chirality,
\begin{equation}
\hQ'=i\kappa\lambda\hat p_4(\hat H-\mu_0\gamma^5)-\hat p_r\gamma^5-(\hat H-\mu_0\gamma^5)\lambda ,
\label{eq:sphQ2}
\end{equation}
with $\hat H$ as in \eqref{eq:sphQ}, $B,\kappa,\lambda>0$, the chemical potential $\mu=0$ and $\mu_0$ fixed, in the flat
angular limit of Sec.~\ref{sec:CMEsphere} and with the reservations stated there\reviewfix{; the doubler-free
periodic implementation \eqref{eq:sphperiodic} applies unchanged, since the twist $W$ commutes with $\hat p_r\gamma^5$}. In the block of chirality $s$, where
$\gamma^5\to s$ and $\hat H\to s\hat H_2$,
\[
Q'_s=s\,\big[i\kappa\lambda p_4(\hat H_2-\mu_0)-p_r-\lambda(\hat H_2-\mu_0)\big]:
\]
the two sectors coincide up to the overall sign $s$, which drops out of $G\,dQ\,dG\,dQ$ ($Q'_s=sq$ gives
$G'_s=sq^{-1}$). On the slice $p_4=\epsilon$, $Q'_s=s\lambda(1-i\kappa\epsilon)(z-\hat H_2)$ with
$z=\mu_0-p_r/\lambda(1-i\kappa\epsilon)$ for both $s$: the two sectors follow the same ray. In Steps 1 and 2 of Sec.~\ref{sec:CMEsphere} the only change is
$\partial_{p_r}Q'_s=-s$, which cancels the sign of the block, so that each chirality contributes the integral
$A(\epsilon)$ along the ray $\ell_+$ of \eqref{eq:sphrays}. Everything else is read off
from Sec.~\ref{sec:CMEsphere}.

\emph{Fermi surfaces.} In the sector of the level $n$, $Q'_{n,s}=s\kappa\lambda(E_n-\mu_0)(ip_4-\varepsilon_n)$ with
$\varepsilon_n=\frac1\kappa+\frac{p_r}{\kappa\lambda(E_n-\mu_0)}$, the same for both chiralities: a sphere $p_r=\lambda(\mu_0-E_n)$
for every level below $\mu_0$, now in both sectors -- two coincident spheres of opposite chirality per level -- and no
sphere for the levels above $\mu_0$ (Fig.~\ref{fig:sph}). For $\mu_0=\delta\to0$ the level $n=0$ gives, for $\delta>0$, the pair of coincident spheres of radius
$\lambda\delta$, and the levels $n<0$ the spheres of radii $\lambda(\delta+\sqrt{2B|n|})\to\lambda\sqrt{2B|n|}$.
Unlike in \eqref{eq:sphQ}, the zero-level spheres exist for $\delta>0$ only: their common frequency coefficient, after
the sign $s$ is factored out, is $X_0=-\kappa\lambda\delta$ for both chiralities, so for $\delta<0$ the zero level
has no sphere in either sector. At $\delta=0$ itself $Q'_{0,s}=-sp_r$ is independent of the frequency and vanishes at
$p_r=0$ on every slice: the transition passes through a singular operator, the junction of the model, at which the
slice invariant is undefined. In the symmetric prescription below $N_3^{(1)}$ changes from $+1$ to $-1$ as $\delta$
decreases through zero. The jump $-2$ is fixed by the level crossing and is the invariant of the pair that
disappears -- the per-piece invariant of \cite{ZubJETP}, its Eq.~(58), on a hypersurface enclosing the small pair is
the residue of the Hall function at $E_0$, one unit per sector (see the residue count below) -- whereas the absolute
value $-1$ on the negative side includes the sea subtraction and cannot be attributed to the remaining spheres. The
pair is thus stable for $\delta\ne0$ and is lost only through the junction, where the operator leaves its class: the
Fermi point is a threshold, not a protected object -- the counterpart, for equal orientations of the two sectors, of
the chirality flip of \eqref{eq:sphQ}. The coefficient of $ip_4$ in $Q'_{n,s}$ is $sX_n$ with
$X_n=\kappa\lambda(E_n-\mu_0)$, constant within a sector and changing sign across the thresholds; on a sphere,
$X_n<0$, the pole of $G'_{n,s}=(sX_n)^{-1}(ip_4-\varepsilon_n)^{-1}$ has negative residue for $s=+1$ and positive
residue for $s=-1$. The model is not of Hamiltonian form, exactly as \eqref{eq:sphQ}, and Bloch's theorem does not
apply to it.}

\reviewfix{\emph{Invariants.} Both sectors supply the same ray integral $A(\epsilon)$ of Sec.~\ref{sec:CMEsphere}, with
a common trace normalization and regulator and in the symmetric prescription $C(\mu_0)$ of that section,
$\mu_0\notin\{E_n\}$. The contour argument needs the regulated Hall function to satisfy $F(\bar z)=-\overline{F(z)}$,
to have its poles on the real axis only and to decay at infinity. The symmetric truncation of
Sec.~\ref{sec:CMEsphere} (Steps 3--5) has these properties and gives ${\rm Re}\,A(\epsilon)=\frac12C(\mu_0)$ for
$\epsilon>0$ and $\mu_0$ between the endpoint levels. For
$\hat T=1$ the slice is $2A$ and, with $A(-\epsilon)=-\overline{A(\epsilon)}$,
\begin{equation}
N_3^{(1)}=\lim_{\epsilon\downarrow0}2\big[A(\epsilon)-A(-\epsilon)\big]=4\,{\rm Re}\,A=2C(\mu_0)
=2\sgn\mu_0\Big(\frac12+\sum_{n>0}\theta(|\mu_0|-\sqrt{2Bn})\Big)
\label{eq:N3sph2}
\end{equation}
in the symmetric prescription of \eqref{eq:N3sph}, while for $\hat T=\gamma^5$ the weights $s$ give $A-A=0$,
$N_3^{(\gamma^5)}=0$ -- exactly, the two blocks having identical domains, couplings and regulator. The conserved
current of \eqref{eq:sphQ2} has the invariant that the $\gamma^5$ current has in \eqref{eq:sphQ}, within the same
prescription, and the $\gamma^5$ current has none: under the assumptions of \eqref{eq:N3T},
$\partial\langle J_i\rangle/\partial\mu\,|_{\mu=0,\,\mu_0\ {\rm fixed}}=N_3^{(1)}\mathcal B_i/4\pi^2$ for a probe field
$\mathcal B$ and $\partial\langle J^{(5)}_i\rangle/\partial\mu=0$ -- an equilibrium current of the conserved charge, a
chiral magnetic like effect with spherical Fermi surfaces, and no chiral separation effect -- the closest of these models to the
chiral magnetic effect proper, $\mu_0$ playing the role of $\mu_5$ (end of Sec.~\ref{sec:FS}). For $0<|\delta|<\sqrt{2B}$,
$N_3^{(1)}=\sgn\delta$, the small spheres existing on the side $\delta>0$ only. The slice values $2A(\pm\epsilon)$ are
complex; only their difference is the invariant.

\emph{The non-Abelian chiral magnetic like effect.} The response \eqref{eq:N3T} is derived in Sec.~\ref{sec:CSEder} for the
non-Abelian field strength, the Fermi-surface structure of the model being colour-blind: for a fermion that carries a
colour index, with the symbol \eqref{eq:sphQ2} multiplied by the unit matrix in colour space and the Fermi spheres
degenerate in colour, the current linear in a colour-magnetic probe $\mathcal B^a_i=\frac12\epsilon_{ijk}F^a_{jk}$ is
obtained exactly as the non-Abelian CSE \eqref{eq:CSE}, the colour trace supplying the index of the adjoint
current:
\[
\frac{\partial\langle J^a_i\rangle}{\partial\mu}\Big|_{\mu=0,\,\mu_0\ {\rm fixed}}
=\frac{N_3^{(1)}}{4\pi^2}\,\trG\big(t^a\boldsymbol{\mathcal B}_i\big)=\frac{T_R\,N_3^{(1)}}{4\pi^2}\,\mathcal B^a_i ,
\qquad \boldsymbol{\mathcal B}_i=\mathcal B^b_it^b,\quad \trG(t^at^b)=T_R\delta^{ab},
\]
with $N_3^{(1)}$ of \eqref{eq:N3sph2} and $T_R=\frac12$ for the fundamental representation -- the factor is the
normalization of the adjoint component implicit in the trace of \eqref{eq:CSE}, not the colour multiplicity -- a
non-Abelian chiral magnetic like effect, the vector counterpart of the
non-Abelian CSE of Sec.~\ref{sec:CSE}, carried by spherical Fermi surfaces. The same holds for the plane model of
Sec.~\ref{sec:FS} with $N_3^{(T)}$ of \eqref{eq:N3T}, and the non-Abelian version of the CSE of
Sec.~\ref{sec:CMEsphere} is \eqref{eq:N3T} with $\hat T=\gamma^5$ and $N_3^{(\gamma^5)}$ of \eqref{eq:N3sph}; in all
three cases the colour structure enters only through the trace, the invariant is that of the colourless model, and
the conditions of Sec.~\ref{sec:CSEder} on the response are the same: zero temperature, a static background, linear
response to a slowly varying probe, and the convention of that section for the adjoint components, which are not
gauge covariant when averaged -- a colour-magnetic probe constant over the supercell in a fixed gauge.

\emph{Static-gauge divergence.} In the static gauge of Sec.~\ref{sec:nonell} ($\mu(\mathbf x)=\mu+\mathbf E\cdot\mathbf x$, $\mu_0$
fixed, constant probe) the divergence of the conserved current is
$\partial_iJ_i=N_3^{(1)}\,\mathbf E\cdot\boldsymbol{\mathcal B}/4\pi^2$ to the indicated order. For a conserved charge
this cannot be a production: by $\partial_t\rho+\nabla\cdot\mathbf J=0$ it is the outflow through the boundary of the
sample of Sec.~\ref{sec:nonell}(e), the enclosed charge decreasing at that rate -- the state is not stationary, or
sources at the ends of the sample must be specified. The $\gamma^5$ current has no response and no static-gauge
divergence; as in Sec.~\ref{sec:nonell}, the static identity is not a derivation of the real-time Ward identity. The
radial branch count of Sec.~\ref{sec:CMEsphere} applies with the chirality weight removed: every sphere has inward
velocity and both sectors count alike, $-2\#\{E_n<\mu_0\}$; with the cutoff $-L\le n\le L$ and the reference $2L+1$
subtracted this is $-2C(\mu_0)=-N_3^{(1)}$, the full invariant in magnitude -- a subtracted count of radial branches,
with the reservations of Sec.~\ref{sec:CMEsphere} about its identification with a Cartesian current.

\emph{Two variants.} Projecting \eqref{eq:sphQ} onto a single two-component block,
$Q_{\rm one}=i\kappa\lambda p_4(\hat H_2-\mu_0)-p_r-\lambda(\hat H_2-\mu_0)$, leaves a single sector and a single ray:
$N_3^{(1)}=2\lim_{\epsilon\downarrow0}{\rm Re}\,A(\epsilon)=C(\mu_0)$, half-integer at the Fermi point
($\frac12\sgn\delta$) -- admissible in the continuum, where slice pairings need not be integers
(Sec.~\ref{sec:NCG}), but less clean. (Merely replacing $\gamma^5$ by $1$ in \eqref{eq:sphQ} does not do this: with
$\hat H={\rm diag}(\hat H_2,-\hat H_2)$ two blocks remain, and since $-\hat H_2$ is unitarily equivalent to $\hat H_2$
they are two copies of $Q_{\rm one}$ with the same ray -- the model \eqref{eq:sphQ2} again, $N_3^{(1)}=2C(\mu_0)$.)
The single-block model with a bounded-below angular Hamiltonian, nonrelativistic Landau levels $\omega(n+\frac12)$,
$\omega>0$, on the angular variables, needs no Dirac-sea subtraction and has
$N_3^{(1)}=n_{\rm filled}=\#\{n\ge0:\omega(n+\frac12)<\mu_0\}$ from the convergent paired Hall sum: the spherical
counterpart of the half $p_3>0$ of the plane model (the radial dispersion remains unbounded below on the
sphere-carrying branches). The model \eqref{eq:sphQ2} keeps the shrinking Fermi sphere and the integer invariant, and
shows in one line that the chiral separation and chiral magnetic like effects of these models are the two orientations of
one and the same Hall-function integral.}

%======================================================================
\section{The non-Abelian quantum Hall effect}
\label{sec:QHE}

\subsection{Derivation}

For the vector current, \eqref{eq:Jvec} with \eqref{eq:Gexp} gives [Ref.~\cite{XZPRD}, Eqs.~(142)--(143)]
\begin{align}
J_\mu&=S^v_\mu+M^v_{\mu\alpha\beta}F_{\alpha\beta},\qquad \corr{S^v_\mu=-\sum_iC_i}\int\frac{d^4p}{(2\pi)^4}\trD(\partial_{p_\mu}QG^{(0)}),\\
M^{(v)}_{\mu\alpha\beta}&=\frac i4\sum_iC_i\int\frac{d^4p}{(2\pi)^4}\trD\big(\partial_{p_\mu}Q\,G^{(0)}\star\partial_{p_{[\alpha}}Q\star\partial_{p_{\beta]}}G^{(0)}\big).
\end{align}
\corr{For the conserved $U(1)$ electric current, the Bloch argument \cite{Bloch} excludes a bulk equilibrium transport current in the thermodynamic limit under its usual locality and boundary assumptions. It does not by itself exclude arbitrary non-Abelian component currents or local magnetization currents.} \reviewfix{After the cyclic spatial trace and the removal of the momentum and spatial boundary terms only the
totally antisymmetric part of $M^{(v)}_{\mu\alpha\beta}$ contributes to the bulk Hall coefficient (the symmetric parts
are total derivatives; for a local current they are magnetization and divergence terms),} so
$J_\mu=\epsilon_{\nu\mu\alpha\beta}M^{(v)}_\nu F_{\alpha\beta}$ with $M^{(v)}_\nu=\frac{i}{12}\epsilon_{\nu\mu\alpha\beta}
\sum_iC_i\int\frac{d^4p}{(2\pi)^4}\trD(\partial_\mu QG\star\partial_\alpha Q\star\partial_\beta G)$ [Ref.~\cite{XZPRD}, Eqs.~(144)--(145)].
For slowly varying fields and $F_{4i}=-iE_i$ [Ref.~\cite{XZPRD}, Eqs.~(146)--(148)]
\begin{equation}
\bar J^{v,\rm QHE}_i=-\frac1{4\pi^2}\epsilon_{ijk}\tilde{\cM}^{(v)}_jE_k,\qquad
\tilde{\cM}^{(v)}_i=-\frac{1}{24\pi^2V}\Big[\int d^3x\wedge dp_i\wedge\trD\big(G^{(0)}\star dQ\star\wedge dG^{(0)}\star\wedge dQ\big)\Big]_{\rm reg},
\label{eq:QHE3}
\end{equation}
``reg'' meaning the Pauli--Villars subtraction of the Dirac sea.\footnote{In Ref.~\cite{XZPRD}, Eq.~(147) the relation between
$\tilde{\cM}^{(v)}$ and $M^{(v)}$ is written as $\tilde{\cM}^{(v)}=2iM^{(v)}$; consistency between (144)--(146) and
(170), (178) requires $\tilde{\cM}^{(v)}=8\pi^2i\,M^{(v)}$.} Similarly
$\bar\rho^v=\frac1{4\pi^2}\tilde{\cM}^{(v)}_iB_i$ and $\frac1{4\pi^2}\tilde{\cM}^{(v)}_i=\partial\cM_i/\partial\mu$
with $\cM$ the magnetization. \reviewfix{The precise statement, with the conditions of Sec.~\ref{sec:logZ}, is given below in two versions (paragraph
``The two versions of the non-Abelian QHE'').} In $2+1$D the same steps give [Ref.~\cite{XZPRD}, Eqs.~(153)--(160)]
\begin{equation}
\bar J^{v,\rm QHE}_i=\frac1{2\pi}\epsilon_{ij}\cM^v_3E_j,\qquad
\cM^v_3=-\frac{1}{24\pi^2S}\Big[\int d^2x\wedge\trD\big(G^{(0)}\star dQ\star\wedge dG^{(0)}\star\wedge dQ\big)\Big]_{\rm reg},
\label{eq:QHE2}
\end{equation}
\begin{GPTcorrection}
This is an integer invariant for a smooth invertible Green symbol with an appropriate closed momentum/frequency
space, or a fixed ultraviolet completion, and a closed trace. The physical integer Hall interpretation also
requires a gapped phase and the current-response hypotheses. In the gapless continuum examples below the
$3+1$D coefficient is an integral of slice invariants and varies with $\mu$ or node separation $b$.
This does not exclude the quantized reciprocal-lattice Hall vector of a fully gapped periodic three-dimensional
system, whose two-dimensional Chern numbers are constant as the remaining crystal momentum varies.
\end{GPTcorrection}

\paragraph{\reviewfix{The two versions of the non-Abelian QHE.}}
\reviewfix{As for the CSE (Sec.~\ref{sec:CSEder}), the results \eqref{eq:QHE3}, \eqref{eq:QHE2} admit two
formulations.}

\reviewfix{\emph{Version I: truncated current, arbitrary generator.} The current is represented by the leading term
\eqref{eq:Jvec}, and $G_W$ by \eqref{eq:Gexp}: $J_\mu(x)=S^v_\mu+M^v_{\mu\alpha\beta}(x)F_{\alpha\beta}(x)$ as a local
relation between colour matrices, and for every generator $t^a$
\begin{align}
&J^a_i(x)=\trG\big(t^aJ_i(x)\big),\qquad E^a_k=i\,\trG\big(t^aF_{4k}\big),\notag\\
&\bar J^{a}_i=-\frac1{4\pi^2}\epsilon_{ijk}\tilde{\cM}^{(v)}_jE^a_k\ \ (3+1\mathrm D),\qquad
\bar J^{a}_i=\frac1{2\pi}\epsilon_{ij}\cM^v_3E^a_j\ \ (2+1\mathrm D).
\label{eq:QHEI}
\end{align}
No condition is imposed on $t^a$. The conditions are: (a) linear response in $F$; (b) for the local relation, the
slow variation \eqref{eq:slowcond}, $(\ell/L)^2/24\ll1$, or one of the cases (i), (ii) of Sec.~\ref{sec:logZ}, and
the magnetization terms mentioned above are to be kept; (c) for the averaged relation \eqref{eq:QHEI}, a
covariantly constant field with $E^a_k$ constant over the supercell in the chosen gauge, for which the omitted terms
of the series of the current average to zero at the linear order in $F$. The components $\bar J^a$ and $E^a$ refer to
the chosen gauge.}

\reviewfix{\emph{Version II: exact current, covariantly constant weight.} Let $n(x)$ be a covariantly constant colour
matrix, $\cD_\mu n=0$ ($n=1$ for an Abelian field or for the gauge-singlet current; $n=t^a$ with $[A_\mu,t^a]=0$ for
a field directed along $t^a$ or lying in the Cartan subalgebra, in the gauge where $A_\mu$ lies there;
$n(x)=\Omega(x)t^a\Omega^\dagger(x)$ in a general gauge). By \eqref{eq:Javexact} the averaged vector current is given
by the leading term exactly,
\begin{equation}
\bar J_i[n]=-\frac1V\int d^4x\,\trG\Big[n\,\trD\int\frac{d^4p}{(2\pi)^4}\,G_W\,\partial_{p_i}Q_W\Big],\qquad\cD n=0,
\label{eq:QHEexact}
\end{equation}
with the regulator contributions understood, to all orders of the derivative expansion and for any strength of the
field: no truncation of the current is made. With the expansion of the exact $G_W$ in the field,
\begin{align}
&\bar J_i[n]=-\frac1{4\pi^2}\epsilon_{ijk}\tilde{\cM}^{(v)}_jE^{(n)}_k+O(F^2)\ \ (3+1\mathrm D),\qquad
E^{(n)}_k=i\,\trG\big(nF_{4k}\big),\notag\\
&\bar J_i[n]=\frac1{2\pi}\epsilon_{ij}\cM^v_3E^{(n)}_j+O(F^2)\ \ (2+1\mathrm D),
\label{eq:QHEII}
\end{align}
for the gauge-invariant field $E^{(n)}_k$ constant over the supercell (for a varying $E^{(n)}$ its derivatives enter
only together with an inhomogeneity of the background). Both sides are gauge invariant. The Hall coefficients are
exact as the coefficients of the first power of the field, without gradient corrections; the terms of the order
$F^2$ and higher (the nonlinear response, including the dependence of the Hall coefficient on the magnetic field)
are not covered.}

\reviewfix{\emph{Relation of the two versions.} It is the same as for the CSE: for a covariantly constant field the
component with a nonzero response in Version~I is directed along $F$ in colour space and can be taken as $n$, and
the two versions coincide at the linear order in the field. If the field admits no covariantly constant weight other
than $n=1$, only Version~I is available for the adjoint components, as a local relation.}

\begin{remark}[Bloch's theorem and the meaning of the average]
\corr{For a static field $E=-\nabla A_0$ with periodic $A_0$, $\int d^3x\,E=0$. This implies a vanishing integrated Hall current when the response coefficient is uniform. For a spatially varying coefficient, $\int E=0$ alone is insufficient; the full equilibrium current, including magnetization contributions, must be used. Equations \eqref{eq:QHE3}--\eqref{eq:QHE2} specify the bulk linear response under the averaging assumptions above.}
\end{remark}

\subsection{Examples and prior work}

$2+1$D: non-relativistic fermions in an $SU(N)$ field $A$ and a constant $U(1)$ field $C_{12}=H$,
$\hQ=-\partial_\tau+\mu-A_0-C_0+\vec D^2/2M$: $\cM^v_3=n$, the number of occupied Landau levels [Ref.~\cite{XZPRD}, Eqs.~(161)--(162)];
relativistic fermions with vierbein: the unregularized value counts the infinite Dirac sea, with Pauli--Villars
$\cM^v_3=\sum_iC_in_i=n_{>0}$ (occupied levels above zero) for $\mu>0$ \corrB{and for the sign of the regulator mass
for which the zero-energy level is not counted; the opposite sign gives $n_{>0}+1$, the two choices differing by the
parity-anomaly contribution discussed below} [Ref.~\cite{XZPRD}, Eqs.~(163)--(165)]. $3+1$D: the Weyl
semimetal $\hQ=\gamma^\mu iD_\mu+\gamma^\mu\gamma^5b_\mu$, for which $\tilde N^{(0)(3)}_3(P)=1$ for $P\in(-|b|,|b|)$,
$\tilde{\cM}^{(v)}=2b$, $\bar J_i=-\frac{\epsilon_{ijk}}{2\pi^2}b_jE_k$\reviewfix{ (for a two-node model whose slice Chern number is zero outside the nodes and one between them,
$\int_{-|b|}^{|b|}dP=2|b|$; a filled-band contribution of the lattice completion would add to it)} [Ref.~\cite{XZPRD}, Eqs.~(166)--(170)] -- the known anomalous Hall
conductivity, now for the non-Abelian current; \reviewfix{and massless Dirac fermions in a magnetic field $H>0$ along $x_3$, for which the lowest Landau level
gives $\tilde{\cM}^{(v)}=2\mu\,\hat e_3$ [Ref.~\cite{XZPRD}, Eq.~(178)] -- the complete result for $\mu^2<2H$; for
larger $\mu$ the levels $E_{n,\pm}(p_3)=\pm\sqrt{p_3^2+2Hn}$ with $2Hn<\mu^2$ are occupied for
$|p_3|<\sqrt{\mu^2-2Hn}$ and contribute as well}\footnote{In Ref.~\cite{XZPRD}, Eq.~(171) the term $\gamma^\mu\gamma^5b_\mu$ is to be added and $\kappa$ defined;
in (176) $B$ stands for $H$. The results are unaffected.}

\paragraph{Prior work.} In $2+1$D the response of a non-Abelian current to a non-Abelian electric field is the
induced non-Abelian Chern--Simons term, with level $\pm\frac12$ per two-component massive fermion: Niemi and Semenoff
\cite{NiemiSemenoff}, Redlich \cite{Redlich1,Redlich2}, Alvarez-Gaum\'e, Della Pietra and Moore \cite{AGDPM}; for
non-relativistic electrons with spin as an $SU(2)$ gauge field, the $U(1)\times SU(2)$ generalized Hall effects of
Fr\"ohlich and Studer \cite{FrohlichStuder1,FrohlichStuder2}. In $3+1$D the non-Abelian generalization of the anomalous
Hall effect of (multi-)Weyl semimetals was obtained in \cite{Dantas}; the four-dimensional QHE with an $SU(2)$ Yang
monopole \cite{ZhangHu} is a different setting. \begin{GPTcorrection}
The parity-odd contribution of one continuum two-component massive Dirac fermion is a half-level before a
gauge-invariant ultraviolet completion is fixed. With one Pauli--Villars subtraction the level is, up to the
overall orientation convention, $k=\tfrac12[\sgn(m)-\sgn(M)]$, hence integer for nonzero masses. The regulator
choice supplies the parity anomaly: gauge invariance and parity cannot both be retained for a single such
fermion. Consequently a total Hall coefficient cannot be assigned by counting only physical positive-energy
levels without specifying the regulator or lattice completion \cite{Redlich1,Redlich2}.
\end{GPTcorrection}

\subsection{Non-renormalization by interactions: the renormalized velocity}
\label{sec:QHEint}

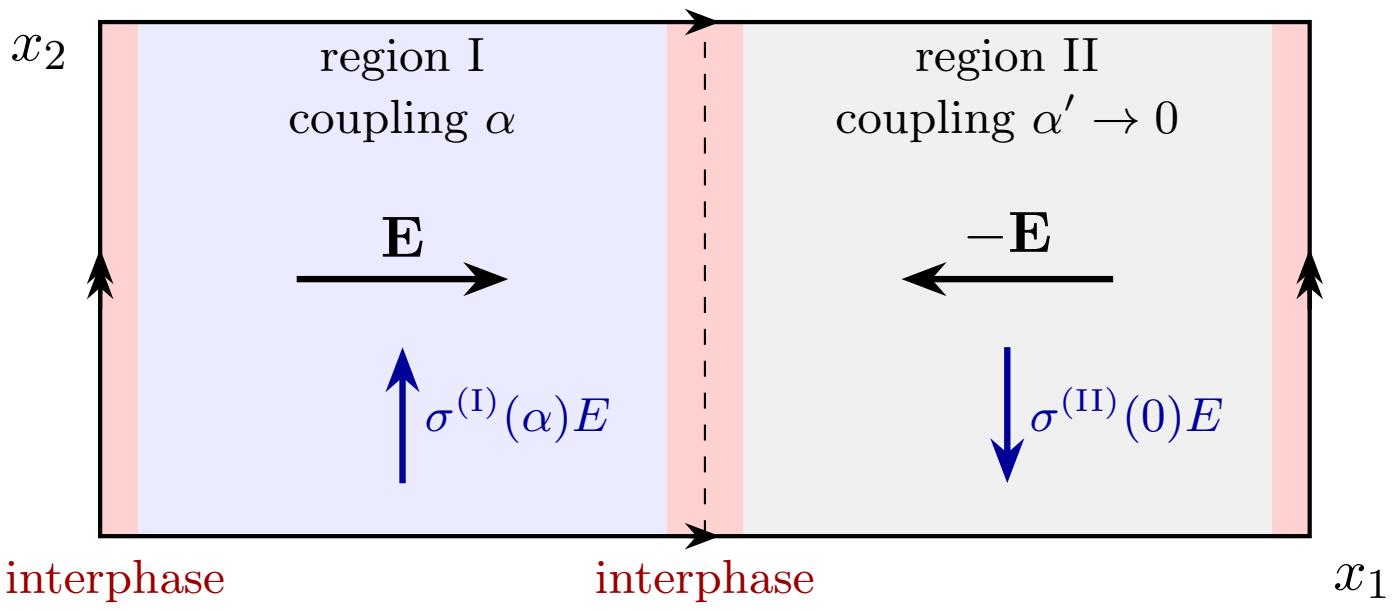
\begin{figure}[htb]
\centering
\begin{tikzpicture}[scale=1.0,>=Stealth]
\fill[blue!8] (0,0) rectangle (4,3.4);
\fill[gray!12] (4,0) rectangle (8,3.4);
\fill[red!18] (3.75,0) rectangle (4.25,3.4);
\fill[red!18] (0,0) rectangle (0.25,3.4);
\fill[red!18] (7.75,0) rectangle (8,3.4);
\draw[thick] (0,0) rectangle (8,3.4);
\draw[dashed] (4,0)--(4,3.4);
% identification marks
\draw[->,thick] (3.9,3.4)--(4.1,3.4); \draw[->,thick] (3.9,0)--(4.1,0);
\draw[->,thick] (0,1.6)--(0,1.8); \draw[->,thick] (8,1.6)--(8,1.8);
\draw[->,thick] (0,1.5)--(0,1.9); \draw[->,thick] (8,1.5)--(8,1.9);
% fields: along x_1, the gradient of a periodic zigzag potential A_0(x_1) (corrected)
\draw[->,very thick] (1.3,1.7)--(2.7,1.7) node[midway,above]{$\mathbf E$};
\draw[->,very thick] (6.7,1.7)--(5.3,1.7) node[midway,above]{$-\mathbf E$};
% Hall currents: along x_2, parallel to the interfaces (corrected)
\draw[->,very thick,blue!60!black] (2.0,0.35)--(2.0,1.25) node[midway,right]{\footnotesize$\sigma^{(\rm I)}(\alpha)E$};
\draw[->,very thick,blue!60!black] (6.0,1.25)--(6.0,0.35) node[midway,right]{\footnotesize$\sigma^{(\rm II)}(0)E$};
\node[align=center] at (2,2.95) {\footnotesize region I\\[-2pt]\footnotesize coupling $\alpha$};
\node[align=center] at (6,2.95) {\footnotesize region II\\[-2pt]\footnotesize coupling $\alpha'\to0$};
\node[red!60!black] at (4,-0.3) {\footnotesize interphase};
\node[red!60!black] at (0.1,-0.3) {\footnotesize interphase};
\node at (8.35,-0.3) {$x_1$}; \node at (-0.4,3.2) {$x_2$};
\end{tikzpicture}
\caption{The two-region gedankenexperiment of Ref.~\cite{ZZAnn} on the torus $T^2$ (opposite sides identified). In
region I the interaction has coupling $\alpha$ and the electric field is $\mathbf E$\corrB{, directed along $x_1$}; in region II the coupling is
switched off and the field is $-\mathbf E$, so that the field is periodic\corrB{ (it is the gradient of a periodic zigzag
potential $A_0(x_1)$; a field along $x_2$ reversing between the regions would have a nonzero curl at the interfaces and
could not be static). The Hall currents flow along $x_2$, parallel to the interfaces, so that no charge accumulates at
them}. The total current along \corrB{$x_2$ (through a cut $x_2={\rm const}$)} is
$I_{\rm pers}+\sigma^{(\rm I)}(\alpha)E-\sigma^{(\rm II)}(0)E$, \reviewfix{and the equality of the two bulk conductivities follows once the interface currents are independent
of the interaction, which is the case for thin interphases and short-ranged interactions (a thin interface can carry
a finite integrated current, so its small area alone does not justify discarding it).} The device is clean when the two regions are
coupled only through the thin interphases (shaded), i.e.\ for short-ranged interactions; for Coulomb interactions the
regions remain coupled, and the renormalized-velocity formulation of Theorem~\ref{thm:LW} replaces the device.}
\label{fig:torus}
\end{figure}

Ref.~\cite{ZZAnn} (and \cite{ZZJPA} for the homogeneous case) states that with contact, Yukawa and Coulomb
interactions the Hall conductivity is given by the same expression \eqref{eq:QHE2} with the non-interacting
$G_W$ replaced by the complete two-point function $\bG_W$, proven to two loops and argued to all orders. The
mechanism is the current with the renormalized velocity $\partial_p(Q_0-\Sigma)=\partial_p\bQ$: the difference
from the bare-velocity current, $\cK=-\int\Tr\,\bG\star\partial_{p_k}\Sigma$, vanishes order by order because the
boson propagator is even in momentum and every term is a total momentum derivative over the Brillouin zone (in the
continuum: over $\mathbb R^D$, with the conditions of Sec.~\ref{sec:latcont} at $|p|\to\infty$; the lattice derivation
uses the approximate Wigner--Weyl calculus, which becomes exact in the continuum). For the
inhomogeneous case \cite{ZZAnn} uses a gedankenexperiment: a torus cut into a region with coupling $\alpha$ and a
region with the coupling switched off by $\theta$-functions, with opposite electric fields, so that the total current
is $I_{\rm pers}+\sigma^{(I)}(\alpha)E-\sigma^{(II)}(0)E$ and the absence of corrections to the total current gives
$\sigma^{(I)}(\alpha)=\sigma^{(I)}(0)$. \begin{GPTcorrection}
For density--density interactions the global identity derived below (Theorem~\ref{thm:LW}) does not depend on the
range of the interaction. The two-region device itself, however, is clean only for short-ranged interactions: for
Coulomb interactions the two regions remain coupled through the interaction and through their electrostatic response,
so that the step-function profile of the coupling has to be supplemented by an interphase (Fig.~\ref{fig:torus}). \reviewfix{The renormalized-velocity formulation of Theorem~\ref{thm:LW} avoids the spatial separation
altogether.}
\end{GPTcorrection}

\begin{GPTcorrection}
\paragraph{A total derivative within the renormalized-current expression.}
Write $\tilde J_k=-\Tr(\bG_W\star\partial_{p_k}\bQ_W)$, where $\Tr$ includes the coordinate and momentum
integrals and the internal trace. The field-dependent inverse correction $\bQ^{(1)}F$ contributes
\begin{equation}
\delta\tilde J_k\big|_{\bQ^{(1)}}=
\Tr\!\left(\bG^{(0)}\star\bQ^{(1)}\star\bG^{(0)}\star\partial_{p_k}\bQ^{(0)}
-\bG^{(0)}\star\partial_{p_k}\bQ^{(1)}\right)F
=-\Tr\,\partial_{p_k}(\bG^{(0)}\star\bQ^{(1)})F=0,
\end{equation}
for a uniform field to this order, a cyclic trace and vanishing momentum boundary terms. This uses
$\partial_k\bG^{(0)}=-\bG^{(0)}\star\partial_k\bQ^{(0)}\star\bG^{(0)}$. It removes the vertex terms from the
integrated current without any use of chiral symmetry; what remains is to show that $\tilde J_k$ coincides with the
bare-velocity current, which is Theorem~\ref{thm:LW}.
\end{GPTcorrection}

\begin{theorem}[Luttinger--Ward identity for the renormalized velocity]\label{thm:LW}
Let the theory be $\Phi$-derivable, i.e.\ let the self-energy be the functional derivative
$\Sigma=\delta\Phi[\bG]/\delta\bG$ of the Luttinger--Ward functional $\Phi[\bG]$ (the sum of the closed
two-particle-irreducible skeleton diagrams built from the full $\bG$; Appendix~\ref{app:manybody}), on a
differentiable branch of $\Phi$; let the interaction vertices be local densities (contact, Yukawa, Coulomb -- of any
range); and let the trace (over the whole system, or per volume in the thermodynamic limit) be regularized so that
commutators with the coordinate operator are traceless and integration by parts in momentum is legitimate (compact
Brillouin zone on a lattice; condition (C2) of Sec.~\ref{sec:latcont} in the continuum). Then, for
any external $A(x)$ compatible with these assumptions,
\begin{equation}
\Tr\big[\bG\,\partial_{p_k}\Sigma\big]=0,\qquad\text{i.e.}\qquad
-\Tr[\bG\,\partial_{p_k}Q_0]=-\Tr[\bG\,\partial_{p_k}\bQ].
\end{equation}
\end{theorem}
\begin{proof}
Under $\bG\to\bG_\varepsilon$, $\bG_\varepsilon(x,y)=e^{i\varepsilon x}\bG(x,y)e^{-i\varepsilon y}$ (Wigner symbol
$\bG_W(x,p-\varepsilon)$), every closed fermion loop of $\Phi$ is invariant (the phases cancel at each density vertex)
and the interaction lines carry no phase; hence $\Phi[\bG_\varepsilon]=\Phi[\bG]$ and
$0=d\Phi/d\varepsilon_k=\Tr[\Sigma\,d\bG_\varepsilon/d\varepsilon_k]=-\Tr[\Sigma\,\partial_{p_k}\bG]$. Then
$\Tr[\bG\partial_k\Sigma]=\Tr[\partial_k(\bG\Sigma)]-\Tr[\partial_k\bG\,\Sigma]=\Tr[\partial_k(\bG\Sigma)]
=-i\Tr[[x_k,\bG\Sigma]]=0$.
\end{proof}
\begin{GPTcorrection}
This is the integrated form of the phase (shift) invariance of a $\Phi$-derivable theory \cite{LW,BaymKadanoff},
\reviewfix{i.e.\ the integrated counterpart of the Ward--Takahashi identity
$q_\mu\Gamma_\mu(p+q,p)=\bQ(p+q)-\bQ(p)$, whose zero-transfer form $\Gamma_k(p,p)=\partial_{p_k}\bQ$ requires a
regular $q\to0$ limit (at a Fermi surface the static and dynamic limits differ). The proof concerns the $U(1)$ phase
and the total trace; an untraced non-Abelian component current is not covered.} The existence
of a differentiable physical branch of $\Phi$ is an assumption; it is known to fail in some strongly correlated
regimes \cite{Kozik}. For Coulomb interactions $A$ is the total (Hartree-screened) field. For the singlet axial
current no regulator preserves both chiral and gauge symmetry, so the renormalized axial current of \cite{ZA} is a
definition supplemented by the chiral-symmetry argument of Sec.~\ref{sec:CSEint}, not a consequence of the vector
identity.
\end{GPTcorrection}

\reviewfix{\paragraph{A short proof of non-renormalization.} Theorem~\ref{thm:LW} and the total-derivative identity
above combine into a proof that does not need the two-region device. By Theorem~\ref{thm:LW} the electric current is
the renormalized-velocity current $\tilde J_k=-\Tr(\bG_W\star\partial_{p_k}\bQ_W)$. Its linear response to a uniform
field follows from the expansion \eqref{eq:G1} with the complete functions,
$\bQ_W=\bQ^{(0)}(p-A)+\bQ^{(1)}_{ij}F_{ij}+\dots$ and
\[
\bG^{(1)}=\Big(\frac i2\,\bG^{(0)}\star\partial_{p_i}\bQ^{(0)}\star\bG^{(0)}\star\partial_{p_j}\bQ^{(0)}\star\bG^{(0)}
-\bG^{(0)}\star\bQ^{(1)}_{ij}\star\bG^{(0)}\Big)F_{ij} .
\]
The terms with $\bQ^{(1)}$, i.e.\ the vertex corrections, drop out as a total momentum derivative, as shown above.
The remaining term is the one that leads from \eqref{eq:Gexp} to \eqref{eq:QHE2}, so that in $2+1$D
\[
\bar J_i=\frac1{2\pi}\epsilon_{ij}\,\boldsymbol{\cM}_3E_j,\qquad
\boldsymbol{\cM}_3=-\frac{1}{24\pi^2S}\int d^2x\wedge\trD\big(\bG\star d\bQ\star\wedge d\bG\star\wedge d\bQ\big),
\]
the invariant \eqref{eq:QHE2} composed of the complete $\bG_W$ and $\bQ_W$ (Ishikawa and Matsuyama \cite{Ishikawa}
obtained the homogeneous form). $\boldsymbol{\cM}_3$ is a topological invariant: its variation under a variation of
$\bQ_W$ is a total derivative, so it does not change under a smooth deformation of $\bQ_W$ that keeps $\bQ_W$ and
$\bG_W$ smooth on the integration domain and keeps $\bQ_W\simeq ip_4$ at $|p_4|\to\infty$. Let the coupling
$\alpha$ be switched off smoothly. Then $\bG_W(\alpha)$ is deformed smoothly into the bare $G^{(0)}_W$, and
$\boldsymbol{\cM}_3(\alpha)=\boldsymbol{\cM}_3(0)=\cM^v_3$. The Hall conductivity is therefore not renormalized by the
interactions, for contact, Yukawa and Coulomb interactions alike.

The conditions are those of Theorem~\ref{thm:LW} together with adiabatic continuity along the path
$\alpha\to0$: the gap must stay open, so that $\bG_W$ has no poles on the integration domain, and $\bG_W$ must not
develop zeros there, at which $\bQ_W$ is singular. A zero of $\bG_W$ is not a phase transition, but it can change
the invariant built from $\bG_W$, and the Green-function invariant then need not equal the Hall conductivity
\cite{Gurarie}. The argument holds for linear response to a uniform field at $T=0$, and, as stated above, it does not
cover phases with intrinsic topological order.}

\begin{GPTcorrection}
\paragraph{Alternatives to the two-region device.} The device is needed only because a static uniform electric
field is not periodic on the torus. It can be avoided: (i) in a gapped phase the Streda relation
$\sigma_{xy}=\partial\rho/\partial B|_\mu$ (Appendix~\ref{app:manybody}) expresses the Hall conductivity through the density
\reviewfix{$\rho=-i\Tr[\bG\,\partial_{p_4}\bQ]$ (per volume, with $\partial_\mu=-i\partial_{p_4}$ and the equal-time
prescription understood; the frequency-shift analogue of Theorem~\ref{thm:LW}, i.e.\ the identity behind the
Luttinger theorem, replaces the bare by the renormalized vertex and requires $T=0$, a differentiable $\Phi$ and
vanishing frequency boundary terms)}, and the Wigner--Weyl expansion of $\rho$ in $B$
gives $\delta\rho=\frac{\cN}{2\pi}\delta B$ directly; (ii) the electric field may be introduced through a
time-dependent vector potential \reviewfix{$A_k(\tau)=-iE_k\tau$ in the stated Euclidean convention $F_{4k}=-iE_k$}, periodic in the spatial directions; (iii) a spatially modulated
field $E(y)$ may be used, with the local response extracted at wave number $q\to0$. In each case the thermodynamic
and zero-temperature limits and their order have to be specified; none of the routes extends the Green-function
formula to phases with intrinsic topological order, in which the single-particle $\bG$ does not carry the Hall
response.
\end{GPTcorrection}

\subsection{Lattice versus continuum: validity of the approximate calculus and the conditions at
$|p|\to\infty$}
\label{sec:latcont}

The non-renormalization results of Secs.~\ref{sec:CSEint} and \ref{sec:QHEint} were obtained in
\cite{ZZJPA,ZZAnn,ZA} (the inhomogeneous non-interacting expressions in
\cite{ZubkovWu,SuleymanovZubkov,SuleymanovZubkovNPB}, and the lattice form of the chiral anomaly in
\cite{ZubkovKhaidukov2017}) for
lattice-regularized models, using the \emph{approximate} lattice Wigner--Weyl calculus (Appendix~\ref{app:ww}): the
lattice Weyl symbol of a product of operators equals the Moyal product of the symbols only up to terms suppressed by
powers of the lattice spacing $a$ times the gradients of the external fields, so that the Groenewold equation, the
gradient expansion of Sec.~\ref{sec:groenewold} and all identities derived from them hold to the leading order in
$a\,\partial_x$, i.e.\ for external fields and interaction-induced inhomogeneities that vary slowly on the lattice
scale ($a^2|F|\ll1$). The present review is formulated in continuum field theory, where the Weyl calculus and its
covariant extension of Sec.~\ref{sec:WW} are exact: the Moyal (star) product is the exact composition law of
symbols, $\bQ_W\star\bG_W=1$ is an identity, and the gradient expansion is the expansion in $\lambda^2|F|$ with no
lattice scale. Every algebraic step of the lattice derivations -- the expansion
$\bQ_W=\bQ^{(0)}(p-\mathcal A)+\bQ^{(1)}F+\dots$, the total-derivative form of the vertex terms, the evenness of the
boson propagator, the order-by-order vanishing of $\cK$, and the Luttinger--Ward identity of Theorem~\ref{thm:LW} --
uses only associativity, the derivations $\partial_\mu$ and the closed trace of the star-product algebra, and
carries over to the continuum verbatim, with the approximate equalities of the lattice calculus replaced by exact
ones\reviewfix{ (the transport coefficients remain, as on the lattice, leading-order response formulas in the
field, while the Luttinger--Ward identity is exact)}. Hence all results of \cite{ZZJPA,ZZAnn,ZA} obtained with
the approximate Wigner--Weyl calculus are valid in continuum field theory, with one qualification: the role played
on the lattice by the compactness of the Brillouin zone has to be taken over by conditions at $|p|\to\infty$.

On the lattice the momentum integrals run over the torus $T^{D-1}$ (times $\mathbb R$ for the continuous Matsubara
frequency at $T=0$), so that (a) the spatial momentum integrals of smooth symbols run over a compact domain, (b) every total
spatial momentum derivative integrates to zero, and (c) the integer-valuedness of the (normalized) invariants
follows from compactness; \reviewfix{the frequency direction is non-compact already on the lattice and needs the
decay of the symbols in $p_4$.} In the continuum the momentum space is $\mathbb R^D$ and (a)--(c) become the
following requirements on the complete symbols, uniformly in $x$.

\begin{enumerate}
\item[(C1)] \emph{Ultraviolet asymptotics of the complete symbol.} $\bQ_W(x,p)$ is a classical elliptic symbol
(Appendix~\ref{app:ops}) whose principal symbol is that of the free operator up to an invertible, homotopically
trivial factor:
\begin{equation}
\begin{gathered}
\bQ_W(x,p)=Z(x,p)\,Q_0(p)+r(x,p),\qquad \corrB{r(x,p)=O(|p|^{1-\delta})},\\
Z\ \text{invertible, homotopic to }1.
\end{gathered}
\label{eq:C1}
\end{equation}
equivalently $\Sigma_W(x,p)=(1-Z)Q_0(p)+\corrB{O(|p|^{1-\delta})}$\corrB{, for some $\delta>0$ and with the corresponding
bounds on the $p$-derivatives (for a classical symbol $\delta=1$; the weaker condition $r=o(|p|)$ would make the
\reviewfix{correction to the slice density only $o(|\vec p|^{-3})$, which does not guarantee integrability over
$\R^3$})}. \reviewfix{Here $Z$ is a symbol of order zero with uniformly bounded inverse and derivatives, and its
deformation to $1$ preserves the chiral structure. Then $\bG_W=O(|p|^{-1})$, $\partial_p\bQ_W=O(1)$, and the
integrands of the topological expressions \eqref{eq:CSE}, \eqref{eq:QHE2} and \eqref{eq:PLB13} are integrable on
every slice: the leading term, built from $\mathsf A=Q_0^{-1}dQ_0$ with $Q_0$ homogeneous of degree $m$, is
horizontal under radial dilation, $\iota_{p\cdot\partial_p}\tr\mathsf A^3=3m\tr\mathsf A^2=0$, i.e.\ the pull-back of
a form on the sphere, so that on the hyperplane at distance $p_4$ from the origin it decays as $|p_4|/|\vec p|^{4}$,
and the remainder is $O(|\vec p|^{-3-\delta})$. Hence the integrals converge, the side faces of the slab of
Fig.~\ref{fig:slab} do not contribute, and the ultraviolet flux of the invariant equals the free one
(Sec.~\ref{sec:stokes}); the analogous estimate holds for the Hall integrals in $D=3$. This replaces (a) and (c).
In a renormalizable theory the self-energy satisfies these bounds up to logarithms of $|p|$, which are admissible as
long as the inverse and derivative estimates and the homotopy of $Z$ survive; the external field enters through
$p-\mathcal A(x)$ and $F$, i.e.\ through terms of lower order in $|p|$, which is why the condition is uniform in $x$.}

\item[(C2)] \emph{Vanishing of the boundary terms.} Each total momentum derivative discarded in
Secs.~\ref{sec:CSEint}, \ref{sec:QHEint} -- the vertex term $\partial_{p_k}(\bG^{(0)}\star\bQ^{(1)})$, the terms of
$\cK$ order by order, and $\partial_{p_k}(\bG\Sigma)$ in the proof of Theorem~\ref{thm:LW} -- leaves the surface
integral $\oint_{S^{D-1}_R}dS_k(\cdots)$, $R\to\infty$, \reviewfix{which vanishes if the traced integrand is uniformly $o(|p|^{1-D})$ (sufficient; angular cancellations
can also make the flux vanish).} Condition (C1) alone does not guarantee this: \corrB{$\bG\Sigma\to Q_0^{-1}(Z^{-1}-1)Q_0$ tends to a
function of degree zero in $p$ (the constant $Z^{-1}-1$ when $Z$ is $p$-independent and commutes with $Q_0$), whose
surface integral $\propto R^{D-1}$ vanishes only if its angular average against $\hat p_k$ does -- which is the case
for a constant --, and the next term, of the type $Q_0^{-1}(p)\,c_\nu(x)\gamma_\nu=O(|p|^{-1})$ for a vector-like
part $c_\nu\gamma_\nu$ of the self-energy (\reviewfix{a constant shift such as that of the chemical potential; a velocity renormalization instead sits in $Z$}), gives after
the spinor trace $\oint dS_k\,c_jp_j/p^2\propto c_kR^{D-2}$, which diverges for $D\ge3$ (a mass-like part $m(x)$
drops out of the trace at this order, $\tr(m\gamma_jp_j/p^2)=0$, and enters only at $O(|p|^{-2})$).}
This is the familiar ambiguity of a shift of the integration momentum in a divergent integral, and it is removed in
the same way: the theory has to be regularized so that shifts of the loop momentum are legitimate -- a lattice,
dimensional regularization, or Pauli--Villars regularization with enough regulator fields (Appendix~\ref{app:ww}),
\reviewfix{the complete regulated expressions, regulator contributions and counterterms included, having vanishing
boundary flux} -- and the regularization is removed
only after the total derivatives have been discarded. In operator language (C2) is the statement that the regularized
trace is closed, $\Tr[\hx_k,\,\cdot\,]=0$, which is the form in which it enters Theorem~\ref{thm:LW}. This replaces
(b). The lattice derivations already needed (C2) in the frequency direction, \reviewfix{where it is provided by the decay of the complete product in each boundary term ($\bG\sim1/ip_4$ alone
does not control $\bG\Sigma$ for an unrestricted self-energy).}
\item[(C3)] \emph{Compatibility of the regularization with the symmetry used.} For the QHE the regularization must
preserve the vector gauge invariance behind the Ward--Takahashi identity (so that $\Gamma_k(p,p)=\partial_{p_k}\bQ$
and the Hartree part of the interaction combines with the external field into the total field). For the CSE it must
in addition not introduce singularities of $\bG_W$ off the hyperplane $p_4=0$ and must leave the chiral
anticommutation $\{\gamma^5,\bQ_W\}=0$ intact where $\bG_W$ is singular, since this is what cancels the
$\bQ^{(1)}$ terms in Sec.~\ref{sec:CSEint}; the regulator contributions then enter only through the flux through
$S^{3}_\infty$, which by (C1) is the free one. \reviewfix{A massive Pauli--Villars operator breaks the chiral anticommutation at every momentum,
$\{\gamma^5,Q+iM_i\}=2iM_i\gamma^5$, but its contribution at finite momenta decouples as $M_i\to\infty$ and only its
ultraviolet flux remains;}\corrB{ this is compatible with the requirement, which concerns the singular set of $\bG_W$ at
finite momenta.} A lattice regularization with doublers requires the
separate analysis of \cite{KhaikinZubkov,ZubkovKhaidukov2017} (the Wigner--Weyl form of the Wilson-fermion
propagator in a varying external field is given in \cite{SuleymanovZubkovNPB}; for the lattice regularization of the
Standard Model itself see \cite{BVZ}).
\end{enumerate}
\reviewfix{Under (C1)--(C3), together with the infrared, response and gap assumptions stated in the derivations, the
statements of Secs.}~\ref{sec:CSEint} and \ref{sec:QHEint} hold in the continuum with the
Green function $\bG_W$ of the regularized theory, and the limit of removed regularization is taken in the final
topological expression, where it affects only the ultraviolet boundary. Conversely, when (C1) fails -- a self-energy
growing as fast as $Q_0$ with a non-trivial matrix structure, or a complete Green function whose large-$|p|$ symbol
is not homotopic to the free one -- the invariant of $\bG$ need not coincide with the free value, even though the
star-product identities remain exact; and when (C2) fails the renormalized-velocity current and the bare-velocity
current differ by a surface term at infinity. The non-elliptic models of Secs.~\ref{sec:FS} and \ref{sec:nonell}
violate (C1) in the $p_4$ direction, which is why they have no ultraviolet degree; for them only the static response
of the Fermi-level (window) current is obtained, Sec.~\ref{sec:nonell}.

%======================================================================
\section{The chiral anomaly}
\label{sec:anomaly}

\subsection{Fujikawa's argument and the index}
\label{sec:Fujikawa}

\reviewfix{Throughout this section the space-time is a closed four-manifold $M$. The physical case $\R^4$ is reached
by compactification, and only in part: the local anomaly density and, under the homotopy hypothesis, the integrated
anomaly hold on $\R^4$ as they stand, whereas the index terms of the general factorization theorem require a
condition at infinity -- a background trivial at infinity, closed to $S^4$, or a periodic one, closed to $T^4$ -- and
depend on it; the two closures and what they share are compared at the end of Sec.~\ref{sec:factor}.}
\corr{On a closed oriented four-manifold, for $Q=\begin{pmatrix}0&O^\dagger\\O&0\end{pmatrix}$ with $O$ elliptic between the specified chiral bundles,} a chiral rotation with constant
$\alpha\in\R$ changes the measure by $\exp(-2i\alpha\Tr\gamma^5)$ and the action by $\alpha\int\tr\cD_\mu J_\mu$;
invariance of $Z$ gives $\sA:=\int\langle\tr\cD_\mu J_\mu\rangle=2i\Tr\gamma^5$, and \corr{with the trace regulated by $e^{-Q^2/\Lambda^2}$, nonzero modes have zero total chiral trace, so} $\Tr\gamma^5=n_+-n_-=\ind O$ [Ref.~\cite{XZPLB}, Eqs.~(5)--(8)]. By the
Atiyah--Singer theorem in Fedosov's form \cite{Fedosov,AS},
\begin{equation}
\corr{\sA=2i\,\ind O=2i\int_{T^*M}\!\left[\ch(\xi)\wedge\pi^*\Td(TM\otimes\C)\right]_{8},}
\label{eq:Fuji}
\end{equation}

\begin{reviewchange}
Strictly, $\sA$ in \eqref{eq:Fuji} is the integrated anomaly density of the regulated axial Ward identity: for
nonzero index the massless partition function without zero-mode insertions vanishes, so that no normalized
expectation value is obtained by dividing by it, and the heat-kernel supertrace defines $2i\,\ind O$ independently of
that division (a small mass or suitable insertions add mass or contact terms to the Ward identity).
\end{reviewchange}
(For lattice models the relation between the chiral anomaly, the Wigner transformation and the momentum-space
invariant $N_3$ was established in \cite{ZubkovKhaidukov2017}.) Here (Appendix~\ref{app:ops}, \ref{app:bundles},
\ref{app:index}): $\ind O=\dim\ker O-\dim\ker O^\dagger$ is
the Fredholm index of the chiral block $O:\Gamma(V_+)\to\Gamma(V_-)$ between the two chiral bundles; $O$ is
elliptic, i.e.\ its principal symbol $o(x,p)$ (the highest-order part of its symbol, $\sigma^\mu p_\mu$ for the
Weyl operator) is invertible for $p\ne0$; $T^*M$ is the cotangent bundle of $M$, with coordinates $(x,p)$ and
projection $\pi:T^*M\to M$, and $\pi^*$ is the pull-back of differential forms from $M$ to $T^*M$;
$\xi=[\pi^*V_+,\pi^*V_-;o]\in K_c(T^*M)$ is the symbol class, the compactly supported $K$-theory class defined by
the two bundles $\pi^*V_\pm$ identified by $o$ away from the zero section; $\ch$ is the Chern character,
$\ch(E)=\tr\exp(F_E/2\pi)=\sum_k\ch_k(E)$ for a bundle with curvature two-form $F_E$, extended to $K_c(T^*M)$ by the
difference construction; $\Td(TM\otimes\C)=1-\frac1{12}p_1(M)+\dots$ is the Todd class of the complexified tangent
bundle, with $p_1(M)=-\frac1{8\pi^2}\tr R\wedge R$ the first Pontryagin form of the Riemann curvature $R$; and
$[\cdot]_8$ selects the component of form degree eight, integrated over the eight-dimensional manifold $T^*M$ with
the orientation of \cite{AS} (the overall sign $(-1)^{\dim M}$ of \cite{AS} is $+1$ here). For $Q=i\gamma^\mu D_\mu$
on flat $M$ this gives $\ind O=-\frac1{8\pi^2}\int\tr F\wedge F$, i.e.\ $\sA=-\frac{i}{4\pi^2}\int\tr F\wedge F$
[Ref.~\cite{XZPLB}, Eqs.~(9)--(11)]; \reviewfix{the number $\frac1{8\pi^2}\int\tr F\wedge F=\int_M\ch_2(E)$ is the instanton number of the gauge bundle
$E$, equal to $-\int_Mc_2(E)$ for $c_1(E)=0$ (the negative of the second Chern number in the convention of
Appendix~\ref{app:bundles}); for a $U(N)$ bundle $\ch_2=\frac12c_1^2-c_2$ and $\int_M\ch_2$ can be half-integral.}
(The anomaly is that of the singlet axial current; the non-singlet one, $\alpha\in\mathfrak g$, has a different
anomaly $\propto\tr(t^aF\tilde F)$.) \begin{GPTcorrection}
For a massless Dirac operator on $\R^4$ with coefficients approaching the free ones, zero lies in the essential
spectrum (the part of the spectrum that does not consist of isolated eigenvalues of finite multiplicity), so the ordinary $H^1\to L^2$ realization is not Fredholm. Compactification is one way to formulate
the index; boundary, weighted-space and relative-index formulations require their own hypotheses. A zero of
the full symbol at finite momentum is not, by itself, a criterion for failure of Fredholmness. Below we use a
closed manifold; when $p_1(M)=0$ in rational cohomology, the Todd factor in \eqref{eq:Fuji} contributes no
nonconstant term in four dimensions.
\end{GPTcorrection}

\subsection{Point splitting and the Wigner--Weyl form of the anomaly}
\label{sec:pointsplit}

The ultraviolet singularity of $\langle x|\hat\rho|x\rangle$ is regularized by $\hat\rho^\epsilon=e^{i\hpi\epsilon}
\hat\rho e^{i\hpi\epsilon}$, whose matrix element $U(x,x+\epsilon)\psi(x+\epsilon)\bar\psi(x-\epsilon)U(x-\epsilon,x)$
is gauge covariant at $x$; $(e^{i\hpi\epsilon})_W=e^{ip\epsilon}$ [Ref.~\cite{XZPLB}, Eq.~(35)]. Repeating the Noether procedure with
$\rho^\epsilon_W=e^{ip\epsilon}\star\rho_W\star e^{ip\epsilon}$ gives $\Gamma^\epsilon=\cD_\mu J^\epsilon_\mu$ with
$J^\epsilon_\mu=-\frac12\trD\gamma^5\int\frac{d^4p}{(2\pi)^4}(\partial_{p_\mu}Q_W\rho^\epsilon_W-\rho^\epsilon_W
\partial_{p_\mu}Q_W)+\dots$ [Ref.~\cite{XZPLB}, Eqs.~(40)--(47)], and with $\langle\hat\rho\rangle=-\hG$ [Ref.~\cite{XZPLB}, Eqs.~(49)--(50)]
\begin{equation}
\begin{aligned}
\trG\langle\cD_\mu J^\epsilon_\mu\rangle={}&i\trD\trG\gamma^5\int\frac{d^4p}{(2\pi)^4}\Big((Q_W\star e^{ip\epsilon}-e^{ip\epsilon}\star Q_W)\star G_W\star e^{ip\epsilon}
\\&\qquad{}-e^{ip\epsilon}\star G_W\star(Q_W\star e^{ip\epsilon}-e^{ip\epsilon}\star Q_W)\Big).
\end{aligned}
\label{eq:PS1}
\end{equation}
\reviewfix{The conditions of Sec.~\ref{sec:logZ} (paragraph ``When the leading term is sufficient'') enter the anomaly
as follows. The dots in $J^\epsilon_\mu$ are the terms with $k\ge1$ of the series of the type \eqref{eq:Jcorr}. The
argument that removes these terms from the averaged currents, \eqref{eq:Javexact} and \eqref{eq:Jaxavexact}, does not
apply to the anomaly in the same form. The anomaly is the integral of the divergence
$\trG\cD_\mu J^\epsilon_\mu=\partial_\mu\trG J^\epsilon_\mu$, and under $\int d^4x$ every term of the series, the
leading one included, is formally a total derivative. The result is different from zero because the limit
$\epsilon\to0$ does not commute with the momentum integral. The anomaly is therefore computed from $\Gamma^\epsilon$
itself: \eqref{eq:PS1} is exact and does not use the truncated current. The counterpart of the series of the current
appears in \eqref{eq:54}, and is discussed after it. For the local anomaly equation written with the leading term of
the current, the condition \eqref{eq:slowcond} is required.}
\begin{GPTcorrection}
Unregulated cyclicity of the trace would set this expression to zero; the ultraviolet limit is singular, and the
manipulations below are performed at fixed $\epsilon$, with the surface terms in $p$ controlled by the decay of the
integrand. \reviewfix{For nonzero index the massless $Q$ has no two-sided inverse: with the pseudoinverse,
$QG=GQ=1-P_0$, the zero-mode projector $P_0$ enters \eqref{eq:PS1}, or one uses an invertible massive regulator
and keeps its chiral-breaking terms until the massless limit; \eqref{eq:PS1} as displayed is the identity of the
invertible sector;} the heat-kernel formula
\eqref{eq:Fuji} defines the integrated anomaly independently.
\end{GPTcorrection}

The key step is the
commutator of $Q_W$ with the point-splitting factor. From \eqref{eq:stardiff} with one factor $e^{ip\epsilon}$
(whose Fourier representation localizes $y$ or $y'$ at $\epsilon$) and the expansion \eqref{eq:Wexp} of the Wilson
loop one finds [Ref.~\cite{XZPLB}, Eqs.~(51)--(54)]
\begin{equation}
e^{ip\epsilon}\star Q_W-Q_W\star e^{ip\epsilon}=e^{ip\epsilon}\Big(\epsilon_\mu\big(\partial_{x_\mu}Q_W-F_{\mu\nu}\partial_{p_\nu}Q_W
+\tfrac1{24}\cD_\alpha\cD_\beta F_{\mu\nu}\partial_{p_\alpha}\partial_{p_\beta}\partial_{p_\nu}Q_W+O(\cD^5)\big)+O(\epsilon^2)\Big),
\label{eq:54}
\end{equation}
where the $-F\partial_pQ$ term comes from $-\frac i2y^\mu y'^\nu F_{\mu\nu}$ with one of $y,y'$ equal to $\epsilon$
and the other converted into $i\partial_p$, and the $\frac1{24}$ term from the $-\frac i{48}(y^\alpha y^\beta+y'^\alpha
y'^\beta)y^\mu y'^\nu\cD\cD F$ term (the $\cD F$ term of \eqref{eq:Wexp} cancels between the two orderings). We have
verified \eqref{eq:54} independently \corr{at the displayed order in the point separation for Weyl-ordered cubic monomials in an Abelian background; this check does not prove the all-orders non-Abelian remainder estimates.}

\reviewfix{The expansion \eqref{eq:54} follows to all orders in $\cD$ from Proposition~\ref{prop:dQW}. To first order
in $\epsilon$, $e^{i\epsilon\hpi}\hQ e^{-i\epsilon\hpi}-\hQ=i\epsilon_\mu[\hpi_\mu,\hQ]$, and $i[\hpi_\mu,\,\cdot\,]$
is the derivation that acts as $\partial_{x_\mu}$ on the coefficient functions and as the shift
$\hpi_\nu\to\hpi_\nu-F_{\mu\nu}$ on the covariant momenta, since $i[\hpi_\mu,\hpi_\nu]=-F_{\mu\nu}$. It is the
variation of $\hQ$ under the translation of the background by $\epsilon$ combined with the variation of the gauge
field $\delta A_\nu=\epsilon_\mu F_{\mu\nu}$, and \eqref{eq:dQWsinglet} gives
\begin{align}
e^{ip\epsilon}\star Q_W-Q_W\star e^{ip\epsilon}
&=e^{ip\epsilon}\epsilon_\mu\Big[\partial_{x_\mu}Q_W-\int_{-1/2}^{1/2}dt\,\Big(e^{it\,\cD\cdot\partial_p}F_{\mu\nu}\Big)\partial_{p_\nu}Q_W\Big]+O(\epsilon^2)\notag\\
&=e^{ip\epsilon}\epsilon_\mu\Big[\partial_{x_\mu}Q_W-\sum_{k\ge0}\frac{(-1)^k}{4^k(2k+1)!}\big(\cD_{\alpha_1}\cdots\cD_{\alpha_{2k}}F_{\mu\nu}\big)\notag\\
&\qquad\qquad\qquad\qquad\times\partial_{p_{\alpha_1}}\cdots\partial_{p_{\alpha_{2k}}}\partial_{p_\nu}Q_W\Big]+O(\epsilon^2),
\label{eq:54all}
\end{align}
where $\cD$ acts on $F$ and $\partial_p$ on $Q_W$. The terms with $k=0,1$ are those of \eqref{eq:54}; this confirms
the coefficient $\frac1{24}$ for an arbitrary non-Abelian field. The field strength is averaged over the same
segment as in \eqref{eq:Jsegment}.}

\reviewfix{In contrast to the averaged currents \eqref{eq:Javexact}, the integration over $x$ does not remove the
terms with $k\ge1$ from the anomaly. In \eqref{eq:61} below they are contracted with the weight $F_{\mu\nu}(x)$, which
is not covariantly constant in general, and $\int\trG[(\cD^{2k}F_{\mu\nu})X]=\int\trG[F_{\mu\nu}\,\cD^{2k}X]\ne0$. They
are removed in other ways. (a) $\partial_p^3Q_W=0$; in particular, for the first-order Dirac operator \eqref{eq:54}
is exact at the order $\epsilon$. (b) A covariantly constant field, $\cD F=0$. (c) For an elliptic operator, by
ultraviolet power counting. The anomaly is the flux \eqref{eq:63} through the sphere at $|p|\to\infty$, to which only
the terms decaying as $|p|^{-3}$ contribute. A term that contains the field factors of total dimension $d$ (two for
each $F$, one for each $\cD$ or $\partial_x$) decays as $|p|^{1-d}$. Only $d=4$ survives: the terms with $F^2$,
which give the anomaly, and the term with $\cD\cD F$, i.e.\ the $k=1$ term multiplied by $G^{(0)}$, whose gauge trace
$\partial\partial\,\trG F$ is a total derivative (and vanishes identically for $SU(N)$). The terms with $k\ge2$ and
all other terms with $d>4$ vanish on the sphere; the terms with $d<4$ vanish by the traces. The integrated anomaly of
an elliptic operator therefore requires no condition of slow variation of the field, as it must for an index. For
the non-elliptic models of Sec.~\ref{sec:nonell} the argument (c) is not available, and the truncation is justified
by (a), (b), or by the condition \eqref{eq:slowcond}. For quasi-elliptic operators the counting (c) has to be done
with the weights of the anisotropic scaling, Sec.~\ref{sec:quasiell}.}
\begin{GPTcorrection}
The truncation at $O(\cD^5)$ is justified when the omitted terms vanish after the momentum integral and the limit
$\epsilon\to0$; since the integral samples $|p|\sim|\epsilon|^{-1}$, this is a statement about the ultraviolet decay
of the integrand (satisfied for the Dirac operator, where each additional derivative of $F$ is accompanied by a power
of $|p|^{-1}$), not merely a count of powers of $\epsilon$.
\end{GPTcorrection}

Inserting \eqref{eq:54} into \eqref{eq:PS1}, \corrB{using at fixed $x$ the identity
\[
\int d^4p\,X_W\star_MY_W=\int d^4p\,e^{-\frac i2\partial_p\partial_x}\big(X_W(x,p)Y_W(x,k)\big)\big|_{k=p},
\]
in which $\partial_p$ acts on $X_W$ and $\partial_x$ on the product (it follows from $\star_M$ by discarding total
$p$-derivatives; the terms with $\partial_x$ are total $x$-derivatives and drop out after the $x$-integration), i.e.\ }\reviewfix{the trace identity $\int d^4x\,d^4p\,\tr(X\star Y)=\int d^4x\,d^4p\,\tr(XY)$ of the fully
integrated, regulated trace (a momentum integral at fixed $x$ alone is not cyclic)}, and
$e^{ip\epsilon}\star G_W+G_W\star e^{ip\epsilon}=e^{ip\epsilon}(2G_W+O(\epsilon))$ [Ref.~\cite{XZPLB}, Eqs.~(55)--(61)]:
\begin{align}
\int d^4x\,\trG\langle\cD_\mu J^\epsilon_\mu\rangle&=-2i\trD\trG\gamma^5\int\frac{d^4x\,d^4p}{(2\pi)^4}e^{2ip\epsilon}\epsilon_\mu\notag\\
&\quad\times\Big(\partial_{x_\mu}Q_W-F_{\mu\nu}\partial_{p_\nu}Q_W+\tfrac1{24}\cD_\alpha\cD_\beta F_{\mu\nu}\partial^3_{\alpha\beta\nu}Q_W\Big)G_W .
\label{eq:61}
\end{align}

\begin{GPTcorrection}
\begin{theorem}[A sufficient hypothesis for the direction-averaged limit]
Let $f\in C^\infty(\R^4)$ be tempered and, outside a ball, let
$f(p)=f_{-3}(p)+r(p)$, with $f_{-3}$ smooth and homogeneous of degree $-3$,
$r=O(|p|^{-3-\delta})$ and $\partial r=O(|p|^{-4-\delta})$ for some $\delta>0$.
Interpret Fourier integrals distributionally, or with a compatible radial cutoff removed first.
For the normalized average over directions of $\epsilon$,
\begin{equation}
\lim_{|\epsilon|\to0}\left\langle\int d^4p\,e^{ip\epsilon}\epsilon_\mu f(p)\right\rangle
=i\lim_{R\to\infty}\int_{|p|<R}d^4p\,\partial_{p_\mu}f(p).
\end{equation}
\end{theorem}
\begin{proof}
Distributional integration by parts gives $\epsilon_\mu\mathrm{FT}(f)=i\mathrm{FT}(\partial_\mu f)$.
Angular averaging of the exponential gives a radial kernel bounded by one and tending pointwise to one.
The spherical mean of $\partial_\mu f_{-3}$ vanishes: its integral over any annulus is the difference of
two equal homogeneous surface fluxes. The remaining spherical mean is integrable with radial measure
$r^3dr$ by the assumed remainder estimate. Dominated convergence after angular integration proves the
formula, and the divergence theorem identifies the right side with the spherical flux.
\end{proof}
In the application the theorem is applied to the traced regulated integrand of \eqref{eq:61}, which for the
Dirac operator is of the required form; smoothness alone would not suffice.
\end{GPTcorrection}

\corr{When these estimates hold, applying the theorem to \eqref{eq:61}} (the exponent is $e^{2ip\epsilon}$; rescaling $2\epsilon\to\epsilon$ gives a
factor $\frac12$, and $-2i\cdot\frac i2=1$) [Ref.~\cite{XZPLB}, Eq.~(63)]:
\begin{equation}
\sA=\trD\trG\gamma^5\int\frac{d^4x\,d^4p}{(2\pi)^4}\partial_{p_\mu}\Big(\big(\partial_{x_\mu}Q_W-F_{\mu\nu}\partial_{p_\nu}Q_W
+\tfrac1{24}\cD_\alpha\cD_\beta F_{\mu\nu}\partial^3_{\alpha\beta\nu}Q_W\big)G_W\Big).
\label{eq:63}
\end{equation}
\corr{The integrated anomaly is fixed by the elliptic symbol class. A local anomaly density may contain lower-order or derivative terms whose integrated variation is exact; the index does not determine that density pointwise.}

\subsection{Evaluation under the homotopy hypothesis}
\label{sec:PLBhom}

Ref.~\cite{XZPLB} assumes $Q_W(x,p)$ homotopic to an $x$-independent $\tilde Q(p)$ and replaces $Q_W$ by $\tilde Q$ in \eqref{eq:63}
(invariance of the index). With the expansion of $\tilde G_W$ to first order in $F$ [Ref.~\cite{XZPLB}, Eqs.~(64)--(69)]
\begin{align}
\sA&=-\int d^4x\,\tr F_{\mu\nu}\int\frac{d^4p}{(2\pi)^4}\trD(\gamma^5\partial_\nu\tilde Q\partial_\mu\tilde G^{(0)})
+\frac i2\int d^4x\,\tr(F_{\mu\nu}F_{\alpha\beta})\int\frac{d^4p}{(2\pi)^4}\partial_{p_\mu}S_{\alpha\beta\nu}\notag\\
&\quad+\frac1{24}\int d^4x\,\partial_\alpha\partial_\beta\tr F_{\mu\nu}\int\frac{d^4p}{(2\pi)^4}\trD(\gamma^5\partial^3_{\alpha\beta\nu}\tilde Q\partial_\mu\tilde G^{(0)}),
\end{align}
with $S_{\alpha\beta\nu}=\frac12\trD(\gamma^5\tilde G^{(0)}\partial_\alpha\tilde Q\tilde G^{(0)}\partial_\beta\tilde Q
\tilde G^{(0)}\partial_\nu\tilde Q)-(\alpha\leftrightarrow\beta)$. The first term vanishes because
$\trD(\gamma^5\partial_\nu\tilde Q\partial_\mu\tilde G^{(0)})$ is symmetric in $\mu\nu$: insert $1=\tilde Q\tilde G^{(0)}$,
use $\partial_\nu\tilde Q\tilde G^{(0)}=-\tilde Q\partial_\nu\tilde G^{(0)}$, cyclicity and $\{\gamma^5,\tilde Q\}=
\{\gamma^5,\tilde G^{(0)}\}=0$ [Ref.~\cite{XZPLB}, Eqs.~(70)--(72)]. The third term is a total $x$-derivative of $F$ and vanishes with $F$
at the boundary. By the same manipulations $S_{\alpha\beta\nu}$ is totally antisymmetric, a 3-form in momentum
space, and \corr{$\tr(F_{\mu\nu}F_{\alpha\beta}+F_{\mu\alpha}F_{\beta\nu}+F_{\mu\beta}F_{\nu\alpha})=\frac14\epsilon_{\mu\nu\alpha\beta}\tr(F_{\rho\sigma}F^\star_{\rho\sigma})$, where $F^\star_{\rho\sigma}=\frac12\epsilon_{\rho\sigma\lambda\eta}F_{\lambda\eta}$}. Hence [Ref.~\cite{XZPLB}, Eqs.~(73)--(74)]
\begin{equation}
\sA=-2iN_3^0\int\frac{d^4x}{16\pi^2}\tr(FF^\star),\qquad \corr{N_3^0=-\frac1{8\pi^2}\int_{\R^4}d_pS} .
\label{eq:anomR4}
\end{equation}
(Excising the singularities and applying Stokes' theorem at fixed regulator identifies the integral over a surface enclosing the singularities with the ultraviolet flux; a small regulator mass breaks the chiral anticommutation and its contribution is kept until the regulator is removed.)
\corr{Deforming the enclosing surface to the two slices with the CSE orientation stated above (the negative of the outward slab orientation;}
the deformation is possible when all singularities lie at $p_4=0$, Sec.~\ref{sec:CSEder}) gives $N_3^0=\frac1{48\pi^2}\int_\Sigma\trD(\gamma^5\tilde G^{(0)}
d\tilde Q\tilde G^{(0)}\wedge d\tilde Q\tilde G^{(0)}\wedge d\tilde Q)$ [Ref.~\cite{XZPLB}, Eqs.~(75)--(76)]. Inserting Moyal products
(free for $x$-independent symbols, for which $N_3^0=N_3$) and $1=\frac1{|V|}\int d^3x$, and then replacing $\tilde Q\to Q_W$, $\tilde G^{(0)}
\to G^{(0)}$ by the topological invariance of the Moyal expression (Sec.~\ref{sec:stokes}) [Ref.~\cite{XZPLB}, Eqs.~(77)--(78)]:
\begin{equation}
\begin{gathered}
\sA=-N_3\,\frac{i}{4\pi^2}\int\tr F\wedge F,\\
N_3=\frac{1}{48\pi^2|V|}\int d^3x\int_\Sigma\trD\big(\gamma^5G^{(0)}\star dQ_W\star G^{(0)}\star\wedge dQ_W\star G^{(0)}\star\wedge dQ_W\big),
\end{gathered}
\label{eq:PLB13}
\end{equation}
and after the inverse Wick rotation $\sA=N_3\frac1{2\pi^2}\int d^4x\,\tr(\mathbf E\cdot\mathbf B)$ [Ref.~\cite{XZPLB}, Eqs.~(79)--(80)].
\begin{GPTcorrection}
For the block convention $Q=\left(\begin{smallmatrix}0&O^\dagger\\O&0\end{smallmatrix}\right)$ and
$\gamma^5=\mathrm{diag}(1,-1)$, $G\,dQ=\mathrm{diag}(O^{-1}dO,(O^\dagger)^{-1}dO^\dagger)$.
After polar deformation to a unitary invertible symbol, the two integrated winding forms have opposite signs.
Thus the chiral trace gives twice the block winding. With the CSE (inward) orientation this is
\[
N_3^0=-\frac1{24\pi^2}\int_{S^3_{\rm out}}\tr(o^{-1}d_po)^3,
\]
which equals $+1$ for $o=p_4+i\vec\sigma\cdot\vec p$.
\reviewfix{Here and below $S^3_{\rm out}$ is the large sphere $|p|=R$ in the four-momentum fibre $\R^4$ at fixed $x$,
oriented as the boundary of the ball $|p|\le R$ (outward normal $\hat p=p/R$), with $R\to\infty$; it is the
sphere at infinity $S^3_\infty$ of Sec.~\ref{sec:CSEder}, and the subscript only records the orientation. The integral is independent of $R$ as soon as
$R$ exceeds all singularities of $o^{-1}$ (exactly so for a symbol homogeneous in $p$), so the limit is a
finite-radius integral of a closed form on $\R^4\setminus\{\text{singularities}\}$.}
\end{GPTcorrection}

\begin{example}[Sec.~5 of Ref.~\cite{XZPLB}]\label{ex:multiWeyl}
$\hO=\hpi_4+i\begin{pmatrix}\hpi_3&\kappa(\hpi_1-i\hpi_2)^n\\ \kappa(\hpi_1+i\hpi_2)^n&-\hpi_3\end{pmatrix}$ is
$x$-independent and in the Weyl-ordered class (powers of the single operators $\hpi_1\pm i\hpi_2$). With
$|O|^2=p_4^2+p_3^2+\kappa^2(p_1^2+p_2^2)^n$, $S_{\alpha\beta\nu}=\epsilon_{\alpha\beta\nu\delta}\tilde p_\delta
4\kappa^2n^2(p_1^2+p_2^2)^{n-1}/|O|^4$ with $\tilde p=(p_1/n,p_2/n,p_3,p_4)$, and with $u=\kappa(p_1^2+p_2^2)^{n/2}$,
$u\,du=n\kappa^2(p_1^2+p_2^2)^{n-1}\rho\,d\rho$, the slice integral reduces to $n$ times the $n=1$ one,
$\int\frac{d^3p}{8\pi^2}S_{123}=\frac n2\sgn p_4$, hence $N_3=N_3^0=n$ and $\sA=-n\frac{i}{4\pi^2}\int\tr F\wedge F$
[Ref.~\cite{XZPLB}, Eqs.~(83)--(86)]. For $n\ge2$ the operator is quasi-elliptic rather than elliptic (its principal symbol in the \corrB{standard isotropic sense of Appendix~\ref{app:ops}} vanishes at $p_1=p_2=0$, while the symbol is invertible for $p\ne0$ and homogeneous with respect to the
anisotropic weights $(1/n,1/n,1,1)$ of $(p_1,p_2,p_3,p_4)$; Appendix~\ref{app:ops}); the integral is still well defined, but its interpretation as a Fredholm index requires the anisotropic (quasi-elliptic) symbol calculus rather than the elliptic theorem used below.\footnote{Sec.~5 of Ref.~\cite{XZPLB} presents the example in the elliptic setting of its
introduction; for $n\ge2$ the appropriate notion is quasi-ellipticity.}
\reviewfix{What happens to the derivations of the anomaly for quasi-elliptic operators, the Fredholm property of
this operator and an exact check of the coefficient $n$ are given in Sec.~\ref{sec:quasiell}.}
\end{example}

\subsection{\corr{General ordering and the integrated anomaly}}
\label{sec:homotopy}

\begin{GPTcorrection}
On a closed manifold two chiral elliptic operators with the same principal symbol differ by an operator of lower
order (for a reordering of the derivatives in an operator of order $m$, of order at most $m-2$), which is compact as
a map $H^m\to L^2$ ($H^m$ the Sobolev space of sections with $m$ derivatives in $L^2$; a compact perturbation does
not change the Fredholm index, Appendix~\ref{app:ops}); the linear interpolation between them stays elliptic and, if both endpoints are Hermitian and
chiral, Hermitian and chiral. Hence the index of the chiral block, and with it the integrated anomaly, is the same
for all orderings, and \eqref{eq:PLB13} extends to any ordering within this class. (A left-coefficient
symmetrization need not be Hermitian by itself; Hermiticity is imposed on the comparison operator. On a non-compact
manifold relative compactness requires decay hypotheses.) The statement concerns the integrated anomaly, not the
local anomaly density or the finite-field transport coefficients, which depend on the ordering through the $O(F)$
terms of Sec.~\ref{sec:class}.
\end{GPTcorrection}

\subsection{Factorization beyond the homotopy hypothesis}
\label{sec:factor}

\paragraph{Setting and notation.} Let $M$ be a closed, connected, oriented four-manifold. The operator is the chiral
block $\hO_A$ of $\hQ=\sum_\alpha c_\alpha(x)\Sym[(-iD)^\alpha]$ (Theorem~\ref{thm:class}), whose coefficient
functions $c_\alpha(x)$ act on the spinor and internal (flavour, orbital) indices and are gauge singlets, the gauge
field entering only through $D_\mu=\partial_\mu-iA_\mu$. The objects entering the theorem are the following.

\emph{The bundles $V_+$, $V_-$.} These are the bundles over $M$ in which the two chiralities of the fermion take
their values \emph{before} the gauge indices are attached: $V_\pm=S_\pm\otimes I$, where $S_\pm$ are the two chiral
spinor bundles (the $\pm1$ eigenbundles of $\gamma^5$ in the spinor bundle $S=S_+\oplus S_-$; rank two each, trivial
on $T^4$) and $I$ is the internal space on which the $c_\alpha$ act ($I=\C$ for the Weyl operator, $I=\C^K$ for the
texture example below). The chiral block of $\hQ$ at $A=0$ for a \emph{single} gauge component is
$\hO_0^{(1)}:\Gamma(V_+)\to\Gamma(V_-)$; with $N$ gauge components and $A=0$ it is $\hO_0=\hO_0^{(1)}\otimes1_{\C^N}$
acting on $V_+\otimes\C^N$; with the gauge field it is $\hO_A:\Gamma(V_+\otimes E)\to\Gamma(V_-\otimes E)$, where
$E$ is the gauge bundle of rank $N$ with connection $A$ (Appendix~\ref{app:index}).

\emph{The symbol classes $\xi_0^{(1)}$, $\xi_0$, $\xi_A$.} Let $\pi:T^*M\to M$, $(x,p)\mapsto x$, be the
projection of the cotangent bundle, and $\pi^*V_\pm$ the \emph{pull-backs} of $V_\pm$ to $T^*M$: the bundles over
$T^*M$ whose fibre over $(x,p)$ is $(V_\pm)_x$, the same space for every momentum $p$ at the point $x$ (for trivial
$V_\pm=M\times\C^r$ simply $T^*M\times\C^r$). They are needed because the principal symbol $o(x,p)$ is a function on
$T^*M$ which at each $(x,p)$ maps $(V_+)_x$ to $(V_-)_x$, i.e.\ a bundle map $o:\pi^*V_+\to\pi^*V_-$ over $T^*M$,
invertible away from the zero section $p=0$. Then $\xi_0^{(1)}=[\pi^*V_+,\pi^*V_-;o]\in K_c(T^*M)$ is the symbol
class (Appendix~\ref{app:index}) of the single-component operator $\hO_0^{(1)}$: the formal difference of the two
pulled-back bundles, identified with each other by $o$ outside a neighbourhood of the zero section. It carries all the information about the background (the $x$-dependence of the $c_\alpha$) and
none about the gauge field. $\xi_0=\xi_0^{(1)}\otimes\C^N=N\xi_0^{(1)}$ is the class of $\hO_0$. Since the
$c_\alpha$ are gauge singlets, the principal symbol of $\hO_A$ is $o\otimes1_E$ -- $\hO_A$ is $\hO_0^{(1)}$ twisted
by $(E,A)$ -- and the twisting formula gives $\xi_A=\xi_0^{(1)}\cdot\pi^*[E]$ (the connection $A$ is of lower order
and does not enter the class).

\emph{The condition $p_1(M)=0$ in rational cohomology.} $p_1(M)=-\frac1{8\pi^2}\tr R\wedge R$ is the first
Pontryagin form of the Levi-Civita curvature (Appendix~\ref{app:bundles}); ``$p_1(M)=0$ in rational cohomology''
means that its de Rham class vanishes, i.e.\ that the form $p_1(M)$ is exact, equivalently that its integral over
every closed four-cycle is zero. On a closed connected oriented four-manifold $H^4(M;\mathbb Q)\cong\mathbb Q$ (and
$H^4(M;\Z)\cong\Z$, so that the rational and the integral conditions coincide), and the condition is simply
$\int_Mp_1(M)=0$; by the Hirzebruch signature theorem, $\sigma(M)=\frac13\int_Mp_1(M)$, it is the vanishing of the
signature of the intersection form on $H^2(M)$. It holds for $T^4$ (flat, $p_1\equiv0$), $S^4$, $S^2\times S^2$ and
$S^1\times S^3$, and fails for $\C P^2$ ($\sigma=1$) and the K3 surface ($\sigma=-16$). The condition is used in one
place only: the Todd factor of \eqref{eq:Fuji}, $\Td(TM\otimes\C)=1-\frac1{12}p_1(M)$ through degree four, contributes
$-\frac1{12}\int_Mp_1(M)\wedge\pi_*\ch_2(\xi_A)=\frac{NN_3^0}{12}\int_Mp_1(M)$ to the index (with $N_3^0$ defined below),
which vanishes exactly when $\int_Mp_1=0$; without the condition this gravitational term is simply to be added to
the formula of Theorem~\ref{thm:main}.

With this notation, by multiplicativity of the Chern character,
\begin{equation}
\ind\hO_A=\int_{T^*M}\ch\big(\xi_0^{(1)}\big)\wedge\pi^*\ch(E),\qquad \ch(E)=\tr\exp(F/2\pi)=N+c_1(E)+\ch_2(E),
\label{eq:twisted}
\end{equation}
with $c_1(E)=\frac1{2\pi}\tr F$ the first Chern class and $\ch_2(E)=\frac1{8\pi^2}\tr F\wedge F$
(with $D=d-iA$, Hermitian $A$, the curvature is $D^2=-iF$ and $\ch=\tr\exp(iD^2/2\pi)$). Integration over the fibre $\R^4_p$ of $\pi:T^*M\to M$,
$\pi_*:H^{2k}_c(T^*M;\mathbb{Q})\to H^{2k-4}(M;\mathbb{Q})$ (it lowers the form degree by four, satisfies
$\pi_*(\pi^*a\wedge b)=a\wedge\pi_*b$ and $\int_{T^*M}b=\int_M\pi_*b$, and realizes the cohomological Thom
isomorphism; Appendix~\ref{app:index}), defines the three invariants of the background
\begin{equation}
N_3^0:=-\pi_*\ch_2(\xi_0^{(1)})\in\Z,\qquad \omega_2:=\pi_*\ch_3(\xi_0^{(1)})\in H^2(M),\qquad
\nu:=\int_{T^*M}\ch_4(\xi_0^{(1)})=\ind\hO_0^{(1)},
\end{equation}
where $N_3^0=-\frac1{24\pi^2}\int_{S^3_\infty}\tr(o^{-1}d_po)^3$ at fixed $x$ is the signed fibre winding \reviewfix{(star-free; the superscript distinguishes it from the Moyal expression $N_3$ of \eqref{eq:PLB13})}, the sign
fixed so that $N_3^0=+1$ for $o=p_4+i\vec\sigma\cdot\vec p$ in agreement with Ref.~\cite{XZPLB}, Eq.~(11) (for $u=p_4+i\vec\sigma\cdot
\vec p$, $\tr(u^{-1}du)^3=12\,dp_1dp_2dp_3$ at the north pole while the outward orientation there is
$-dp_1dp_2dp_3$). $N_3^0$ is automatically independent of $x$: $\ch_2(\xi_0^{(1)})$ is closed with compact support
in the fibres. \reviewfix{In elementary terms: $N_3^0(x)$ is the degree of the map $p\mapsto o(x,p)$ from the sphere
$|p|=R$ to the invertible matrices, an integer. Ellipticity means that $o(x,p)$ is invertible for all $p\ne0$ and,
by homogeneity and continuity, that the sphere of a fixed radius $R$ avoids the zeros of $o$ for every $x$ of the compact $M$; the
family of maps is therefore continuous in $x$, and an integer-valued continuous function on a connected $M$ is
constant. The continuity of the degree is the statement that it cannot jump at any $x_0$: the degree is a
homotopy invariant, and for $x$ close to $x_0$ the straight-line homotopy $o_t=(1-t)\,o(x_0,\cdot)+t\,o(x,\cdot)$,
$t\in[0,1]$, stays invertible on the sphere because
$\|o(x,\cdot)-o(x_0,\cdot)\|_{S^3}<\|o(x_0,\cdot)^{-1}\|_{S^3}^{-1}$ by uniform continuity, so $N_3^0(x)=N_3^0(x_0)$
in a neighbourhood of $x_0$. A jump at $x_0$ would require this homotopy to fail for $x$ arbitrarily close to
$x_0$, i.e.\ a zero of $o(x_0,\cdot)$ on the sphere -- the breakdown of ellipticity at $x_0$. Thus two regions with
different $N_3^0$ cannot be joined by a smooth elliptic symbol: any interpolation, e.g.\
$o_t=p_4+i(1-2t)\vec\sigma\cdot\vec p$ between $N_3^0=1$ and $N_3^0=-1$, passes through a symbol ($o_{1/2}=p_4$)
that is not invertible on the sphere, and such a domain wall is not a lower-order (slowly varying background)
modification of the operator but a degeneration of its principal symbol. The same follows from Stokes' theorem: with $d=d_x+d_p$ the form $\tr(o^{-1}do)^3$ is closed wherever $o$
is invertible, $d\tr(o^{-1}do)^3=-\tr(o^{-1}do)^4=0$ by cyclicity, so its component with one $dx$ and three $dp$'s
obeys $\partial_{x^\mu}\big[\tr(o^{-1}d_po)^3\big]=-d_p\big[\ldots\big]_{\mu}$, an exact form on the sphere whose
integral vanishes. The degree could change only if a zero of $o$ crossed the sphere as $x$ varies, i.e.\ if a
singularity escaped to infinite momentum, which uniform ellipticity forbids; for the non-elliptic models, where $o$
is not invertible on any large sphere, no such degree exists.}

\paragraph{The content of the twisted-index formula and of the three invariants.} We spell out the objects
entering \eqref{eq:twisted} and the display following it.

\emph{$\pi^*\ch(E)$.} $\ch(E)=\tr\exp(F/2\pi)=N+\frac1{2\pi}\tr F+\frac1{8\pi^2}\tr F\wedge F$ is a closed,
inhomogeneous differential form on $M$ (a $0$-form, a $2$-form and a $4$-form; higher terms vanish since $\dim M=4$),
built from the gauge field alone. $\pi^*\ch(E)$ is the same form regarded as a form on the eight-dimensional
$T^*M$: in the coordinates $(x,p)$ its components contain only $dx$'s and depend only on $x$. In a wedge product with
a form on $T^*M$ it therefore acts as a coefficient which is constant along the fibres, which is what allows the
fibre integration below to be performed first.

\emph{Why \eqref{eq:twisted} holds.} The Atiyah--Singer formula \eqref{eq:Fuji} with $\Td=1$ (the $p_1$ condition)
gives $\ind\hO_A=\int_{T^*M}\ch(\xi_A)$. The symbol class of the twisted operator is represented by the triple
$(\pi^*V_+\otimes\pi^*E,\ \pi^*V_-\otimes\pi^*E;\ o\otimes1_E)$. Choosing on $\pi^*V_\pm$ connections $\nabla_\pm$
identified by $o$ outside a neighbourhood of the zero section, and on $\pi^*E$ the pull-back of the gauge connection,
the curvature of a tensor product is $F_\pm\otimes1+1\otimes F_E$ with commuting summands, so that
$\tr\exp\big((F_\pm\otimes1+1\otimes F_E)/2\pi\big)=\ch(\nabla_\pm)\wedge\pi^*\ch(E)$ and
\[
\ch(\xi_A)=\big(\ch(\nabla_+)-\ch(\nabla_-)\big)\wedge\pi^*\ch(E)=\ch(\xi_0^{(1)})\wedge\pi^*\ch(E),
\]
a form with compact support in the fibres because the first factor has. This is the multiplicativity used in
\eqref{eq:twisted}; the integral over $T^*M$ selects the part of total degree eight, the orientation of $T^*M$ being
$dx^1\wedge\dots\wedge dx^4\wedge dp_1\wedge\dots\wedge dp_4$ (for $n=4$ this coincides with the symplectic
orientation of \cite{AS}, which is why the sign $(-1)^n$ of the index formula is $+1$).

\emph{Degree bookkeeping.} $\ch(\xi_0^{(1)})=\ch_2+\ch_3+\ch_4$ as a compactly supported form on $T^*M$
($\ch_0$ is the rank difference, zero; $\ch_1$ would have to be paired with a $6$-form from $\ch(E)$, which does not
exist on a four-manifold; forms of degree $>8$ vanish). The degree-eight part of
$\ch(\xi_0^{(1)})\wedge\pi^*\ch(E)$ is therefore
\[
\ch_2(\xi_0^{(1)})\wedge\pi^*\ch_2(E)+\ch_3(\xi_0^{(1)})\wedge\pi^*c_1(E)+\ch_4(\xi_0^{(1)})\cdot N ,
\]
and the integral over $T^*M$ is performed as a fibre integral followed by an integral over $M$ (Fubini), using
$\int_{T^*M}\pi^*a\wedge b=\int_Ma\wedge\pi_*b$ (projection formula), which turns the three terms into
$\int_M\pi_*\ch_2(\xi_0^{(1)})\wedge\ch_2(E)$, $\int_M\pi_*\ch_3(\xi_0^{(1)})\wedge c_1(E)$ and
$N\int_M\pi_*\ch_4(\xi_0^{(1)})$: this is Fig.~\ref{fig:degrees} and, after inserting the definitions of the second
display, Theorem~\ref{thm:main}.

\emph{$\pi_*\ch_2(\xi_0^{(1)})$.} $\ch_2(\xi_0^{(1)})$ is a compactly supported $4$-form on $T^*M$. Decompose it
according to the number of $dp$'s; the fibre integral $\pi_*$ keeps only the component with four $dp$'s,
$f(x,p)\,dp_1\wedge dp_2\wedge dp_3\wedge dp_4$, and integrates it over the fibre:
$(\pi_*\ch_2)(x)=\int_{\R^4}d^4p\,f(x,p)$, a function ($0$-form) on $M$. Fibre integration commutes with $d$ for
forms with compact support in the fibres, $d\,\pi_*b=\pi_*db$, so $\pi_*\ch_2$ is closed, i.e.\ locally constant,
hence constant on connected $M$. Its value is the winding of the symbol at infinity: with $\nabla_+=d_p$ and
$\nabla_-=d_p+\chi(|p|)\,o\,d_p(o^{-1})$ at fixed $x$ (Appendix~\ref{app:index}),
$\ch_2(\nabla_+)-\ch_2(\nabla_-)$ is the exterior derivative of a Chern--Simons three-form which on the sphere
$|p|=R\to\infty$ equals $\pm\frac1{24\pi^2}\tr(o^{-1}d_po)^3$, and Stokes' theorem gives
$\pi_*\ch_2(\xi_0^{(1)})=\frac1{24\pi^2}\int_{S^3_{\rm out}}\tr(o^{-1}d_po)^3=-N_3^0$ in the orientation of this
review ($-1$ for the Bott class). Thus $N_3^0$ is an integer, the same at every $x$, which depends only on the
principal symbol on a large sphere in the fibre: the quantity which the CSE expression \eqref{eq:PLB13} computes
under the conditions of Sec.~\ref{sec:stokes}. Its coefficient in \eqref{eq:twisted} is
$\int_M\ch_2(E)=\frac1{8\pi^2}\int\tr F\wedge F$, the instanton number: the gauge-quadratic term of the anomaly.

\emph{$\pi_*\ch_3(\xi_0^{(1)})$.} $\ch_3(\xi_0^{(1)})$ is a compactly supported $6$-form on $T^*M$; the component
with four $dp$'s and two $dx$'s, $\frac12g_{\mu\nu}(x,p)\,dx^\mu\wedge dx^\nu\wedge d^4p$, integrated over the fibre
gives the $2$-form $\omega_2=\frac12\big(\int d^4p\,g_{\mu\nu}\big)dx^\mu\wedge dx^\nu$ on $M$, closed for the same
reason, with a rational cohomology class.

\emph{Explicit form of the fibre integrals.} Take $\nabla_+=d$ and $\nabla_-=d-\chi(|p|)\,\theta$, where
\[
\theta=(do)\,o^{-1}=\theta_x+\theta_p,\qquad \theta_x=\partial_{x^\mu}o\,o^{-1}\,dx^\mu,\qquad
\theta_p=\partial_{p_a}o\,o^{-1}\,dp_a,
\]
$d=d_x+d_p$ is the exterior derivative on $T^*M$ and $\chi$ a cutoff equal to $0$ near $p=0$ and to $1$ for large
$|p|$ (so that $\nabla_-=o\,\nabla_+o^{-1}$ there). Using $d\theta=\theta\wedge\theta$, the curvature of $\nabla_-$ is
$F_-=-d\chi\wedge\theta-\chi(1-\chi)\,\theta\wedge\theta$, and since $\tr\theta^{2k}=0$ and $d\chi\wedge d\chi=0$,
\[
\ch_k(\xi_0^{(1)})=\ch_k(\nabla_+)-\ch_k(\nabla_-)
=-\frac1{(k-1)!}\Big(\frac{-i}{2\pi}\Big)^k\chi^{k-1}(1-\chi)^{k-1}\,d\chi\wedge\tr\theta^{2k-1},\qquad k\ge1 .
\]
The fibre integral keeps the component of $\tr\theta^{2k-1}$ with three $dp$'s (the $dx$'s being moved to the left),
the radial integral gives $\int_0^1\chi^{k-1}(1-\chi)^{k-1}d\chi=((k-1)!)^2/(2k-1)!$, and therefore
\begin{equation}
\pi_*\ch_k(\xi_0^{(1)})=-\frac{(k-1)!}{(2k-1)!}\Big(\frac{-i}{2\pi}\Big)^k\oint_{S^3_{\rm out}}
\big[\tr\theta^{2k-1}\big]_{(2k-4,\,3)},
\label{eq:pistarchk}
\end{equation}
the subscript denoting the component with $2k-4$ $dx$'s and three $dp$'s, integrated over a large sphere in the
fibre (for a symbol homogeneous in $p$ the result does not depend on the radius; in general it changes by an exact
form on $M$). For $k=2$ this is $\pi_*\ch_2=\frac1{24\pi^2}\oint\tr(o^{-1}d_po)^3=-N_3^0$, as above. For $k=3$,
\begin{equation}
\omega_2=-\frac{i}{480\pi^3}\oint_{S^3_{\rm out}}\big[\tr(\theta\wedge\theta\wedge\theta\wedge\theta\wedge\theta)\big]_{(2,3)},
\qquad
\big[\tr\theta^5\big]_{(2,3)}=5\,\tr(\theta_x\theta_x\theta_p\theta_p\theta_p)+5\,\tr(\theta_x\theta_p\theta_x\theta_p\theta_p),
\label{eq:omega2}
\end{equation}
the two terms being the two cyclic classes of words with two $\theta_x$ and three $\theta_p$ (products are wedge
products, and cyclic permutations of five one-forms under the trace carry no sign). In components,
$\omega_2=\frac12\Omega_{\mu\nu}dx^\mu\wedge dx^\nu$ with
\[
\Omega_{\mu\nu}=-\frac{i}{96\pi^3}\oint_{|p|=R}dS\,\hat p_d\,\epsilon_{dabc}\,
\tr\big(\theta_\mu\theta_\nu\theta_a\theta_b\theta_c-\theta_\mu\theta_a\theta_\nu\theta_b\theta_c
+\theta_\mu\theta_a\theta_b\theta_\nu\theta_c-\theta_\mu\theta_a\theta_b\theta_c\theta_\nu\big),
\]
$\theta_\mu=\partial_{x^\mu}o\,o^{-1}$, $\theta_a=\partial_{p_a}o\,o^{-1}$, $dS$ the area element of the sphere and
$\hat p=p/R$ its outward normal. We have checked \eqref{eq:omega2} numerically on the texture symbol
$o=\sigma^\mu p_\mu\otimes P+\bar\sigma^\mu p_\mu\otimes(1-P)$ with a generic rank-one projector field $P(x)$ in
$\C^2$: the right-hand side reproduces $-2c_1(W)=-\frac i\pi\tr(P\,dP\wedge dP)$ pointwise to the accuracy of the
quadrature. For $k=4$, \eqref{eq:pistarchk} integrated over $M$ gives
$\nu=-\frac1{13440\pi^4}\int_{S^*M}\tr\theta^7$, the index of $\hO_0^{(1)}$ as a winding number on the cosphere
bundle $S^*M$ (on which a seven-form has exactly four $dx$'s and three $dp$'s); this is the formula
$\ind=-\frac{(n-1)!}{(2\pi i)^n(2n-1)!}\int_{S^*M}\tr(o^{-1}do)^{2n-1}$ of Fedosov \cite{Fedosov} for $n=4$, which
confirms the normalization of \eqref{eq:pistarchk}. It is a \emph{mixed} invariant: it vanishes when the symbol does not
depend on $x$ (then $\ch(\xi_0^{(1)})$ contains no $dx$ at all) and, more generally, when the $x$-dependence is
topologically trivial; it is nonzero when the symbol winds jointly in $x$ and $p$, as in the texture example below,
where $\omega_2=-2c_1(W)$ is the first Chern class of the subbundle selected by the projector field. In
\eqref{eq:twisted} it is paired with $c_1(E)=\frac1{2\pi}\tr F$, so it contributes only for a gauge group with a
$U(1)$ factor, $\int_M\omega_2\wedge c_1(E)$ being $2\pi$ times the flux quanta of $\tr F$ through the two-cycles dual
to $\omega_2$; for $SU(N)$ it drops out.

\emph{$\nu=\int_{T^*M}\ch_4(\xi_0^{(1)})$.} The compactly supported $8$-form $\ch_4(\xi_0^{(1)})$ integrated over
all of $T^*M$ (equivalently $\int_M\pi_*\ch_4$) is, by the Atiyah--Singer formula with $\Td=1$, the index of the
untwisted single-component operator $\hO_0^{(1)}$; it is paired with $\ch_0(E)=N$, each gauge component
contributing the same untwisted index. It, too, vanishes for $x$-independent symbols.

In words: the index of the gauged operator is a sum over the three ways of distributing the eight form degrees
between the background (fibre winding, mixed class, untwisted index) and the gauge field (instanton number, first
Chern class, rank), and $N_3^0$ is the only one of the three background invariants that survives for a homogeneous
symbol or for an $SU(N)$ gauge field. \begin{GPTcorrection}
(If the symbol extends invertibly over the whole fibre at one point $x$, its boundary winding vanishes and $N_3^0=0$
everywhere.) On $S^4$ with trivial chiral bundles the restriction of the symbol class to the zero section vanishes,
so that $e(T^*M)\,\pi_*\ch_2(\xi)=0$, with $e$ the Euler class and $\int_{S^4}e=\chi(S^4)=2$: a nonzero fibre charge requires nontrivial chiral bundles,
as for the spinor bundles. The bookkeeping of the three invariants is summarized in Fig.~\ref{fig:degrees}.
\end{GPTcorrection}

On $T^4$ the group of symbol classes, $K_c(T^*T^4)\cong K^0(T^4)\cong\Z^8$ (compactly supported $K$-theory of the cotangent bundle; the first isomorphism is the Thom isomorphism and $8=1+6+1$ is the rank of the even cohomology of $T^4$; Appendix~\ref{app:index}), is torsion free and $\xi_0^{(1)}$ is determined by
$(N_3^0,\omega_2,\nu)$; an $x$-independent symbol has $\omega_2=\nu=0$; hence the \corr{stable symbol class equals that of an
$x$-independent representative (allowing pseudodifferential stabilization) iff} $\omega_2=0$ and $\nu=0$ (a $K$-theory criterion: homotopy implies it, not conversely --
$e^{ix_1}(p_4+i\vec\sigma\cdot\vec p)$ has the class of the Weyl symbol but is not homotopic to it).

\begin{theorem}[General factorization]\label{thm:main}
With no assumption on the $x$-dependence of the symbol,
\begin{align}
\sA(A)&=\sA(0)+2i\int_M\omega_2\wedge c_1(E)-2iN_3^0\int_M\ch_2(E)\notag\\
&=\sA(0)+2i\int_M\omega_2\wedge c_1(E)-N_3^0\frac{i}{4\pi^2}\int_M\tr F\wedge F,\qquad
\sA(0)=2i\,\ind\hO_0=2iN\nu .
\end{align}
For $SU(N)$ (or any $E$ with $c_1(E)=0$), $\sA(A)-\sA(0)=-N_3^0\frac{i}{4\pi^2}\int\tr F\wedge F$ for every elliptic
$Q(x,p)$, and Ref.~\cite{XZPLB}, Eq.~(13) holds verbatim if and only if $\ind\hO_0=0$ -- strictly weaker than the homotopy hypothesis,
\corr{which implies also $\omega_2=0$ on $T^4$ with fixed trivializations.}
\end{theorem}
\begin{proof}
Insert $\ch(E)$ into \eqref{eq:twisted}, keep the degree-eight part and integrate over the fibres first in the last
two terms (Fig.~\ref{fig:degrees}).
\end{proof}

\begin{figure}[htb]
\centering
\begin{tikzpicture}[>=Stealth,every node/.style={font=\small}]
\def\w{3.1} \def\h{1.05}
% headers
\node at (0.5*\w,3.5*\h) {$\ch_0(E)=N$};
\node at (1.5*\w,3.5*\h) {$\ch_1(E)=c_1(E)$};
\node at (2.5*\w,3.5*\h) {$\ch_2(E)=\frac1{8\pi^2}\tr F\wedge F$};
\node[anchor=east] at (-0.15,2.5*\h) {$\ch_2(\xi_0^{(1)})\ \xrightarrow{\ \pi_*\ }\ -N_3^0$};
\node[anchor=east] at (-0.15,1.5*\h) {$\ch_3(\xi_0^{(1)})\ \xrightarrow{\ \pi_*\ }\ \omega_2$};
\node[anchor=east] at (-0.15,0.5*\h) {$\ch_4(\xi_0^{(1)})\ \xrightarrow{\ \pi_*\ }\ \nu$};
% grid
\draw (0,0) grid[xstep=\w,ystep=\h] (3*\w,3*\h);
% anti-diagonal (total degree 8 on T^*M)
\fill[gray!20] (2*\w,2*\h) rectangle (3*\w,3*\h);
\fill[gray!20] (\w,\h) rectangle (2*\w,2*\h);
\fill[gray!20] (0,0) rectangle (\w,\h);
\draw (0,0) grid[xstep=\w,ystep=\h] (3*\w,3*\h);
\node at (2.5*\w,2.5*\h) {$-N_3^0\int_M\ch_2(E)$};
\node at (1.5*\w,1.5*\h) {$\int_M\omega_2\wedge c_1(E)$};
\node at (0.5*\w,0.5*\h) {$N\nu=\ind\hO_0$};
\node[gray] at (0.5*\w,2.5*\h) {degree $<8$};
\node[gray] at (1.5*\w,2.5*\h) {degree $<8$};
\node[gray] at (0.5*\w,1.5*\h) {degree $<8$};
\node[gray] at (2.5*\w,1.5*\h) {degree $>8$};
\node[gray] at (1.5*\w,0.5*\h) {degree $>8$};
\node[gray] at (2.5*\w,0.5*\h) {degree $>8$};
\node at (1.5*\w,-0.6) {$\ind\hO_A=\displaystyle\int_{T^*M}\ch(\xi_0^{(1)})\wedge\pi^*\ch(E)=
N\nu+\int_M\omega_2\wedge c_1(E)-N_3^0\int_M\ch_2(E),\qquad \sA=2i\,\ind\hO_A$};
\end{tikzpicture}
\caption{Degree bookkeeping of the twisted index \eqref{eq:twisted} on a four-manifold. Only the products of total
degree eight on $T^*M$ (shaded) survive; fibre integration $\pi_*$ turns them into the three invariants of the
background, $N_3^0$ (fibre winding), $\omega_2\in H^2(M)$ and $\nu=\ind\hO_0^{(1)}$, paired with the components of
$\ch(E)$ of complementary degree. The gauge-quadratic term is universal; the mixed term needs $c_1(E)\ne0$; the
$A$-independent term is the untwisted index.}
\label{fig:degrees}
\end{figure}
Here $N_3^0$ is the fibre degree, not the Moyal expression $N_3$ of \eqref{eq:PLB13}, which equals it under the conditions of
Sec.~\ref{sec:stokes}; background gauge fields enter through $\omega_2,\nu$ (equivalently through $E\to L\otimes E$),
not as separate terms. \corr{$\ind\hO_0=0$ follows from an actual invertible symmetry identifying the two chiral kernels. On the flat torus it also follows when the principal symbol is independent of $x$ and only lower-order coefficients vary; the analogous assertion on a general curved manifold requires the bundle and gravitational contributions to vanish.} \corr{For $U(N)$ with $c_1(E)\ne0$ a mixed anomaly term remains; identifying it with a specific transport coefficient requires a separate response derivation.} Which classes $[E]$, and hence which of the three terms, can occur is fixed by the global form of the gauge group rather than by its Lie algebra: for the Standard Model group $(SU(3)\times SU(2)\times U(1))/Z_6$ the admissible bundles carry correlated fractional fluxes of the three factors, which is the origin of the additional $Z_6$ symmetry of its lattice formulation \cite{BVZ}. \corr{On general $M$ the omitted Todd factor contributes $-\frac1{12}\int_M p_1(M)\wedge\pi_*\ch_2(\xi_A)$ to the index, since $\Td(TM\otimes\C)=1-p_1(M)/12$ through degree four.}

\begin{example}[Texture not homotopic to any $Q(p)$]
$M=T^4$, internal space $V\otimes\C^K$, $P(x)$ a rank-$k$ projector field, $W=\mathrm{im}P$,
$o=\sigma^\mu p_\mu\otimes P+\bar\sigma^\mu p_\mu\otimes(1-P)$. With $\beta$ the Bott class (the symbol class of the Weyl operator $p_4+i\vec\sigma\cdot\vec p$ on $\R^4$, the
generator of $K_c(\R^4)\cong\Z$; Appendix~\ref{app:index}), $\ch(\xi_0^{(1)})=\ch(\beta)\wedge(2\ch(W)-K)$,
$\int_{\R^4}\ch_2(\beta)=-1$ in our orientation, and $c_1(W)=\frac{i}{2\pi}\tr(P\,dP\wedge dP)$ the first Chern
class of the subbundle $W$: $N_3^0=2k-K$, $\omega_2=-2c_1(W)$, $\nu=-2\int\ch_2(W)$. For $k=1$, $K=2$ one has
$c_1(W)^2=0$; for $P=P_{12}\otimes P_{34}$ in $\C^4$ (unit flux on each 2-torus) $\int\ch_2(W)=1$, $N_3^0=-2$, $\nu=-2$,
and, for a $U(N)$ field $E$, $\sA(A)=-4iN\int\ch_2(W)-4i\int c_1(W)\wedge c_1(E)-(2k-K)\frac{i}{4\pi^2}\int\tr F\wedge F$ (the mixed term vanishes for $SU(N)$). For $P=P_{12}\oplus0$ in $\C^3$: $N_3^0=-1$,
$\nu=0$, $\omega_2\neq0$: not homotopic to a $Q(p)$, yet \corrB{Ref.~\cite{XZPLB}, Eq.~(13)} holds for $SU(N)$.
\end{example}

\begin{example}[Background magnetic field]
A background $U(1)$ field \corrB{$A^{\rm bg}$ (field strength $F_0$; $A_0$ denotes the time component elsewhere)} with flux (constant or not) is the translation-invariant symbol twisted by a line
bundle $L$: $N_3^0$ unchanged, \corr{$\omega_2=-N_3^0c_1(L)$, $\nu=-N_3^0\int\ch_2(L)$}, and
$\sA=-N_3^0\frac{i}{4\pi^2}\int\tr(F+F_0)\wedge(F+F_0)$. For an $SU(N)$ field and a static magnetic background \corrB{Ref.~\cite{XZPLB}, Eq.~(13)}
holds verbatim; for a $U(1)$ field the cross term $-N_3^0\frac{i}{2\pi^2}\int\tr F\wedge F_0$ appears. Only the flux
enters the topology; the slices are Landau-level systems whose Hall-type invariants differ by the chiral zero modes
\corr{(on a compact transverse surface the Dirac index fixes the net chiral zero-mode count; on the plane the Aharonov--Casher count also depends on flux and normalizability conditions).}
\end{example}

\paragraph{\reviewfix{$\R^4$ and the two compactifications.}}
\reviewfix{The physical setting is often $\R^4$, where the chiral operator with coefficients approaching the free ones is
not Fredholm (Sec.~\ref{sec:anomaly}; Appendix~\ref{app:ops}, open manifolds), and the question is which of the
statements above survive and by which closure of $\R^4$ they are obtained. The local anomaly density of
Sec.~\ref{sec:pointsplit} is a point-splitting statement and holds pointwise on $\R^4$; under the homotopy hypothesis
the integrated anomaly, \eqref{eq:anomR4}, follows by Stokes in momentum space, again without compactness. What
requires a closed manifold is Theorem~\ref{thm:main}, whose content is an index: the terms $\sA(0)=2i\,\ind\hO_0$ and
$2i\int\omega_2\wedge c_1(E)$ are defined only after a condition at infinity has been imposed, and the two natural
conditions are the two compactifications.

\emph{$S^4$.} The one-point compactification is the closure for backgrounds trivial at infinity: a finite-action gauge
field, $A\to$ pure gauge and $F=O(|x|^{-4})$, extends to a bundle $E$ over $S^4$ with $\int_{S^4}\ch_2(E)$ the
instanton number, and the coefficients $c_\alpha(x)$ of the symbol must have a limit as $|x|\to\infty$, uniform in the
direction, so that the principal symbol is defined at the added point. On $S^4$, $p_1=0$ and $H^2(S^4)=0$, so the mixed
term vanishes identically, and
\[
\sA(A)=2i\,\ind\hO_0-N_3^0\frac{i}{4\pi^2}\int\tr F\wedge F ,
\]
with $\ind\hO_0$ the untwisted index of the compactified background: zero for the Dirac operator (positive scalar
curvature, no harmonic spinors), so that \eqref{eq:anomR4} is recovered verbatim; possibly nonzero for a symbol with
an $x$-dependent texture in a compact region, which is then a genuine $\R^4$ contribution invisible to the
homotopy-hypothesis route. Two caveats. The chiral bundles on $S^4$ are nontrivial, which is what permits $N_3^0\ne0$
there (above). And the identification of the $S^4$ index with the $L^2$ zero modes on $\R^4$ rests on conformal
covariance, available for the first-order Dirac operator [Appendix~\ref{app:ops}, item (c)]; for an operator of order
$m$ the inversion chart multiplies a homogeneous symbol by $|y|^{2m}$, the $S^4$ operator is the $\R^4$ one conjugated
by conformal weights, its index is given by the Atiyah--Singer formula, but its relation to $\R^4$ zero modes requires
decay conditions, i.e.\ weighted spaces [item (d)].

\emph{$T^4$.} Periodization is the closure the review already uses: the trace per unit volume is defined by periodizing
the symbol over supercells (Sec.~\ref{sec:CSEder}), $T^3$ in space and, with the thermal circle, $T^4$; it is the
natural closure for periodic backgrounds and for everything inherited from lattice regularizations. On $T^4$,
$p_1\equiv0$, the chiral bundles are trivial, the symbol class is torsion free and determined by $(N_3^0,\omega_2,\nu)$,
and the theorem holds with all three terms, the mixed one included for $U(N)$, since $H^2(T^4)\ne0$. For fermions in
the fundamental representation the gauge bundle is a genuine $SU(N)$ bundle and the instanton number is an integer;
't~Hooft's twisted boundary conditions, which admit fractional charges, are not available with fundamental fermions.
A finite-action configuration on $\R^4$ is not periodic: the same background cannot be put on both closures, each
being a different completion of $\R^4$.

\emph{Comparison.} The local density and the coefficient of $\int\tr F\wedge F$ are the same on either closure and agree
with the $\R^4$ statement. The background terms are not interchangeable: $\omega_2$ lives in $H^2(M)$ and vanishes on
$S^4$ but not on $T^4$, and $\ind\hO_0$ depends on how the symbol is closed up, trivially at one point or periodically.
This is as it should be -- on $\R^4$ these terms are not defined without a condition at infinity. When the background
is nontrivial at spatial infinity, for instance a texture winding on the sphere at infinity, neither closure applies:
that is the Callias or scattering-calculus setting of Appendix~\ref{app:ops}, items (a) and (b), where the index
acquires a contribution from the sphere at infinity for which the closed-manifold theorem has no term.}

\paragraph{Prior work.} The twisting formula is Atiyah--Singer III \cite{AS}; the homogeneous case of the theorem --
the anomaly of a general translation-invariant higher-derivative Dirac operator multiplied by its momentum-space
winding -- is due to Yee and Yi \cite{YeeYi}; the same idea is implicit in Nielsen--Ninomiya and Volovik
\cite{NielsenNinomiya,Volovik} and, for multi-Weyl nodes, in \cite{Dantas}, where $N_3^0$ is expressed through the Berry
curvature (non-interacting) and the background is homogeneous. The decomposition into three invariants and the criterion $\ind\hO_0=0$ are direct four-dimensional consequences of the twisting formula; we are not aware of their explicit statement in the literature.

\subsection{The Moyal $N_3$ equals the degree $N_3^0$: the same charge in the CSE and in the anomaly}
\label{sec:stokes}

\begin{GPTcorrection}
\reviewfix{Recall that $N_3$ denotes the Moyal expression for the charge, the second line of \eqref{eq:PLB13}
[Eq.~(78) of Ref.~\cite{XZPLB}],
\[
N_3=\frac{1}{48\pi^2|V|}\int d^3x\int_\Sigma\trD\big(\gamma^5G^{(0)}\star dQ_W\star G^{(0)}\star\wedge dQ_W\star G^{(0)}\star\wedge dQ_W\big),
\]
as distinct from the star-free degree $N_3^0$ of the principal symbol on the sphere $S^3_{\rm out}$ (Sec.~\ref{sec:factor});
the purpose of this subsection is to show that the two coincide under stated assumptions.}
Let $\Sigma(p_4)$ be the slice integral of \eqref{eq:PLB13} in a static background, with the orientation fixed in
the conventions. Assume (i) that $G_W=Q_W^{\star-1}$ exists and is smooth for every $p_4\ne0$ (Hamiltonian form: all
singularities at $p_4=0$); (ii) chiral anticommutation, $\{\gamma^5,Q_W\}=0$; (iii) that the spatial phase-space trace
$\int d^3x\,d^3p$ is closed under the Moyal product, the surface terms in $x$ and $\vec p$ being discarded. With
$\mathsf A=G\star dQ$ (exterior products in $(p_1,\dots,p_4)$ understood) one has
$d\,\trD\gamma^5\mathsf A^3=-\trD\gamma^5\mathsf A^4$ modulo traced star commutators, and the traced four-form
vanishes: moving one one-form around the other three changes its sign while $\gamma^5$ commutes with $\mathsf A$.
Hence $d\Sigma/dp_4=0$ on every interval free of singularities, and
\begin{equation}
N_3=\Sigma(0^+)-\Sigma(0^-)=\Sigma(+\infty)-\Sigma(-\infty),
\label{eq:slices}
\end{equation}
\reviewfix{the invariant expressed through the ultraviolet slices. For the same unnormalized
three-form the outward-sphere flux has the opposite sign to the inward-slice difference
(Fig.~\ref{fig:slab}); the minus sign in the sphere formula below accounts for this.}

\reviewfix{\emph{The surface term at $|\vec p|\to\infty$ behind \eqref{eq:slices}.} The first equality in
\eqref{eq:slices} is the definition \eqref{eq:CSE} with the slices at $p_4=\pm\epsilon$; the second one uses
$d\Sigma/dp_4=0$, and this step discards a surface term which deserves to be made explicit. Write
$\omega=\trD\gamma^5\mathsf A^3$ for the three-form in $p$-space (its coefficients depend on $x$ and are integrated
with $\int d^3x$), so that $\Sigma(p_4)=\int d^3x\int d^3p\,\omega_{123}(x,p_4,\vec p)$. Since
$(d\omega)_{4123}=\partial_4\omega_{123}-\partial_1\omega_{423}+\partial_2\omega_{413}-\partial_3\omega_{412}$,
\begin{equation}
\frac{d\Sigma}{dp_4}=\int d^3x\int d^3p\,(d\omega)_{4123}
+\lim_{R\to\infty}\int d^3x\oint_{|\vec p|=R}dS\,n_j\,\tfrac12\epsilon_{jkl}\,\omega_{4kl}(x,p_4,\vec p).
\label{eq:dSigma}
\end{equation}
The first term vanishes by the algebraic identity above: $d\omega$ is a sum of traced star commutators, i.e.\ of
total derivatives in $\vec x$ and $\vec p$ (for a static background the Moyal product involves $(\vec x,\vec p)$
only, $p_4$ being a parameter); their $\vec x$ part is removed by the trace per unit volume and their $\vec p$ part
is again a surface term of the same kind as the second one. The second term is the flux, through the large
two-sphere of the slice, of the components of $\omega$ with one leg along $p_4$; its vanishing is the assumption
behind \eqref{eq:slices} (assumption (iii)), and it holds whenever $\omega_{4kl}=o(R^{-2})$ uniformly on
$|\vec p|=R$ at fixed $p_4\ne0$. For an elliptic symbol of order $m$ this is automatic: $Q_W=O(|p|^{m})$,
$G_W=O(|p|^{-m})$ and $dQ_W=O(|p|^{m-1})$ give $\mathsf A=G_W\star dQ_W=O(|p|^{-1})$ in every direction, hence
$\omega=O(|p|^{-3})$ against the area $4\pi R^2$ of the sphere, and the flux is $O(R^{-1})$; the Moyal corrections
to $G_W=Q_W^{\star-1}$ carry additional powers of $|p|^{-1}$ (each $\partial_x\partial_p$ pair lowers the order by
one), so the estimate survives under the uniform symbol bounds of condition (C1). The margin is one power of $R$
whatever the order $m$, and the condition concerns the behaviour in $\vec p$ at fixed $p_4$ only. For the
non-elliptic Fermi-surface models of Sec.~\ref{sec:FS} the symbol still grows along every direction of $\vec p$ at
fixed $p_4\neq0$ (linearly in $p_3$, quadratically in $p_\perp$), the slice integrals converge and
\eqref{eq:slices} holds, $\Sigma(p_4)$ being constant for $p_4\gtrless0$; what fails for them is the next step
-- the absence of decay along $p_4$ -- which prevents the identification of the ultraviolet slices with the flux
through a large sphere. Were the decay in $\vec p$ slower than $|p|^{-2}$, the flux would be finite, $\Sigma$ would
depend on $p_4$, and the slice integral itself would converge only conditionally.}

A convenient chiral realization of a
Hamiltonian-form operator is
\[
O=p_4+ih,\quad h=h^\dagger,\qquad Q=\begin{pmatrix}0&O^\dagger\\O&0\end{pmatrix}
\]
(the resolvent $ip_4-h$ alone does not anticommute with a fixed $\gamma^5$ and must not be substituted for the full
chiral $Q$ in the trace formula).

To identify the ultraviolet flux with the fibre winding assume in addition that $Q_W$ is a classical elliptic symbol
of positive order (an asymptotic sum of terms homogeneous in $p$ of decreasing degree with invertible leading term;
Appendix~\ref{app:ops}): at large four-momentum the Moyal inverse is then given by the parametrix expansion (the
recursive inversion of the symbol starting from the inverse of the principal symbol), after radial
normalization the inverse approaches the inverse of the principal symbol, \reviewfix{and, with uniform symbol estimates for the inverse, integrable remainders on the slices and a vanishing
side boundary in the iterated limits (condition (C1) of Sec.~\ref{sec:latcont}), the subprincipal and Moyal
corrections to the integrated three-form vanish in the limit, so that}
\begin{equation}
N_3=-\frac{1}{24\pi^2|V|}\int d^3x\int_{S^3_{\rm out}}\tr(o^{-1}d_po)^3=N_3^0,
\label{eq:sphere}
\end{equation}
with the sign fixed in Sec.~\ref{sec:factor}.
\reviewfix{In short, \eqref{eq:sphere} holds because the star products disappear at $|p|\to\infty$, and the reason
is the following. The Moyal product is the series
$f\star g=fg+\sum_{n\ge1}\frac{1}{n!}\big(\tfrac{i}{2}\big)^n f\big(\overleftarrow{\partial}_x\cdot\overrightarrow{\partial}_p
-\overleftarrow{\partial}_p\cdot\overrightarrow{\partial}_x\big)^n g$, whose $n$-th term carries $n$ derivatives with
respect to $p$. For symbols with the estimates $|\partial_x^\alpha\partial_p^\beta f|\le C_{\alpha\beta}|p|^{m-|\beta|}$
(the class of Appendix~\ref{app:ops}; for the covariant symbol the $x$-derivatives bring the field strength and its
covariant derivatives, which are bounded) each $\partial_p$ lowers the growth by one power of $|p|$ while
$\partial_x$ does not raise it, so the $n$-th term is $O(|p|^{m_f+m_g-n})$: the star product is the pointwise product
plus corrections suppressed by powers of $1/|p|$, i.e.\ an asymptotic expansion in $1/|p|$ -- the same fact that
makes the composition formula of pseudodifferential operators asymptotic. Physically, the Moyal terms are powers of
(field gradient)/(momentum), and a fermion of large momentum does not resolve the inhomogeneity. Consequently the
star inverse $G_W=Q_W^{\star-1}$ is $o^{-1}$ plus $O(|p|^{-m-1})$, $G_W\star dQ_W=o^{-1}do+O(|p|^{-2})$, and the
Moyal three-form on the sphere $|p|=R$ is $\trD\gamma^5(o^{-1}d_po)^3+O(R^{-4})$ against the volume $2\pi^2R^3$ of
the sphere: the leading term gives the $R$-independent flux of \eqref{eq:sphere}, the corrections give $O(R^{-1})$.
The same applies to the far slices $p_4\to\pm\infty$ of \eqref{eq:slices}, on which $|p|\to\infty$ as well; the
star-free three-form of the principal symbol is closed outside its singularities, and Stokes' theorem on the region
between the far slices and the sphere (its side boundary at $|\vec p|\to\infty$ vanishing by the estimate of
\eqref{eq:dSigma}) converts the slice difference into the sphere flux. Uniformity of the symbol estimates in the
direction of $p$ is essential -- this is condition (C1) -- and it is exactly what the non-elliptic models lack along
$p_4$.} Thus the CSE expression and the gauge-quadratic term of the integrated
anomaly contain the same signed charge whenever both derivations meet their assumptions. Ellipticity by itself neither
places the singularities at $p_4=0$ nor provides chirality, so (i)--(iii) are separate hypotheses. The background and
mixed terms of Theorem~\ref{thm:main} are additional index data which do not enter the CSE.
\end{GPTcorrection}

\subsection{\reviewfix{Quasi-elliptic operators: the anomaly with anisotropic scaling}}
\label{sec:quasiell}

\begin{GPTcorrection}
\reviewfix{The multi-Weyl operator of Example~\ref{ex:multiWeyl} is not elliptic for $n\ge2$: its part of the highest
order $n$ consists of the off-diagonal entries $\kappa(p_1\mp ip_2)^n$ only, and these vanish on the plane
$p_1=p_2=0$. It is \emph{quasi-elliptic} (Appendix~\ref{app:ops}). This class lies between the elliptic operators,
for which the derivations of Secs.~\ref{sec:pointsplit}--\ref{sec:stokes} are formulated, and the non-elliptic
Fermi-surface models of Sec.~\ref{sec:nonell}. This subsection states which steps of these derivations hold for
quasi-elliptic operators, which have to be modified, and what remains unproven. \GPT{For the multi-Weyl family the index calculation below gives the coefficient $n$.
For a general matrix-valued quasi-elliptic symbol, the identification of the heat-kernel index with the weighted
symbol winding requires the additional analytic input stated below. Weighted power counting alone does not
establish this identification or justify every point-splitting remainder.}}

\paragraph{\reviewfix{The formula for the anomaly and the conditions of its validity.}}
\reviewfix{\GPT{The formula below holds for the multi-Weyl family under the heat-kernel and locality assumptions specified
here; for a general quasi-elliptic chiral block it additionally requires condition (vi). The integrated anomaly
means the regulated axial Ward-identity density, as in Sec.~\ref{sec:Fujikawa}, rather than the integral of the
ordinary divergence of a globally smooth current on a closed manifold:}
\begin{equation}
\begin{gathered}
\sA(A)=\int_{T^4}d^4x\,\langle\tr\cD_\mu J_\mu\rangle=2i\,\ind\hO_A=-N_3^0\,\frac{i}{4\pi^2}\int_{T^4}\tr F\wedge F,\\
N_3^0=-\frac1{24\pi^2}\oint_{S}\tr\big(o_w^{-1}d_po_w\big)^3\in\Z ,
\end{gathered}
\label{eq:qeanomaly}
\end{equation}
where $o_w(p)$ is the weighted principal symbol \eqref{eq:qedef} of the chiral block and $S$ is any closed
three-surface in momentum space that encloses the point $p=0$, with the outward orientation (the conventions of
Sec.~\ref{sec:factor}). Equivalently, $\ind\hO_A=-N_3^0\,\frac1{8\pi^2}\int\tr F\wedge F$, and after the inverse Wick
rotation $\sA=N_3^0\,\frac1{2\pi^2}\int d^4x\,\tr(\mathbf E\cdot\mathbf B)$. For the multi-Weyl operator of
Example~\ref{ex:multiWeyl}, $N_3^0=n$ and
\begin{equation}
\sA(A)=-n\,\frac{i}{4\pi^2}\int_{T^4}\tr F\wedge F .
\label{eq:qeanomalyn}
\end{equation}
The formula has the same form as \eqref{eq:PLB13}; the difference is in the conditions.}

\reviewfix{\emph{Conditions.} (i) Chirality: $\hQ=\left(\begin{smallmatrix}0&\hO_A^\dagger\\ \hO_A&0\end{smallmatrix}\right)$,
$\{\gamma^5,\hQ\}=0$. (ii) Minimal coupling with constant coefficients: $\hQ=\sum_\alpha c_\alpha\Sym[(-iD)^\alpha]$,
where the $c_\alpha$ do not depend on $x$ and are gauge singlets, and the gauge field enters only through
$D_\mu=\partial_\mu-iA_\mu$. (Another ordering of the covariant derivatives changes the operator by terms of lower
weighted degree and does not change the integrated anomaly.) (iii) Quasi-ellipticity: the symbol of the chiral block
has the form \eqref{eq:qedef} with some weights $w_\mu>0$, and $o_w(p)$ is invertible for all $p\ne0$. (iv) The
space-time is the Euclidean torus $T^4$, and $A$ is a smooth connection on a $U(N)$ or $SU(N)$ bundle over it; the
fermions are in the fundamental representation, and no interaction other than with the external field $A$ is
included. \GPT{(v) The heat kernel admits a complete gauge-covariant small-$t$ expansion with local polynomial
coefficients of definite weighted dimension and remainders allowing integration over $T^4$; paragraph (e)
also uses the local classification of gauge-invariant integrated densities independent of smooth changes of
the connection. (vi) For a general matrix-valued $o_w$, an anisotropic analytic index theorem identifies the
Fredholm index with the class $[\pi^*V_+,\pi^*V_-;o_w]$, or an equivalent comparison of the full regulated
point-split Ward identity with the heat-kernel supertrace is established. Condition (vi) is not proved here
by calculating only the displayed $F^2$ term. For the multi-Weyl family its coefficient is instead fixed by
the signed calculation in (f), together with the local classification in (e).}}

\reviewfix{\emph{No condition of weak inhomogeneity.}
\GPT{The heat-kernel index uses the complete axial current and does not require the slow-variation condition
\eqref{eq:slowcond} for a smooth fixed connection on $T^4$. This statement does not follow merely by truncating
the point-split current. For $\partial_p^3Q=0$ the $k\ge1$ terms of \eqref{eq:54all} are absent; otherwise the
weighted analysis in (d) and control of the regulator are required. In particular, terms with more than two
field strengths need not have dimension greater than $|w|$ when the weights differ.
The relevant scaling is $p=R^w q$, $\epsilon=R^{-w}\eta$ with a controlled weighted direction $\eta$, rather
than the isotropic rule $[p]\sim1/|\epsilon|$. A uniform estimate on the remainder is needed before taking
an arbitrary direction-dependent point-splitting limit. Replacing the complete current by its leading term
in a local response requires separate estimates; the factor $1/24$ in the derivative series is not a universal
relative error bound for its divergence. Nor may the ultraviolet anomaly be discarded solely because a formal
current correction is a total derivative before removal of the regulator.}}

\reviewfix{\emph{Status.}
\GPT{Under the Fredholm, heat-kernel and local-classification assumptions, (e) gives
$\sA(A)=-c\,i(4\pi^2)^{-1}\int\tr F\wedge F$. The coefficient is integral when the formula is valid for
bundles with unit $\int\tr F\wedge F/(8\pi^2)$. On the trivial bundle at $A=0$, the Fourier blocks are square
matrices, so their kernels and adjoint kernels have equal dimension and $\sA(0)=0$.
For the multi-Weyl family, (f) fixes $c=n$, including the sign. Extending that constant-field check to every
smooth connection uses (e), not just deformation of one connection: connections on different bundles cannot
in general be joined by such a deformation. For a general matrix-valued symbol, the coefficient $N_3^0$
found in the isolated $F^2$ term of (c) yields the full integrated anomaly only after condition (vi) or the
required regulator comparison and omitted-term estimates have been supplied.}}

\reviewfix{\emph{What is not covered.} (1) Coefficients depending on $x$: the expected formula is that of
Theorem~\ref{thm:main}, with the terms containing $\omega_2$ and $\nu$, \GPT{but its analytic index identification is not established in this subsection [paragraph (g)]; the cases in which
condition (ii) can be weakened are listed under ``Coefficients depending on $x$'' below.
(2) $\R^4$: Fredholmness is not automatic and requires conditions at infinity; the gapless multi-Weyl operator
on the unweighted whole-space domain is not Fredholm. Without such conditions \eqref{eq:qeanomaly} cannot be
interpreted as a whole-space Fredholm index;} what remains is the point-splitting result of paragraph (c), the density
$-N_3^0\frac{i}{8\pi^2}\tr(F_{\mu\nu}F^\star_{\mu\nu})$ modulo total derivatives, with the reservations of paragraph
(d). (3) A general four-manifold: the weights are attached to coordinate directions. (4) The local anomaly density
pointwise: the formula is for the integral. (5) The real-time response: the Minkowski form is the formal Wick
rotation of the Euclidean result, as in the elliptic case.}

\reviewfix{\emph{The mixed term.} In Theorem~\ref{thm:main} the anomaly contains the mixed term
$2i\int\omega_2\wedge c_1(E)$. In \eqref{eq:qeanomaly} it is absent, and this is a consequence of condition (ii): for
constant coefficients $\omega_2=0$ and $\nu=0$. This is seen in three ways. Topologically, $\omega_2$ of
\eqref{eq:omega2} is built from the part of $\tr\theta^5$ with two factors
$\theta_x=\partial_{x^\mu}o\,o^{-1}dx^\mu$, which vanish for a symbol that does not depend on $x$. In the
heat-kernel argument of paragraph (e) a term linear in the field, $c_{\mu\nu}\tr F_{\mu\nu}$, has the weighted
dimension $w_\mu+w_\nu<|w|$, and its integral -- the flux of $\tr F$, which does not vanish for $U(N)$ -- drops out
together with the other terms of $d_w\ne|w|$. In the exact computation of paragraph (f) the index vanishes for
$F_{34}=0$ and an arbitrary flux $N_{12}$. The mixed term appears in two cases. (1) A background $U(1)$ field
$A^{\rm bg}$ with the field strength $F_0$, in addition to the field $A$. The operator is still of the form (ii),
with the total connection $A+A^{\rm bg}$, and \eqref{eq:qeanomaly} gives \GPT{under the corresponding conditions above, including (vi) when needed}
\begin{equation}
\sA=-N_3^0\frac{i}{4\pi^2}\int\tr F\wedge F-N_3^0\frac{i}{2\pi^2}\int\tr F\wedge F_0
-N\,N_3^0\frac{i}{4\pi^2}\int F_0\wedge F_0 .
\label{eq:qemixed}
\end{equation}
The second term is the mixed term $2i\int\omega_2\wedge c_1(E)$ with $\omega_2=-N_3^0\,F_0/2\pi$, the third is
$2iN\nu$ with $\nu=-N_3^0\frac1{8\pi^2}\int F_0\wedge F_0$, as in the example of the background magnetic field of
Sec.~\ref{sec:factor}. The mixed term differs from zero only for a gauge group with a $U(1)$ factor, $\tr F\ne0$;
for the multi-Weyl operator it is confirmed by paragraph (f) with $F_{12}$ taken from the background and $F_{34}$
from the field $A$. (2) Coefficients $c_\alpha(x)$ that depend on $x$ (a texture). The expected term is
$2i\int\omega_2\wedge c_1(E)$ with $\omega_2$ given by \eqref{eq:omega2}, in which $o$ is replaced by the weighted
principal symbol $o_w$; the topological part of the derivation is unchanged [paragraph (g)], but the index theorem
\GPT{for $x$-dependent quasi-elliptic symbols has not been supplied here; this is a conditional extension, not a claim that no such index theorem exists.}}

\reviewfix{\emph{Coefficients depending on $x$.} For elliptic operators the coefficients $c_\alpha(x)$ may depend
on $x$, because the Atiyah--Singer theorem holds for any such dependence; the anomaly then has the three terms of
Theorem~\ref{thm:main}. For quasi-elliptic operators condition (ii) is used in the classification step of paragraph
(e). It can be weakened in three cases without any new theorem, and in all of them \eqref{eq:qeanomaly} holds
unchanged, under the remaining conditions. (1) Only the terms of lower weighted degree depend on $x$, the weighted
principal symbol $o_w(p)$ being $x$-independent. These terms are relatively compact with respect to the principal
part [paragraph (b)] and do not change the index. (2) The weighted homotopy hypothesis: $o_w(x,p)$ is joined to an
$x$-independent symbol by a continuous family $o_w^{(t)}(x,p)$, $t\in[0,1]$, of quasi-elliptic symbols with the same
weights and degree, invertible for $p\ne0$ uniformly in $x$ and $t$. The operators then form a norm-continuous
family of Fredholm maps $\mathcal H_w\to L^2$ on one and the same domain, and the index is constant along it. This
is the hypothesis of Sec.~\ref{sec:PLBhom} in the weighted form. An example is the multi-Weyl operator with a
position-dependent $\kappa(x)>0$: the family $\kappa_t=(1-t)\kappa(x)+t\kappa_0$ stays quasi-elliptic, and the
anomaly is \eqref{eq:qeanomalyn}. (3) A background $U(1)$ field, which is the case \eqref{eq:qemixed}.}

\reviewfix{For a dependence on $x$ that cannot be removed by such a homotopy (a texture) the expected result is the
formula of Theorem~\ref{thm:main},
\begin{equation}
\sA(A)=2iN\nu+2i\int\omega_2\wedge c_1(E)-N_3^0\frac{i}{4\pi^2}\int\tr F\wedge F ,
\label{eq:qetexture}
\end{equation}
with $N_3^0$, $\omega_2$, $\nu$ computed from $o_w(x,p)$. The reasoning of paragraph (e) gives its structure but not
its coefficients. Assume that (v) extends to $x$-dependent coefficients, the local density being a gauge trace of a
polynomial in $F$, its covariant derivatives and the derivatives of the $c_\alpha$. The integral does not change
under variations of $A$ at a fixed background. For the terms quadratic in the field,
$g_{\mu\nu\rho\sigma}(x)\tr F_{\mu\nu}F_{\rho\sigma}$, the variation is
$-4\int\tr\big(\delta A_\nu\cD_\mu(g_{\mu\nu\rho\sigma}F_{\rho\sigma})\big)$; it vanishes for all $\delta A$ and all
$F$ only if $g$ is totally antisymmetric and constant. The quadratic term is therefore a constant multiple of
$\int\tr F\wedge F$; the constant is determined by the part of the density without derivatives of the coefficients,
i.e.\ by the symbol at one point, which gives $N_3^0$ under the conditions stated above for constant coefficients.
For the terms linear in the field, $f_{\mu\nu}(x)\tr F_{\mu\nu}$, the variation is
$-2\int\partial_\mu f_{\mu\nu}\tr\delta A_\nu$, so that $f_{\mu\nu}$ is divergence-free, i.e.\ dual to a closed
two-form: this is the structure of the mixed term. The term without the field is $\sA(0)$. What this reasoning does
not give is the identification of the closed two-form with $\omega_2$ of \eqref{eq:omega2} and of $\sA(0)$ with
$2iN\nu$; it requires the index theorem for $x$-dependent quasi-elliptic symbols or an explicit heat-kernel
computation, and neither is supplied here. For $SU(N)$ the mixed term is absent, and under these assumptions
$\sA(A)-\sA(0)=-N_3^0\frac{i}{4\pi^2}\int\tr F\wedge F$.}

\paragraph{\reviewfix{(a) Anisotropic scaling and the symbol estimates.}}
\reviewfix{Let $w_\mu>0$ be the weights, $\lambda^wp=(\lambda^{w_1}p_1,\dots,\lambda^{w_4}p_4)$,
$|w|=\sum_\mu w_\mu$, and let $[p]$ be a weighted norm, $[\lambda^wp]=\lambda[p]$. The chiral block is quasi-elliptic
of weighted degree $m$ if
\begin{equation}
o(x,p)=o_w(x,p)+\dots,\qquad o_w(x,\lambda^wp)=\lambda^mo_w(x,p),\qquad o_w(x,p)\ \text{invertible for}\ p\ne0,
\label{eq:qedef}
\end{equation}
the dots being terms of lower weighted degree. For the multi-Weyl operator $w=(\frac1n,\frac1n,1,1)$, $m=1$,
$|w|=2+\frac2n$, and $o^\dagger o=[p]^2$ with $[p]^2=p_4^2+p_3^2+\kappa^2(p_1^2+p_2^2)^n$. Three properties follow.
(i) The inverse decays in every direction of the momentum, $o^{-1}=O([p]^{-m})$, and each derivative
$\partial_{p_\mu}$ lowers the weighted degree by $w_\mu$. The Moyal series and the parametrix expansion therefore
\GPT{remain asymptotic in the anisotropic calculus. Derivative corrections gain at least $w_{\min}$;
if the full symbol also contains lower weighted orders, the first relative gain is
$\delta=\min(w_{\min},m-m_{\rm next})>0$, where $m_{\rm next}<m$ is the largest such order
(omit the second entry if no lower term is present). It need not equal $w_{\min}$.} (ii) The gauge field enters at a lower
weighted degree: in $\hpi_\mu=\hat p_\mu-A_\mu$ the momentum of weight $w_\mu$ is replaced by a function of weight
zero. The weighted principal symbol of $\hO_A$ is $o_w\otimes1_E$ and does not contain $A$. (iii) This is the
difference from the plane model of Sec.~\ref{sec:FS}, whose symbol does not grow along the $p_4$ axis; no choice of
weights makes it quasi-elliptic.}

\paragraph{\reviewfix{(b) The Fredholm property.}}
\reviewfix{On $T^4$ the operator $\hO_A$, closed in $L^2$ on its natural domain $\mathcal H_w$ \GPT{(defined using local bundle trivializations and anisotropic
Sobolev norms; on a trivial bundle it is the space with $\langle p\rangle_w^m\hat\psi$ square summable;} for the multi-Weyl operator those with
$\partial_3\psi,\partial_4\psi,\partial_1^n\psi,\partial_2^n\psi\in L^2$), is Fredholm. The operator with the symbol
$\chi(p)\,o_w^{-1}$, $\chi$ a cutoff of small momenta, inverts it up to operators whose symbols are
\GPT{$O([p]^{-\delta})$, with $\delta>0$ as in (a),} and these are compact on $T^4$; the kernel and the cokernel are finite-dimensional and consist
of smooth sections. For the multi-Weyl operator at $A=0$ this is explicit: $\hO^\dagger\hO=[p]^2$ on the lattice of
momenta has a compact resolvent. These are the standard properties of quasi-elliptic (semi-elliptic) operators, and
we do not reproduce the proofs. Three remarks. First, as a map $H^n\to L^2$ of the isotropic Sobolev space, which is
the setting of Sec.~\ref{sec:homotopy}, the operator is bounded but not Fredholm for $n\ge2$: its inverse gains only
one derivative along $x_3,x_4$, and the range is not closed. \GPT{The smooth kernel and the kernel of the formal adjoint are unchanged, but a nonclosed range does not
have the same finite-dimensional Fredholm cokernel. The number $\dim\ker\hO-\dim\ker\hO^\dagger$ is the
index of the anisotropic realization, not a Fredholm index of $H^n\to L^2$,} and the compactness argument of
Sec.~\ref{sec:homotopy} holds with $H^m$ replaced by $\mathcal H_w$. Second, on $\R^4$ the operator is not Fredholm
without conditions at infinity, as for the Dirac operator. Third, the weights are attached to coordinate
directions; the statements hold on $T^4$, or on a manifold with a fixed splitting of the directions, but not on a
general four-manifold.}

\paragraph{\reviewfix{(c) Point splitting: the three terms before the limit.}}
\reviewfix{\GPT{First take the weighted-homogeneous chiral block $o=o_w$, smooth and invertible for $p\ne0$,
and set $\tilde Q=\left(\begin{smallmatrix}0&o^\dagger\\o&0\end{smallmatrix}\right)$,
$\tilde G^{(0)}=\tilde Q^{-1}$ off the origin. The three displayed contributions in \eqref{eq:61} can be treated as regulated Fourier distributions.
This does not assert that \eqref{eq:61}, including its omitted orders in the point separation and in the field,
is exact at finite $\epsilon$. Extension to a nonhomogeneous symbol or a larger compact singular set requires
a distributional extension across that set, control of boundary terms, and, in the first identity below,
a globally single-valued logarithmic phase. Invertibility outside a compact set alone is insufficient:
the scalar block $o=(p_1^2+p_2^2-1)+i(p_3^2+p_4^2-1)$ has a compact zero torus, and its phase winds around a
small normal loop linking that torus, so no single-valued logarithm exists on its full invertible region.} \emph{The term with one $F$.} In the block form,
$\partial_\nu\tilde Q\,\tilde G^{(0)}=\mathrm{diag}(\partial_\nu o^\dagger\,o^{\dagger-1},\partial_\nu o\,o^{-1})$,
hence
\begin{equation}
\begin{gathered}
\trD\big(\gamma^5\partial_{p_\nu}\tilde Q\,\tilde G^{(0)}\big)=\partial_{p_\nu}\varphi,\qquad
\varphi=\log\det o^\dagger-\log\det o,\\
\int d^4p\,e^{2ip\epsilon}\partial_{p_\nu}\varphi=-2i\epsilon_\nu\int d^4p\,e^{2ip\epsilon}\varphi ,
\end{gathered}
\label{eq:qeoneF}
\end{equation}
\GPT{Here the logarithm can be chosen on $\R^4\setminus\{0\}$, which is simply connected;
the identities are understood distributionally, with a compatible extension at the origin and no omitted
branch-cut terms.} The term is proportional to $\epsilon_\mu\epsilon_\nu F_{\mu\nu}=0$ (for the multi-Weyl symbol $\det o=[p]^2$ is
real and $\varphi=0$). \emph{The term with $\cD\cD F$} is a total $x$-derivative \GPT{for this $x$-independent symbol at the displayed order.} \emph{The term with
$F^2$.} By the trace identity below \eqref{eq:63} its momentum integrand is proportional to
$\epsilon_\mu j_\mu(p)\,e^{2ip\epsilon}\,\tr(FF^\star)$, where
$j_\mu=\frac1{3!}\epsilon_{\mu\alpha\beta\nu}S_{\alpha\beta\nu}$ is the vector dual to the three-form $S$. \GPT{For the weighted-homogeneous symbol, $j_\mu$ has weighted degree $w_\mu-|w|$ and is locally
integrable at the origin, since $w_\mu>0$. Its locally integrable extension is tempered and its divergence
is supported at $p=0$. With this specified extension, the Fourier-distribution identity is}
\begin{equation}
\int d^4p\,e^{2ip\epsilon}\epsilon_\mu j_\mu(p)=\frac i2\int d^4p\,e^{2ip\epsilon}\,\partial_{p_\mu}j_\mu(p)
\ \xrightarrow[\ \epsilon\to0\ ]{}\ \frac i2\int d^4p\,\partial_{p_\mu}j_\mu(p) .
\label{eq:qeFsq}
\end{equation}
\GPT{The last integral is the flux of $S$ through a surface enclosing the origin. It reproduces the
coefficient of the displayed $F^2$ term in \eqref{eq:anomR4}; identifying it with the complete regulated anomaly
still requires the hypotheses stated above.}
\GPT{Weighted homogeneity gives $\partial_\mu j_\mu=c\,\delta^4(p)$: derivatives of the delta function
have different weighted degrees, and $c$ is fixed by the flux. Thus the contracted Fourier distribution in
\eqref{eq:qeFsq} is constant. This is not a claim of absolute convergence of the unregulated Fourier integral
for every nonzero $\epsilon$.} For the multi-Weyl operator $j_\mu\propto\tilde p_\mu s(p)$ with
$\tilde p_\mu=w_\mu p_\mu$ and $s$ of weighted degree $-|w|$: $\tilde p_\mu\partial_\mu$ is the generator of the
anisotropic scaling, and $\partial_\mu(\tilde p_\mu s)=(|w|-|w|)s=0$ for $p\ne0$. \GPT{Thus for this homogeneous symbol the specified $F^2$ distribution requires no direction average.
For more general symbols the same conclusion requires the extension and boundary hypotheses stated above.} \GPT{The isotropic theorem of Sec.~\ref{sec:pointsplit}, with its degree-$-3$ hypothesis, is not itself
a theorem for arbitrary weighted symbols. For $x$-dependent quasi-elliptic symbols, the corresponding regulated
version of \eqref{eq:63} has not been established here.}}

\paragraph{\reviewfix{(d) The omitted terms: weighted power counting.}}
\reviewfix{The terms omitted in \eqref{eq:61} were removed in Sec.~\ref{sec:pointsplit} by the isotropic counting
(c). With weights it reads as follows. The field strength $F_{\mu\nu}$ has the weighted dimension $w_\mu+w_\nu$,
the derivative $\cD_\mu$ the dimension $w_\mu$, the anomaly density the dimension $|w|$. \GPT{For the homogeneous principal model, a homogeneous term whose field factors have total dimension
$d_w$ has a coefficient with the scaling below. Lower-order symbols require an asymptotic expansion rather
than this exact relation for the whole coefficient:}
\begin{equation}
C(\lambda^{-w}\epsilon)=\lambda^{|w|-d_w}\,C(\epsilon) .
\label{eq:qecount}
\end{equation}
Terms with $d_w>|w|$ vanish at $\epsilon\to0$ (provided $C$ is bounded as a function of the weighted direction of
$\epsilon$), the terms with $d_w=|w|$ are marginal, \GPT{whereas terms with $d_w<|w|$ are potentially divergent. Their cancellation requires traces,
symmetry, counterterms or the integrated heat-kernel identity; it does not follow from power counting alone.} \GPT{For $w_\mu=1$, only $d=4$ survives after the required cancellations of lower-dimensional contributions.} With unequal weights more terms are superficially
relevant: for the multi-Weyl operator $F_{12}$ has the dimension $\frac2n$, and $\tr F_{12}^k$ is relevant or
marginal for $k\le n+1$. For this operator such terms are excluded by symmetry:
\begin{equation}
\begin{gathered}
o(R^{(12)}_\phi p)=U\,o(p)\,U^{-1},\quad U=e^{-in\phi\sigma_3/2};\qquad
o(R^{(34)}_\alpha p)=V\,o(p)\,V,\quad V=e^{i\alpha\sigma_3/2};\\
o\big|_{p_4\to-p_4}=-o^\dagger;\qquad o\big|_{p_3\to-p_3}=\sigma_3\,o^\dagger\sigma_3;\\
\GPT{o\big|_{p_2\to-p_2}=\sigma_2\,o^\dagger\sigma_2.}
\end{gathered}
\label{eq:qesym}
\end{equation}
where $R^{(12)}_\phi$: $p_1+ip_2\to e^{i\phi}(p_1+ip_2)$ and $R^{(34)}_\alpha$: $p_4+ip_3\to e^{i\alpha}(p_4+ip_3)$.
\GPT{These are local symbol symmetries; arbitrary rotations need not preserve the global torus lattice.
The rotations are implemented by unitary transformations of $Q$ commuting with $\gamma^5$; the displayed
longitudinal and transverse reflections (and, combined with rotations, the reflection of any single coordinate)} are implemented by matrices anticommuting with $\gamma^5$ and
exchange the chiral blocks. The anomaly density is therefore invariant under the rotations in the planes $(1,2)$
and $(3,4)$ and odd under the reflection of any single coordinate, so that each of the indices $1,2,3,4$ enters
every term an odd number of times. Its weighted dimension is then at least $w_1+w_2+w_3+w_4=|w|$, with equality only
if each index enters once. Hence there are no relevant terms, and the marginal terms are the components of
gauge-invariant four-forms: $\tr F_{12}F_{34}$, $\epsilon_{ij}\epsilon_{ab}\tr F_{ia}F_{jb}$ ($i,j\in\{1,2\}$,
$a,b\in\{3,4\}$) and total derivatives. The terms quadratic in $F$ are those computed in (c), which gives the two
structures with equal coefficients, i.e.\ $\tr(FF^\star)$ with the coefficient $N_3^0=n$. Two reservations. The
symmetries act on $\epsilon$ as well, and the argument applies to the limit when it does not depend on the
direction of $\epsilon$, or to the average over the directions related by \eqref{eq:qesym}. And this is a counting
argument: the remainders have not been estimated. For a general quasi-elliptic operator, without such symmetries,
we have no argument within point splitting that removes the superficially relevant terms.}

\paragraph{\reviewfix{(e) The Fujikawa form: the relevant terms integrate to zero.}}
\reviewfix{In the heat-kernel regularization of Sec.~\ref{sec:Fujikawa} no separate argument is required for the
integrated anomaly. \GPT{The anisotropic estimates in (b) give compact resolvent and polynomial spectral growth, hence
trace-class heat operators for $t>0$, for
$\hQ^2=\mathrm{diag}(\hO^\dagger\hO,\hO\hO^\dagger)$. Therefore} $\sA=2i\,\ind\hO_A=2i\,\Tr\gamma^5e^{-t\hQ^2}$ for every $t>0$. In the expansion of the diagonal of the heat
kernel at small $t$ (it follows from the symbol calculus of (a); we do not prove it here) a local term of the
weighted dimension $d_w$ enters with the power $t^{(d_w-|w|)/2m}$. \GPT{Since the supertrace is independent of $t$, the sum of the integrated coefficients at each
nonzero power of $t$ vanishes (and any logarithmic coefficients must cancel). This does not require each
local invariant or each unintegrated coefficient to vanish separately. The constant term gives the index.} The terms of lower weighted degree in the symbol are relatively compact perturbations by (b) and do not
change the index, so that one may take $o=o_w$; the coefficients of the density are then pure numbers, and no
dimensionful constant can raise the dimension of a term. For an $x$-independent symbol this density is a single
gauge trace of a polynomial in $F$ and its covariant derivatives with constant coefficients, and its integral is
invariant under deformations of $A$. A term linear in the field, $c_{\mu\nu}\tr F_{\mu\nu}$, has the dimension
$w_\mu+w_\nu<|w|$ and is not marginal; the marginal terms linear in $F$ contain derivatives and are total
derivatives. \GPT{Using the local classification of gauge-invariant polynomial densities whose integrals are
unchanged by arbitrary connection variations, modulo total derivatives, the surviving connection-dependent
functional here is a multiple of $\int\tr F\wedge F$. This classification is an additional ingredient beyond
weighted counting; in particular it excludes combinations with covariant derivatives as well as higher
powers of $F$. As a check on the derivative-free quadratic sector,} for
$c_{\mu\nu\rho\sigma}\tr F_{\mu\nu}F_{\rho\sigma}$ the variation
$-4\int c_{\mu\nu\rho\sigma}\tr(\delta A_\nu\cD_\mu F_{\rho\sigma})$ vanishes for all $\delta A$ only if $c$ is
totally antisymmetric (Bianchi identity), and the terms of higher order in $F$ are not invariant. Thus
\begin{equation}
\sA(A)-\sA(0)=-c\,\frac{i}{4\pi^2}\int\tr F\wedge F
\label{eq:qeindex}
\end{equation}
for every $x$-independent quasi-elliptic chiral operator on $T^4$, with a constant $c$ fixed by the symbol. It may
be evaluated in any convenient field. \GPT{For the multi-Weyl operator (f) fixes $c=n=N_3^0$. For a general matrix-valued symbol, the
isolated $F^2$ computation (c) agrees with that coefficient if the regulator comparison or anisotropic index
identification in (vi) is available. It is not a substitute for that step. The multi-Weyl factor $n$ is also obtained
by the independent heat-kernel calculation of Ref.~\cite{YeeYi}, where the operator of
Example~\ref{ex:multiWeyl} is considered explicitly.}}

\paragraph{\reviewfix{(f) An exact check for the multi-Weyl operator.}}
\reviewfix{\GPT{For this exact check use a charge-one $U(1)$ line bundle on $T^4$, with
$F_{12}=B>0$, $F_{34}=E$ and all other components zero. The flux numbers
$N_{12}=BL_1L_2/(2\pi)>0$ and $N_{34}=EL_3L_4/(2\pi)$ are integers. For a diagonal non-Abelian embedding the
indices of the line-bundle summands must be added; a scalar flux product cannot silently replace a colour trace.} Then $[\hpi_1+i\hpi_2,\hpi_1-i\hpi_2]=2B$, i.e.\ $\hpi_1+i\hpi_2=\sqrt{2B}\,a$ with
$[a,a^\dagger]=1$, the pair $(\hpi_1,\hpi_2)$ commutes with $(\hpi_3,\hpi_4)$, and
\[
\hO=\hpi_4+i\begin{pmatrix}\hpi_3&\kappa(2B)^{n/2}(a^\dagger)^n\\ \kappa(2B)^{n/2}a^n&-\hpi_3\end{pmatrix}.
\]
On the states $(|l\rangle,0)$ with $l=0,\dots,n-1$ the off-diagonal part vanishes, $a^n|l\rangle=0$, and
$\hO=\hpi_4+i\hpi_3$: these are $n$ branches of $1+1$-dimensional chiral fermions in the field $E$, each with the
degeneracy $N_{12}$ and with \GPT{$\ind(\hpi_4+i\hpi_3)=-N_{34}$}, since
$(\hpi_4+i\hpi_3)^\dagger(\hpi_4+i\hpi_3)=\hpi_4^2+\hpi_3^2+E$ and the opposite order gives $-E$. The remaining
states form the pairs $(|l+n\rangle,|l\rangle)$, on which $\hO=\hpi_4+i(\sigma_3\hpi_3+c_l\sigma_1)$ with $c_l\ne0$
and $\hO^\dagger\hO=\hpi_4^2+\hpi_3^2+c_l^2+\sigma_3E\ge c_l^2>0$, and likewise for $\hO\hO^\dagger$; they carry no
zero modes. Therefore
\begin{equation}
\GPT{\ind\hO_A=-nN_{12}N_{34}=-\frac{n}{8\pi^2}\int F\wedge F,\qquad
N_3^0=n ,}
\label{eq:qeLL}
\end{equation}
in agreement with \eqref{eq:qeindex} for $c=N_3^0=n$ and with Example~\ref{ex:multiWeyl}. This computation uses
neither ellipticity nor an expansion in the field.}

\paragraph{\reviewfix{(g) The factorization theorem and the Moyal form of $N_3$.}}
\reviewfix{In Sec.~\ref{sec:factor} the topological part holds without change. The isotropic principal symbol is
not invertible away from the zero section, but the weighted principal symbol $o_w$ is, and this is all that the
definition of the class $\xi_0^{(1)}=[\pi^*V_+,\pi^*V_-;o_w]\in K_c(T^*M)$ requires. By (a) the symbol of the gauged
operator is $o_w\otimes1_E$, so that $\xi_A=\xi_0^{(1)}\cdot\pi^*[E]$; the multiplicativity of the Chern character,
the fibre integration, the degree bookkeeping and the expressions of $N_3^0$, $\omega_2$, $\nu$ are unchanged, with
the sphere $S^3_{\rm out}$ replaced by any surface enclosing $p=0$ in the fibre; $N_3^0$ does not depend on $x$ by the
same argument. What does not hold as stated is the analytic input: the Atiyah--Singer formula \eqref{eq:Fuji} is a
theorem about elliptic operators. For $x$-independent symbols on $T^4$ it is replaced by \eqref{eq:qeindex}. For
$x$-dependent quasi-elliptic symbols Theorem~\ref{thm:main} requires an index theorem for this class. \GPT{The weighted derivative gain cannot simply be identified with the isotropic H\"ormander
parameter $\rho$ without checking that symbol class and its coordinate calculus. In particular a restriction of
the type $\rho>1/2$, which appears in the coordinate-invariant calculus of the isotropic classes on a manifold, does
not prove an obstruction to an anisotropic index theorem in the fixed global directions of $T^4$. No such analytic identification is proved here for the $x$-dependent case; the topological
twisting formula is conditional on it, rather than evidence that the required theorem is unavailable.} The Todd factor plays no role, since the setting is restricted
to $T^4$ by (b). In Sec.~\ref{sec:stokes} the estimates leading to \eqref{eq:sphere} hold with $|p|$ replaced by
$[p]$ and the sphere by the surface $[p]=R$: \GPT{after pull-back to a fixed weighted unit surface, derivative-generated Moyal corrections gain
at least $R^{-w_{\min}}$. Lower-order symbol terms may instead give $R^{-\delta}$ as in (a). Uniform bounds
on these pulled-back forms, the inverse and the trace are required; pointwise decay alone does not control
the surface integral or justify discarding boundaries.}}

\paragraph{\reviewfix{(h) Summary: the three classes.}}
\reviewfix{The comparison of the elliptic operators, the quasi-elliptic operators and the non-elliptic
Fermi-surface models of Sec.~\ref{sec:nonell} is given in Table~\ref{tab:classes}.}
\begin{table}[htb]
\centering
\small
\begin{tabular}{p{3.3cm}|p{3.2cm}|p{4.0cm}|p{3.7cm}}
& elliptic & quasi-elliptic & non-elliptic (Sec.~\ref{sec:nonell})\\ \hline
symbol at large momentum & invertible, $G=O(|p|^{-m})$ & invertible, $G=O([p]^{-m})$ & not invertible, no decay along $p_4$\\
Fredholm on $T^4$ & $H^m\to L^2$ & $\mathcal H_w\to L^2$ & \GPT{no elliptic Fredholm conclusion}\\
ultraviolet degree $N_3^0$ & yes & yes (surface around $p=0$) & no\\
term with $F^2$, \eqref{eq:qeFsq} & \GPT{regulated term} & \GPT{distributional term under (c)} & \GPT{no ultraviolet flux derivation}\\
omitted terms & isotropic counting, $d>4$ & weighted counting, $d_w>|w|$, and symmetry or (e) & $\partial_p^3Q=0$, $\cD F=0$ or \eqref{eq:slowcond}\\
integrated anomaly & index theorem, Theorem~\ref{thm:main} & \GPT{multi-Weyl: (e)--(f); general: analytic identification needed} & \GPT{static response only; no Ward identity derived}\\
\end{tabular}
\caption{\reviewfix{The derivations of the anomaly for the three classes of operators.}}
\label{tab:classes}
\end{table}
\end{GPTcorrection}

\subsection{\GPT{Non-elliptic models: static responses and the limits of the anomaly argument}}
\label{sec:nonell}

\begin{GPTcorrection}
\reviewfix{The Fermi-surface models of Sec.~\ref{sec:FS} have a nontrivial Moyal invariant but no ultraviolet
degree. \GPT{This subsection evaluates formal static response coefficients and the spatial divergence of the
associated constitutive terms. Such a calculation does not, without a regulated Ward identity and a specified
state, establish a quantum anomaly. A window of states near a Fermi sheet is an open subsystem and its charge
can exchange with states outside that window. The full conserved charge obeys a continuity equation if the
regularization preserves its symmetry.}
(The lattice regularization of the model is in Sec.~\ref{sec:FS}, the current of the conserved charge along $B$ in
Sec.~\ref{sec:CME}.)
Only static fields are considered: the real-time response, and with it the spectral flow in
a time-dependent gauge, requires a separate analysis because of the form $\kappa p_3\cdot ip_4$ of the frequency
dependence of $Q$, and is not attempted here. The model is \eqref{eq:FSmodel},
$Q=i\kappa p_4p_3-p_3-H_\perp+\mu_0$ with $H_\perp=\pi_\perp^2/2m$ and Landau levels $E_n$, in the notation of
Proposition~\ref{prop:FS}; paragraphs (f) and (g) apply the same analysis to the spherical models of
Secs.~\ref{sec:CMEsphere} and~\ref{sec:CMEsphere2}.}

\paragraph{\reviewfix{(a) What survives: the infrared half of the argument.}}
\reviewfix{\GPT{For this model the explicit regulated residue calculation of Proposition~\ref{prop:FS} gives
$\Sigma^{(T)}(p_4)=\sgn(p_4)n_{\rm filled}$ for $p_4\ne0$, and hence $d\Sigma/dp_4=0$ on each half-axis.
This calculation establishes the slice identity without invoking an isotropic elliptic estimate for this
non-elliptic symbol.} Thus $N_3^{(T)}=2n_{\rm filled}$
\corrB{($n_{\rm filled}$ in the normalization of \eqref{eq:CSE})}, with
$n_{\rm filled}$ the number of Landau levels below $\mu_0$.
\GPT{The formal response \eqref{eq:N3T},
$\partial\langle J^{(T)}\rangle_i/\partial\mu=N_3^{(T)}B_i/(4\pi^2)$, additionally requires its stated current,
trace, regulator and boundary hypotheses. The residue count is unchanged between Landau thresholds, with the
regulator and asymptotic reference fixed. Integer lattice pairings require the specific Fredholm-module
hypotheses of Sec.~\ref{sec:NCG}; discretizing one coordinate alone does not establish them.}}

\paragraph{\reviewfix{(b) What fails: the ultraviolet half.}}
\reviewfix{The passage from the slices to the sphere, \eqref{eq:sphere}, is not available. The operator is not
elliptic: the principal symbol $i\kappa p_4p_3-\pi_\perp^2/2m$ vanishes on non-compact sets of the fibre.
\GPT{In a Landau sector $Q_n(p_4,0)=\mu_0-E_n$, independently of $p_4$, so the inverse has no uniform
high-frequency decay.} Condition (C1) fails there. \GPT{Taking $|p_4|\to\infty$ is a nonuniform ultraviolet
limit: a window $p_3=O((\kappa|p_4|)^{-1})$ retains a finite product $\kappa p_4p_3$.}
The far-slice values remain $\pm n_{\rm filled}$ by the residue calculation, \GPT{rather than becoming a
winding of an invertible principal symbol. No compactly supported cotangent symbol class of the type used in
Theorem~\ref{thm:main} is supplied by this principal symbol. That theorem therefore does not apply; the computed
$n_{\rm filled}$ depends on $\mu_0$ and $B$ and is not the ultraviolet degree $N_3^0$.}}

\paragraph{\reviewfix{(c) The chiral structure of the Fermi sheets.}}
\reviewfix{For $p_3\ne0$, $Q=\kappa p_3(ip_4-H_{\rm eff})$ with
\GPT{$H_{\rm eff}=(p_3+H_\perp-\mu_0)/(\kappa p_3)$}: the
Landau level $n$ has the effective one-dimensional dispersion
\GPT{$\varepsilon_n(p_3)=(p_3-p_{3,n})/(\kappa p_3)$}, with the
sheet at $p_{3,n}=\mu_0-E_n$, the slope \GPT{$v_n=1/(\kappa p_{3,n})$} and the degeneracy $B/2\pi$ per unit area
(Fig.~\ref{fig:contour}a). The $n_{\rm filled}$ levels with $E_n<\mu_0$ have $v_n>0$ -- the sign selected by the
contour of Fig.~\ref{fig:contour}b -- the infinitely many others $v_n<0$.
\GPT{The formal negative-energy set at physical $\mu=0$ is $(0,p_{3,n})$ for $E_n<\mu_0$ and
$(p_{3,n},0)$ for $E_n>\mu_0$. Calling it an occupied set presupposes a state and a sea prescription.}
There are two boundaries: the sheet, where $\varepsilon_n$ crosses zero,
and the junction \GPT{$p_3=0$}, where the coefficient \GPT{$\kappa p_3$} of $ip_4$ vanishes and
$\varepsilon_n$ passes through infinity \GPT{(for $E_n<\mu_0$, $-\infty$ at $0^+$ and $+\infty$ at $0^-$)}.
The junction is not a zero of $Q$, $Q_n(p_4,0)=\mu_0-E_n$.
\GPT{It is nevertheless a singularity of the effective dispersion and of its frequency normalization.
The residue $1/(\kappa p_3)$ is negative on $p_3<0$, so the full propagator is not a canonical positive-spectral-weight
Green function as written. A formal analogy with a Weyl Landau branch does not specify transport through this junction.}}

\paragraph{\reviewfix{\GPT{(d) Static divergence of the Fermi-window constitutive current.}}}
\reviewfix{In \eqref{eq:N3T} $\mu$ is the ordinary chemical potential, $ip_4\to ip_4+\mu$, which shifts all
$\varepsilon_n$ uniformly and is distinct from the parameter $\mu_0$ that places the sheets. Introduce a static
electric field through the scalar potential $A_0=-\mathbf E\cdot\mathbf x$.
\GPT{If a local constitutive approximation is applicable with fixed $\mu_0$, constant $B$ and
$\mu(\mathbf x)=\mu+\mathbf E\cdot\mathbf x$, its current term $J_i^{\rm loc}(\mathbf x)=J_i(\mu(\mathbf x))$
obeys the chain rule}
\begin{equation}
\GPT{\partial_iJ_i^{\rm loc}=\frac{\partial J_i}{\partial\mu}\,E_i.}
\label{eq:staticgauge}
\end{equation}
\GPT{A static external field does not by itself make the state stationary or establish local equilibrium;
these require dynamics and boundary or reservoir conditions. Equation~\eqref{eq:staticgauge} is not a general
derivation of an anomalous Ward identity, even when its coefficient agrees with a known Dirac anomaly.}

For the Fermi-surface model the $\mu$-derivative of the \GPT{formal leading current integrand} is supported at
the Fermi points after frequency integration: per mode,
\GPT{for $p_3\ne0$,}
\[
\GPT{\partial_\mu(-G_n\partial_{p_3}Q_n)
=-\kappa p_{3,n}G_n^2
=-\frac{\partial\varepsilon_n}{\partial p_3}(ip_4+\mu-\varepsilon_n)^{-2}.}
\]
\GPT{With the zero-temperature Matsubara prescription and a fixed window avoiding $p_3=0$, this gives
$(\partial\varepsilon_n/\partial p_3)\delta(\mu-\varepsilon_n)$ under the remaining momentum integral.}
Thus a level contributes $\partial j_n/\partial\mu=\sgn v_n/(2\pi)$ per orbital at its sheet.
Let $J^{(T)}_{\rm w}$ be the $\hat T$-current of the states in windows of fixed width in \GPT{$p_3$} around the sheets
of the levels selected by the contour, $E_n<\mu_0$ -- the Fermi-level current.
\GPT{For the plane model $T=1$, at physical $\mu=0$ and for sufficiently small shifts keeping each root inside
its window, the formal sheet response is}
\[
\GPT{\frac{\partial J_{{\rm w},3}^{(T)}}{\partial\mu}
=\frac{n_{\rm filled}B}{4\pi^2}=\frac{N_3^{(T)}B}{8\pi^2}.}
\]
Consequently \eqref{eq:staticgauge} gives \GPT{for this local constitutive term}
\begin{equation}
\GPT{\partial_iJ^{(T),\rm loc}_{{\rm w},i}
=\frac{N_3^{(T)}}{8\pi^2}\,\mathbf E\cdot\mathbf B.}
\label{eq:FSanomaly}
\end{equation}
\GPT{This is a static window-response identity. To turn it into a balance law one must determine
$\partial_t\rho_{\rm w}+\boldsymbol\nabla\cdot\mathbf J_{\rm w}=S_{\rm w}$, including exchange across the momentum
window and any source terms. Neither $\partial_t\rho_{\rm w}=0$ nor $S_{\rm w}$ has been derived here.
The coefficient per selected sheet equals the familiar one-dimensional chiral-branch coefficient times the
Landau degeneracy, but that agreement does not establish a regulator-independent quantum anomaly of the window
charge or extend the ultraviolet index factorization theorem to this model.}}

\paragraph{\reviewfix{\GPT{(e) The conserved current and the role of the reference subtraction.}}}
\reviewfix{\GPT{If the regulator preserves the conserved $T=1$ symmetry, the full current satisfies
$\partial_t\rho+\partial_iJ_i=0$ in a source-free region. For a slab $a<x_3<b$ with cross-section $S$, inserting
the isolated constitutive term would give}
\[
\GPT{\oint_{\partial V}\mathbf J^{\rm loc}\cdot d\mathbf S
=S(b-a)\frac{\partial J_3}{\partial\mu}E_3,\qquad
\frac{d}{dt}\int_V\rho\,d^3x=-\oint_{\partial V}\mathbf J\cdot d\mathbf S.}
\]
\GPT{A nonzero net outflow implies depletion, an explicit source, or failure of the isolated constitutive term
to represent the full current in that state. Boundary outflow does not permit a stationary source-free state
with nonzero bulk divergence. The equations above do not select the deep states as the carriers of the loss,
or prove that the window population remains in local equilibrium; that would require the real-time analysis
explicitly left open here.}

\GPT{The factor of two between \eqref{eq:N3T} and the selected-sheet coefficient has a precise bookkeeping
aspect. A cutoff including levels $0\le n\le L$, with $E_L>\mu_0$, gives the signed sheet count
\[
 S_L(\mu_0)=\sum_{n=0}^{L}\sgn(\mu_0-E_n)=2n_{\rm filled}-(L+1).
\]
For a reference coupling $\mu_0^{\rm ref}<E_0$, $S_L(\mu_0^{\rm ref})=-(L+1)$; the relative count is
$S_L(\mu_0)-S_L(\mu_0^{\rm ref})=2n_{\rm filled}$, whereas the selected positive-slope windows count
$n_{\rm filled}$. Thus one is comparing a subtracted full count with a restricted count. This observation
does not fix a physical sea prescription or a time-dependent flow. Likewise, the equal real parts of the two
momentum half-ray contour integrals are properties of the regulated contour calculation, not an assignment of
half of a real-time current to a prescribed set of deep states.}}

\paragraph{\reviewfix{(f) Example: the model with spherical Fermi surfaces (Sec.~\ref{sec:CMEsphere}).}}
\reviewfix{For the model \eqref{eq:sphQ} the symmetry current is that of $\hat T=\gamma^5$.
In the sector of the level $n$ and chirality $s$, $Q_{n,s}=X_{n,s}(ip_4-\varepsilon_{n,s})$
with the constant $X_{n,s}=s\kappa\lambda(E_n-\mu_0)$ and
\[
\varepsilon_{n,s}(p_r)=\frac1\kappa+\frac{p_r}{X_{n,s}} ,\qquad p_r\ge0 .
\]
If $X_{n,s}<0$ the dispersion falls from $1/\kappa>0$ at $p_r=0$ through zero at the sphere
$p_r=-X_{n,s}/\kappa=s\lambda(\mu_0-E_n)$ to $-\infty$.
\GPT{The negative-energy region is the exterior of the sphere; the pole residue $1/X_{n,s}$ is negative.
If $X_{n,s}>0$, the energy stays positive. There is no radial junction inside a fixed sector away from the
thresholds $\mu_0=E_n$, but these facts do not define a canonical Weyl sea. At the level of a one-dimensional
radial count only, a sphere contributes $\sgn(\partial\varepsilon/\partial p_r)/(2\pi)=-1/(2\pi)$ per orbital,
before weighting by $T$. A radial momentum branch is not a Cartesian current channel.}

\GPT{If the static probe-response assumptions and the symmetric angular sea subtraction of
Sec.~\ref{sec:CMEsphere} hold, applying the local constitutive rule gives}
\begin{equation}
\GPT{\begin{gathered}
\partial_iJ_i^{(5),\rm loc}
=\frac{N_3^{(\gamma^5)}}{4\pi^2}\,\mathbf E\cdot\boldsymbol{\mathcal B},\\
N_3^{(\gamma^5)}=2C(\mu_0)
=2\sgn\mu_0\Big(\frac12+\sum_{n>0}\theta(|\mu_0|-\sqrt{2Bn})\Big).
\end{gathered}}
\label{eq:sphanomaly}
\end{equation}
\GPT{The probe $\mathcal B$ is distinct from the parameter $B$ of the angular operator. For
$0<|\delta|<\sqrt{2B}$ the coefficient is $\sgn\delta/(4\pi^2)$, but identifying it with a single-Weyl
anomaly requires a regulated Ward identity, a specified state and a valid global realization. At $\delta=0$
itself, $Q_{0,s}=-p_r$ loses its frequency dependence, so the limiting object is not a regular Weyl propagator.}

\GPT{The chirality-weighted radial count is, with $-L\le n\le L$,
\[
 \lim_{L\to\infty}\big[\#\{E_n>\mu_0\}-\#\{E_n<\mu_0\}\big]
 =-2C(\mu_0)=-N_3^{(\gamma^5)}.
\]
Its magnitude is $2|C(\mu_0)|=|N_3^{(\gamma^5)}|$, not $2C(\mu_0)$ for both signs of $\mu_0$.
An identification with a Cartesian response needs the measure, velocity projections, trace normalization and
probe coupling. It cannot be obtained by discarding the minus sign or merely calling it an orientation choice.
Nor does the full subtracted slice invariant automatically equal a charge localized inside a small momentum ball.
For $T=1$, the regulated result $N_3^{(1)}=0$ removes this particular linear constitutive term; it does not by
itself prove the absence of every flow or a quantum anomaly. The periodic implementations discussed in
Sec.~\ref{sec:CMEsphere} concern the angular spectrum and its regulated count; an angular doubling or its removal
does not supply the missing Ward-identity derivation.} The coefficient $\sgn\delta/(4\pi^2)$ refers to the flat
model and to its doubler-free periodic implementation \eqref{eq:sphperiodic}; with the single-valued $f_a$ of
\cite{ZubJETP} it would be doubled.}

\paragraph{\reviewfix{(g) Example: the spherical model with the chiral magnetic like effect (Sec.~\ref{sec:CMEsphere2}).}}
\reviewfix{For the model \eqref{eq:sphQ2} the two chiral sectors of a level coincide up to the overall sign $s$: both
have the sphere $p_r=\lambda(\mu_0-E_n)$ for $E_n<\mu_0$, with the same dispersion
\GPT{$\varepsilon_n=1/\kappa+p_r/[\kappa\lambda(E_n-\mu_0)]$} and the same inward velocity, and none for $E_n>\mu_0$.
The $\gamma^5$ current weights the two coincident spheres of a level with $s=\pm1$: their contributions to
$\partial J^{(5)}/\partial\mu$ cancel level by level, $(+1)(-1/2\pi)+(-1)(-1/2\pi)=0$ per orbital,
in agreement with $N_3^{(\gamma^5)}=0$ of \eqref{eq:N3sph2}.
\GPT{More directly, $Q'_s=sq$ implies $G'_s=sg$ and identical block current and three-form integrands.
Their $\gamma^5$-weighted difference vanishes when the domains, probe couplings and regulators agree. Thus}
\[
\GPT{\partial_iJ_i^{(5),\rm loc}=0}
\]
\GPT{for this constitutive response. The two blocks have the same dispersion, so their labels $s=\pm1$ alone
do not make them a pair of opposite-propagating Weyl branches. A statement about the complete quantum Ward
identity still requires a symmetry-compatible regulator and the specified current and state.}
\GPT{The conserved-current slice count sums the two copies, $N_3^{(1)}=2C(\mu_0)$.}
\GPT{Its formal local response yields
$\partial_iJ_i^{\rm loc}=N_3^{(1)}\mathbf E\cdot\boldsymbol{\mathcal B}/(4\pi^2)$ under the same assumptions.
The continuity constraints of (e) apply: this term cannot describe a stationary source-free conserved current
with nonzero divergence. The two spherical constructions therefore illustrate addition and cancellation of
regulated Hall-ray contributions. By themselves these response coefficients do not establish the presence or the
absence of a quantum anomaly; the current of the conserved charge along the field, with its conditions, is the
subject of Secs.~\ref{sec:CME} and~\ref{sec:CMEsphere2}.} In terms of the static responses the two spherical models
are the two halves of the plane model taken separately: \eqref{eq:sphQ} has the response of the symmetry current
and none of the conserved one, \eqref{eq:sphQ2} has the response of the conserved current and none of the symmetry
one; the plane model, with $\hat T=1$ its only current, has both the window response \eqref{eq:FSanomaly} and the
continuity constraint of (e).}

\end{GPTcorrection}

%======================================================================
\section{Noncommutative-geometric form of the invariants}
\label{sec:NCG}

For 2D lattice systems with the precise lattice Wigner--Weyl calculus, the trace per unit volume $\cT$ (defined by
periodizing a finite lattice; a closed trace: $\cT(a\star b)=\cT(b\star a)$, $\cT(\partial_\mu a)=0$) and the Hall
conductivity $\sigma_H=\frac{e^2}{h}N_3$, Refs.~\cite{KMMZZ,KMMZZindex} write
\begin{align}
N_3&=\frac{\epsilon^{\mu\nu\rho}}{24\pi^2}\cT\big(G\star\partial_\mu Q\star G\star\partial_\nu Q\star G\star\partial_\rho Q\big)
=\big\langle[G],[\varphi]\big\rangle,\\
\varphi(a_0,\dots,a_3)&=\frac{\epsilon^{\mu\nu\rho}}{24\pi^2}\cT\big(a_0\star\partial_\mu a_1\star\partial_\nu a_2\star\partial_\rho a_3\big),
\end{align}
\reviewfix{Related cohomological constructions -- foliations of spaces associated with vertex operator algebras,
categories of $V$-structures for foliations, and the reduction cohomology of Riemann surfaces -- are developed in
Refs.~\cite{Z1,Z3,Z5}.}
\begin{GPTcorrection}
Here (Appendix~\ref{app:ncg}) the symbols form an algebra $\mathcal A$ under the star product, closed under the
derivations $\partial_\mu$ and under inversion, and $\cT$ is cyclic and closed; $\varphi$ is then a cyclic
3-cocycle, i.e.\ a 4-linear functional with $\varphi(a_1,a_2,a_3,a_0)=-\varphi(a_0,a_1,a_2,a_3)$ and vanishing
Hochschild coboundary; $[G]$ is the class of the invertible element $G$ in $K_1(\mathcal A)$ (the group of homotopy
classes of invertible matrices over $\mathcal A$); and the pairing $\langle[u],[\varphi]\rangle=\varphi(u^{-1},u,u^{-1},u)$
depends only on the classes of $u$ and $\varphi$. Thus $N_3=\varphi(G^{-1},G,G^{-1},G)$ (the sign of the winding of
$Q$, opposite to that of $G$, is absorbed in the pairing convention). Closedness alone does not give integrality;
integrality follows from the identification of $\varphi$ with the Chern character of a Fredholm module. For a finite
spatial period and discretized imaginary time the symbol is a smooth invertible function on the frequency--Bloch
three-torus with an integer winding; in the infinite lattice the module is built from the position Dirac operator
$D=\gamma^0\hat\tau+\gamma^1\hat x_1+\gamma^2\hat x_2$, $F=\operatorname{sgn}D$ (its finite-dimensional kernel treated
separately), $P=(1+F)/2$ -- an odd Fredholm module, $F=F^\dagger$, $F^2=1$, $[F,a]$ compact for every symbol operator $a$ -- which is $(3,\infty)$-summable (the singular values of $[F,a]$ decay as $k^{-1/3}$), and $N_3=\operatorname{Index}(PGP)$ by the Connes--Moscovici
local index formula \cite{Connes,CM}; the normalization constants of the residue (Dixmier-trace) formula are fixed by
the lattice cell, the Clifford trace and \reviewfix{the trace per unit volume (the identification requires a sufficiently local representation of the
symbol algebra and the frequency integration in $\cT$; cyclicity alone does not construct the module). Continuous
imaginary time requires} a suspension or a
frequency compactification with a fixed ultraviolet reference.

For a Matsubara Green function with positive spectral weight and nonsingular $G(0)$, $Q(0)$, a smooth deformation
keeping $G$ and $Q$ invertible gives the topological-Hamiltonian reduction $N_3[Q]=N_3[\corrB{ip_4}-H_t]$, $H_t=-Q(0)$
\cite{WangZhang} (a spectral gap alone does not exclude zeros of $G$ at zero frequency), and the resulting projector
pairing is the integer Hall index of \cite{BES}. The mobility-gap case requires the localization (Sobolev-algebra)
analysis of \cite{BES} and is not covered by the invertible-symbol argument.
\end{GPTcorrection}

\paragraph{The CSE invariant in the same language.} \begin{GPTcorrection}
The slice expression of \eqref{eq:PLB13} has the form of an odd three-dimensional winding in
$(x_1,x_2,x_3;p_1,p_2,p_3)$ at fixed $p_4$. If the slice defines a smooth bounded invertible element $O_{p_4}$ of a
lattice algebra of the type just described (with $x_3$ in the role of $\tau$), the chiral decomposition
$\trD\gamma^5(G\,dQ)^3=\tr(O^{-1}dO)^3-\tr(O^{\dagger-1}dO^\dagger)^3$ and
$\langle[O^\dagger],\varphi\rangle=-\langle[O],\varphi\rangle$ give
\begin{equation}
\reviewfix{\Sigma(p_4)=-\langle[O_{p_4}],[\varphi]\rangle,\qquad}
\reviewfix{N_3=-\langle[O_{0^+}],[\varphi]\rangle+\langle[O_{0^-}],[\varphi]\rangle}
\end{equation}
\reviewfix{(with the pairing convention above $\varphi(u^{-1},u,u^{-1},u)=-(24\pi^2)^{-1}\cT\tr(u^{-1}du)^3$,
because $d(u^{-1})=-u^{-1}(du)u^{-1}$, which fixes the minus sign of the slice pairing for the sign convention of
\eqref{eq:PLB13}; the opposite windings of $O$ and $O^\dagger$ account for the factor of two} between $1/48\pi^2$ and $1/24\pi^2$.
Under these hypotheses each slice pairing is an integer, and the constancy in $p_4$ of Sec.~\ref{sec:stokes} is the
homotopy invariance of the pairing. The hypotheses are not automatic: global invertibility of the slice and its
normalization at infinity have to be established; unregularized continuum slices can give $\pm n/2$ (the example of
Sec.~5 of Ref.~\cite{XZPLB}), which are not individual integer pairings although their difference is the integer $n$; \reviewfix{and no lattice completion satisfying these hypotheses has been specified for the continuum model of
Sec.~\ref{sec:FS} (the lattice regularization of Sec.~\ref{sec:FS} discretizes $x_3$ only).}
\end{GPTcorrection}

%======================================================================
\section{Summary and open problems}
\label{sec:summary}

\begin{GPTcorrection}
The covariant Wigner--Weyl transform and its inverse form an exact operator calculus; the expansion in the field
strength is controlled by the gradient expansion. The covariant symbol of an operator with gauge-singlet coefficients
is independent of the gauge field exactly when the covariant derivatives are completely symmetrized; a change of
ordering changes the symbol by $O(F)$ terms but not the integrated anomaly, which depends only on the elliptic chiral
symbol class.

\reviewfix{The non-Abelian CSE and QHE are given, at leading order in the slowly varying external field and under the
stated current and averaging assumptions, by the Abelian topological expressions with $F\to F^a$ and the generator
trace, the Moyal product accounting exactly for the inhomogeneity.} On a closed four-manifold with vanishing rational $p_1$ the anomaly
decomposes into the untwisted index, a mixed term $\omega_2\wedge c_1(E)$ and the gauge-quadratic term
$-N_3^0\frac{i}{4\pi^2}\int\tr F\wedge F$; for $c_1(E)=0$ the factorization into the instanton number and $N_3^0$ holds
precisely when the untwisted index vanishes, which is weaker than the homotopy hypothesis. The fibre charge $N_3^0$
coincides with the Moyal CSE expression $N_3$ under the Hamiltonian-form, chirality, trace and ellipticity assumptions of
Sec.~\ref{sec:stokes}; \reviewfix{for the non-elliptic Fermi-surface models the CSE invariant survives as a contour invariant
($N_3^{(T)}=2n_{\rm filled}$), in a static electric field the local constitutive term of the Fermi-level (window) current, obtained from the CSE response, has the divergence
$\partial_iJ^{(T),\rm loc}_{{\rm w},i}=N_3^{(T)}\mathbf E\cdot\mathbf B/8\pi^2$: a static response identity whose coefficient
factorizes into the Fermi-surface invariant of the CSE and $\mathbf E\cdot\mathbf B$, but which is not a derivation of an
anomalous Ward identity (Sec.~\ref{sec:nonell}); the conserved charges obey the continuity equation
(a divergence of their constitutive term in the static gauge means an outflow through the boundary), and the model carries an equilibrium
current of the conserved charge along $B$ -- a chiral magnetic like effect, even in $B$ and allowed because the model breaks parity and is
not of Hamiltonian form at the junction; in the lattice regularization of Sec.~\ref{sec:FS} the invariant is
$N_3^{(T)}=2[n_{\rm filled}(\mu_0)-n_{\rm filled}(\mu_0-2\alpha)]$, counting the Landau levels between the effective
chemical potentials of the two junctions of the Brillouin zone, and the equilibrium current is $N_3^{(T)}B/4\pi^2$;
in the model with spherical Fermi surfaces of \cite{ZubJETP} the two chiral sectors cancel and the equilibrium current
of the conserved charge vanishes, while its spherical Fermi point carries the CSE coefficient of a single Weyl fermion; correlating the radial term with
the chirality exchanges the two effects and gives a chiral magnetic like effect with spherical Fermi surfaces
(Sec.~\ref{sec:CME}).}

\reviewfix{Interactions do not change the topological expressions when the current is defined with the renormalized
velocity: the vertex terms are total momentum derivatives, and the Luttinger--Ward identity equates this current to
the bare-velocity one for density interactions of any range on a differentiable branch of $\Phi$, with a
shift-invariant regulated trace (the identity concerns the total traced $U(1)$ current; the axial current requires
the chiral-symmetry argument). These statements were originally derived for lattice models with the approximate
lattice Wigner--Weyl calculus; in continuum field theory the calculus is exact and the derivations hold verbatim,
the compactness of the Brillouin zone being replaced by the ultraviolet conditions (C1)--(C3) of
Sec.~\ref{sec:latcont} on the complete symbol at $|p|\to\infty$: free asymptotics up to a homotopically trivial
factor, vanishing of the surface terms of the discarded total momentum derivatives (a regularization in which
loop-momentum shifts are legitimate), and compatibility of the regularization with the gauge (and, for the CSE,
chiral) symmetry used. For lattice symbols the invariants are index pairings of the $K_1$ class of $G$ with a cyclic
3-cocycle; the CSE invariant is a difference of two such slice pairings (with the sign convention of
\eqref{eq:PLB13}, the negative of the $O$ pairing). The mobility-gap extension remains open within this framework.}
\end{GPTcorrection}

\paragraph{Open problems.} The explicit Wigner--Weyl computation of the $F^0$ term of the anomaly for $x$-dependent
symbols (it should reproduce $2i\,\ind\hO_0$); the extension of the non-renormalization of the CSE to non-Abelian
external fields in the covariant calculus; the mobility-gap case of the index theorem; for the non-elliptic models
of \cite{ZubJETP}, the equilibrium current itself at the junctions and the Luttinger--Ward identity for an
effective $G^{-1}=ip_4X-Y$ with sign-changing $X$; gravitational contributions on general $M$.

\appendix
\section{Definitions of the notions used}
\label{app:defs}

This appendix collects, in the conventions of the main text, the definitions of the notions that are used without
further explanation.

\subsection{Operators, symbols, index}
\label{app:ops}

\paragraph{Symbol, principal symbol.} A differential operator of order $m$ on $\R^D$ (or on a manifold $M$),
$\hQ=\sum_{|\alpha|\le m}a_\alpha(x)(-i\partial)^\alpha$ with matrix-valued coefficients, has the Weyl symbol
$Q(x,p)$ of Sec.~\ref{sec:WW}, a polynomial of degree $m$ in $p$. Its \emph{principal symbol} is the top-degree part
$q_m(x,p)=\sum_{|\alpha|=m}a_\alpha(x)p^\alpha$, homogeneous of degree $m$ in $p$; it is independent of the ordering
of the derivatives (\reviewfix{permuting covariant derivatives with fixed left coefficients changes the order by at most
$m-2$; moving an $x$-dependent coefficient across a derivative can already produce order $m-1$}) and transforms as a function on
the cotangent bundle $T^*M$. For a pseudodifferential operator the symbol is a function $Q(x,p)$ with prescribed
growth in $p$, and the principal symbol is its leading homogeneous term. \emph{Classical symbols} of order $m$ are
those admitting an asymptotic expansion $Q\sim\sum_{j\ge0}q_{m-j}(x,p)$ with $q_{m-j}$ homogeneous of degree $m-j$
for $|p|\ge1$.

\paragraph{Ellipticity, quasi-ellipticity.} $\hQ$ is \emph{elliptic} if $q_m(x,p)$ is invertible for all $x$ and
all $p\ne0$. The Dirac operator is elliptic since $(\gamma^\mu p_\mu)^2=p^2$. $\hQ$ is \emph{quasi-elliptic}
(elliptic with respect to weights $w_\mu>0$) if its symbol is a sum of terms homogeneous of the same degree under
$p_\mu\to\lambda^{w_\mu}p_\mu$ and the leading such part is invertible for $p\ne0$; the multi-Weyl operator of
Sec.~5 of Ref.~\cite{XZPLB} is quasi-elliptic with weights $(1/n,1/n,1,1)$ \reviewfix{(the anomaly for this class:
Sec.~\ref{sec:quasiell})}.

\paragraph{Parametrix.} Let $\hQ$ be an elliptic classical pseudodifferential operator of order $m$ with symbol
$Q\sim\sum_{j\ge0}q_{m-j}$. A \emph{parametrix} is an operator $\hat P$ of order $-m$ with
\[
\hat P\hQ=1+\hat R_L,\qquad \hQ\hat P=1+\hat R_R,
\]
where $\hat R_{L,R}$ are \emph{smoothing}: their symbols belong to $S^{-\infty}$ (they decay faster than any power
of $|p|$ together with all derivatives), equivalently their integral kernels are $C^\infty$, so that they map
$H^{-\infty}=\bigcup_sH^s$ into $C^\infty$. The symbol of $\hat P$ is constructed recursively from the composition
rule. In the Weyl calculus the symbol of a product is the Moyal product (Appendix~\ref{app:ww}), which for symbols
with asymptotic expansions reads
$a\star b\sim\sum_{n\ge0}\frac1{n!}\big(\frac i2\big)^na\,(\overleftarrow\partial_x\overrightarrow\partial_p
-\overleftarrow\partial_p\overrightarrow\partial_x)^n\,b$, each power lowering the degree of homogeneity in $p$ by
one. Writing $P\sim\sum_{j\ge0}p_{-m-j}$ and requiring $P\star Q\sim1$ degree by degree gives
\[
p_{-m}=q_m^{-1},\qquad
p_{-m-j}=-\Big[\sum_{\substack{k+l+n=j\\ k<j}}\frac1{n!}\Big(\frac i2\Big)^n
p_{-m-k}\,(\overleftarrow\partial_x\overrightarrow\partial_p-\overleftarrow\partial_p\overrightarrow\partial_x)^n\,q_{m-l}\Big]q_m^{-1},
\qquad j\ge1,
\]
each term in the bracket being homogeneous of degree $-j$. A symbol with this asymptotic expansion exists (Borel
summation), and the operator it defines is a left parametrix; a right parametrix is constructed in the same way, and
the two differ by a smoothing operator. Consequences: (i) \emph{Elliptic regularity}: if $\hQ u\in H^s$ for some
$u\in H^{-\infty}$, then $u=\hat P\hQ u-\hat R_Lu\in H^{s+m}$; in particular $\ker\hQ\subset C^\infty$, and the kernel
does not depend on the Sobolev space on which $\hQ$ is considered. (ii) On a closed manifold the smoothing remainders
are compact, so that $\hQ$ is invertible modulo compact operators and hence Fredholm (below). (iii) If $\hQ$ is
invertible, $\hQ^{-1}-\hat P$ is smoothing, so that the symbol of $\hQ^{-1}$ has the same asymptotic expansion as
$P$:
\[
G_W(x,p)=(\hQ^{-1})_W=q_m(x,p)^{-1}+O(|p|^{-m-1}),\qquad |p|\to\infty,\qquad
\partial_x^\alpha\partial_p^\beta G_W=O(|p|^{-m-|\beta|}).
\]
This is the \emph{parametrix expansion} used in Sec.~\ref{sec:stokes}: after radial normalization the Moyal inverse
approaches the inverse of the principal symbol, and the subprincipal terms $p_{-m-j}$, $j\ge1$, together with the
Moyal corrections contribute to the integrated three-form only through terms that vanish as the radius of the sphere
tends to infinity. The Weyl symbol of a differential operator differs from its left (Kohn--Nirenberg) symbol
$\sum_\alpha a_\alpha(x)p^\alpha$ by terms of lower order, so that the principal symbol, the ellipticity and the
leading term of the parametrix expansion do not depend on which symbol is used.

\paragraph{Sobolev spaces, compact operators.} On $\R^D$ the Sobolev space $H^s(\R^D)$, $s\in\R$, consists of the
tempered distributions $u$ whose Fourier transform $\hat u$ satisfies
$\|u\|_s^2=\int d^Dp\,(1+|p|^2)^s|\hat u(p)|^2<\infty$. For integer $s\ge0$ this is equivalent to
$\sum_{|\alpha|\le s}\|\partial^\alpha u\|_{L^2}<\infty$ (square-integrable weak derivatives through order $s$);
$H^0=L^2$; $H^{s'}\subset H^s$ continuously for $s'>s$; $(H^s)'=H^{-s}$ under the $L^2$ pairing; $H^s\subset C^k$
for $s>k+D/2$ (Sobolev embedding); $H^\infty=\bigcap_sH^s$ consists of the smooth functions with all derivatives in
$L^2$, and $H^{-\infty}=\bigcup_sH^s$ of the tempered distributions of finite order. On a closed manifold $M$
(sections of a vector bundle) $H^s(M)$ is defined through a finite atlas $\{(U_i,\kappa_i)\}$ and a subordinate
partition of unity $\{\chi_i\}$, $\|u\|_s^2=\sum_i\|(\chi_iu)\circ\kappa_i^{-1}\|^2_{H^s(\R^D)}$; different choices
give equivalent norms, as does $\|(1+\Delta)^{s/2}u\|_{L^2}$ for any positive elliptic second-order operator $\Delta$.
A pseudodifferential operator of order $m$ is bounded $H^{s+m}\to H^s$ for every $s$. A \emph{compact operator}
between Banach spaces maps bounded sets to relatively compact sets; between Hilbert spaces the compact operators are
the norm limits of finite-rank operators, and they form a two-sided ideal (products with bounded operators are
compact). \emph{Rellich's lemma}: on a closed manifold the inclusion $H^{s'}\hookrightarrow H^s$ is compact for
$s'>s$. Hence an operator of order $m'<m$ is compact as a map $H^m\to L^2$ (it is bounded $H^m\to H^{m-m'}$ and
$H^{m-m'}\hookrightarrow L^2$ is compact), and smoothing operators are compact on every $H^s$. On a non-compact
manifold Rellich's lemma fails: on $\R^D$ the translates $u(\cdot-a)$, $|a|\to\infty$, of a fixed bump function form
a bounded sequence in every $H^{s'}$ without a convergent subsequence in any $H^s$; accordingly a smoothing operator
such as the convolution with a Schwartz function is bounded on $L^2(\R^D)$ but not compact (it is a Fourier
multiplier with continuous spectrum). Compactness on $\R^D$ requires decay in $x$ as well as in $p$, e.g.\ for the
product of multiplication by a function $V(x)\to0$ with $(1+\Delta)^{-s/2}$, $s>0$.

\paragraph{Fredholm operators, index.} A bounded operator $O:\mathcal H_1\to\mathcal H_2$ is \emph{Fredholm} if
its kernel and cokernel are finite dimensional and its range is closed; its \emph{index} is
$\ind O=\dim\ker O-\dim\mathrm{coker}\,O=\dim\ker O-\dim\ker O^\dagger$. Equivalently (Atkinson's theorem) $O$ is
Fredholm iff it is invertible modulo compact operators, i.e.\ iff there is a bounded $P$ with $PO-1$ and $OP-1$
compact. The index is unchanged by compact perturbations and by small norm perturbations, hence constant along
continuous families of Fredholm operators (homotopy invariance); it is additive, $\ind(O_1O_2)=\ind O_1+\ind O_2$,
and $\ind O^\dagger=-\ind O$. The \emph{essential spectrum} of a self-adjoint operator is the complement of the
isolated eigenvalues of finite multiplicity; $O$ is Fredholm iff $0$ is not in the essential spectrum of
$O^\dagger O$ and of $OO^\dagger$.

\emph{Closed manifolds.} An elliptic operator $\hQ$ of order $m$ on a closed manifold is Fredholm as a map
$H^{s+m}\to H^s$ for every $s$: the parametrix inverts it modulo smoothing operators, which are compact by Rellich's
lemma. By elliptic regularity the kernel and the cokernel ($=\ker\hQ^\dagger$, $\hQ^\dagger$ being elliptic as
well) consist of smooth sections and do not depend on $s$, so neither does the index. Lower-order perturbations are
compact and do not change the index, and neither does a continuous deformation of the principal symbol through
elliptic symbols (homotopy invariance); the index therefore depends only on the homotopy class of the principal
symbol, i.e.\ on the symbol class $\xi\in K_c(T^*M)$ (Appendix~\ref{app:index}), and is given by the Atiyah--Singer
formula. This is the setting of Secs.~\ref{sec:anomaly}, \ref{sec:homotopy} and \ref{sec:factor}.

\emph{Open (non-compact) manifolds.} On a non-compact manifold the ellipticity of the principal symbol does not
imply the Fredholm property, because smoothing operators are no longer compact (previous paragraph): what is missing
is control of the operator at spatial infinity. The simplest example is the free massless Dirac operator
$i\gamma^\mu\partial_\mu$ on $\R^4$: it is elliptic and self-adjoint on $H^1$, its spectrum is the whole real line,
$0$ belongs to the essential spectrum, and it is not Fredholm $H^1\to L^2$; a gauge field decaying at infinity (a
finite-action configuration) is a relatively compact perturbation and changes neither the essential spectrum nor
this conclusion (Weyl's theorem). Likewise a Hamiltonian-form operator $ip_4-H$ with a Fermi point or a Fermi
surface is not Fredholm on $\R^D$. The massive operator $i\gamma^\mu\partial_\mu+im$, on the other hand, is
invertible, hence Fredholm, with index zero. The settings in which an index is defined on open manifolds are the
following. (a) \emph{Uniform ellipticity at infinity.} If the full symbol is invertible with
$|Q(x,p)^{-1}|\le C(1+|p|)^{-m}$ uniformly for $|x|+|p|$ large (the ``SG'' or scattering calculus on $\R^D$
\cite{Parenti,Cordes}; e.g.\ a Dirac operator with a mass term $m(x)$ bounded away from zero outside a compact set,
or with a confining potential), the parametrix remainders decay in $x$ as well as in $p$, are compact, and the
operator is Fredholm on the standard Sobolev spaces; the index formula then contains a contribution from the sphere
at spatial infinity. (b) \emph{Callias-type operators.} For $D_\Phi=i\gamma^\mu\partial_\mu\otimes1+1\otimes\Phi(x)$
on $\R^{2k+1}$ with a Hermitian matrix field $\Phi$ invertible outside a compact set, $D_\Phi$ is Fredholm and its
index is a topological invariant of the restriction of $\Phi$ to the sphere at infinity -- the Chern number of the
positive spectral subbundle of $\Phi/|\Phi|$ over $S^{2k}$, for $k=1$ and traceless $2\times2$ $\Phi$ the degree of
$\Phi/|\Phi|:S^2\to S^2$, i.e.\ the monopole charge \cite{Callias,BottSeeley}; this is the index theorem behind the
fermion zero modes on monopoles and domain walls, the topological charge sitting at spatial infinity rather than in
the momentum fibre. (c) \emph{$L^2$-index.} When $0$ lies in the essential spectrum one may still define the
\emph{$L^2$-index} $\dim\ker_{L^2}O-\dim\ker_{L^2}O^\dagger$ whenever both kernels are finite dimensional; it is not
stable under arbitrary compact perturbations, and its evaluation needs information at infinity. For the massless
Dirac operator on $\R^4$ in an instanton background of charge $k$ (field strength decaying as $|x|^{-4}$) the
$L^2$-index of the Weyl operator in the fundamental representation equals $k$: the Dirac operator is conformally
covariant, and the conformal compactification $\R^4\to S^4$ maps the $L^2$ zero modes bijectively onto the zero modes
of the elliptic, hence Fredholm, Dirac operator on $S^4$, whose index is given by the Atiyah--Singer formula (this is
the mechanism behind the 't~Hooft and Jackiw--Rebbi zero modes). The heat-kernel expression
$\Tr\gamma^5e^{-t\hQ^2}$ of Fujikawa's argument then equals $n_+-n_-$; in general, on an open manifold, it may
acquire a $t$-dependent contribution from the continuous spectrum (the spectral densities of $O^\dagger O$ and
$OO^\dagger$ away from zero need not coincide), which is the origin of the boundary terms in (a) and (b).
(d) \emph{Weighted Sobolev spaces, manifolds with ends.} On manifolds with cylindrical or conical ends the operator
is Fredholm between weighted spaces $H^{s+m}_\delta\to H^s_\delta$ for all weights $\delta$ outside a discrete set
(the indicial roots), and the index jumps when $\delta$ crosses such a value \cite{LockhartMcOwen}; the
Atiyah--Patodi--Singer theorem on a compact manifold with boundary, $\ind=\int_M(\text{local index density})
-\frac12(\eta+h)$ with the spectral asymmetry $\eta$ of the boundary operator, is the prototype \cite{APS}.
(e) \emph{Relative index.} For two operators which coincide outside a compact set and are invertible at infinity
the difference of the indices is the integral of the difference of the local index densities \cite{GromovLawson}.
In this review the index statements are made on closed manifolds (compactification being the route of (c)), and
the open-manifold notions enter only through the remarks on $\R^4$ in Sec.~\ref{sec:anomaly}: a zero of the full
symbol at finite momentum (a Fermi point or Fermi surface at $p_4=0$) is not a failure of ellipticity, but on $\R^D$
it places $0$ in the essential spectrum, which is why the Fermi-surface invariants of Sec.~\ref{sec:FS} are not
indices.

\subsection{Degree, winding number, Fermi points}
\label{app:degree}

\paragraph{Degree of a map.} Let $f:X\to Y$ be a smooth map between closed oriented manifolds of the same
dimension $n$, with $Y$ connected. By Sard's theorem the regular values $y\in Y$ (points at which $Df(x)$ is
invertible for every $x\in f^{-1}(y)$) are dense; for such $y$ the preimage $f^{-1}(y)$ is finite (discrete by the
inverse function theorem, compact since $X$ is), and
\[
\deg f=\sum_{x\in f^{-1}(y)}\sgn\det Df(x)\in\Z,
\]
the sign being $+1$ when $Df(x)$ carries the orientation of $T_xX$ to that of $T_yY$. The number does not depend on
the regular value (two regular values are joined by a path whose preimage is a compact one-manifold, the boundary
points of which carry opposite signs) and is invariant under homotopy of $f$ (the same argument applied to the
homotopy). Equivalently, in de Rham terms,
\[
\int_Xf^*\omega=\deg f\int_Y\omega\qquad\text{for every $n$-form $\omega$ on $Y$},
\]
so that $\deg f=\int_Xf^*\omega/\int_Y\omega$ for any $\omega$ with $\int_Y\omega\ne0$; this is the form in which
the degree appears as the integral of a closed form (a winding number or a Chern number). Examples: $z\mapsto z^k$
on $S^1$ has degree $k$; the antipodal map of $S^n$ has degree $(-1)^{n+1}$; the Chern number of a two-band
Hamiltonian $H(\vec p)=\vec d(\vec p)\cdot\vec\sigma$ on the two-torus is the degree of $\hat d:T^2\to S^2$. The case
used in this review is $X=S^3$ and $Y=SU(2)\cong S^3$ or $Y=U(N)$: for $g:S^3\to U(N)$ the three-form
$\frac1{24\pi^2}\tr(g^{-1}dg)^3$ is the pull-back of the normalized generator of $H^3(U(N);\Z)$ -- on $SU(2)$ it is
the volume form of the unit three-sphere divided by its volume $2\pi^2$, since $\tr(g^{-1}dg)^3$ equals $12$ times
that volume form -- so that its integral is the degree of $g$, followed for $N>2$ by the identification
$\pi_3(U(N))\cong\pi_3(SU(2))\cong\Z$. A degree is defined only for a map between closed manifolds of equal
dimension: a symbol which is not invertible on large spheres (the non-elliptic models of Sec.~\ref{sec:FS}) defines
no map $S^3\to GL(N,\C)$ and hence no degree, and a scalar invertible symbol $S^3\to\C^\times\simeq S^1$ has no
degree either, its three-form $(g^{-1}dg)^3$ vanishing identically.

\paragraph{Winding number of a matrix function.} For $g:S^3\to GL(N,\C)$ (or $U(N)$), $N\ge2$, the homotopy class
in $\pi_3(GL(N,\C))=\pi_3(U(N))\cong\Z$ is measured by
$W[g]=\frac1{24\pi^2}\int_{S^3}\tr(g^{-1}dg)^3$ up to the orientation convention. In this review the sign is fixed
by $N_3^0=-\frac1{24\pi^2}\int_{S^3_{\rm out}}\tr(o^{-1}d_po)^3=+1$ for $o=p_4+i\vec\sigma\cdot\vec p$ (outward
orientation of $S^3$, $\epsilon_{1234}=1$), \reviewfix{and \eqref{eq:N3hom} uses the inverse chiral block $G=o^{-1}$, not the full Dirac
Green function. For the full Dirac operator,} $\tr\gamma^5(GdQ)^3=\tr(O^{-1}dO)^3-\tr(O^{\dagger-1}dO^\dagger)^3$ accounting for the factor
$\frac1{48\pi^2}$ instead of $\frac1{24\pi^2}$. Since $\tr(g^{-1}dg)^3$ is closed, the integral does not depend on
the surface as long as it is deformed without crossing singularities of $g$ (Stokes), which is the topological
invariance used throughout.

\paragraph{Fermi points and Fermi surfaces.} A \emph{Fermi point} is an isolated zero of $Q=G^{-1}$ in the
four-dimensional Euclidean momentum space (for a Hamiltonian-form $Q=ip_4-H$: a point $\vec p_0$ with
$H(\vec p_0)=0$ at $p_4=0$, a node of the spectrum); a \emph{Fermi surface} is a two-dimensional surface of zeros at
$p_4=0$. \reviewfix{A nonzero winding cannot disappear while the Green function and its inverse remain
invertible on the enclosing surface and the defining symmetry is preserved. A chiral-weighted Dirac
charge does not prohibit a symmetry-breaking mass from opening a gap} \cite{Volovik}; the
analogous invariant for Fermi surfaces is the winding around the surface in the $(p_4,p_\perp)$ plane, \reviewfix{whereas the three-form considered in Sec.~\ref{sec:FS} is distinct from that local
$\pi_1$ charge: an ordinary scalar commuting symbol has $(G\,dQ)^3=0$ but can have nonzero
$\pi_1$ winding around a Fermi surface. Noncommutativity of the transverse Landau problem is essential
to the scalar three-form calculated here.}

\paragraph{Homotopy hypothesis.} Two symbols are \emph{homotopic} if they can be connected by a continuous family
of symbols within the admissible class (here: elliptic, chiral); the homotopy hypothesis of Ref.~\cite{XZPLB} is that $Q_W(x,p)$
is homotopic to an $x$-independent $\tilde Q(p)$.

\subsection{Bundles and characteristic classes}
\label{app:bundles}

\paragraph{Vector bundles, connection, curvature.} A complex vector bundle $E\to M$ of rank $N$ assigns to every
$x\in M$ a copy of $\C^N$ (the gauge indices); a gauge field $A=A_\mu dx^\mu$ (Hermitian, $\mathfrak u(N)$-valued in
a trivialization) defines the covariant derivative $D=d-iA$ and the curvature $D^2=-iF$,
$F=dA-iA\wedge A$, which is the field-strength two-form $\frac12F_{\mu\nu}dx^\mu\wedge dx^\nu$ of the main text. A
\emph{line bundle} $L$ is the case $N=1$; a \emph{subbundle} $W=\mathrm{im}\,P$ of a trivial bundle is defined by a
\reviewfix{projector field $P(x)$, with induced connection $\nabla s=P\,ds$ on sections satisfying
$Ps=s$. Its anti-Hermitian curvature is $P\,dP\wedge dP\,P$; the Hermitian curvature in the convention
$\nabla^2=-iF$ is $F=iP\,dP\wedge dP\,P$.}
Over the torus $T^4$ all bundles used in the text are given explicitly in this way.

\paragraph{Chern classes, Chern character.} With the normalization $F/2\pi$ (so that the flux of a line bundle
through a closed surface is $2\pi\times$integer),
\begin{align*}
c(E)&=\det\Big(1+\frac F{2\pi}\Big)=1+c_1(E)+c_2(E)+\dots,\\
c_1(E)&=\frac1{2\pi}\tr F,\qquad
c_2(E)=\frac12\Big[\Big(\frac{\tr F}{2\pi}\Big)^2-\tr\Big(\frac F{2\pi}\Big)^2\Big],
\end{align*}
\[
\ch(E)=\tr\exp\Big(\frac F{2\pi}\Big)=N+c_1(E)+\ch_2(E)+\dots,\qquad \ch_k(E)=\frac1{k!}\tr\Big(\frac F{2\pi}\Big)^k,
\quad \ch_2=\tfrac12c_1^2-c_2 .
\]
These are closed forms whose de Rham classes do not depend on the connection; the integrals of $c_k$ over closed
$2k$-cycles are integers, those of $\ch_k$ rational. The Chern character is additive on direct sums and
multiplicative on tensor products, $\ch(E\otimes E')=\ch(E)\wedge\ch(E')$ (the property used in
Sec.~\ref{sec:factor}); for a line bundle $\ch(L)=e^{c_1(L)}$, $\ch_2(L)=\frac12c_1(L)^2$.

\paragraph{Instanton number.} For a gauge bundle on a closed four-manifold,
$\int_M\ch_2(E)=\frac1{8\pi^2}\int_M\tr F\wedge F\in\Z$ when $c_1(E)=0$ (it equals $-\int c_2(E)$); this integer is
\reviewfix{the chosen \emph{instanton number}, with sign opposite to the second Chern number in these conventions.} In the main text the anomaly is written through
$\frac{i}{4\pi^2}\int\tr F\wedge F=2i\int\ch_2(E)$.

\paragraph{Chern roots and the splitting principle.} For a complex vector bundle $V$ of rank $r$ over $M$ there
is a space $Y$ with a map $\phi:Y\to M$, injective on cohomology, such that $\phi^*V=L_1\oplus\dots\oplus L_r$
splits into line bundles (the flag bundle of $V$). The \emph{Chern roots} $x_i=c_1(L_i)\in H^2(Y)$ are formal
variables in terms of which every characteristic class of $V$ is a symmetric polynomial:
$c(V)=\prod_i(1+x_i)$, i.e.\ $c_k=e_k(x_1,\dots,x_r)$ (elementary symmetric functions), and
$\ch(V)=\sum_ie^{x_i}$, $\ch_k=\frac1{k!}\sum_ix_i^k$ (power sums). Conversely every symmetric formal power series
in the $x_i$ defines a characteristic class (it is a polynomial in the $c_k$ by Newton's identities), and the
classes of the form $\prod_if(x_i)$ (multiplicative sequences) are multiplicative on direct sums. In terms of the
curvature, the $x_i$ are the eigenvalues of $F/2\pi$ treated as formal variables, which is how the formulas of the
previous paragraph arise. For a real oriented bundle $V$ of rank $2k$ (such as $TM$) the Chern roots of the
complexification $V\otimes\C$ come in pairs $\pm x_i$, $i=1,\dots,k$; the odd Chern classes of $V\otimes\C$ vanish
rationally and the \emph{Pontryagin classes} are $p(V)=\prod_i(1+x_i^2)$, $p_j=e_j(x_1^2,\dots,x_k^2)
=(-1)^jc_{2j}(V\otimes\C)$, in particular $p_1=\sum_ix_i^2=-c_2(V\otimes\C)$.

\paragraph{Pontryagin class, Todd class.} For the tangent bundle $TM$ of an oriented Riemannian four-manifold with
curvature two-form $R$ (an $\mathfrak{so}(4)$-valued matrix) the first Pontryagin form is
$p_1(M)=-\frac1{8\pi^2}\tr R\wedge R$; it vanishes identically for flat $T^4$, and its integral vanishes on $S^4$
(the signature $\frac13\int p_1$ is zero). The \emph{Todd class} of a complex bundle $V$ with Chern roots $x_i$ is
\[
\Td(V)=\prod_i\frac{x_i}{1-e^{-x_i}},\qquad
\frac{x}{1-e^{-x}}=1+\frac x2+\frac{x^2}{12}-\frac{x^4}{720}+\dots,
\]
so that $\Td(V)=1+\frac12c_1+\frac1{12}(c_1^2+c_2)+\frac1{24}c_1c_2+\dots$ (the degree-two term is
$\sum_ix_i^2/12+\sum_{i<j}x_ix_j/4=(c_1^2-2c_2)/12+c_2/4$). It is multiplicative, $\Td(V\oplus V')=\Td(V)\Td(V')$,
and it is the class which converts the $K$-theoretic push-forward into the cohomological one (Riemann--Roch; next
appendix), which is why it enters the Atiyah--Singer formula. For the complexified tangent bundle the roots come in
pairs $\pm x_i$, and
\begin{gather*}
\Td(TM\otimes\C)=\prod_i\frac{x_i}{1-e^{-x_i}}\cdot\frac{-x_i}{1-e^{x_i}}
=\prod_i\Big(\frac{x_i/2}{\sinh(x_i/2)}\Big)^2=\hat A(M)^2,\\
\hat A(M)=1-\frac{p_1}{24}+\frac{7p_1^2-4p_2}{5760}+\dots,
\end{gather*}
so that $\Td(TM\otimes\C)=1-\frac1{12}p_1(M)$ through degree four, in agreement with $c_1=0$, $c_2=-p_1$ in the
general formula. $\hat A(M)$ is the index density of the Dirac operator, $\ind D_E=\int_M\hat A(M)\wedge\ch(E)$,
which is the form the Atiyah--Singer formula takes after the fibre integration of the spinor symbol class.

\paragraph{Pfaffian and Euler class.} For a real antisymmetric $2k\times2k$ matrix $A$ the \emph{Pfaffian} is
\[
\mathrm{Pf}(A)=\frac1{2^kk!}\sum_{\sigma\in S_{2k}}\sgn(\sigma)\prod_{i=1}^kA_{\sigma(2i-1)\sigma(2i)}
=\frac1{2^kk!}\,\epsilon^{i_1\dots i_{2k}}A_{i_1i_2}\cdots A_{i_{2k-1}i_{2k}},
\]
e.g.\ $\mathrm{Pf}(A)=A_{12}$ for $k=1$ and $\mathrm{Pf}(A)=A_{12}A_{34}-A_{13}A_{24}+A_{14}A_{23}$ for $k=2$; it
satisfies $\mathrm{Pf}(A)^2=\det A$ and $\mathrm{Pf}(BAB^T)=\det B\,\mathrm{Pf}(A)$, so that it is invariant under
$SO(2k)$ and changes sign under orientation reversal. The \emph{Euler class} of an oriented real bundle of rank $2k$
with an $SO(2k)$ connection of curvature $\Omega$ (an antisymmetric matrix of two-forms, for which the formula makes
sense because two-forms commute) is $e=\mathrm{Pf}(\Omega/2\pi)$; for rank four,
\[
e=\frac1{4\pi^2}\big(\Omega_{12}\wedge\Omega_{34}-\Omega_{13}\wedge\Omega_{24}+\Omega_{14}\wedge\Omega_{23}\big)
=\frac1{32\pi^2}\epsilon^{abcd}\,\Omega_{ab}\wedge\Omega_{cd}.
\]
It is multiplicative on direct sums, satisfies $e^2=p_k$ (the top Pontryagin class), equals $c_k$ for a complex
bundle of rank $k$ regarded as a real oriented bundle of rank $2k$, vanishes for a bundle admitting a
nowhere-vanishing section, and for $TM$ integrates to the Euler characteristic (Gauss--Bonnet--Chern):
$\chi(S^4)=2$, $\chi(T^4)=0$; $e(T^*M)=e(TM)$ for rank four. In the $K$-theoretic language of the next appendix the
Euler class is the restriction of the Thom class to the zero section, which is how it enters the argument of
Sec.~\ref{sec:factor} that a nonzero fibre charge on $S^4$ requires nontrivial chiral bundles.

\subsection{$K$-theory, symbol class, index theorem}
\label{app:index}

\paragraph{$K$-groups, stable isomorphism.} Two vector bundles $E,E'$ over a compact space $X$ are \emph{stably
isomorphic} if $E\oplus\theta^n\cong E'\oplus\theta^n$ for some trivial bundle $\theta^n=X\times\C^n$. Stable
isomorphism is coarser than isomorphism (the tangent bundle of $S^2$ is stably trivial, $TS^2\oplus\theta^1\cong
\theta^3$, but not trivial) and is the equivalence relation detected by characteristic classes and by homotopy
invariants. $K^0(X)$ is the Grothendieck group of the monoid of isomorphism classes of bundles under $\oplus$:
formal differences $[E]-[E']$, with $[E]-[E']=[F]-[F']$ iff $E\oplus F'\oplus\theta^n\cong E'\oplus F\oplus\theta^n$
for some $n$; the reduced group $\tilde K^0(X)=\ker(\mathrm{rank}:K^0(X)\to\Z)$ (for connected $X$) is the group of
stable isomorphism classes. $K^0$ is a contravariant homotopy functor and a ring under $\otimes$, and
$\ch:K^0(X)\otimes\mathbb Q\to H^{\rm even}(X;\mathbb Q)$ is an isomorphism, so that rationally a class is determined
by its Chern character; integrally, torsion is lost, which is why the torsion-freeness of $K^0(T^4)$ is used in
Sec.~\ref{sec:factor}.

\paragraph{Locally compact spaces and compactly supported $K$-theory.} A Hausdorff space is \emph{locally compact}
if every point has a compact neighbourhood: $\R^n$, open subsets of manifolds, and the total spaces of vector bundles
over compact manifolds -- in particular $T^*M$ with its non-compact fibres $\R^4_p$ -- are locally compact but not
compact. Such a space has a one-point compactification $Y^+=Y\cup\{\infty\}$ (for $Y=\R^n$, $Y^+=S^n$; for
$Y=T^*M$, $Y^+$ is the Thom space $D(T^*M)/S(T^*M)$, the unit disc bundle with the unit sphere bundle collapsed to a
point), and the \emph{compactly supported $K$-theory} is
$K_c(Y)=\tilde K^0(Y^+)=\ker\big(K^0(Y^+)\to K^0(\{\infty\})\big)$. Concretely, its elements are represented by
triples $(E_0,E_1;\sigma)$: two bundles over $Y$ and an isomorphism $\sigma:E_0|_{Y\setminus K}\to E_1|_{Y\setminus K}$
defined outside a compact set $K$, modulo homotopy of the data, modulo stabilization
$(E_0\oplus F,E_1\oplus F;\sigma\oplus1)$, and modulo the triples in which $\sigma$ extends to an isomorphism over all
of $Y$ (the \emph{difference construction}; the triple is to be thought of as ``$E_0-E_1$, cancelled outside
$K$''). $K_c$ is covariant under open inclusions and contravariant under proper maps; $K_c(\R^n)=\tilde K^0(S^n)$ is
$\Z$ for even $n$ and $0$ for odd $n$ (Bott periodicity); $K_c(Y)$ is a module over $K^0(Y)$ and, for $Y=T^*M$, over
$K^0(M)$ through $\pi^*$, the product of a triple with a bundle $E$ being $(E_0\otimes E,E_1\otimes E;\sigma\otimes1)$.
The Chern character extends to $K_c(Y)$ with values in compactly supported cohomology: choosing connections
$\nabla_0,\nabla_1$ on $E_0,E_1$ which are intertwined by $\sigma$ outside $K$ ($\nabla_1=\sigma\nabla_0\sigma^{-1}$
there), the form $\ch(E_0,E_1;\sigma)=\ch(\nabla_0)-\ch(\nabla_1)$ vanishes outside $K$, and its class in
$H^{\rm even}_c(Y;\mathbb Q)$ does not depend on the choices.

\paragraph{The bundles $V_\pm$, the projection $\pi$ and the pull-back $\pi^*$.} Throughout, $M$ is a closed
oriented four-manifold and $V_+$, $V_-$ are Hermitian vector bundles over $M$ of equal rank: the spaces in which the
two chiralities of the fermion field take their values. For the Dirac operator $V_\pm=S_\pm$ are the two chiral
spinor bundles, the $\pm1$ eigenbundles of $\gamma^5$ in the spinor bundle $S=S_+\oplus S_-$ (rank two each; on flat
$T^4$ or $\R^4$ the trivial bundles $M\times\C^2$), and in the presence of internal indices and of a gauge field
$V_\pm\to V_\pm\otimes\C^N$ or $V_\pm\otimes E$. A chiral operator
$Q=\big(\begin{smallmatrix}0&O^\dagger\\O&0\end{smallmatrix}\big)$ acts on sections of $S\otimes E$, with
$O:\Gamma(V_+\otimes E)\to\Gamma(V_-\otimes E)$ and $O^\dagger$ its formal adjoint. The cotangent bundle $T^*M$ is
the eight-dimensional manifold of pairs $(x,p)$, $x\in M$ and $p$ a covector at $x$ (the momentum of the
Wigner--Weyl calculus), and $\pi:T^*M\to M$, $\pi(x,p)=x$, is the projection. For a bundle $V\to M$ the
\emph{pull-back} $\pi^*V\to T^*M$ is the bundle whose fibre over $(x,p)$ is $V_x$ -- the bundle $V$ carried along
unchanged in the momentum directions; a connection on $V$ pulls back to a connection on $\pi^*V$ whose curvature
has no $dp$ components. For a differential form $a$ on $M$, $\pi^*a$ is the same form regarded as a form on $T^*M$
(it contains no $dp$'s), and for a class in $K^0(M)$, $\pi^*[E]=[\pi^*E]$; $\pi^*$ is a ring homomorphism in
cohomology and in $K$-theory, and fibre integration is adjoint to it, $\int_{T^*M}\pi^*a\wedge b=\int_Ma\wedge\pi_*b$.

\paragraph{Symbol class.} The principal symbol of $O$ is, for every $(x,p)$, a linear map
$o(x,p):(V_+)_x\to(V_-)_x$, smooth in $(x,p)$ and homogeneous of degree $m$ in $p$ -- i.e.\ a bundle map
$o:\pi^*V_+\to\pi^*V_-$ over $T^*M$ -- and $O$ is elliptic iff $o(x,p)$ is invertible for $p\ne0$. Outside the
compact set $K=\{|p|\le1\}$ (any neighbourhood of the zero section would do) $o$ therefore identifies $\pi^*V_+$
with $\pi^*V_-$, and the triple
\[
\xi(O)=[\pi^*V_+,\pi^*V_-;o]\in K_c(T^*M)
\]
is the \emph{symbol class} of $O$ (difference construction of the previous paragraph). Equivalently, in the
clutching picture, $\xi(O)=[\pi^*V_+\cup_o\pi^*V_-]-[\pi^*V_-\cup_1\pi^*V_-]$ on the double of the disc bundle,
where $\pi^*V_+\cup_o\pi^*V_-$ denotes the bundle obtained by gluing $\pi^*V_+$ over one copy of $D(T^*M)$ to
$\pi^*V_-$ over the other copy along $S(T^*M)$ by means of $o$. Properties: (i) $\xi(O)$ depends only on the homotopy
class of $o$ through elliptic symbols, is unchanged by lower-order terms of $O$ and under stabilization
$V_\pm\to V_\pm\oplus\theta$, $o\to o\oplus1$; two elliptic operators with the same symbol class have the same
index. (ii) $\xi(O^\dagger)=-\xi(O)$, $\xi(O_1\oplus O_2)=\xi(O_1)+\xi(O_2)$, and for the composition of elliptic
operators $\xi(O_1O_2)=\xi(O_1)+\xi(O_2)$ (additivity of the index). (iii) \emph{Twisting:} replacing $\partial$ by
the covariant derivative $D=\partial-iA$ of a bundle $E$ with connection, $V_\pm\to V_\pm\otimes E$, gives the
principal symbol $o\otimes1_E$ and
\[
\xi(O_A)=\xi(O)\cdot\pi^*[E],
\]
the module product of $K_c(T^*M)$ over $K^0(M)$; the connection does not enter the class, being of lower order.
This is the twisting formula used in Sec.~\ref{sec:factor}. (iv) \emph{Chern character:} choosing connections
$\nabla_\pm$ on $\pi^*V_\pm$ with $\nabla_-=o\nabla_+o^{-1}$ for $|p|\ge1$, the form
$\ch(\xi)=\ch(\nabla_+)-\ch(\nabla_-)$ is supported in $|p|\le1$, and its fibre integrals are the invariants of
Sec.~\ref{sec:factor}. For an $x$-independent symbol on $\R^4$ with trivial $V_\pm$ one may take $\nabla_+=d$ and
$\nabla_-=d+\chi(|p|)\,o\,d(o^{-1})$ with a cutoff $\chi$ equal to $1$ for $|p|\ge1$; the transgression formula
$\ch(\nabla_-)-\ch(\nabla_+)=d\,\mathrm{CS}(\nabla_+,\nabla_-)$ and Stokes' theorem then reduce the fibre integral of
$\ch_2(\xi)$ to the boundary term $\frac1{24\pi^2}\int_{S^3}\tr(o^{-1}do)^3$: the fibre integral of the degree-four
part of the Chern character of a symbol class is the winding number of the symbol on the sphere at infinity,
$N_3^0=-\pi_*\ch_2(\xi)=-\frac1{24\pi^2}\int_{S^3_{\rm out}}\tr(o^{-1}d_po)^3$ in the orientation of this review.
Examples: $V_\pm=\C^2$ trivial and $o=p_4+i\vec\sigma\cdot\vec p$ give the Bott class $\beta$ (next paragraph); the
texture $o=\sigma^\mu p_\mu\otimes P+\bar\sigma^\mu p_\mu\otimes(1-P)$ of Sec.~\ref{sec:factor} gives
$\ch(\xi)=\ch(\beta)\wedge(2\ch(W)-K)$.

\paragraph{Bott class and Thom isomorphism.} \emph{Bott periodicity} states that $K_c(\R^{2k})\cong\Z$,
$K_c(\R^{2k+1})=0$, and more generally $K_c(Y\times\R^2)\cong K_c(Y)$ for every locally compact $Y$, the isomorphism
being multiplication by the generator $\beta_1$ of $K_c(\R^2)$. This generator is represented by the triple
$(\C,\C;z)$ with $z=p_1+ip_2$ -- the symbol class of the Cauchy--Riemann operator $\partial_1+i\partial_2$ -- and the
generator of $K_c(\R^{2k})$ is its $k$-fold exterior product $\beta_k=\beta_1^{\boxtimes k}$, represented by the
symbol of the Dirac operator on $\R^{2k}$: the Clifford multiplication $c(p):S_+\to S_-$ between the two half-spinor
spaces, an isomorphism for $p\ne0$ since $c(p)^\dagger c(p)=|p|^2$. For $k=2$ this is the \emph{Bott class}
$\beta=(\C^2,\C^2;p_4+i\vec\sigma\cdot\vec p)$, the symbol class of the Weyl operator; its Chern character has fibre
integral $\pm1$ (orientation convention; $-1$ in the orientation of this review, Sec.~\ref{sec:factor}), and every
class in $K_c(\R^4)$ is an integer multiple of $\beta$, the integer being the winding number of the symbol on
$S^3$. The \emph{Thom isomorphism} extends this from a single fibre to a bundle: for a complex vector bundle $W\to M$
of complex rank $k$ there is a canonical class $\lambda_W\in K_c(W)$, the \emph{Thom class}, restricting to the Bott
class in every fibre -- it is represented by the Koszul complex, $\Lambda^{\rm even}\pi^*W\to\Lambda^{\rm odd}\pi^*W$
by exterior multiplication with the fibre coordinate $w$ (for $k=2$ this is the Clifford multiplication above) --
and the map
\[
K^0(M)\to K_c(W),\qquad a\mapsto\pi^*a\cdot\lambda_W,
\]
is an isomorphism of $K^0(M)$-modules. The same holds for real bundles with a spin$^c$ structure, in particular for
$W=T^*M$ when $M$ is spin$^c$ (every oriented four-manifold is); for $T^4$, whose cotangent bundle is trivial,
$\lambda_{T^*T^4}$ is the Bott class $\beta$ in each fibre, and $K_c(T^*T^4)\cong K^0(T^4)$. In cohomology the
analogous statement is the Thom isomorphism $H^j(M)\to H^{j+2k}_c(W)$, $a\mapsto\pi^*a\wedge u_W$, with the
\emph{Thom form} $u_W$ (closed, compactly supported in the fibres, with fibre integral $1$); its inverse is fibre
integration $\pi_*$, and the restriction of $u_W$ to the zero section is the Euler class $e(W)$. The two
isomorphisms are intertwined by the Chern character up to the Todd class,
\[
\ch(\pi^*a\cdot\lambda_W)=\pi^*\big(\ch(a)\,\Td(W)^{-1}\big)\wedge u_W
\]
(with the conventions of \cite{AS}; other conventions conjugate $W$ or change the sign), which is the
Riemann--Roch relation responsible for the factor $\Td(TM\otimes\C)$ in the Atiyah--Singer formula: the analytic
index is a push-forward in $K$-theory, and converting it into a cohomological integral introduces the Todd class of
the fibre. \emph{Fibre integration} $\pi_*:H^{j}_c(T^*M)\to H^{j-4}(M)$ is the integration of a compactly supported
form over the fibre $\R^4_p$ with the fibre orientation $dp_1\wedge dp_2\wedge dp_3\wedge dp_4$ of the Conventions:
only the components of a form containing all four $dp$'s contribute, the degree is lowered by four, and
$\pi_*(\pi^*a\wedge b)=a\wedge\pi_*b$ (projection formula), $\int_{T^*M}b=\int_M\pi_*b$. For $T^4$,
$K^0(T^4)\cong H^{\rm even}(T^4;\Z)\cong\Z^{1+6+1}=\Z^8$ (K\"unneth), torsion free, so that a class is determined by
its Chern character, i.e.\ by the three numbers/forms $(N_3^0,\omega_2,\nu)$ of Sec.~\ref{sec:factor}.

\paragraph{Atiyah--Singer index theorem.} For an elliptic operator $O$ on a closed manifold $M$ of dimension $n$
\cite{AS},
\[
\ind O=(-1)^n\int_{T^*M}\ch(\xi)\wedge\pi^*\Td(TM\otimes\C),
\]
with $T^*M$ oriented as in \cite{AS}; for $n=4$ the sign is $+1$ and only the degree-eight component of the
integrand contributes, which is formula \eqref{eq:Fuji}. For a twisted operator $\ch(\xi\cdot\pi^*[E])=
\ch(\xi)\wedge\pi^*\ch(E)$ (multiplicativity), which is \eqref{eq:twisted}. Fedosov's form of the theorem
\cite{Fedosov} expresses the index as the trace of a symbol expression in the deformation-quantized (star-product)
algebra, i.e.\ in the language of the Wigner--Weyl calculus used here.

\subsection{Wigner--Weyl calculus}
\label{app:ww}

\paragraph{Weyl symbol, Moyal product.} Without gauge field the Weyl symbol of $\hX$ is
$X_W(x,p)=\int d^Dy\,e^{ipy}\langle x-\frac y2|\hX|x+\frac y2\rangle$ (the $A=0$ case of \eqref{eq:WT});
conversely, the operator with symbol $X_W$ has the integral kernel
$\langle x|\hX|y\rangle=\int\frac{d^Dp}{(2\pi)^D}e^{ip(x-y)}X_W\big(\frac{x+y}2,p\big)$ (Weyl quantization). The
symbol of a product is the \emph{Moyal product}
\begin{align*}
(X_W\star_MY_W)(x,p)&=X_W(x,p)\,e^{\frac i2(\overleftarrow\partial_x\overrightarrow\partial_p-
\overleftarrow\partial_p\overrightarrow\partial_x)}\,Y_W(x,p)\\
&=\frac1{\pi^{2D}}\int dy\,dk\,dy'\,dk'\,e^{2i(k'y-ky')}X_W(x+y,p+k)\,Y_W(x+y',p+k'),
\end{align*}
the integral form being the definition for general symbols and the exponential form its expansion in powers of the
derivatives, which terminates for polynomial symbols and is asymptotic in general (this is the gradient expansion).
The first terms are
\[
X\star_MY=XY+\frac i2\{X,Y\}-\frac18\big(\partial_{x_\mu}\partial_{x_\nu}X\,\partial_{p_\mu}\partial_{p_\nu}Y
-2\,\partial_{x_\mu}\partial_{p_\nu}X\,\partial_{p_\mu}\partial_{x_\nu}Y+\partial_{p_\mu}\partial_{p_\nu}X\,
\partial_{x_\mu}\partial_{x_\nu}Y\big)+O(\partial^3),
\]
with the Poisson bracket $\{X,Y\}=\partial_{x_\mu}X\,\partial_{p_\mu}Y-\partial_{p_\mu}X\,\partial_{x_\mu}Y$, and the
\emph{Moyal bracket} is
$X\star_MY-Y\star_MX=2i\,X\sin\big(\tfrac12(\overleftarrow\partial_x\overrightarrow\partial_p
-\overleftarrow\partial_p\overrightarrow\partial_x)\big)Y=i\{X,Y\}+O(\partial^3)$. Properties: the product is
associative and noncommutative; $(X\star_MY)^\dagger=Y^\dagger\star_MX^\dagger$, the symbol of $\hX^\dagger$ being
$X_W^\dagger$; $\partial_{x_\mu}$ and $\partial_{p_\mu}$ are derivations of it; and
\begin{gather*}
x_\mu\star_MY=x_\mu Y+\tfrac i2\partial_{p_\mu}Y,\qquad Y\star_Mx_\mu=x_\mu Y-\tfrac i2\partial_{p_\mu}Y,\\
p_\mu\star_MY=p_\mu Y-\tfrac i2\partial_{x_\mu}Y,\qquad Y\star_Mp_\mu=p_\mu Y+\tfrac i2\partial_{x_\mu}Y,
\end{gather*}
so that $[x_\mu,Y]_\star=i\partial_{p_\mu}Y$, $[p_\mu,Y]_\star=-i\partial_{x_\mu}Y$ and
$[x_\mu,p_\nu]_\star=i\delta_{\mu\nu}$; e.g.\ $x_\mu\star_Mp_\nu=x_\mu p_\nu+\frac i2\delta_{\mu\nu}$ is the symbol
of $\hx_\mu\hp_\nu$, while the symbol $x_\mu p_\nu$ corresponds to the symmetrized product
$\frac12\{\hx_\mu,\hp_\nu\}$. \emph{Trace property}: $\int dx\,dp\,X\star_MY=\int dx\,dp\,XY=\int dx\,dp\,Y\star_MX$
whenever the integrals converge, all derivative terms being total derivatives; a $p$-integral at fixed $x$ alone is
not cyclic (the terms containing $\partial_x$ survive), which is the identity used in Sec.~\ref{sec:pointsplit}.
The \emph{Weyl ordering} is the quantization rule under which $x^kp^l$ is mapped to the completely symmetrized
product of $\hx$ and $\hp$; for the covariant momenta this is $\Sym(\hpi^\alpha)$ of Sec.~\ref{sec:cons}. The
\emph{gradient (derivative) expansion} is the expansion of $\star_M$ in powers of $\partial_x\partial_p$, i.e.\ in
gradients of the background; in the covariant calculus the expansion is in $F$ and its covariant derivatives, the
triangular Wilson loop of \eqref{eq:starWilson} taking the place of the phase $e^{2i(k'y-ky')}$.

\paragraph{Groenewold equation.} The symbol of the inverse operator is determined by $Q_W\star G_W=1$ (Groenewold
\cite{Wigner}); its iterative solution in powers of gradients gives $G^{(0)},G^{(2)},\dots$ of Sec.~\ref{sec:groenewold}.

\paragraph{Lattice Wigner--Weyl calculus: approximate and precise.} For a tight-binding model the momentum $p$
ranges over the compact Brillouin zone and the coordinate over the lattice. Defining the Weyl symbol by the continuum
formula with the lattice (or interpolated) coordinate and the Brillouin-zone momentum gives the \emph{approximate
lattice Wigner--Weyl calculus} \cite{SuleymanovZubkovNPB,ZubkovWu,SuleymanovZubkov,ZZAnn,ZA} (developed in
\cite{SuleymanovZubkovNPB} for Wilson fermions in a varying electromagnetic field): the symbol of a product equals the Moyal
product of the symbols only up to corrections of relative order $a\,\partial_x$ ($a$ the lattice spacing acting on
the gradients of the external fields and of the inhomogeneities), so that the Groenewold equation and the gradient
expansion hold to leading order in $a\partial_x$, i.e.\ for $a^2|F|\ll1$ and slowly varying backgrounds. The
\emph{precise lattice Wigner--Weyl calculus} \cite{FZ} (rectangular lattices; the honeycomb lattice in
\cite{ChobanyanZubkov}) defines the symbol so that the basic identities of the continuum calculus hold exactly, \reviewfix{by specifying an extended coordinate lattice, an admissible symbol space and its trace}\corrB{, at the price of a star product which is not the Moyal product and of a doubled coordinate lattice}. In the continuum
limit $a\to0$ both reduce to the exact Weyl calculus above; the conditions which then replace the compactness of the
Brillouin zone are listed in Sec.~\ref{sec:latcont}.

\paragraph{Peierls substitution.} In a tight-binding model a hopping term $t\,e^{ib\cdot\hp}$ (translation by the
bond vector $b$) is coupled to the gauge field by $t\,e^{ib\cdot\hp}\to t\,e^{ib\cdot(-iD)}$, i.e.\ by multiplying
the hopping amplitude by the Wilson line along the bond; in the covariant calculus this is the covariant
quantization \eqref{eq:inv} of the field-free dispersion.

\paragraph{Non-Abelian Stokes theorem.} For an Abelian field Stokes' theorem gives $\oint_{\partial\Delta}A=
\int_\Delta F$, hence $\Pexp(i\oint_{\partial\Delta}A)=e^{i\int_\Delta F}$. For a non-Abelian field the holonomy of a
closed path is not determined by the flux alone, but it has a surface representation
\cite{Arefeva,Bralic,FGK}. Let $\Delta$ be parametrized by $z(s,t)$, $(s,t)\in[0,1]^2$, with $z(s,0)=x_0$ and
$z(s,1)=x_1$ for all $s$: the surface is swept by a family of curves $t\mapsto z(s,t)$ from $x_0$ to $x_1$, and
its boundary is the loop $\gamma$ formed by the curve $s=1$ followed by the curve $s=0$ traversed backwards (the
boundary orientation induced by $ds\wedge dt$). Let $U_s(t)=\Pexp\big(i\int_0^tA_\mu\,\partial_{t'}z^\mu dt'\big)$ be
the transporter from $x_0$ to $z(s,t)$ along the curve at fixed $s$, and $W(s)=U_0(1)^{-1}U_s(1)$ the holonomy of
the loop formed by the curves $s$ and $0$. The derivative of a transporter with respect to a deformation of its path
with fixed endpoints is given by the field strength transported to the starting point,
\[
\frac{d}{ds}U_s(1)=i\,U_s(1)\int_0^1dt\,\tilde F(s,t),\qquad
\tilde F(s,t)=U_s(t)^{-1}F_{\mu\nu}(z(s,t))\,\partial_sz^\mu\,\partial_tz^\nu\,U_s(t)
\]
(differentiate the ordered product defining $U_s(1)$; the terms $\partial_sA$ and the commutators produced by the
ordering combine into $F_{\mu\nu}\partial_sz^\mu\partial_tz^\nu$), hence $dW/ds=i\,W(s)\int_0^1dt\,\tilde F(s,t)$, whose
solution is the \emph{surface-ordered exponential}
\[
\Pexp\Big(i\oint_{\partial\Delta}A\Big)=W(1)=\bar P_s\exp\Big(i\int_0^1ds\int_0^1dt\,\tilde F(s,t)\Big),
\]
with later $s$ to the right (the loop starts at $x_0$ along the curve $s=1$, in the convention of
Sec.~\ref{sec:WW} in which the starting point of a transporter stands on the right). In the Abelian limit this is
$\exp(+i\int_\Delta F)$ with $\Delta$ oriented by $(s,t)$ and $\partial\Delta$ in the induced orientation. Expanding,
\[
W=1+i\int_\Delta\tilde F-\frac12\Big(\int_\Delta\tilde F\Big)^2
+\frac12\int\!\!\int_{s>s'}\big[\tilde F(s,t),\tilde F(s',t')\big]+O(F^3),
\]
and the \emph{ordering ambiguity} -- the dependence on the family of curves along which $F$ is transported --
resides entirely in the commutator terms: for an $F$ which is constant after transport to a single point the
commutators vanish, and in general they are of the order $[F,\cD F]\times(\text{area})^2\times(\text{size of the
loop})$, one power of the size higher than the terms retained in \eqref{eq:Wexp}. In the proof of \eqref{eq:Wexp}
the loop is the triangle of Fig.~\ref{fig:triangle}, traversed clockwise in the $(y,y')$ plane, so that its oriented
area is $-\frac12y\wedge y'$ and the first-order term is $-\frac i2y^\mu y'^\nu F_{\mu\nu}$; the parallel transport
of $F$ to $x$ along straight segments produces the covariant Taylor expansion $F(z)=F(x)+(z-x)^\alpha\cD_\alpha
F+\dots$ used there, and the quadratic term $-\frac18(y^\mu y'^\nu F_{\mu\nu})^2$ is $-\frac12(\int_\Delta\tilde F)^2$
with constant $F$.

\paragraph{Pauli--Villars regularization.} The theory is supplemented by heavy regulator fields with masses $M_i$
and weights $C_i$ ($C_0=1$ for the physical field, $\sum_iC_i=0$, and further sum rules as needed), all
contributions being summed with these weights; the regulated Dirac-sea quantities are finite and the limit
$M_i\to\infty$ is taken at the end. Sufficiently many sum rules ($\sum_iC_iM_i^n=0$ for $n=0,\dots,D-1$ suffices for a fermion
line, since \reviewfix{$\sum_iC_i(\gamma^\mu p_\mu+iM_i)^{-1}=O(|p|^{-D-1})$}) make the regulated Green function decay fast enough at $|p|\to\infty$ for the surface terms of total momentum derivatives to vanish (condition
(C2) of Sec.~\ref{sec:latcont}), i.e.\ shifts of the loop momentum become legitimate, as in dimensional
regularization or on a lattice. In the topological expressions the regulators contribute only through the
ultraviolet boundary of momentum space.

\subsection{Many-body notions}
\label{app:manybody}

\paragraph{Matsubara Green function.} In Euclidean time $\tau$, $G(\tau,x;\tau',x')=-\langle T_\tau\psi(\tau,x)
\bar\psi(\tau',x')\rangle$; at $T=0$ its Fourier transform depends on a continuous frequency $p_4=\omega$ (at $T>0$
on the Matsubara frequencies $\omega_n=(2n+1)\pi T$). $\bQ=\bG^{-1}$ is the complete inverse Green function
(Dyson: $\bQ=Q_0-\Sigma$, $\Sigma$ the self-energy). A Green function with the spectral representation
\reviewfix{$G(i\omega)=\int d\varepsilon\,\rho(\varepsilon)/(i\omega-\varepsilon)$, $\rho\ge0$,
$\int\rho\,d\varepsilon=1$, has $\operatorname{Im}G(i\omega)=-\omega\int\rho(\varepsilon)\,d\varepsilon/(\omega^2
+\varepsilon^2)$, strictly definite for $\omega\ne0$ when $\rho$ is a full-rank matrix weight, which excludes poles
and zeros off the hyperplane $\omega=0$ (for the canonical $\psi\psi^\dagger$ propagator; in infinite dimension the
absence of a kernel does not by itself give a bounded inverse, and in an inhomogeneous system operator invertibility
does not imply pointwise invertibility of the Wigner symbol).}

\paragraph{Luttinger--Ward functional, $\Phi$-derivability.} Let $G_0$ be the free and $\bG$ the complete
Matsubara Green function, $\bQ=\bG^{-1}=Q_0-\Sigma$. The \emph{Luttinger--Ward functional} $\Phi[\bG]$ \cite{LW} is
the sum of all closed connected two-particle-irreducible (2PI, ``skeleton'') vacuum diagrams -- those which do not
fall apart when any two fermion lines are cut -- evaluated with the complete $\bG$ on every fermion line and the
bare interaction on every interaction line, each with its symmetry factor; equivalently
$\Phi[\bG]=\sum_{n\ge1}\frac1{2n}\Tr\big(\Sigma^{(n)}[\bG]\,\bG\big)$, where $\Sigma^{(n)}[\bG]$ is the sum of the
skeleton self-energy diagrams with $n$ interaction lines (the factor $1/2n$ compensates the $2n$ ways of opening a
vacuum diagram with $n$ interaction lines into a self-energy diagram). Its defining property is that the exact
self-energy is its functional derivative,
\[
\Sigma(y,x)=\frac{\delta\Phi[\bG]}{\delta\bG(x,y)},\qquad\text{i.e.}\qquad\delta\Phi=\Tr(\Sigma\,\delta\bG),
\]
and that the grand potential is the stationary value of the functional
\[
\Psi[\bG]=\Tr\ln\bG-\Tr(Q_0\bG)+\Phi[\bG]
\]
(up to a convention-dependent overall factor and an additive constant), whose stationarity condition
$\delta\Psi/\delta\bG=\bG^{-1}-Q_0+\Sigma=0$ is the Dyson equation; non-perturbatively $\Phi$ is the
two-particle-irreducible effective action of Cornwall, Jackiw and Tomboulis \cite{CJT}, the Legendre transform of
the generating functional with a bilocal source coupled to $\psi\bar\psi$. $\Phi$ depends on $\bG$ and on the
interaction only, not on $G_0$ (universality), which is what allows the chemical potential and the external fields
to be varied at fixed functional form. An approximation in which $\Sigma=\delta\Phi/\delta\bG$ is computed from a
subset of skeleton diagrams (Hartree--Fock, $GW$, fluctuation exchange, \dots) is \emph{$\Phi$-derivable}
(``conserving''): Baym and Kadanoff showed \cite{BaymKadanoff} that the resulting Green function obeys the
conservation laws of particle number, momentum and energy, because $\Phi$ inherits the symmetries of the interaction.
The symmetry relevant here is the invariance under the phase transformation
$\bG(x,y)\to e^{i\varepsilon x}\bG(x,y)e^{-i\varepsilon y}$ with constant $\varepsilon$ (Wigner symbol
$\bG_W(x,p)\to\bG_W(x,p-\varepsilon)$): for density-type vertices, which couple $\bar\psi\psi$ at one point to the
interaction line, every vertex $x_j$ of a closed fermion loop has one incoming line (phase $e^{-i\varepsilon x_j}$)
and one outgoing line (phase $e^{+i\varepsilon x_j}$), the phases cancel vertex by vertex, and the interaction lines
carry no phase; hence $\Phi[\bG_\varepsilon]=\Phi[\bG]$ for an interaction of any range. The same holds for the
frequency shift $\bG(\tau_x,\tau_y)\to e^{i\omega_0\tau_x}\bG\,e^{-i\omega_0\tau_y}$ when the interaction is
instantaneous, and for a global internal rotation when the interaction is a singlet. These invariances are used in
Theorem~\ref{thm:LW} and in the next-but-one paragraph. The existence of a single differentiable branch of $\Phi$ as
a functional of $\bG$ can fail: the map $G_0\mapsto\bG$ need not be injective, so that $\Phi[\bG]$ is multivalued,
and the skeleton series can converge to an unphysical branch \cite{Kozik}.

\paragraph{Ward--Takahashi identity, renormalized velocity.} Gauge invariance gives
$q_\mu\Gamma_\mu(p,p+q)=\bG^{-1}(p+q)-\bG^{-1}(p)$ for the vector vertex; \reviewfix{if the vertex has a regular and unique zero-transfer limit,
$\Gamma_k(p,p)=\partial_{p_k}\bG^{-1}=\partial_{p_k}\bQ$. Gapless systems require an explicit
order of the frequency and momentum limits.} The \emph{renormalized velocity} is this quantity, in
contrast to the bare velocity $\partial_{p_k}Q_0$ (equal to $\gamma_k$ for Dirac fermions). The frequency analogue,
$\Tr[\Sigma\,\partial_{p_4}\bG]=0$, is the identity from which Luttinger's theorem follows.

\paragraph{The shift identity $\Tr[\bG\,\partial_{p_k}\Sigma]=0$.} This is the identity of Theorem~\ref{thm:LW}; we
spell out its content. Let $\Tr$ denote the full regularized trace, $\Tr X=\int\frac{d^Dx\,d^Dp}{(2\pi)^D}\tr X_W$
(per volume in the thermodynamic limit), under which the star product is cyclic and total momentum derivatives
vanish, $\Tr[\hx_k,\,\cdot\,]=0$ (compact Brillouin zone on a lattice; condition (C2) of Sec.~\ref{sec:latcont} in
the continuum). (i) \emph{Invariance:} for density-type vertices $\Phi[\bG_\varepsilon]=\Phi[\bG]$, with
$\bG_{\varepsilon,W}(x,p)=\bG_W(x,p-\varepsilon)$ (previous paragraphs). (ii) \emph{Differentiation:}
$0=\frac{d}{d\varepsilon_k}\Phi[\bG_\varepsilon]\big|_{\varepsilon=0}=\Tr\big[\Sigma\,\partial_{\varepsilon_k}
\bG_\varepsilon\big]=-\Tr[\Sigma\,\partial_{p_k}\bG]$. (iii) \emph{Integration by parts:}
$\Tr[\bG\,\partial_k\Sigma]=\Tr[\partial_k(\bG\star\Sigma)]-\Tr[\partial_k\bG\star\Sigma]=\Tr[\partial_k(\bG\star\Sigma)]
=-i\,\Tr\big[[\hx_k,\hat{\bG}\hat\Sigma]\big]=0$, where the cyclicity of the trace was used to identify
$\Tr[\partial_k\bG\star\Sigma]$ with $\Tr[\Sigma\star\partial_k\bG]$ and its closedness to discard the total
derivative. Hence
\[
\Tr[\bG\,\partial_{p_k}\Sigma]=\Tr[\Sigma\,\partial_{p_k}\bG]=0,\qquad\text{equivalently}\qquad
-\Tr[\bG\,\partial_{p_k}Q_0]=-\Tr[\bG\,\partial_{p_k}\bQ]:
\]
the integrated current computed with the bare velocity $\partial_{p_k}Q_0$ equals the one computed with the
renormalized velocity $\partial_{p_k}\bQ=\partial_{p_k}(Q_0-\Sigma)$. (iv) \emph{Diagrammatic meaning:}
$\Tr[\bG\,\partial_k\Sigma]$ is the sum over the skeleton diagrams of $\Tr(\Sigma\bG)$ (closed diagrams) with the
momentum derivative inserted on one internal fermion line; summing the insertions over all lines of a closed diagram
gives the derivative of the diagram with respect to a common shift of all loop momenta, which vanishes because a
shift of the integration variables does not change the integral (the interaction lines depend on momentum
differences only). The identity is thus the statement that loop momenta may be shifted, which is precisely what the
regularization has to guarantee ((C2)). (v) \emph{Lowest order:} for the Fock term with an even potential,
$\Sigma(p)\propto\int_{p'}V(p-p')\bG(p')$, an integration by parts in $p'$ gives
$\Tr[\bG\,\partial_k\Sigma]\propto\int_p\int_{p'}V(p-p')\,\bG(p)\,\partial'_k\bG(p')$, and symmetrizing in
$p\leftrightarrow p'$ turns this into $\frac12\int\!\int V(p-p')(\partial_k+\partial'_k)[\bG(p)\bG(p')]
=-\frac12\int\!\int\big[(\partial_k+\partial'_k)V(p-p')\big]\bG(p)\bG(p')=0$ -- the ``evenness of the boson
propagator'' of Refs.~\cite{ZZJPA,ZZAnn} in its simplest instance; the Hartree term is $p$-independent and drops out
trivially. (vi) \emph{Frequency analogue:} for an instantaneous interaction the same argument with the frequency
shift gives $\Tr[\Sigma\,\partial_{p_4}\bG]=0$; inserted into the particle number $\rho=\Tr[\bG\,\partial_\mu\bQ]$ it
yields Luttinger's theorem, the particle number being the volume enclosed by the surface $\bQ(\omega=0,\vec p)=0$
independently of the interaction. (vii) \emph{What the identity does not give:} it concerns the integrated (total)
current, not the local current density nor an untraced non-Abelian component (the trace over the gauge indices is
essential for the cyclicity); it requires a differentiable branch of $\Phi$ and the vanishing of the surface term at
$|p|\to\infty$; and for the axial current no regularization preserves the required invariance together with the
chiral symmetry, so that the renormalized axial current of Sec.~\ref{sec:CSEint} is a definition rather than a
consequence of the identity.

\paragraph{Streda relation; Bloch theorem.} In a gapped (or mobility-gapped) phase the Hall conductivity equals the
derivative of the particle density with respect to the magnetic field at fixed chemical potential,
$\sigma_{xy}=\partial\rho/\partial B|_\mu$ (Streda). The \emph{Bloch theorem} on the absence of equilibrium currents
states that the total electric current in the ground state of a system with short-ranged interactions vanishes in the
thermodynamic limit \cite{Bloch}; it concerns the total conserved current, not its local or component densities.

\paragraph{Topological Hamiltonian, mobility gap.} The \emph{topological Hamiltonian} of an interacting system is
$H_t=-\bG(i\omega=0)^{-1}=-\bQ(0)$ \cite{WangZhang}; \reviewfix{for a physical canonical Green function with a normalized positive spectral representation,
a gap, nonsingular $G(0)$ and fixed ultraviolet behavior, the standard deformation to
$(i\omega-H_t)^{-1}$ stays invertible. Smoothness and invertibility alone, without the spectral or
homotopy assumptions, do not prove this reduction.}

\paragraph{Mobility gap and the quantized Hall conductance without a spectral gap.} A \emph{mobility gap} is an
interval of energies in which all eigenstates are (Anderson) localized although the density of states need not
vanish: for a disordered two-dimensional electron gas in a strong magnetic field the Landau levels are broadened
into bands in which only the states near the centres are extended, and a Fermi level between two centres lies in a
mobility gap but not in a spectral gap. In the noncommutative-geometric formulation of Bellissard, van Elst and
Schulz-Baldes \cite{BES} the observables form the algebra of covariant (homogeneous) operators on $\ell^2(\Z^2)$ or
$L^2(\R^2)$ generated by a translation-invariant ensemble of random Hamiltonians $H_\omega$, with the derivations
$\partial_jX=i[X_j,X]$ ($X_j$ the position operators) and the trace per unit volume $\cT$ (an ergodic average); the
zero-temperature Hall conductivity is given by the Kubo formula in the form
\[
\sigma_{xy}=\frac{e^2}{h}\,\mathrm{Ch}(P_F),\qquad
\mathrm{Ch}(P_F)=2\pi i\,\cT\big(P_F[\partial_1P_F,\partial_2P_F]\big),\qquad P_F=\chi(H_\omega\le E_F),
\]
the \emph{Chern character} of the Fermi projection (for a periodic system it reduces to the TKNN integral, and with
$\partial_j\to\partial_{p_j}$ and the frequency integral performed it is the $N_3$ of Sec.~\ref{sec:NCG}).
Localization at the Fermi level is expressed by the \emph{Sobolev condition} $\cT\big(|[X_j,P_F]|^2\big)<\infty$
(a finite localization length, $\ell^2\propto\cT(|[X,P_F]|^2)$), which holds when the states at $E_F$ are
exponentially localized and follows, e.g., from the fractional-moment bounds of Aizenman and Graf
\cite{AizenmanGraf}. The theorem of \cite{BES} states: if $E_F$ lies in a region in which the Sobolev condition
holds, then (i) $\mathrm{Ch}(P_F)$ is an integer, almost surely independent of the disorder realization, namely the
index of the Fredholm operator $P_FUP_F$ on $P_F\mathcal H$ with $U=(X_1+iX_2)/|X_1+iX_2|$ -- the pairing of the
$K_0$-class of $P_F$ with the Connes--Chern character of the Fredholm module defined by $U$, which requires only
$[U,P_F]$ to lie in the Schatten class $\mathcal L^3$, guaranteed by the Sobolev condition, and not the compactness
that a spectral gap would provide; (ii) $\mathrm{Ch}(P_F)$ is constant as $E_F$ moves inside the region of localized
states (the Hall plateau) and can change only when $E_F$ crosses extended states; (iii) the direct conductivity
$\sigma_{xx}$ vanishes there in the limit of infinite relaxation time. Thus the Fermi level in a mobility gap gives
a quantized Hall conductance without a spectral gap, which is the explanation of the plateaux of the integer quantum
Hall effect in disordered samples \cite{BES,AizenmanGraf,ProdanSB}. This situation is not covered by the
invertible-symbol framework of Sec.~\ref{sec:NCG}: when the density of states at $E_F$ is nonzero, the Matsubara
Green function $\bG(\omega)$ is not invertible at $\omega=0$ in the required (smooth, uniformly bounded) sense, the
frequency integral defining $N_3$ passes through singularities, and the homotopy to a topological Hamiltonian is not
available; the localization length replaces the gap as the quantity controlling the convergence of the index
pairing. The corresponding statement for the interacting Green-function invariant, and for the three-dimensional
slice invariants of this review, is the open problem listed in Sec.~\ref{sec:summary}.

\paragraph{Landau levels, Hall function.} For a particle of charge $1$ and mass $m$ in a uniform magnetic field
\reviewfix{$B$ the transverse kinetic energy $\pi_\perp^2/2m$ has the spectrum
$E_n=|B|(n+\tfrac12)/m$, $n=0,1,\dots$, and each Landau level has degeneracy $|B|/(2\pi)$ per unit area} (one state per flux quantum
$2\pi$). The \emph{Hall function} of Sec.~\ref{sec:FS} is the TKNN/Ishikawa-type integrand of the $2+1$D Hall
conductivity written as a function of a complex chemical potential $z$,
$F(z)=\epsilon^{ab}\int\tr(G_z\star\partial_aH\star G_z\star\partial_bH\star G_z)$, $G_z=(z-H)^{\star-1}$; \reviewfix{with $\int\tr$ interpreted as a transverse trace per area and $B>0$,
$\tfrac12\oint F(z)\,dz=n_{\rm filled}$ for the positively oriented occupied-level contour.
The factor $1/2$ comes from $3(2\pi)^2/(24\pi^2)$ when the three-form is converted to this
operator trace; dropping it would double the Hall number.}

\paragraph{Chiral modes, spectral flow.} A \emph{chiral one-dimensional mode} is a branch of the spectrum crossing
the Fermi level with velocity of definite sign $\chi=\sgn v$. In an electric field $E$ along the mode the momenta
move at the rate $\dot p=E$, so the occupation of the branch changes at the rate $\partial_t\rho=\chi E/2\pi$ per
unit length (the $1+1$D anomaly), while the response of its current to the chemical potential is
$\partial j/\partial\mu=\chi/2\pi$. \emph{Spectral flow} is the net number of eigenvalues crossing the Fermi level
under a change of parameters; the chiral anomaly in a magnetic field is the spectral flow of the lowest Landau
level, a chiral mode with degeneracy $B/2\pi$ per unit area, which gives $\partial_t\rho_R=\frac1{4\pi^2}\mathbf
E\cdot\mathbf B$ per Weyl fermion, i.e.\ $\partial_t\rho_5=\frac1{2\pi^2}\mathbf E\cdot\mathbf B$ for a Dirac
fermion.

\paragraph{Chiral symmetry; vierbein.} \emph{Chiral symmetry} of $\hQ$ means $\{\gamma^5,\hQ\}=0$, \reviewfix{equivalently an off-diagonal block form. The more specific form
$Q=\left(\begin{smallmatrix}0&O^\dagger\\O&0\end{smallmatrix}\right)$ additionally requires $Q=Q^\dagger$} in the basis in which
$\gamma^5=\mathrm{diag}(1,-1)$; it is exact for massless Dirac fermions and is broken by a mass or by a chirally
non-invariant regularization. A \emph{vierbein} $e^\mu_a(x)$ is a field of frames converting the flat $\gamma^a$
into $\gamma^\mu(x)=e^\mu_a\gamma^a$; $|e|=\det e^a_\mu$; in the condensed-matter context it is the emergent
anisotropic and position-dependent velocity of the Dirac quasiparticles.

\subsection{Noncommutative geometry}
\label{app:ncg}

\paragraph{Algebra, trace per unit volume, derivations.} The symbols of the lattice Wigner--Weyl calculus with the
star product form an algebra $\mathcal A$; $\partial_\mu$ (derivatives with respect to the quasi-momentum,
equivalently commutators with the coordinates) are derivations, $\partial(a\star b)=\partial a\star b+a\star\partial
b$; the \emph{trace per unit volume} $\cT$ is defined by periodization \cite{KMMZZ}: the lattice is divided into
supercells, the symbols are taken periodic under supercell translations, the trace is restricted to one supercell
and divided by its volume, and the limit of infinite supercell is taken at the end; it is \emph{cyclic},
$\cT(a\star b)=\cT(b\star a)$, and \emph{closed}, $\cT(\partial_\mu a)=0$. The factor $\frac1V\int d^3x$ in the
continuum expressions \eqref{eq:CSE}, \eqref{eq:QHE2}, \eqref{eq:PLB13} is the same construction with the lattice
replaced by continuous space.

\paragraph{Cyclic cocycles.} A \emph{cyclic $n$-cocycle} on $\mathcal A$ is an $(n+1)$-linear functional $\varphi$
with cyclicity $\varphi(a_1,\dots,a_n,a_0)=(-1)^n\varphi(a_0,\dots,a_n)$.
It also satisfies $b\varphi=0$, where
\[
(b\varphi)(a_0,\dots,a_{n+1})=\sum_{j=0}^n(-1)^j\varphi(a_0,\dots,a_ja_{j+1},\dots,a_{n+1})
+(-1)^{n+1}\varphi(a_{n+1}a_0,a_1,\dots,a_n)
\]
is the Hochschild coboundary. Its class lies in the cyclic cohomology $HC^n(\mathcal A)$ \cite{Connes}. For a
closed cyclic trace and commuting derivations,
$\varphi(a_0,\dots,a_3)=\epsilon^{\mu\nu\rho}\,\cT(a_0\star\partial_\mu a_1\star\partial_\nu a_2\star\partial_\rho a_3)$
is a cyclic 3-cocycle (the lattice counterpart of the fundamental class of the three-torus).

\paragraph{$K_1$ and the pairing.} $K_1(\mathcal A)$ is the group of homotopy classes of invertible elements of the
matrix algebras $M_k(\mathcal A)$ (stabilized in $k$); $[G]$ is the class of the invertible symbol $G$. The
\emph{pairing} of $K_1$ with odd cyclic cohomology is $\langle[u],[\varphi]\rangle=c_n\,\varphi(u^{-1},u,u^{-1},u,
\dots)$ with a normalization constant $c_n$ \cite{Connes}; for the cocycle above the normalization is absorbed in
the prefactor $\frac1{24\pi^2}$ so that $\langle[u],[\varphi]\rangle=\varphi(u^{-1},u,u^{-1},u)$ is the winding
number. The pairing depends only on the classes $[u]$ and $[\varphi]$ (homotopy invariance).

\paragraph{Fredholm modules, summability, Dixmier trace.} An \emph{odd Fredholm module} over $\mathcal A$ is a
Hilbert space $\mathcal H$ carrying a representation of $\mathcal A$ together with a self-adjoint operator $F$ with
$F^2=1$ and $[F,a]$ compact for all $a\in\mathcal A$. It is \emph{$p$-summable} if $[F,a]$ belongs to the Schatten
class $\mathcal L^p$ and \emph{$(p,\infty)$-summable} if $[F,a]\in\mathcal L^{(p,\infty)}$, the ideal of operators
whose singular values satisfy $\mu_k=O(k^{-1/p})$. The \emph{Dixmier trace} $\Tr_\omega$ is a trace on
$\mathcal L^{(1,\infty)}$ (the logarithmic average of the partial sums of singular values), vanishing on trace-class
operators; \reviewfix{for classical pseudodifferential operators of critical order on a closed manifold the Dixmier trace is
proportional to the \emph{Wodzicki residue}, its symbol-level expression.} Here $\mathcal H$ is the lattice Hilbert
space (with Euclidean time) tensored with $\C^2$, $D=\gamma^0\hat\tau+\gamma^1\hat x_1+\gamma^2\hat x_2$ is the
position Dirac operator and $F=\operatorname{sgn}D$; the commutators $[F,a]$ decay as the inverse distance, which
gives $(3,\infty)$-summability in three dimensions.

\paragraph{Index pairing and the local index formula.} For $P=(1+F)/2$ and an invertible $u$, the Toeplitz
compression $PuP:P\mathcal H\to P\mathcal H$ is Fredholm and $\operatorname{Index}(PuP)=\langle[u],[\ch_*(\mathcal
H,F)]\rangle$, where $\ch_*(\mathcal H,F)$ is the Chern character of the module, the cyclic cocycle
$\tau_n(a_0,\dots,a_n)\propto\Tr(F[F,a_0]\cdots[F,a_n])$ \reviewfix{for odd $n$ with $n+1>p$ (a sufficient trace-class condition; endpoint cases require separate care)} \cite{Connes}. The Connes--Moscovici
\emph{local index formula} \cite{CM} expresses this cocycle through residues (Dixmier traces) of the operators
involved, i.e.\ through local expressions of the type $\cT(a_0\partial a_1\partial a_2\partial a_3)$; applied to
the module above it reproduces $\varphi$ and gives $N_3=\operatorname{Index}(PGP)$ \cite{KMMZZindex}.

\end{document}